\documentclass[10pt,onecolumn,oneside,a4paper]{article}

\usepackage[dvips]{graphics,lscape,rotating}
\usepackage{psfrag}
\usepackage{bbm}
\usepackage{amsfonts,amsmath}
\usepackage{mathrsfs}
\usepackage{pstricks,array,longtable}
\newcommand{\lo}[1]{^{#1}\!\!} 

\newcommand{\fcmt}[2]{\fbox{\mbox{\begin{minipage}[t]{#1cm} #2 \end{minipage}}}}
\newcommand{\cmt}[2]{\mbox{\begin{minipage}[t]{#1cm} #2 \end{minipage}}}
\newcommand{\mb}[1]{\mathbbm{#1}}  
\newcommand{\Muserfunction}[1]{A}               %Mathematica-Importe
\newcommand{\Mvariable}[1]{#1}

\makeatletter
\@addtoreset{equation}{section}
\makeatother

\def\ba{\begin{eqnarray}}
\def\ea{\end{eqnarray}}
\def\be{\begin{equation}}
\def\ee{\end{equation}}

\newtheorem{Theorem}{Theorem}[section]                             %define the theoremes to use
\newtheorem{Definition}{Definition}[section]

\DeclareMathOperator{\tr}{tr}
\DeclareMathOperator{\sgn}{sgn}

\begin{document}

%%%%%%%%%%%%%%%%%%%%%%%%%%%%%%%%%%%%%%%%%%%%%%%%%%%%%%%%%%%%%%%%%%%%%%%%%%%%%%%%%%%%%%%%%%%%%%%%%%%%%%%
\vspace{-2cm}
\title{\vspace{-2cm} Unboundedness of Triad -- Like Operators\\ in 
Loop Quantum Gravity}
\author{ 
J.
Brunnemann\thanks{jbrunnem@aei.mpg.de,jbrunnemann@perimeterinstitute.ca},
T.
Thiemann\thanks{thiemann@aei.mpg.de,tthiemann@perimeterinstitute.ca}\\
\\
MPI f. Gravitationsphysik, Albert-Einstein-Institut, \\
           Am M\"uhlenberg 1, 14476 Potsdam, Germany\\
\\
and\\
\\
Perimeter Institute for Theoretical Physics,\\
31 Caroline Street N, Waterloo, ON N2L 2Y5, Canada}
%14424 Potsdam
\date{{\small Preprint AEI-2005-099}}
%\\ PACS code : 04.60.Pp (Loop Quantum Gravity, Quantum geometry, Spin
%Foams), 04.60 (Quantum Theory of Gravitation)\\
%       keywords :}}

\maketitle

\begin{abstract}
In this paper we deliver the proofs for the claims, made in a companion 
paper, concerning the avoidance of cosmological curvature singularities in 
in full Loop Quantum Gravity (LQG).
\end{abstract}

\newpage

\tableofcontents
  
\section{Introduction}

In this paper we display a self -- contained derivation of the results 
reported in our companion paper \cite{0}. Basically that paper consisted 
of three parts: Two claims and a scheme.

The first claim was that Triad Like 
Operators, which are the direct analogon in full Loop Quantum Gravity 
(LQG) (see \cite{1,TT:review} for books and \cite{1a,TT:review_k} for 
reviews), of the inverse 
scale factor in Loop Quantum Cosmology (LQC)\footnote{LQC
are the usual homogeneous minisuperspace models 
quantized by LQG methods. See for instance \cite{4} for a review.},
are unbounded even right at the big bang (zero volume eigenstates) and 
even when restricting to states which are isotropic and homogeneous on 
large scales. This is in contrast to the LQC result. Notice that unbounded 
does not mean ill -- defined: In fact,
our triad operators are well -- defined, that is, they are defined on a 
dense subset of the Hilbert space consisting of the finite linear span 
of spin network states. This is a well -- known result derived almost a 
decade ago \cite{TT:QSD I,TT:QSD V,TT:length-paper}.

Rather, unboundedness means that there are states in the closure of the 
span on which the operator norm diverges. This means that, in contrast to 
bounded operators, not every state is in the domain of the operator and 
therefore it is not granted that the local curvature singularity is absent 
in full LQG because a priori any state could be semiclassically relevant 
in order to describe a collapsing universe. 

The second result we claimed was that with respect to a candidate class of 
semiclassical states\footnote{Subject to the reservation that currently we 
can do this only when artificially substituting $SU(2)$ by $U(1)^3$, which 
seems not to invalidate the qualitative result.} describing a collapsing 
universe which is homogeneous 
and isotropic on large scales, the analogon of the inverse scale factor 
is bounded even at zero volume. This is a calculation within full LQG 
and takes the inhomogeneous and non isotropic fluctuations into account. 

However, that calculation can at best be a first promising hint because, 
as we showed, reliable conclusions can only be drawn when working with 
physical operators and physical states. A crude way of implementing such a 
scheme was also described in \cite{0}. In particular, kinematical 
boundedness does not imply physical boundedness because a Dirac observable 
constructed from a partial observable such as the inverse scale factor 
may have a different spectrum. Secondly, whether or not the quantum 
evolution with respect to an unphysical time parameter breaks down at zero 
volume on certain states or not may or may not be an indication of a 
global initial singularity: On the one hand, if there are restrictions on 
the allowed 
set of states that one can evolve with the Hamiltonian constraint, 
the physical Hilbert space may still 
be large enough to describe a collapsing universe. On the other hand, if 
there are no restrictions, it maybe the case that not a sufficient number 
of these formal solutions are normalizable with respect to the physical 
inner product in order to capture a sector of the physical Hilbert space 
describing a collapsing universe at large scales.\\
\\
This paper is organized as follows:\\
\\
In section two we derive the contribution of the kinetic term of a scalar 
field to the Hamiltonian constraint operator. This is one of the terms 
which become singular classically at the big bang.

In section three we set up the calculation of the norm of the analog of 
the inverse scale factor in LQG, which is a triad like operator that 
couples to the quantum scalar matter, with respect to arbitrary spin 
network states.

In section four we specialize the general framework of section three to 
the simplest possible zero volume states: gauge invariant spin network 
states supported on a trivalent graph. The final result was already 
displayed in \cite{0} and is given in formula 
(\ref{e'(B)^2 Endresultat eichinvarianter 3-Vertex}). We display the 
unboundedness of this expression graphically by selecting simple 
configurations. We also display the rather irregular behaviour of the
norm of the inverse scale factor on different spin network states.

In section five we calculate the norm of the inverse scale factor with 
respect to kinematical coherent states (for $U(1)^3$) and show that 
the norm remains bounded right at the big bang subject to the constraint 
that we restrict the valence of the vertex of the graph underlying the 
coherent states. This seems to be a reasonable restriction on 
(kinematical) semiclassicality because one would not expect that the 
Hausdorff dimension of a graph diverges from the classical dimension.

In appendices A, B, C we give a self -- contained account of various 
identities for the volume operator matrix elements and recoupling schemes.
We also review the complexifier coherent states \cite{2} needed for our 
calculation in section five.

%%%%%%%%%%%%%%%%%%%%%%%%%%%%%%%%%%%%%%%%%%%%%%%%%%%%%%%%%%%%%%%%%%%%%%%%%%%%%%%%%%%%%%%%%%%%%%%%%%%%%%%
\section{Derivation}
%%%%%%%%%%%%%%%%%%%%%%%%%%%%%%%%%%%%%%%%%%%%%%%%%%%%%%%%%%%%%%%%%%%%%%%%%%%%%%%%%%%%%%%%%%%%%%%%%%%%%%%
In this section we will show how to implement an operator version of 
$H_{kin}=\displaystyle\int_\Sigma d^3x\frac{\pi^2(x)}{\sqrt{\det(q)}(x)}$ where $q_{ab}$ is the spatial metric on the hypersurfaces $\Sigma$ with $\det(q) \ne 0$ $\forall~x\in\Sigma$. We will closely follow \cite{TT:QSD V}, section 3.3 .

\begin{figure}[hbt]
    \center
    \cmt{8}{
    \psfrag{x}{${x}$}
    \psfrag{y}{${y}$}
    \psfrag{xy}{${x}-{y}$}
    \psfrag{ep}{$\epsilon$}
    \psfrag{O}{$0$}
    \includegraphics[height=5cm]{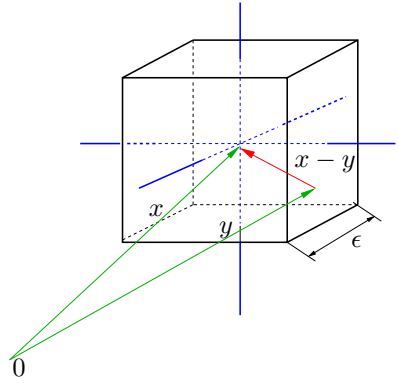} 
    \caption{The regulator $\chi_\epsilon(x,y)$} }
\end{figure}

Recall from \cite{TT:QSD V} the following definitions:

\[\begin{array}{lllll}

    V(R)&=&\displaystyle\int\limits_R d^3x~\sqrt{\det(q)}(x)
    &\ldots&\cmt{6}{volume of a spatial region $R$}
    \\

    \chi_\epsilon(x,y)&=&\displaystyle\prod_{a=1}^{3}\Theta\Big(\frac{\epsilon}{2}-\big|x^a-y^a \big| \Big) &\ldots&\cmt{6}{characteristic function of a cube centered at $x$ with coordinate volume $\epsilon^3$}
    
    \\
    
    V(x,\epsilon)&=&\displaystyle\int d^3y~\chi_\epsilon(x,y)~\sqrt{\det(q)}(y)
    &\ldots&\cmt{6}{volume of that cube as measured by $q_{ab}$}
    
\end{array} \]\be\label{characteristic functions}\ee

here we assume $\det(q) > 0$ $\forall x\in R$ and $\Theta(z)$ is the usual unit step function with $\Theta(z)=0$ if $z< 0$, $\Theta(z)=\frac{1}{2}$ if $z=0$ and $\Theta(z)=1$ if $z> 0$. If we take the limit $\epsilon \rightarrow 0$ we realize that

\[\begin{array}{lllll}

    \lim\limits_{\epsilon \rightarrow 0}~ \frac{1}{\epsilon^3}~ \chi_\epsilon(x,y)= \delta(x,y)
    
    &~~~~~~~~~~&
    \mbox{and}
    &~~~~~~~~~~& 
    
    \lim\limits_{\epsilon \rightarrow 0}~ \frac{1}{\epsilon^3}~V(x,\epsilon)=\sqrt{\det(q)}(x)

\end{array} \] \be\label{limiting properties of the characteristic functions} \ee

such that we have $\forall~x\in R$ and sufficiently small $\epsilon>0~:~\frac{\delta V(R)}{\delta E^a_i(x)}=\frac{\delta V(x,\epsilon)}{\delta E^a_i(x)}$.

By using the classical identities 
\ba
   V(R)&=&\int_R~d^3x~\sqrt{\det q}(x)
   \nonumber
   \\
   &=&\int_R~d^3x~\sqrt{\Big|\frac{1}{3!} \epsilon^{ijk}\epsilon_{abc}E^a_i(x)E^b_j(x)E^c_k(x) \Big|}
\ea
and the key identities

\ba\label{e-identity}
   E^a_i(x)=\sqrt{\det(q)}(x)~ e^a_i(x)
           =\det((e^j_b))~e^a_i(x)
   ~~~~~~\mbox{where}~~~e^j_b(x)=-\frac{2}{\kappa}\big\{A^j_b(x),V(R) \big\}
\ea

Using (\ref{characteristic functions}),(\ref{limiting properties of the characteristic functions}) and (\ref{e-identity}) we can now derive a regulated expression:

\begin{footnotesize}

\[\begin{array}{lllll}
   H_{kin}&=&\displaystyle\int\limits_R d^3x~\frac{\pi^2(x)}{\sqrt{\det(q)}(x)}&&
   
   \\
   
   &=&
   \displaystyle\int\limits_R d^3x~\pi^2(x)\frac{\big[\det(e^i_a)\big]^2}{\big[\det(q)\big]^{\frac{3}{2}}}(x)
   
   \\
   
   &=&\multicolumn{3}{l}{
   \displaystyle\lim_{\epsilon \rightarrow 0}
   \int\limits_R d^3x~\pi(x)
   \int\limits_R d^3y~\pi(y)
   \int\limits_R d^3u~
   \frac{\big[\det(e^i_a)\big]}{\big[\det(q)\big]^{\frac{3}{4}}}(u)
   \int\limits_R d^3w~
   \frac{\big[\det(e^i_a)\big]}{\big[\det(q)\big]^{\frac{3}{4}}}(w)
   ~\frac{1}{\epsilon^3} \chi_\epsilon(x,y)
   ~\frac{1}{\epsilon^3} \chi_\epsilon(u,x)
   ~\frac{1}{\epsilon^3} \chi_\epsilon(w,y)}
   
   \\
   &=&
   
   \displaystyle\lim_{\epsilon \rightarrow 0}~\frac{1}{\epsilon^9}
   \int\limits_R d^3x~\pi(x)
   \int\limits_R d^3y~\pi(y)
   &\displaystyle\hspace{-1cm}
   \int\limits_R d^3u~\frac{1}{3!} \epsilon_{ijk}\epsilon^{abc} \Big[-\frac{8}{\kappa^3}\Big]
   \frac{\big\{A^i_a(u),V(u,\epsilon) \big\}}{\big[\frac{1}{\epsilon^3}V(u,\epsilon)\big]^{\frac{1}{2}}}
   \frac{\big\{A^j_b(u),V(u,\epsilon) \big\}}{\big[\frac{1}{\epsilon^3}V(u,\epsilon)\big]^{\frac{1}{2}}}
   \frac{\big\{A^k_c(u),V(u,\epsilon) \big\}}{\big[\frac{1}{\epsilon^3}V(u,\epsilon)\big]^{\frac{1}{2}}}
   \\
   &&&\displaystyle\hspace{-1cm}
   \int\limits_R d^3w~\frac{1}{3!} \epsilon_{lmn}\epsilon^{def} \Big[-\frac{8}{\kappa^3}\Big]
   \frac{\big\{A^l_d(w),V(w,\epsilon) \big\}}{\big[\frac{1}{\epsilon^3}V(w,\epsilon)\big]^{\frac{1}{2}}}
   \frac{\big\{A^m_e(w),V(w,\epsilon) \big\}}{\big[\frac{1}{\epsilon^3}V(w,\epsilon)\big]^{\frac{1}{2}}}
   \frac{\big\{A^n_f(w),V(w,\epsilon) \big\}}{\big[\frac{1}{\epsilon^3}V(w,\epsilon)\big]^{\frac{1}{2}}}

   \\
   &&&\hspace{-1cm}
   \times\displaystyle~ \chi_\epsilon(x,y)
   ~ \chi_\epsilon(u,x)
   ~ \chi_\epsilon(w,y)
   
   \\\\
   
   &=&
   \displaystyle\frac{64\cdot 2^6}{\kappa^6\cdot6\cdot6}~\lim_{\epsilon \rightarrow 0}~
   \int\limits_R d^3x~\pi(x)
   \int\limits_R d^3y~\pi(y)
   &\displaystyle
   \int\limits_R d^3u~ \epsilon_{ijk}\epsilon^{abc} 
   \big\{A^i_a(u),\big[V(u,\epsilon)\big]^{\frac{1}{2}} \big\}
   \big\{A^j_b(u),\big[V(u,\epsilon)\big]^{\frac{1}{2}} \big\}
   \big\{A^k_c(u),\big[V(u,\epsilon)\big]^{\frac{1}{2}} \big\}
   \\
   &&&\displaystyle
   \int\limits_R d^3w~ \epsilon_{lmn}\epsilon^{def} 
   \big\{A^l_d(w),\big[V(w,\epsilon)\big]^{\frac{1}{2}} \big\}
   \big\{A^m_e(w),\big[V(w,\epsilon)\big]^{\frac{1}{2}} \big\}
   \big\{A^n_f(w),\big[V(w,\epsilon)\big]^{\frac{1}{2}} \big\}

   \\
   &&&
   \times\displaystyle~ \chi_\epsilon(x,y)
   ~ \chi_\epsilon(u,x)
   ~ \chi_\epsilon(w,y) 
\end{array}\]
\end{footnotesize}
 \\[-8mm]\be\label{regularization step I} \ee

Now we introduce a triangulation $T$ of  $R$  adapted to a graph $\gamma$ as follows: at every vertex $v\in V(\gamma)$ choose a triple $e_I, e_J, e_K$ of edges of $\gamma$ and let a tetrahedron $\Delta$ be based at $v$ which is spanned by segments $s_I, s_J, s_K$ of this triple. Each segment $s_I$ is given by the part of the according edge $e_I(t^I)$ for which the curve parameter $t^I \in [0,\epsilon ]$
\footnote{Of course, this $\epsilon$ a priori has nothing to do with the $\epsilon$ we take the limit of. However one can synchronize both quantities \cite{TT:QSD V} , justifying our simplification.} .
We have:
\vfill
\pagebreak

\[\begin{array}{lllll}

   \lefteqn{\displaystyle\int\limits_\Delta~d^3u ~ \epsilon_{ijk}\epsilon^{abc} 
   \big\{A^i_a(u),\big[V(u,\epsilon)\big]^{\frac{1}{2}} \big\}
   \big\{A^j_b(u),\big[V(u,\epsilon)\big]^{\frac{1}{2}} \big\}
   \big\{A^k_c(u),\big[V(u,\epsilon)\big]^{\frac{1}{2}} \big\}=}
   
   \\
   &&\multicolumn{3}{l}{\fcmt{16}{introduce an adapted coordinate sytem:\\
              \[\begin{array}{lclllll}
	       u^a=\sum\limits_{L=I,J,K} s^a_L(t^L)&~~~~~~~~&
	      \rightarrow& d^3u&=&\Big|\det\Big(\frac{\partial(u^1,u^2,u^3)}{\partial(t^I,t^J,t^K)} \Big) \Big|~~dt^Idt^Jdt^K
	      
	      \\
	      &&&&=&\sgn(\det(\dot{s}^a_I))~\frac{1}{3!} \epsilon_{def}\epsilon^{IJK}
	            \dot{s}^d_I(t^I)~\dot{s}^e_J(t^J)~\dot{s}^f_K(t^K)~dt^Idt^Jdt^K
	      \end{array}\]
	      
	      Also notice that:
	      \[\epsilon_{def}\epsilon^{abc} = 3!~\delta_{[d}^a~\delta_e^b~\delta_{f]}^c  \]}}
	      
 \\\\

 &=&\displaystyle\int\limits_{[0,\epsilon]^3} d^3t~\sgn(\det(\dot{s}^a_I)(t))~
    \epsilon^{IJK}\epsilon_{ijk}~\dot{s}^a_I(t^I)~\dot{s}^b_J(t^J)~\dot{s}^c_K(t^K)
  \\[-3mm]
  &&~~~~~~~~  
    \big\{A^i_a(u(t)),\big[V(u(t)),\epsilon)\big]^{\frac{1}{2}} \big\}
   \big\{A^j_b(u(t)),\big[V(u(t)),\epsilon)\big]^{\frac{1}{2}} \big\}
   \big\{A^k_c(u(t)),\big[V(u(t),\epsilon)\big]^{\frac{1}{2}} \big\}
   
 \\[3mm]
 &\approx&\sgn(\det(\dot{s}^a_I)(t))~\epsilon^{IJK}\epsilon_{ijk}
 
   ~\epsilon~ \dot{s}^a_I(v)~\big\{A^i_a(v),\big[V(v),\epsilon)\big]^{\frac{1}{2}} \big\}
   ~\epsilon~ \dot{s}^b_J(v)~\big\{A^j_b(v),\big[V(v),\epsilon)\big]^{\frac{1}{2}} \big\}
   ~\epsilon~ \dot{s}^c_K(v)~\big\{A^k_c(v),\big[V(v,\epsilon)\big]^{\frac{1}{2}} \big\}
   
  \\[2mm]
  &&\multicolumn{3}{l}{\fcmt{16}{Now we are ready to invoke the relation (valid for small $\epsilon$)\\
       \[\epsilon~ \dot{s}^a_I(v)~\big\{A^i_a(v),\big[V(v),\epsilon)\big]^{\frac{1}{2}} \big\}
         \approx
	 \tr\Big[\tau_i~h_{s_I}(\epsilon) \big\{h_{s_I}^{-1}(\epsilon),[V(v,\epsilon)]^{\frac{1}{2}} \big\} \Big]
	 =\tr\Big[\tau_i~h_{I}(\epsilon) \big\{h_{I}^{-1}(\epsilon),[V(v,\epsilon)]^{\frac{1}{2}} \big\} \Big]\]
	 \\
	 and continue}}
  
  \\\\[-1mm]
  
  &\approx&\sgn(\det(\dot{s}^a_I)(v))~\epsilon^{IJK}\epsilon_{ijk}
     \tr\Big[\tau_i~h_{I}(\epsilon) \big\{h_{I}^{-1}(\epsilon),[V(v,\epsilon)]^{\frac{1}{2}} \big\} \Big]
     \\&&\hspace{3.6cm} \times
     \tr\Big[\tau_j~h_{J}(\epsilon) \big\{h_{J}^{-1}(\epsilon),[V(v,\epsilon)]^{\frac{1}{2}} \big\} \Big]
     \\
     &&\hspace{3.8cm} \times
     \tr\Big[\tau_k~h_{K}(\epsilon) \big\{h_{K}^{-1}(\epsilon),[V(v,\epsilon)]^{\frac{1}{2}} \big\} \Big]	        
\end{array}\] \nopagebreak\\[-14mm]\be\label{approximate integral over tetrahedron} \ee

The reason for calculating (\ref{approximate integral over tetrahedron}) is given by the fact that the integration over the spatial region $R$ in (\ref{regularization step I}) can be (symbolically) split  as follows 
\cite{TT:review}:

\ba\label{symbolic integral decomposition of R}
   \int\limits_R
   &=&\int\limits_{\overline{U^\epsilon_\gamma}} 
       + \sum\limits_{v\in V(\gamma)}~ \int\limits_{U^\epsilon_{\gamma v}}
   \nonumber
   \\.
   &=&\int\limits_{\overline{U^\epsilon_\gamma}} 
       + \sum\limits_{v\in V(\gamma)}~\frac{1}{E(v)}\sum_{e_I\cap e_J\cap e_K=v}
        ~\bigg[\int\limits_{U^\epsilon_{\gamma v}( e_I e_J e_K)} 
	+\int\limits_{\overline{U^\epsilon_{\gamma v}}( e_I e_J e_K)}\bigg] 
   \nonumber
   \\
   &\approx&\int\limits_{\overline{U^\epsilon_\gamma}} 
       + \sum\limits_{v\in V(\gamma)}~\frac{1}{E(v)}\sum_{e_I\cap e_J\cap e_K=v}
        ~\bigg[
	8\cdot \int\limits_{\Delta^\epsilon_{\gamma v}( e_I e_J e_K)} 
	+\int\limits_{\overline{U^\epsilon_{\gamma v}}( e_I e_J e_K)}\bigg]       
\ea 

Here we have first decomposed $R$ into a region $\overline{U^\epsilon_\gamma}$ not containig the vertices of the graph $\gamma$ and regions $U^\epsilon_{\gamma v}$ around each vertex of $v\in V(\gamma)$. Now we choose a triple $e_I, e_J, e_K$ of edges outgoing from $v$ and decompose $U^\epsilon_{\gamma v}$ into the region $U^\epsilon_{\gamma v}( e_I e_J e_K)$ covered by the tetrahedron $\Delta^\epsilon_{\gamma v}( e_I e_J e_K)$ spanned by $e_I, e_J, e_K$ and its 8 mirror images and the rest $\overline{U^\epsilon_{\gamma v}}( e_I e_J e_K)$, not containing $v$. In the last step we notice that the integral over $U^\epsilon_{\gamma v}( e_I e_J e_K)$ classically converges to 8 times the integral over the original single tetrahedron $\Delta^\epsilon_{\gamma v}( e_I e_J e_K)$ as we shrink the tetrahedron to zero.

We average over all such triples $e_I\cap e_J\cap e_K=v$ and divide by the number of possible choices of triples for a vertex with $N$ edges $E(v)=\Big({N\atop 3}\Big)$.
\\\\
We can now decompose the $u$ and $w$ integration over $R$ in (\ref{regularization step I}) according to (\ref{symbolic integral decomposition of R}).

%%%%%%%%%%%%%%%%%%%%%%%%%%%%%%%%%%%%%%%%%%%%%%%%%%%%%%%%%%%%%%%%%%%%%%%%%%%%%%%%%%%%%%%%%%%%%%%%%%%%%%
\subsection{Quantization of the Regulated Expression}
%%%%%%%%%%%%%%%%%%%%%%%%%%%%%%%%%%%%%%%%%%%%%%%%%%%%%%%%%%%%%%%%%%%%%%%%%%%%%%%%%%%%%%%%%%%%%%%%%%%%%%

Now we can quantize the thus decomposed expression (\ref{regularization step I}) by changing Poisson brackets to commutators times $\frac{1}{\mb{i}\hbar}$. The holonomies act by multiplication only and  $V\rightarrow \hat{V}$ becomes the volume operator as defined in \cite{TT:Closed ME of V in LQG}. Now the reason for the decomposition (\ref{symbolic integral decomposition of R}) becomes clear, because if $\hat{V}$ is applied to a spin network function $f_\gamma$ we will only have contributions at the vertices $v\in V(\gamma)$ and so only the integration over the tetrahedra $\Delta^\epsilon_{\gamma v}( e_I e_J e_K)$ has to be considered. 
\\\\
Also the smeared momenta $\pi(R)=\int_R d^3x \pi(x) \chi_\epsilon(x,v)$ of the scalar field $\phi$ will in the limit $\epsilon\rightarrow 0$ turn into $-\mb{i}\hbar \kappa X(v)$ where $X(v):=\frac{1}{2}[X_R(v)+X_L(v)]$ is the sum over right and left invariant vectorfields acting on the point holonomies $U(v)$ as defined in \cite{TT:Kin. HS for Ferm and Higgs QFT's}. So we will work in a Hilbert space $\mathcal{H}=\mathcal{H}_{gravity}\otimes\mathcal{H}_{matter}$ as introduced in \cite{TT:Kin. HS for Ferm and Higgs QFT's} and \cite{QFT on CST II}. Here we are only interested in the action of $\hat{H}_{kin}$ on the gravitational part $\mathcal{H}_{gravity}=\mathcal{H}_{AL}$, the usual Ashtekar Lewandowski Hilbert space of Loop Quantum Gravity. 
\\\\
Notice, that we can identify the sections $s_I$ used in (\ref{approximate integral over tetrahedron}) of the edge $e_I$ with edges of the graph $\gamma$ by introducing trivial 2-valent vertices $v_{triv}$ and write $h_{e_I [0,1]}=h_{e_I [0,\epsilon]}h_{e_I [\epsilon,1]}=h_{s_I [0,\epsilon]}h_{e_I [\epsilon,1]}$. Now the vertices $v_{triv}$ do not change the gauge behaviour of $f_\gamma$ and they are annihilated by $\hat{V}$. In what follows we will imply this construction and replace $s_I\rightarrow e_I$. So we can determine the action of $\hat{H}_{kin}$ on a cylindrical function $f_\gamma$ starting from (\ref{regularization step I}) and using (\ref{approximate integral over tetrahedron}), (\ref{symbolic integral decomposition of R}). We will use the shorthand $\epsilon(I J K):= \sgn\big(\det(\dot{e}_I \dot{e}_J \dot{e}_K) \big)(v)$ and as already mentioned $h_I(\epsilon)=h_{e_I}(\epsilon)$.

\[\begin{array}{lllll}
   \hat{H}_{kin}f_\gamma
   
   &=&\multicolumn{3}{l}{\bigg[
   
   \displaystyle\frac{64\cdot2^6\cdot\mb{i}^2\cdot \hbar^2 \kappa^2}{\kappa^6\cdot6\cdot6\cdot\mb{i}^6\cdot\hbar^6}~\lim_{\epsilon \rightarrow 0}~
   \int\limits_R d^3x~X(x)
   \int\limits_R d^3y~X(y)}\hspace{6cm}
   \\
   &&\times\displaystyle
   \sum\limits_{v\in V(\gamma)}\frac{8}{E(v)}\sum\limits_{e_I\cap e_J\cap e_K=v}& 
   \epsilon(e_I e_J e_K)~ \epsilon^{IJK}~ \epsilon_{ijk}
   \\
   &&& 
   ~~\hspace{-1cm}~\times\tr\Big[\tau_i~h_{I}(\epsilon) \big[h_{I}^{-1}(\epsilon),\hat{V}^{\frac{1}{2}} \big] \Big]
   \tr\Big[\tau_j~h_{J}(\epsilon) \big[h_{J}^{-1}(\epsilon),\hat{V}^{\frac{1}{2}}\big] \Big]
   \tr\Big[\tau_k~h_{K}(\epsilon) \big[h_{K}^{-1}(\epsilon),\hat{V}^{\frac{1}{2}}\big]\Big]
   &
   \\
   &&\times\displaystyle
   \sum\limits_{v'\in V(\gamma)}\frac{8}{E(v')}\sum\limits_{e_L\cap e_M\cap e_N=v'}& 
   \epsilon(e_L e_M e_N)~ \epsilon^{LMN}~ \epsilon_{lmn}
   \\
   &&& 
   ~~\hspace{-1cm}~\times\tr\Big[\tau_l~h_{M}(\epsilon) \big[h_{M}^{-1}(\epsilon),\hat{V}^{\frac{1}{2}} \big] \Big]
   \tr\Big[\tau_m~h_{M}(\epsilon) \big[h_{M}^{-1}(\epsilon),\hat{V}^{\frac{1}{2}}\big] \Big]
   \tr\Big[\tau_n~h_{N}(\epsilon) \big[h_{N}^{-1}(\epsilon),\hat{V}^{\frac{1}{2}}\big]\Big]
   &

   \\[3mm]
   &&\multicolumn{2}{l}{
   \times\displaystyle~ \chi_\epsilon(x,y)
   ~ \chi_\epsilon(v,x)
   ~ \chi_\epsilon(v',y) \bigg]~f_\gamma}
\end{array}\]\be \label{quantization step I} \ee

Now we take the limit $\epsilon \rightarrow 0$ and due to the characteristic functions in (\ref{quantization step I}) the $x$ and $y$ integrations drop out and we must have  $x=y=v=v'$. For convenience, we introduce 

\be
   \lo{(r)\,}\hat{e}^i_I(v)
   :=
   \lim\limits_{\epsilon\rightarrow 0}~\tr\Big[{\tau_i h_{I}(\epsilon) \big[h_{I}^{-1}(\epsilon), \hat{V}^r \big]}\Big]\bigg|_{v\in V(\gamma)}
\ee 
It is easy to see that the limit is trivial, i.e. the operator $\lo{(r)\,}\hat{e}^i_I(v)$ is $\epsilon$--independent for sufficiently small $\epsilon$. We will not display the $\epsilon$--dependence any more in what follows. 
Notice, that $\big[\mb{i}\cdot\lo{(r)\,}\hat{e}^i_I(v)  \big]^\dagger=\mb{i}\cdot \lo{(r)\,}\hat{e}^i_I(v)$ is selfadjoint.
Moreover we will abbreviate $\sum\limits_{e_I\cap e_J\cap e_K=v}\hspace{-3mm}\rightarrow \sum\limits_{IJK}$. Finally we find:

\[\begin{array}{lllll}
   \hat{H}_{kin}f_\gamma
   
   &=&\bigg[
   
   \displaystyle\frac{P}{\hbar^4\kappa^4}~
   \sum\limits_{v\in V(\gamma)}&
   \displaystyle \frac{1}{E(v)}X(v) \sum\limits_{IJK}
   \epsilon(IJK)~\epsilon^{IJK}~\epsilon_{ijk}
   ~\lo{(\frac{1}{2})\,}\hat{e}^i_I(v)
   ~\lo{(\frac{1}{2})\,}\hat{e}^j_J(v)
   ~\lo{(\frac{1}{2})\,}\hat{e}^k_K(v)
   \\
   &&&\displaystyle \frac{1}{E(v)} X(v) \sum\limits_{LMN}
   \epsilon(LMN)~\epsilon^{LMN}~\epsilon_{lmn}
   ~\lo{(\frac{1}{2})\,}\hat{e}^l_L(v)
   ~\lo{(\frac{1}{2})\,}\hat{e}^m_M(v)
   ~\lo{(\frac{1}{2})\,}\hat{e}^n_N(v)
   
     \bigg]~f_\gamma
\end{array}\]\be \label{quantization step II}\ee

where $P=\frac{2^{16}}{9}$ and $X(v)=\lim\limits_{\epsilon\rightarrow 0}\int d^3x^\chi_\epsilon(v,x)X(x)$.
This reduces the problem to analyzing the smeared operator: $\tr\Big[{\tau_k h_{e_K} \big[h_{e_K}^{-1}, \hat{V}^r \big]}\Big]=:~\lo{(r)\,}e^k_K$. This will be done in the following.

%%%%%%%%%%%%%%%%%%%%%%%%%%%%%%%%%%%%%%%%%%%%%%%%%%%%%%%%%%%%%%%%%%%%%%%%%%%%%%%%%%%%%%%%%%%%%%%%%%%%%%%
\section{Calculation}
%%%%%%%%%%%%%%%%%%%%%%%%%%%%%%%%%%%%%%%%%%%%%%%%%%%%%%%%%%%%%%%%%%%%%%%%%%%%%%%%%%%%%%%%%%%%%%%%%%%%%%%
In this section we will calculate the action of the  $~\lo{(r)\,}e^k_K$-operators in a recoupling scheme basis. For definitions see section \ref{Action of a Holonomy on a Recoupling Scheme} in the appendix.

%%%%%%%%%%%%%%%%%%%%%%%%%%%%%%%%%%%%%%%%%%%%%%%%%%%%%%%%%%%%%%%%%%%%%%%%%%%%%%%%%%%%%%%%%%%%%%%%%%%%%%%
\subsection{General Case}
%%%%%%%%%%%%%%%%%%%%%%%%%%%%%%%%%%%%%%%%%%%%%%%%%%%%%%%%%%%%%%%%%%%%%%%%%%%%%%%%%%%%%%%%%%%%%%%%%%%%%%%

Let $\big| T_{J'}\big >:=\big|\,\vec{a}'\, J'\, M'\,;\,\vec{j}'\,;\,\vec{n}'\, \big>$ and
    $\big| T_{J}\big >:=\big|\,\vec{a}\, J\, M\,;\,\vec{j}\,;\,\vec{n}\, \big>$.
Then we have:
\ba
  \big<\,T_{J'}\, \big| \,\lo{(r)}e^k_K\, \big|\,T_J\, \big>
  &=&
  \big<\,T_{J'}\, \big| \,
  \tr\Big[{\tau_k~ h_{K} \big[h_{K}^{-1}, \hat{V}^r \big]}\Big]
  \,\big|\,T_J\, \big>
  \nonumber
  \\
  &=&\big<\,T_{J'}\, \big|\,
  \tr\Big[{\tau_k~ h_{K}~ h_{K}^{-1}~ \hat{V}^r } - {\tau_k~ h_{K}~ \hat{V}^r~ h_{K}^{-1} } \Big]
  \,\big|\,T_J\, \big>
  \nonumber
  \\
  &&\fcmt{12}{By linearity of $\tr$ we have
          \[\begin{array}{rcl}
	      \tr\Big[{\tau_k~ h_{K}~ h_{K}^{-1}~ \hat{V}^r } - {\tau_k~ h_{K}~ \hat{V}^r~ h_{K}^{-1} } \Big]
	      &=&\underbrace{\tr\Big[\tau_k~ \mb{1}\big]}_{\displaystyle 0} \hat{V}^r  -\tr\Big[ {\tau_k~ h_{e_K} \hat{V}^r h_{e_K}^{-1} } \Big] 
	  \end{array}\] since the $\tau_k$ are trace free}
  \nonumber
  \\
  &=&-\big<\,T_{J'}\, \big|\,
     \tr\Big[ {\tau_k~ h_{e_K} \hat{V}^r h_{e_K}^{-1} } \Big] 
     \,\big|\,T_J\, \big>
  \nonumber
  \\
  &&\fcmt{10}{Using the properties of $SU(2)$ given in appendix \ref{SU(2) properties} we have\\
             \[\begin{array}{rcl}
	         h_{K }&=&\epsilon^{-1}~ \overline{h}_{K}~\epsilon = \epsilon ~\overline{h}_{K}~ \epsilon^{-1} \\
		 
		 h_{K}^{-1}&=&\epsilon~[h_{K}]^T~\epsilon^{-1}
	     \end{array}\] here the overline denotes complex conjugation} 
  \nonumber
  \\
  &=&-\big<\,T_{J'}\, \big|\,
     \tr\Big[ {\tau_k~\epsilon ~\overline{h}_{K}~ \not\epsilon^{-1}~  \hat{V}^r ~\not\epsilon~[h_{K}]^T~\epsilon^{-1} } \Big]
     \,\big|\,T_J\, \big>
  \nonumber
  \\
  &&\fcmt{10}{Use cyclicity of $\tr$ and remove the slashed inner $\epsilon$-terms}
  \nonumber
  \\
  &=&-\big<\,T_{J'}\, \big|\,
     \tr\Big[ {\underbrace{\epsilon^{-1} ~\tau_k~\epsilon}_{\displaystyle \overline{\tau_k}} ~\overline{h}_{K}~\hat{V}^r ~[h_{K}]^T~ } \Big]
     \,\big|\,T_J\, \big> 
  \nonumber
  \\
  &=&-\sum_{A\,B\,C} \big[\overline{\tau}_k \big]_{CA}~
     \big<\,T_{J'}\, \big|\,
     \big[\overline{h}_{K} \big]_{AB}
                    ~ \hat{V}^r ~
		    \big[[h_{K}]^T \big]_{BC}~
     \,\big|\,T_J\, \big> 		    
  \nonumber\\
  &=&-\sum_{A\,B\,C} \big[\overline{\tau}_k \big]_{CA}~
     \big<\,T_{J'}\, \big|\,
     \big[\overline{h}_{K} \big]_{AB}
                    ~ \hat{V}^r ~
		    \big[h_{K} \big]_{CB}
     \,\big|\,T_J\, \big>
  \nonumber
  \\   
  &=&-\sum_{A\,B\,C}~~ \big[\overline{\tau}_k \big]_{CA}~\sum_{\tilde{T}_{J'}\,\tilde{T}_J}
     \overline{} ~
     ~~ \big<\,T_{J'}\, \big|\,
     \big[\overline{h}_{K} \big]_{AB}
                    ~\big|\,\tilde{T}_{J'}\, \big>\,
		    \big<\,\tilde{T}_{J'}\, \big|\, \hat{V}^r \,\big|\,\tilde{T}_J\, \big>
		    \,\big<\,\tilde{T}_J\, \big|\big[h_{K} \big]_{CB}
     \,\big|\,T_J\, \big> 		    
  \nonumber
  \\   
&=&-\sum_{A\,B\,C}~~ \big[\overline{\tau}_k \big]_{CA}~\sum_{\tilde{T}_{J'}\,\tilde{T}_J}
     \overline{} ~
     ~~ \overline{\big<\,\tilde{T}_{J'}\, \big|\,
     \big[{h}_{K} \big]_{AB}
                    ~\big|\,{T}_{J'}\, \big>}\,
		    \big<\,\tilde{T}_{J'}\, \big|\, \hat{V}^r \,\big|\,\tilde{T}_J\, \big>
		    \,\big<\,\tilde{T}_J\, \big|\big[h_{K} \big]_{CB}
     \,\big|\,T_J\, \big> 		    
  \nonumber
  \\   
  &=&-\sum_{A\,B\,C}~~ \big[\overline{\tau}_k \big]_{CA}~\sum_{\tilde{T}_{J'}\,\tilde{T}_J}
     \overline{C^{T_{J'}}_{\tilde{T}_{J'}}(K,A,B)} ~
     C^{T_{J}}_{\tilde{T}_{J}}(K,C,B)~~
     \big<\,\tilde{T}_{J'}\, \big|\, \hat{V}^r  \,\big|\,\tilde{T}_J\, \big> 		    
\ea

In the last line we have used the expansion coefficients $C^{T_{J}}_{\tilde{T}_{J}}(K,A,B)$ defined by
\ba
   \big[h_K\big]_{AB}~\big|T_J \big> 
   &=& \sum_{\tilde{T}_J} \big<\,\tilde{T}_J \,\big|\,\big [h_K \big]_{AB}\,\big|\, T_J\, \big>\cdot |\,\tilde{T}_J\, \big>
   \nonumber
   \\
   &=& \sum_{\tilde{T}_J} C^{T_{J}}_{\tilde{T}_{J}}(K,A,B) \cdot |\,\tilde{T}_J\, \big>
\ea
and explicitely calculated in (\ref{Holonomieaktion auf allgemeine Kante Endresultat}). Moreover we sum over all quantum numbers of the possible intermediate states
$\big| \tilde{T}_{J'}\big >:=\big|\,\tilde{\vec{a}}'\, \tilde{J}'\, \tilde{M}'\,;\,\tilde{\vec{j}}'\,;\,\tilde{\vec{n}}'\, \big>$ and
    $\big| \tilde{T}_{J}\big >:=\big|\,\tilde{\vec{a}}\, \tilde{J}\, \tilde{M}\,;\,\tilde{\vec{j}}\,;\,\tilde{\vec{n}}\, \big>$~~.
\\\\
If we now insert the explicit result of (\ref{Holonomieaktion auf allgemeine Kante Endresultat}) we find as a general expression for the matrix element:

\[\begin{array}{ccllllll}
      \lefteqn{\big<\,T_{J'}\, \big| \,\lo{(r)}e^k_K\, \big|\,T_J\, \big>=}
      \\ 
      \\
      &=&-\displaystyle\sum_{A\,B\,C}~~ \big[\overline{\tau}_k \big]_{CA}
      &\multicolumn{4}{l}{\displaystyle\sum_{\tilde{T}_{J'}\,\tilde{T}_J}
     \overline{C^{T_{J'}}_{\tilde{T}_{J'}}(K,A,B)} ~
     C^{T_{J}}_{\tilde{T}_{J}}(K,C,B)~~
     \big<\,\tilde{T}_{J'}\, \big|\, \hat{V}^r  \,\big|\,\tilde{T}_J\, \big> }
     \\
     \\
     &&\multicolumn{4}{l}{\fcmt{15}{Now the Volume Operator only affects the intermediate
      recouplings $\tilde{\vec{a}}',~\tilde{\vec{a}}$ 
      so we must have (in order to get non vanishing contributions): 
     
     \[\begin{array}{lclcll}
          \tilde{J}'=\tilde{J} 
	  
	  &\leadsto& J'=\left\{\begin{array}{ll}
	                         J-1\\J\\J+1
			      \end{array} \right.
	  &\mbox{because}&\tilde{J}'=J'\pm\frac{1}{2} ,~\tilde{J}=J\pm\frac{1}{2}&
	  \\
	  \\
	  \tilde{M}'=\tilde{M} &\leadsto&M'=M+C-A 
	  &\mbox{because}&\tilde{M}'=M'+A,~\tilde{M}=M+C
	  \\
	  \tilde{\vec{j}}'=\tilde{\vec{j}} &\leadsto& 
	  j_L'=j_L~\forall L\ne K,~j_K'=\left\{\begin{array}{ll}
	                         j_K-1\\j_K\\j_K+1
			      \end{array} \right.
	  &\mbox{because}&\tilde{j}_K'=j_K'\pm\frac{1}{2},~\tilde{j_K}=j_K\pm\frac{1}{2}
	  \\
	  &\leadsto&\tilde{\vec{j}}=[j_1,\ldots\,j_{K-1},\tilde{j}_K,j_{K+1},\ldots,j_N]
	  \\
	  \\
	  \tilde{\vec{n}}'=\tilde{\vec{n}}&\leadsto&n_L'=n_L~\forall L\ne K,~n_K'=n_K
	  &\mbox{because}&\tilde{n}_K'=n_K'+B,~\tilde{n}_K=n_K+B 
	  \\
	  \\
	  &\leadsto&\tilde{\vec{n}}=[n_1,\ldots\,n_{K-1},n_K+B,n_{K+1},\ldots,n_N]
     \end{array}\]}} 
      \\
      \\
      &=&-\displaystyle\sum_{A\,B\,C}\big[\overline{\tau}_k \big]_{CA}&\displaystyle\sum_{\tilde{\vec{a}}'\,\tilde{\vec{a}}}
                 \displaystyle\sum_{\tilde{J}=J\pm\frac{1}{2}\atop\tilde{J}'=J'\pm\frac{1}{2}}

       \displaystyle\sum_{\tilde{j}_K'=j_K'\pm\frac{1}{2}\atop \tilde{j}_K=j_K\pm\frac{1}{2}}
                & C^{j_K'}_{\tilde{j}_K'}(B,n_K')
       \displaystyle\sum_{g_{N-1}'} 
                 C^{J'~j_K'}_{\tilde{J}'~ \tilde{j}_K'}(A,M',g_{N-1}')
       \displaystyle\sum_{g_K'\ldots g_{N-2}'}
		 
                 C^{\vec{a}'~J'~j_K'}_{\tilde{\vec{a}}'\tilde{J}'~\tilde{j}_K'}(g_K',\ldots,g_{N-1}')
	
    \\&&&&\!
     \times\,      C^{j_K}_{\tilde{j}_K}(B,n_K)
       \displaystyle\sum_{g_{N-1}} 
                 C^{J~j_K}_{\tilde{J}~ \tilde{j}_K}(C,M,g_{N-1})
       \displaystyle\sum_{g_K\ldots g_{N-2}}
		 
                 C^{\vec{a}~J~j_K}_{\tilde{\vec{a}}~\tilde{J}~\tilde{j}_K}(g_K,\ldots,g_{N-1})

     \\&&&&
     
     \multicolumn{4}{l}{	 
		 \times~
			       
    \Big[\displaystyle\prod_{L=1 \atop L\ne K}^N  \delta_{n_L'n_L}\delta_{{j}_L'j_L}\Big]
	         \times\delta_{{n}_K'+B~ n_K+B}~\delta_{\tilde{j}_K'\tilde{j}_K}
		 		
    \times\delta_{M'~M+C-A}~\delta_{\tilde{J}'\tilde{J}}}~

    \\&&&&\multicolumn{4}{l}{\!
    \times~
    
    \displaystyle\prod_{R=2}^{K-1}\delta_{\tilde{a}_R'a_R'} \delta_{\tilde{a}_R a_R}} 
    
    \\ \\
    
    &&&&\multicolumn{4}{l}{\times \big<\,\tilde{\vec{a}}'~\tilde{J}~M'\!+\!A~;~
                           \tilde{\vec{j}}'~;~\tilde{\vec{n}}
			   \,\big|\, \hat{V}^r \,\big|\,\tilde{\vec{a}}~\tilde{J}~M+C~;~
                           \tilde{\vec{j}}~;~\tilde{\vec{n}}\,\big> }
   
    \\    
\end{array}\] \\[-17mm]

\be \label{zwI} ~\ee

By inspection of (\ref{zwI}) we can see that the sum over the index $B=\pm\frac{1}{2}$ will only affect the factors \linebreak $C^{j_K}_{\tilde{j}_K}(B,n_K)~C^{j_K'}_{\tilde{j}_K}(B,n_K')$. Since $n_K'=n_K$ and always $\tilde{j}_K'=\tilde{j}_K$ we can explicitely evaluate $\sum_{B=\pm\frac{1}{2}}C^{j_K}_{\tilde{j}_K}(B,n_K)~C^{j_K'}_{\tilde{j}_K}(B,n_K)$. In order to get non vanishing contributions we have one of the following cases:

\begin{footnotesize}

\[\begin{array}{llll}
   \mbox{\fbox{\mbox{$j_K'=j_K-1$}}} & \rightarrow \tilde{j}_K=j_K-\frac{1}{2} 
   ~~&\cmt{1.8}{$\tilde{j_K}=j_K'+\frac{1}{2}$\\
              $\tilde{j_K}=j_K-\frac{1}{2}$}
   &\begin{array}{lcl}
                  \displaystyle\sum_{B=\pm\frac{1}{2}}C^{j_K}_{\tilde{j}_K}(B,n_K)~C^{j_K'}_{\tilde{j}_K}(B,n_K)
		&=&\displaystyle\sum_{B=\pm\frac{1}{2}}
		   -2B\Big[\frac{j_K'+2Bn_K+1}{2(j_K'+1)}\cdot \frac{j_K-2Bn_K}{2j_K}\Big]^{\frac{1}{2}} 
		=0    
		      
	       \end{array}      
   
   \\
   \\
   
   \mbox{\fbox{\mbox{$j_K'=j_K$}}} &  
   ~~&\cmt{1.8}{$\tilde{j_K}=j_K-\frac{1}{2}$}
   &\begin{array}{lll}
                  \displaystyle\sum_{B=\pm\frac{1}{2}}C^{j_K}_{\tilde{j}_K}(B,n_K)~C^{j_K'}_{\tilde{j}_K}(B,n_K)
		&=&\displaystyle\sum_{B=\pm\frac{1}{2}}
		   4B^2~ \frac{j_K-2Bn_K}{2j_K} 
		=1    
		      
	       \end{array} 
   \\	        	      
   &~~&\cmt{1.8}{$\tilde{j_K}=j_K+\frac{1}{2}$}
   &\begin{array}{lll}
                  \displaystyle\sum_{B=\pm\frac{1}{2}}C^{j_K}_{\tilde{j}_K}(B,n_K)~C^{j_K'}_{\tilde{j}_K}(B,n_K)
		&=&\displaystyle\sum_{B=\pm\frac{1}{2}}
		   4B^2~ \frac{j_K+2Bn_K+1}{2(j_K+1)} 
		=1    
		      
	       \end{array}
\\\\
   \mbox{\fbox{\mbox{$j_K'=j_K+1$}}} & \rightarrow \tilde{j}_K=j_K+\frac{1}{2} 
   ~~&\cmt{1.8}{$\tilde{j_K}=j_K'-\frac{1}{2}$\\
              $\tilde{j_K}=j_K+\frac{1}{2}$}
   &\begin{array}{lcl}
                  \displaystyle\sum_{B=\pm\frac{1}{2}}C^{j_K}_{\tilde{j}_K}(B,n_K)~C^{j_K'}_{\tilde{j}_K}(B,n_K)
		&=&\displaystyle\sum_{B=\pm\frac{1}{2}}
		   -2B\Big[\frac{j_K'-2Bn_K}{2j_K'}\cdot \frac{j_K+2Bn_K+1}{2(j_K+1)}\Big]^{\frac{1}{2}} 
		=0    
		      
	       \end{array}           
\end{array}\]

\end{footnotesize}

So we can carry out the sum over $B$ and have as additional selection rule \fbox{$j_K'\stackrel{!}{=}j_K$}. This is a consistency check for the following reason: During the regularization process of the operator $\hat{H}_{kin}$ as defined in (\ref{quantization step II}) we had to introduce trivial 2-valent vertices when decomposing the edges $e\rightarrow s[0,\epsilon]\circ e[\epsilon,1]$, which do not change the gauge behaviour of the spin network function over the graph $\gamma$. In order to be gauge invariant the spin $j$ of the sergment $s$ must not be modified by the $\lo{(r)}e^k_K$-operators.  Moreover the $\lo{(r)}e^k_K$ transform in the $J=1$ representation. So we get \fbox{$J'\stackrel{!}{=}J\pm 1$} .
\\
%%%%%%%%%%%%%%%%%%%%%%%%%%%%%%%%%%%%%%%%%%%%%%%%%%%%%%%%%%%%%%%%%%%%%%%%%%%%%%%%%%%%%%%%%%%%%%%%%%%%%%%
The result is a simplified expression for the matrix element of $\lo{(r)}e^k_K$:

\fcmt{18}{~\\
\[\begin{array}{ccllllll}
      \big<\,T_{J'}\, \big| \,\lo{(r)}e^k_K\, \big|\,T_J\, \big>
      
      &=&\multicolumn{5}{l}{\big<\,\vec{a}'\, J'\, M'\,;\,\vec{j}'\,;\,\vec{n}'\, \big|
          \,\lo{(r)}e^k_K\,
         \big|\,\vec{a}\, J\, M\,;\,\vec{j}\,;\,\vec{n}\, \big>}
      \\

      \\
      &=&\multicolumn{5}{l}{-\displaystyle\sum_{A\,C} \big[\overline{\tau}_k \big]_{CA}~\delta_{M'~M+C-A}
            
      \displaystyle\sum_{\tilde{j}_K=j_K\pm\frac{1}{2}}
      
      \displaystyle\sum_{\tilde{J}=J\pm\frac{1}{2}\atop\tilde{J}'=J'\pm\frac{1}{2}}
      \delta_{\tilde{J}'\tilde{J}}      	  
           
      \displaystyle\sum_{\tilde{\vec{a}}'\,\tilde{\vec{a}}}
      \bigg[\displaystyle\prod_{R=2}^{K-1}\delta_{\tilde{a}_R'a_R'} \delta_{\tilde{a}_R a_R}\bigg]

      \Bigg\{}         
		 
     \\\\
     
     &&~\hspace{3cm}~&~~\displaystyle\sum_{g_K'\ldots g_{N-1}'} 
                 C^{J'~j_K}_{\tilde{J}~ \tilde{j}_K}(A,M',g_{N-1}')
       	        ~C^{\vec{a}'~J'~j_K}_{\tilde{\vec{a}}'~\tilde{J}~\tilde{j}_K}(g_K',\ldots,g_{N-1}')
     \\\\		
     && & \times\displaystyle\sum_{g_K\ldots g_{N-1}}		
		~C^{J~j_K}_{\tilde{J}~ \tilde{j}_K}(C,M,g_{N-1})
		~C^{\vec{a}~J~j_K}_{\tilde{\vec{a}}~\tilde{J}~\tilde{j}_K}(g_K,\ldots,g_{N-1})

     \\\\
     &&&\times~ \big<\,\tilde{\vec{a}}'~\tilde{J}~M'\!+\!A~;~
                           \tilde{\vec{j}}~;~\tilde{\vec{n}}
			   \,\big|\, \hat{V}^r \,\big|\,\tilde{\vec{a}}~\tilde{J}~M\!+\!C~;~
                           \tilde{\vec{j}}~;~\tilde{\vec{n}}\,\big>
     \\\\
     &&& \times~\delta_{\vec{j}'\vec{j}}~\delta_{\vec{n}'\vec{n}}~~ \Bigg\}

\end{array}\] \\[-15mm]
\be\label{Endresult e-Op I} \ee \\[-4mm]}
\\[0.5cm]                
Here we have used the abbreviations (according to (\ref{Holonomieaktion auf allgemeine Kante Endresultat}) and (\ref{Entwicklung in anderes Recoupling Scheme})~)
\begin{footnotesize}

\[\begin{array}{llllllllll}	   
    \multicolumn{6}{l}{
    
        \begin{array}{lcllllll}
         C^{J~j_K}_{\tilde{J}~ \tilde{j}_K}(A,M,g_{N-1})
  	      &=&\multicolumn{5}{l}{\displaystyle\sum_m
               \big<\,g_{N\!-\!1}\,m\,;\,j_K\,M\!-\!m\,\big|\,J(g_{N\!-\!1}\,j_K)~M\big> }
	       \\ [-1.5mm]
               &&&\multicolumn{5}{l}{~~~~~\big<\,g_{N\!-\!1}\,m\,;\,\tilde{j}_K\,M\!-\!m\!+\!A_0\,\big|\,\tilde{J}(g_{N\!-\!1}\,\tilde{j}_K)~M\!+\!A\big> }
           \\
              & &&\multicolumn{5}{l}{
	        ~~~~~\big<~j_K~M\!-\!m~;~\frac{1}{2}~A~\big|~\tilde{j}_K~M\!-m\!+\!A~\big>}
	
	\end{array}}

    \\\\\\

    \multicolumn{6}{l}{C^{\vec{a}~J~j_K}_{\tilde{\vec{a}}~\tilde{J}~\tilde{j}_K}(g_K,\ldots,g_{N-1})=}
    \\
    \\	   
    =&
      \big<\tilde{a}_K(a_{K\!-\!1}\,\tilde{j}_k)~\tilde{a}_{K\!+\!1}(\tilde{a}_K\,j_{K\!+\!1})
	  \,\big|\,
	  g_K(a_{K-1}\,j_{K+1})~\tilde{a}_{K+1}(g_K\,\tilde{j}_K)\,\big>
      &
      \big<a_K(a_{K\!-\!1}\,j_K)~a_{K\!+\!1}(a_K\,j_{K\!+\!1})
	  \,\big|\,
	  g_K(a_{K-1}\,j_{K+1})~a_{K+1}(g_K\,j_K)\,\big>&&&~	  
      \\
      
      &\big<\tilde{a}_{K\!+\!1}(g_{K}\,\tilde{j}_K)~\tilde{a}_{K\!+\!2}(\tilde{a}_{K\!+\!1}\,j_{K\!+\!2})
	  \,\big|\,
	  g_{K\!+\!1}(g_{K}\,j_{K+2})~\tilde{a}_{K+2}(g_{K\!+\!1}\,\tilde{j}_K)\,\big>
      &\big<a_{K\!+\!1}(g_{K}\,j_K)~a_{K\!+\!2}(a_{K\!+\!1}\,j_{K\!+\!2})
	  \,\big|\,
	  g_{K\!+\!1}(g_{K}\,j_{K+2})~a_{K+2}(g_{K\!+\!1}\,j_K)\,\big>
	  	  
      \\
      
     & ~~~~~~~~~\vdots&~~~~~~~~~\vdots
      
      \\
      
      &\big<\tilde{a}_{N\!-\!3}(g_{N-4}\,\tilde{j}_K)~\tilde{a}_{N\!-\!2}(\tilde{a}_{N\!-\!3}\,j_{N\!-\!2})
	  \,\big|\,
	  g_{N\!-\!3}(g_{N-4}\,j_{N-2})~\tilde{a}_{N-2}(g_{N\!-\!3}\,\tilde{j}_K)\,\big>
      &\big<a_{N\!-\!3}(g_{N-4}\,j_K)~a_{N\!-\!2}(a_{N\!-\!3}\,j_{N\!-\!2})
	  \,\big|\,
	  g_{N\!-\!3}(g_{N-4}\,j_{N-2})~a_{N-2}(g_{N\!-\!3}\,j_K)\,\big>
	  	  
      \\
      
      &\big<\tilde{a}_{N\!-\!2}(g_{N-3}\,\tilde{j}_K)~\tilde{a}_{N\!-\!1}(\tilde{a}_{N\!-\!2}\,j_{N\!-\!1})
	  \,\big|\,
	  g_{N\!-\!2}(g_{N-3}\,j_{N-1})~\tilde{a}_{N-1}(g_{N\!-\!2}\,\tilde{j}_K)\,\big>
      &\big<a_{N\!-\!2}(g_{N-3}\,j_K)~a_{N\!-\!1}(a_{N\!-\!2}\,j_{N\!-\!1})
	  \,\big|\,
	  g_{N\!-\!2}(g_{N-3}\,j_{N-1})~a_{N-1}(g_{N\!-\!2}\,j_K)\,\big>
	  	  
      \\
      
     &\big<\tilde{a}_{N\!-\!1}(g_{N-2}\,\tilde{j}_K)~\tilde{J}(\tilde{a}_{N\!-\!1}\,j_{N})
	  \,\big|\,
	  g_{N\!-\!1}(g_{N-2}\,j_{N})~\tilde{J}(g_{N\!-\!1}\,\tilde{j}_K)\,\big>
      &\big<a_{N\!-\!1}(g_{N-2}\,j_K)~J(a_{N\!-\!1}\,j_{N})
	  \,\big|\,
	  g_{N\!-\!1}(g_{N-2}\,j_{N})~J(g_{N\!-\!1}\,j_K)\,\big>

\\
\\\multicolumn{6}{l}{\mbox{where the individual $g_K\ldots g_{N-1}$ can take all values allowed by their arguments due to the Clebsch Gordan theorem and (\ref{Definitions for the action of holonomy on a recoupling scheme}})}

\\\\\\  \multicolumn{6}{l}{\vec{j}=\big\{j_1,\ldots,j_{K-1},j_K,j_{K+1},\ldots,j_N \big\}}
  \\  \multicolumn{6}{l}{\tilde{\vec{j}}=\big\{j_1,\ldots,j_{K-1},\tilde{j}_K,j_{K+1},\ldots,j_N \big\}}

\\\\  \multicolumn{6}{l}{{\vec{a}}~=
              \big\{{a}_2(j_1\,j_2)~{a}_3({a}_2\,j_3)
	            ~\ldots~ 
	            {a}_{K\!-\!1}({a}_{K\!-\!2}\,j_{K\!-\!1})~
		    {a}_K({a}_{K\!-\!1}\,{j}_K) ~
		    {a}_{K\!+\!1}({a}_{K}\,j_{K\!+\!1})~
		    \ldots~
		    {a}_{N\!-\!1}({a}_{N\!-\!2}\,j_{N\!-\!1})~
		    {J}({a}_{N\!-\!1}\,j_{N})~
		        \big\}}
\\  \multicolumn{6}{l}{{\vec{a}}'=
              \big\{{a}_2'(j_1\,j_2)~{a}_3'({a}_2'\,j_3)
	            ~\ldots~ 
	            {a}_{K\!-\!1}'({a}_{K\!-\!2}'\,j_{K\!-\!1})~
		    {a}_K'({a}_{K\!-\!1}'\,{j}_K) ~
		    {a}_{K\!+\!1}'({a}_{K}'\,j_{K\!+\!1})~
		    \ldots~
		    {a}_{N\!-\!1}'({a}_{N\!-\!2}'\,j_{N\!-\!1})~
		    {J'}({a}_{N\!-\!1}'\,j_{N})~
		     \big\}}

\\\\  \multicolumn{6}{l}{\tilde{\vec{a}}~=
              \big\{\tilde{a}_2(j_1\,j_2)~\tilde{a}_3(\tilde{a}_2\,j_3)
	            ~\ldots~ 
	            \tilde{a}_{K\!-\!1}(\tilde{a}_{K\!-\!2}\,j_{K\!-\!1})~
		    \tilde{a}_K(\tilde{a}_{K\!-\!1}\,\tilde{j}_K) ~
		    \tilde{a}_{K\!+\!1}(\tilde{a}_{K}\,j_{K\!+\!1})~
		    \ldots~
		    \tilde{a}_{N\!-\!1}(\tilde{a}_{N\!-\!2}\,j_{N\!-\!1})~
		    \tilde{J}(\tilde{a}_{N\!-\!1}\,j_{N})~
		        \big\}}
\\  \multicolumn{6}{l}{\tilde{\vec{a}}'=
              \big\{\tilde{a}_2'(j_1\,j_2)~\tilde{a}_3'(\tilde{a}_2'\,j_3)
	            ~\ldots~ 
	            \tilde{a}_{K\!-\!1}'(\tilde{a}_{K\!-\!2}'\,j_{K\!-\!1})~
		    \tilde{a}_K'(\tilde{a}_{K\!-\!1}'\,\tilde{j}_K) ~
		    \tilde{a}_{K\!+\!1}'(\tilde{a}_{K}'\,j_{K\!+\!1})~
		    \ldots~
		    \tilde{a}_{N\!-\!1}'(\tilde{a}_{N\!-\!2}'\,j_{N\!-\!1})~
		    \tilde{J}(\tilde{a}_{N\!-\!1}'\,j_{N})~
		        \big\}}

\\\\ \multicolumn{6}{l}{{\vec{g}}~=
              \big\{{g}_2(j_1\,j_2)~{g}_3({g}_2\,j_3)
	            ~\ldots~ 
	            {g}_{K\!-\!1}({g}_{K\!-\!2}\,j_{K\!-\!1})~
		    {g}_K({g}_{K\!-\!1}\,{j}_{K\!+\!1}) ~
		    {g}_{K\!+\!1}({g}_{K}\,j_{K\!+\!2})~
		    \ldots~
		    {g}_{N\!-\!1}({g}_{N\!-\!2}\,j_{N})~
		    {J}({g}_{N\!-\!1}\,j_{K})~
		        \big\}} 
			
\end{array}\]

\end{footnotesize}

%%%%%%%%%%%%%%%%%%%%%%%%%%%%%%%%%%%%%%%%%%%%%%%%%%%%%%%%%%%%%%%%%%%%%%%%%%%%%%%%%%%%%%%%%%%%%%%%%%%%%%%
\subsection{Special Cases}
%%%%%%%%%%%%%%%%%%%%%%%%%%%%%%%%%%%%%%%%%%%%%%%%%%%%%%%%%%%%%%%%%%%%%%%%%%%%%%%%%%%%%%%%%%%%%%%%%%%%%%%

Furthermore for the {\bf\underline{special cases}} \fbox{$K=1$}, \fbox{$K=2$} and \fbox{$K=N$} we find:

%%%%%%%%%%%%%%%%%%%%%%%%%%%%%%%%%%%%%%%%%%%%%%%%%%%%%%%%%%%%%%%%%%%%%%%%%%%%%%%%%%%%%%%%%%%%%%%%%%%%%
\subsubsection{\fbox{$K = 1$}} 
%%%%%%%%%%%%%%%%%%%%%%%%%%%%%%%%%%%%%%%%%%%%%%%%%%%%%%%%%%%%%%%%%%%%%%%%%%%%%%%%%%%%%%%%%%%%%%%%%%%%%

\cmt{18}{~\\
\[\begin{array}{ccllllll}
      \big<\,T_{J'}\, \big| \,\lo{(r)}e^k_{K=1}\, \big|\,T_J\, \big>
      
      &=&\multicolumn{5}{l}{\big<\,\vec{a}'\, J'\, M'\,;\,\vec{j}'\,;\,\vec{n}'\, \big|
          \,\lo{(r)}e^k_{K=1}\,
         \big|\,\vec{a}\, J\, M\,;\,\vec{j}\,;\,\vec{n}\, \big>}
      \\

      \\
      &=&\multicolumn{5}{l}{-\displaystyle\sum_{A\,C} \big[\overline{\tau}_k \big]_{CA}~\delta_{M'~M+C-A}
            
      \displaystyle\sum_{\tilde{j}_1=j_1\pm\frac{1}{2}}
      
      \displaystyle\sum_{\tilde{J}=J\pm\frac{1}{2}\atop\tilde{J}'=J'\pm\frac{1}{2}}
      \delta_{\tilde{J}'\tilde{J}}      	  
           
      \displaystyle\sum_{\tilde{\vec{a}}'\,\tilde{\vec{a}}}
      
      \Bigg\{} ~~~~~~~~~~~~~~~~~~~~~~~~~~~~~~~~~~~         
		 
     \\\\
     
     &&~\hspace{3cm}~&~~\displaystyle\sum_{g_2'\ldots g_{N-1}'} 
                 C^{J'~j_1}_{\tilde{J}~ \tilde{j}_1}(A,M',g_{N-1}')
       	        ~C^{\vec{a}'~J'~j_1}_{\tilde{\vec{a}}'~\tilde{J}~\tilde{j}_1}(g_2',\ldots,g_{N-1}')
     \\\\		
     && & \times\displaystyle\sum_{g_2\ldots g_{N-1}}		
		~C^{J~j_1}_{\tilde{J}~ \tilde{j}_1}(C,M,g_{N-1})
		~C^{\vec{a}~J~j_1}_{\tilde{\vec{a}}~\tilde{J}~\tilde{j}_1}(g_2,\ldots,g_{N-1})

     \\\\
     &&&\times~ \big<\,\tilde{\vec{a}}'~\tilde{J}~M'\!+\!A~;~
                           \tilde{\vec{j}}~;~\tilde{\vec{n}}
			   \,\big|\, \hat{V}^r \,\big|\,\tilde{\vec{a}}~\tilde{J}~M\!+\!C~;~
                           \tilde{\vec{j}}~;~\tilde{\vec{n}}\,\big>
     \\\\
     &&& \times~\delta_{\vec{j}'\vec{j}}~\delta_{\vec{n}'\vec{n}}~~ \Bigg\}

\end{array}\] \\[-15mm]
\be\label{Endresult e-Op K=1} \ee \\}
\\[1cm]                
Here we have used the  same shorthands as in (\ref{Endresult e-Op I})  but must use (\ref{3nj Expansion: special case K=1}) instead of (\ref{3nj Expansion I}) in the definition of the coefficients
$C^{\vec{a}~J~j_1}_{\tilde{\vec{a}}~\tilde{J}~\tilde{j}_1}(g_2,\ldots,g_{N-1})$ and the intermediate recoupling scheme $\vec{g}$:

\begin{footnotesize}

\[\begin{array}{llllllllll}	   
    \multicolumn{6}{l}{C^{\vec{a}~J~j_1}_{\tilde{\vec{a}}~\tilde{J}~\tilde{j}_1}(g_2,\ldots,g_{N-1})=}
    \\
    \\	   
    =&
      \big<\tilde{a}_2(\tilde{j}_1\,j_2)~\tilde{a}_{3}(\tilde{a}_2\,j_{3})
	  \,\big|\,
	  g_2(j_2\,j_{3})~\tilde{a}_{3}(g_2\,\tilde{j}_1)\,\big>
      &
      \big<a_2(j_1\,j_2)~a_{3}(a_2\,j_{3})
	  \,\big|\,
	  g_2(j_2\,j_{3})~a_{3}(g_2\,j_1)\,\big>&&&~	  
      \\
      
      &\big<\tilde{a}_{3}(g_{2}\,\tilde{j}_1)~\tilde{a}_{4}(\tilde{a}_{3}\,j_{4})
	  \,\big|\,
	  g_{3}(g_{2}\,j_{4})~\tilde{a}_{4}(g_{3}\,\tilde{j}_1)\,\big>
      &\big<a_{3}(g_{2}\,j_1)~a_{4}(a_{3}\,j_{4})
	  \,\big|\,
	  g_{3}(g_{2}\,j_{4})~a_{4}(g_{3}\,j_1)\,\big>
	  	  
      \\
      
     & ~~~~~~~~~\vdots&~~~~~~~~~\vdots
      
      \\
      
      &\big<\tilde{a}_{N\!-\!3}(g_{N\!-\!4}\,\tilde{j}_1)~\tilde{a}_{N\!-\!2}(\tilde{a}_{N\!-\!3}\,j_{N\!-\!2})
	  \,\big|\,
	  g_{N\!-\!3}(g_{N\!-\!4}\,j_{N\!-\!2})~\tilde{a}_{N\!-\!2}(g_{N\!-\!3}\,\tilde{j}_1)\,\big>
      &\big<a_{N\!-\!3}(g_{N\!-\!4}\,j_1)~a_{N\!-\!2}(a_{N\!-\!3}\,j_{N\!-\!2})
	  \,\big|\,
	  g_{N\!-\!3}(g_{N\!-\!4}\,j_{N\!-\!2})~a_{N\!-\!2}(g_{N\!-\!3}\,j_1)\,\big>
	  	  
      \\
      
      &\big<\tilde{a}_{N\!-\!2}(g_{N\!-\!3}\,\tilde{j}_1)~\tilde{a}_{N\!-\!1}(\tilde{a}_{N\!-\!2}\,j_{N\!-\!1})
	  \,\big|\,
	  g_{N\!-\!2}(g_{N\!-\!3}\,j_{N\!-\!1})~\tilde{a}_{N\!-\!1}(g_{N\!-\!2}\,\tilde{j}_1)\,\big>
      &\big<a_{N\!-\!2}(g_{N\!-\!3}\,j_1)~a_{N\!-\!1}(a_{N\!-\!2}\,j_{N\!-\!1})
	  \,\big|\,
	  g_{N\!-\!2}(g_{N\!-\!3}\,j_{N\!-\!1})~a_{N\!-\!1}(g_{N\!-\!2}\,j_1)\,\big>
	  	  
      \\
      
     &\big<\tilde{a}_{N\!-\!1}(g_{N\!-\!2}\,\tilde{j}_1)~\tilde{J}(\tilde{a}_{N\!-\!1}\,j_{N})
	  \,\big|\,
	  g_{N\!-\!1}(g_{N\!-\!2}\,j_{N})~\tilde{J}(g_{N\!-\!1}\,\tilde{j}_1)\,\big>
      &\big<a_{N\!-\!1}(g_{N\!-\!2}\,j_1)~J(a_{N\!-\!1}\,j_{N})
	  \,\big|\,
	  g_{N\!-\!1}(g_{N\!-\!2}\,j_{N})~J(g_{N\!-\!1}\,j_1)\,\big>

\\
\\\multicolumn{6}{l}{\mbox{where the individual $g_2\ldots g_{N-1}$ can take all values allowed by their arguments due to the Clebsch Gordan theorem and (\ref{Definitions for the action of holonomy on a recoupling scheme}})}

\\\\ \multicolumn{6}{l}{{\vec{g}}~=
              \big\{{g}_2(j_2\,j_3)~{g}_3({g}_2\,j_4)
	            ~\ldots~ 
	            {g}_{N\!-\!1}({g}_{N\!-\!2}\,j_{N})~
		    {J}({g}_{N\!-\!1}\,j_{1})~
		        \big\}} 
			
\end{array}\]

\end{footnotesize}

%%%%%%%%%%%%%%%%%%%%%%%%%%%%%%%%%%%%%%%%%%%%%%%%%%%%%%%%%%%%%%%%%%%%%%%%%%%%%%%%%%%%%%%%%%%%%%%%%%%%%
\subsubsection{\fbox{$K = 2$}} 
%%%%%%%%%%%%%%%%%%%%%%%%%%%%%%%%%%%%%%%%%%%%%%%%%%%%%%%%%%%%%%%%%%%%%%%%%%%%%%%%%%%%%%%%%%%%%%%%%%%%%
This case can be handled by (\ref{Endresult e-Op I}). One only has to leave out the prefactor  $\displaystyle\prod_{R=2}^{K-1}\delta_{\tilde{a}_R'a_R'} \delta_{\tilde{a}_R a_R}$ .

\pagebreak
%%%%%%%%%%%%%%%%%%%%%%%%%%%%%%%%%%%%%%%%%%%%%%%%%%%%%%%%%%%%%%%%%%%%%%%%%%%%%%%%%%%%%%%%%%%%%%%%%%%%%
\subsubsection{\fbox{$K = N$}} 
%%%%%%%%%%%%%%%%%%%%%%%%%%%%%%%%%%%%%%%%%%%%%%%%%%%%%%%%%%%%%%%%%%%%%%%%%%%%%%%%%%%%%%%%%%%%%%%%%%%%%

In this case the expansion (\ref{3nj Expansion I}) into intermediate recoupling schemes  is not necessary because the spin the holonomies in $\lo{(r)}e^k_{K=N}$ act on is already the last one and we can work directly with 
(\ref{Holonomieaktion auf allgemeine Kante}) instead to obtain:

\cmt{18}{ 
 \[\begin{array}{ccllllll}
      \big<\,T_{J'}\, \big| \,\lo{(r)}e^k_{K=N}\, \big|\,T_J\, \big>
      
      &=&\multicolumn{5}{l}{\big<\,\vec{a}'\, J'\, M'\,;\,\vec{j}'\,;\,\vec{n}'\, \big|
          \,\lo{(r)}e^k_{K=N}\,
         \big|\,\vec{a}\, J\, M\,;\,\vec{j}\,;\,\vec{n}\, \big>}
      \\

      \\
      &=&\multicolumn{5}{l}{-\displaystyle\sum_{A\,C} \big[\overline{\tau}_k \big]_{CA}~\delta_{M'~M+C-A}
            
      \displaystyle\sum_{\tilde{j}_N=j_N\pm\frac{1}{2}}
      
      \displaystyle\sum_{\tilde{J}=J\pm\frac{1}{2}\atop\tilde{J}'=J'\pm\frac{1}{2}}
      \delta_{\tilde{J}'\tilde{J}}

      \Bigg\{} ~~~~~~~~~~~~~~~~~~~~~~~~~~~~~~~~~~~         
		 
     \\\\
     
     &&~\hspace{3cm}~&~~
                 C^{J'~j_N}_{\tilde{J}~ \tilde{j}_N}(A,M',a_{N-1}')
       	        ~
     \\\\		
     && & \times		
		~C^{J~j_N}_{\tilde{J}~ \tilde{j}_N}(C,M,a_{N-1})
		~

     \\\\
     &&&\times~ \big<\,\vec{a}'~\tilde{J}~M'\!+\!A~;~
                           \tilde{\vec{j}}~;~\tilde{\vec{n}}
			   \,\big|\, \hat{V}^r \,\big|\,\vec{a}~\tilde{J}~M\!+\!C~;~
                           \tilde{\vec{j}}~;~\tilde{\vec{n}}\,\big>
     \\\\
     &&& \times~\delta_{\vec{j}'\vec{j}}~\delta_{\vec{n}'\vec{n}}~~ \Bigg\}

 \end{array}\] \\[-15mm]

\be \label{Endresult e-Op K=N} \ee }

%%%%%%%%%%%%%%%%%%%%%%%%%%%%%%%%%%%%%%%%%%%%%%%%%%%%%%%%%%%%%%%%%%%%%%%%%%%%%%%%%%%%%%%%%%%%%%%%%%%%%%%
\subsection{Evaluation of the Matrix Elements $\big<\cdot\big|\hat{V}^r\big|\cdot\big>$}
%%%%%%%%%%%%%%%%%%%%%%%%%%%%%%%%%%%%%%%%%%%%%%%%%%%%%%%%%%%%%%%%%%%%%%%%%%%%%%%%%%%%%%%%%%%%%%%%%%%%%%%

In order to evaluate the matrix elements 
$\big<\,\tilde{\vec{a}}'~\tilde{J}~M'\!+\!A~;~
                           \tilde{\vec{j}}~;~\tilde{\vec{n}}
			   \,\big|\, \hat{V}^r \,\big|\,\tilde{\vec{a}}~\tilde{J}~M\!+\!C~;~
                           \tilde{\vec{j}}~;~\tilde{\vec{n}}\,\big>=:
\big<\,\tilde{T}_{J'}\,\big|\,\hat{V}^r\,\big|\,\tilde{T}_J\,\big>$ 
in (\ref{Endresult e-Op I}) containing the $r^th$ power of the Volume Operator
we have to recall the definition of the Volume operator as $\hat{V}:=(\ell_P)^3\sqrt[4]{\hat{q}^\dagger \hat{q}}=\sqrt{|\hat{q}|}$ with $\hat{q}$ beeing the matrix analyzed in \cite{Volume_Article_I} and $\ell_P$ the Planck length. It is in particular a real antisymmetric matrix times $\mb{i}$.  So the eigenvalues $\lambda_{\hat{q}}$ of $\hat{q}$ are real and come in pairs $\pm\lambda_{\hat{q}}$ or $\lambda_{\hat{q}}=0$ if the dimension of $\hat{q}$ is odd. Moreover its eigenvectors $|\gamma(\lambda_{\hat{q}})>$ are orthogonal. 

The definition of $\hat{V}$ is to be understood as follows: 
For each of its eigenvalues $\lambda_{\hat{q}}$ the the matrix $\hat{q}$ has an eigenvector $|\gamma(\lambda_{\hat{q}})>$.  

$\hat{V}$ has the same eigenvectors  as $\hat{q}$ but its eigenvalues are given by $(\ell_P)^3\sqrt{|\lambda_{\hat{q}}|}$ .
\\

We have to insert a $\mb{1}=\sum_{\lambda_{\hat{q}}}|\gamma(\lambda_{\hat{q}})><\gamma(\lambda_{\hat{q}})|$ in terms of normalized eigenstates $|\gamma(\lambda_{\hat{q}})>$ belonging to each eigenvalue (of course counting multiplicity!)  $\lambda_{\hat{q}}$ of $\hat{q}$ in order to evaluate $\hat{V}^r$:
   
   \be\label{Eigenstate expansion q}\begin{array}{ccl}
     \big<\,\tilde{T}_{J'}\,\big|\,\hat{V}^r\,\big|\,\tilde{T}_J\,\big>
     &=&
     \displaystyle\sum_{\lambda_{\hat{q}}, \lambda'_{\hat{q}}} 
     \big<~\tilde{T}_{J'}~\big|\gamma(\lambda_{\hat{q}})~\big>
     \underbrace{\big<\gamma(\lambda_{\hat{q}})\big|~\hat{V}^r~ \big|\gamma(\lambda'_{\hat{q}})\big>}
     _{\displaystyle{\Big[\sqrt{|\lambda_{\hat{q}}|}\,\Big]^r\cdot
           \delta_{\lambda_{\hat{q}} \lambda'_{\hat{q}}}}}
         \big<\gamma(\lambda'_{\hat{q}})\big|~\tilde{T}_J~\big>\\
	
     \\
     &=&(\ell_P)^{3r}
     \displaystyle\sum_{\lambda_{\hat{q}}}\big|\lambda_{\hat{q}}\big|^{\frac{r}{2}}\cdot
     ~\big< ~\tilde{T}_{J'}~\big|\gamma(\lambda_{\hat{q}})~\big>
                                \overline{\big<~\tilde{T}_J~\big|\gamma(\lambda_{\hat{q}})\big>}\\	  
   \end{array}\ee
   %\\\\

Sometimes it might be more convenient to use the matrix $\hat{q}^{\dagger}\hat{q}$ instead, because as a real symmetric matrix its eigenvalues $\lambda_{\hat{q}^{\dagger}\hat{q}}$ are real and $\ge0$ from the beginning and we do not need to introduce by hand the modulus from the definition of $\hat{V}$ in terms of $\hat{q}$ as required in (\ref{Eigenstate expansion q}) but can write:  

   \be\label{Eigenstate expansion q+q}\begin{array}{ccl}
     \big<\,\tilde{T}_{J'}\,\big|\,\hat{V}^r\,\big|\,\tilde{T}_J\,\big>
     &=&(\ell_P)^{3r}
     \displaystyle\sum_{\lambda_{\hat{q}^{\dagger}\hat{q}}}
     \big(\lambda_{\hat{q}^{\dagger}\hat{q}}\big)^{\frac{r}{4}}\cdot
     ~\big< ~\tilde{T}_{J'}~\big|\gamma(\lambda_{\hat{q}^{\dagger}\hat{q}})~\big>
     \overline{\big<~\tilde{T}_J~\big|\gamma(\lambda_{\hat{q}^{\dagger}\hat{q}})\big>}\\	  
   \end{array}\ee
   %\\\\

Anyway, we need to know explicitely the spectrum and the eigenstates of $\hat{V}$. This will only be possible to do exactly in very special cases.

\pagebreak
%%%%%%%%%%%%%%%%%%%%%%%%%%%%%%%%%%%%%%%%%%%%%%%%%%%%%%%%%%%%%%%%%%%%%%%%%%%%%%%%%%%%%%%%%%%%%%%%%%%%%%%
\section{Gauge Invariant 3-Vertex}
%%%%%%%%%%%%%%%%%%%%%%%%%%%%%%%%%%%%%%%%%%%%%%%%%%%%%%%%%%%%%%%%%%%%%%%%%%%%%%%%%%%%%%%%%%%%%%%%%%%%%%%
\subsection{General Properties}
%%%%%%%%%%%%%%%%%%%%%%%%%%%%%%%%%%%%%%%%%%%%%%%%%%%%%%%%%%%%%%%%%%%%%%%%%%%%%%%%%%%%%%%%%%%%%%%%%%%%%%%

As we will see in the simple case of the gauge invariant 3-vertex it is possible to explicitely evaluate the action of $\hat{H}_{kin}$ as derived in (\ref{quantization step II}). We notice that $E(3)=1$ and look at one 3-vertex only. Then we realize that the sign-factors $\epsilon(IJK)\epsilon(LMN)\stackrel{3-vertex}{\longrightarrow}\epsilon(123)\epsilon(123)=1$, because every permutation of the edges changing the sign of $\epsilon(123)$ will also change the sign of $\epsilon^{IJK}$. So we have to consider only the triple $(123)$. Then (\ref{quantization step II}) reads as (we get an additional multiplicity factor of $6\cdot6$ while summing over all permutations of the edge triple $1,2,3$):

\[\begin{array}{lllll}
   \hat{H}_{kin}f_{\gamma}^v
   
   &=&\bigg[
   
   \displaystyle\frac{P\cdot6^2}{\hbar^4\kappa^4}\,
   
   \displaystyle X(v)X(v) 
   \epsilon^{IJK}\,\epsilon^{LMN}\,\epsilon_{ijk}\,\epsilon_{lmn}\,
   \lo{(\frac{1}{2})\,}\hat{e}^i_I(v)
   \,\lo{(\frac{1}{2})\,}\hat{e}^j_J(v)
   \,\lo{(\frac{1}{2})\,}\hat{e}^k_K(v)
   
   \,\lo{(\frac{1}{2})\,}\hat{e}^l_L(v)
   \,\lo{(\frac{1}{2})\,}\hat{e}^m_M(v)
   \,\lo{(\frac{1}{2})\,}\hat{e}^n_N(v)
   
     \bigg]~f_{\gamma}^v
   \\[5mm]
   &&\fcmt{10}{Now we again take advantage of the identity
        \\[1mm]
    ~~~~~~~~~~$\epsilon_{ijk}\epsilon_{lmn}=6\cdot\delta_{[l}^i\delta_m^j\delta_{n]}^k$ 
        \\[1mm]
    and introduce the manifestly {\bf gauge invariant} quantities
        \\[1mm] 
   ~~~~~~~~~~$\hat{q}_{IL}(r,v)= \lo{(r)\,}\hat{e}^i_I(v)~\lo{(r)\,}\hat{e}^l_L(v)~\delta_{il} $ 
   \\[1mm]
   }
  
  \\\\

   &=&\bigg[
   
   \displaystyle\frac{P\cdot6^3}{\kappa^4\cdot6\cdot6\cdot\hbar^4}\,
   
   \displaystyle X(v)X(v) 
   \underbrace{\epsilon^{IJK}\,\epsilon^{LMN}\,
   \hat{q}_{IL}\Big(\frac{1}{2},v\Big)\,
   \hat{q}_{JM}\Big(\frac{1}{2},v\Big)\,
   \hat{q}_{KN}\Big(\frac{1}{2},v\Big)}_{\displaystyle e'(v)^2}
     \bigg]~f_{\gamma}^v

\end{array}\] \\[-2cm]\be \ee\\[0mm]

We have introduced the quantity

\be
   e'(v)^2= \epsilon^{IJK}\epsilon^{LMN} 
   q_{IL}\big(\frac{1}{2}\big)~  q_{JM}\big(\frac{1}{2}\big)~q_{KN}\big(\frac{1}{2}\big)
\ee

which is the gravitational part of $\hat{H}_{kin}$ and up to a constant the according operator-version "$\widehat{\frac{1}{\sqrt{\det(q)}(v)}}$". In what follows we will explicitely calculate its expectation values wrt. to gauge invariant states $f^v_\gamma$ at an unspecified 3-vertex, therefore we drop the argument $v$ of $e'(v)^2$.
\\\\  
At the gauge invariant 3-vertex we have

\ba
   \big\|~e'^2 ~|~ 0 >\big\|^2&=& < 0\, |\, [e']^2 \,[e']^2 \,|\, 0 >\nonumber\\
   &=&< 0\, |\, [e']^2\,|\, 0 >~< 0\, |\,  [e']^2\, |\, 0\, >  \nonumber\\
   &=&\big[ < 0 \,| \,[e']^2\,|\, 0\, >\big]^2
\ea

Furthermore
\ba\label{Definition of e'(B)^2}
   < 0 \,| \,[e']^2\,|\, 0\, >
   &=&
       \epsilon^{IJK}\epsilon^{LMN}~<0\,|\,  
   q_{IL}\big(\frac{1}{2}\big)~  q_{JM}\big(\frac{1}{2}\big)~q_{KN}\big(\frac{1}{2}\big)\,|\,0> \nonumber \\
   &=&
   \epsilon^{IJK}\epsilon^{LMN}~
   <0\,|\,q_{IL}\big(\frac{1}{2}\big) \,|\,0>\,
   <0\,|\,q_{JM}\big(\frac{1}{2}\big) \,|\,0>\,
   <0\,|\,q_{KN}\big(\frac{1}{2}\big) \,|\,0> 
\ea

We can evaluate this as follows (the $e^i_I$'s transform in the $J=1$ representation of $SU(2)$):

\ba\label{Definition of the q_{IL} at the 3-vertex}
   <0\,|\,q_{IL}\big(\frac{1}{2}\big) \,|\,0> 
   &=&
   \sum_{|\,1\,>}  <0\,|\, \lo{(\frac{1}{2})}e^i_I \,|\,1\,>\,<1\,|\, \lo{(\frac{1}{2})}e^l_L\,|\,0>~~\delta_{il} \nonumber\\
   &=&\delta_{il}
   \sum_{a_2'=\left\{ \tiny\begin{array}{l} \!\!j_3\!-\!1\\\!\!j_3\\\!\!j_3\!+\!1\end{array}\right.}
   \sum_{M'=\left\{ \tiny\begin{array}{r} \!\!-1\\\!\!0\\\!\!1\end{array}\right.}~~~ \overline{\big<\,a_2'(j_1\,j_2)~J'(a_2'\,j_3)\!=\!1~M'\,
   \big|\,\lo{(\frac{1}{2})}e^i_I\,\big| 
   \,a_2(j_1\,j_2)\!=\!j_3~J(a_2\,j_3)\!=\!0~M\!=\!0\,\big>} \nonumber\\[-5mm]
   &&\hspace{3.2cm} \times\big<\,a_2'(j_1\,j_2)~J'(a_2\,j_3)\!=\!1~M'\,
   \big|\,\lo{(\frac{1}{2})}e^l_L\,\big| 
   \,a_2(j_1\,j_2)\!=\!j_3~J(a_2\,j_3)\!=\!0~M\!=\!0\,\big> \nonumber\\
\ea
Here the overline denotes complex conjugation.
As we can see the remaining task is to calculate matrix elements of the form $<0\,|\, \lo{(r)}e^i_I \,|\,1\,>$ at the gauge invariant 3-vertex. We will apply the general expression 
(\ref{Endresult e-Op I}) separately for every $I=1,2,3$.

%%%%%%%%%%%%%%%%%%%%%%%%%%%%%%%%%%%%%%%%%%%%%%%%%%%%%%%%%%%%%%%%%%%%%%%%%%%%%%%%%%%%%%%%%%%%%%%%%%%%%%%
\subsubsection{The Coefficients $~C^{J~j_K}_{\tilde{J}~ \tilde{j}_K}(A,M,g_{N-1})$}

It is difficult to find a general expression for coefficients of the form $~C^{J~j_K}_{\tilde{J}~ \tilde{j}_K}(A,M,g_{N-1})$ occuring in equation (\ref{Endresult e-Op I}). However one can in principle carry out the summation over 
$-g_{N-1} \le m \le g_{N-1}$ where the Clebsch Gordan coefficients are only contributing for with 
$j_K$ compatible values of $m$. \\
For the given case of the matrix element of $<~T_{J=1}~|{\lo{(r)}e_K^k}~|~T_{J=0}~>$ the relevant non-vanishing coefficients $~C^{J~j_K}_{\tilde{J}~ \tilde{j}_K}(A,M,g_{N-1})$ can be calculated to be ({\it MATHEMATICA}):\\

\begin{tabular}{|lllll||r|r|}\hline\
 &&&&&&\\
 &&&&& \cmt{3.8}{$\tilde{j}_K=j_K+\frac{1}{2}$ }  
 &    \cmt{3.8}{$\tilde{j}_K=j_K-\frac{1}{2}$ }
 \\&&&&&&
 \\
 \hline\hline
 
 $J=0$ & $\tilde{J}=\frac{1}{2}$ & $M=0$ & $A=\pm\frac{1}{2}$ & $g_{N-1}=j_K$ 
 &$(-1)^{2j_K} \left[\displaystyle \frac{j_K+1}{2j_K+1} \right]^{\frac{1}{2}}$
 &$-(-1)^{2j_K} \left[\displaystyle \frac{j_K}{2j_K+1} \right]^{\frac{1}{2}}$
 \\
 \hline\hline
 $J=1$&$\tilde{J}=\frac{1}{2}$ & $M=-1$ & $A=+\frac{1}{2}$ & $g_{N-1}=j_K$
 &$-\sqrt{\frac{2}{3}} \left[\displaystyle\frac{j_K}{2j_K+1} \right]^{\frac{1}{2}}$
 &$-\sqrt{\frac{2}{3}} \left[\displaystyle\frac{j_K+1}{2j_K+1} \right]^{\frac{1}{2}}$
 \\
 \hline
 $J=1$&$\tilde{J}=\frac{1}{2}$ & $M=0$ & $A=\pm\frac{1}{2}$ & $g_{N-1}=j_K$
 &$-2A(-1)^{2j_K}\frac{1}{\sqrt{3}} \left[\displaystyle\frac{j_K}{2j_K+1} \right]^{\frac{1}{2}}$
 &$-2A(-1)^{2j_K}\frac{1}{\sqrt{3}} \left[\displaystyle\frac{j_K+1}{2j_K+1} \right]^{\frac{1}{2}}$
 \\
 \hline
 $J=1$&$\tilde{J}=\frac{1}{2}$ & $M=1$ & $A=-\frac{1}{2}$ & $g_{N-1}=j_K$
 &$(-1)^{2j_K}\sqrt{\frac{2}{3}} \left[\displaystyle\frac{j_K}{2j_K+1} \right]^{\frac{1}{2}}$
 &$(-1)^{2j_K}\sqrt{\frac{2}{3}} \left[\displaystyle\frac{j_K+1}{2j_K+1} \right]^{\frac{1}{2}}$
 \\
 \hline
 
\end{tabular}\\[1cm]

%%%%%%%%%%%%%%%%%%%%%%%%%%%%%%%%%%%%%%%%%%%%%%%%%%%%%%%%%%%%%%%%%%%%%%%%%%%%%%%%%%%%%%%%%%%%%%%%%%%%%%%
\subsubsection{The Matrix Elements $\big<\cdot\big|\hat{V}^r\big|\cdot\big> $}
%%%%%%%%%%%%%%%%%%%%%%%%%%%%%%%%%%%%%%%%%%%%%%%%%%%%%%%%%%%%%%%%%%%%%%%%%%%%%%%%%%%%%%%%%%%%%%%%%%%%%%%
As one can easily see, for $N=3, J'=1, J=0$ the only allowed intermediate total angular momentum in the matrix elements $\big<\cdot\big|\hat{V}^r\big|\cdot\big> $ of (\ref{Endresult e-Op I}) is given by $\tilde{J}=\frac{1}{2}$. Therefore we have to evaluate expressions of the general form:

\ba
     \big<\,a_2'(j_1\,j_2)~\tilde{J}(a_2'\,j_3)=\frac{1}{2}~M'\,
            \big|\,\hat{V}^r\,\big|\,
	    a_2(j_1\,j_2)~\tilde{J}(a_2'\,j_3)=\frac{1}{2}~M\,\big>
\ea  
Note that we have to perform an eigenvector expansion according to (\ref{Eigenstate expansion q+q}).
In our special case $\hat{V}$ takes the form: 

\ba
   \hat{V}&=&(\ell_P)^3\sqrt{\big| \tilde{Z} \cdot \big[(J_{12})^2,(J_{23})^2\big]\big|}  ~~~~~~~~~\mbox{with}~J_{12}=J_1+J_2~\mbox{and}~J_{23}=J_2+J_3
   \nonumber
   \\
   &=&(\ell_P)^3\big|\tilde{Z} \big|^{\frac{1}{2}}\cdot\sqrt{\big| \hat{q}_{123} \big|}
\ea

Here $\ell_P$ denotes the Planck length, $\tilde{Z}$ is a constant prefactor dependend on the regularization, on the Immirzi-parameter $\beta$ and on the relative orientation of the edges $e_1,e_2,e_3$, see \cite{TT:Closed ME of V in LQG} for details\footnote{There $\tilde{Z}$ takes the value $\tilde{Z}=\frac{\mb{i}}{4}\cdot\beta^3\cdot\Big(\frac{3}{4}\Big)^3$, $\beta$ being the Immirzi parameter.  }. The matrix elements of $\hat{q}_{123}=\big[(J_{12})^2,(J_{23})^2\big]$ have the general form at the N-vertex (see \cite{Volume_Article_I} for details\footnote{a similar expression was also derived in \cite{de Pietri} using a different method.}):
   
   \be\label{ME of V}\begin{array}{llll}
      \multicolumn{2}{l}{<a_2~|~\hat{q}_{123}~|~a_2-1>=-<a_2-1~|~\hat{q}_{123}~|~a_2>=}~~~~~\\
      =\frac{\mb{i}}{\sqrt{(2a_2-1)(2a_2+1)}}
                          &\big[(j_1+j_2+a_2+1)(-j_1+j_2+a_2)(j_1-j_2+a_2)(j_1+j_2-a_2+1)\\ 
                          &~(j_3+a_3+a_2+1)(-j_3+a_3+a_2)(j_3-a_3+a_2)(j_3+a_3-a_2+1)\big]^{\frac{1}{2}}\\  
    \end{array}\ee

For $N=3$ we have to set $a_3 = a_N = \tilde{J}=\frac{1}{2}$. Now for $\tilde{J}=\frac{1}{2}$ we have
$a_2, a_2' = j_3\pm\frac{1}{2}$ hence the Hilbert space $\hat{V}$ is represented on is twodimensional and $\hat{q}_{123}$ is of the form:
   
\be
    \hat{q}_{123}=   \begin{array}{cc}
                  \left(\begin{array}{cc}
	             0&-\mb{i}A_1 \\
	             \mb{i} A_1&0
		  \end{array} \right)
	          \begin{array}{l}
	            \leftarrow a_2=j_3-\frac{1}{2} \\
		    \leftarrow a_2=j_3+\frac{1}{2}
	          \end{array}	
                \end{array}
\ee   
   
Here $A_1$ is (\ref{ME of V}) evaluated at $a_3=\tilde{J}=\frac{1}{2}$ and $a_2=j_3+\frac{1}{2}$:

\be
   A_1=A_1(j_1,j_2,j_3,\tilde{J},a_2)=:MEV2D[j_1,j_2,j_3,\tilde{J},a_2]=(-\mb{i})\cdot<j_3+\frac{1}{2}~|~\hat{q}_{123}~|~j_3-\frac{1}{2}>
\ee  

So we can immediately see, that 

\be 
   \hat{q}^\dagger\hat{q}=   
                \left(\begin{array}{cc}
	             |A_1|^2&0 \\
	             0&|A_1|^2
	        \end{array} \right)
\ee	          
is already diagonal with eigenvalue $|A_1|^2$ and eigenvectors $\left(1\atop 0 \right)$ and 
$\left( 0 \atop 1 \right)$. So	  

\fcmt{17}{\ba\label{explicit V-ME 3-vertex}
     \big<\,a_2'(j_1\,j_2)~\tilde{J}(a_2'\,j_3)\!=\!\frac{1}{2}~M'\,
            \big|\,\hat{V}^r\,\big|\,
	    a_2(j_1\,j_2)~\tilde{J}(a_2'\,j_3)\!=\!\frac{1}{2}~M\,\big>
     &~~=~~& (\ell_P)^{3r}\big|\tilde{Z} \big|^{\frac{r}{2}}\cdot\big|A_1\big|^{\frac{r}{2}}~\delta_{a_2'a_2}
\ea } 
~
\\\\
In the following calculations we will frequantly use (\ref{explicit V-ME 3-vertex}). Keep in mind that 
(\ref{explicit V-ME 3-vertex}) is derived for general $j_1,j_2,j_3$, which will be modified due to the action of holonomies in the $e$-Operators! In order to avoid confusion we will always write down all the arguments of $A_1=MEV2D[j_1,j_2,j_3,\tilde{J},a_2]$

%%%%%%%%%%%%%%%%%%%%%%%%%%%%%%%%%%%%%%%%%%%%%%%%%%%%%%%%%%%%%%%%%%%%%%%%%%%%%%%%%%%%%%%%%%%%%%%%%%%%%%%
\subsection{Calculation of the different cases}
%%%%%%%%%%%%%%%%%%%%%%%%%%%%%%%%%%%%%%%%%%%%%%%%%%%%%%%%%%%%%%%%%%%%%%%%%%%%%%%%%%%%%%%%%%%%%%%%%%%%%%%

\subsubsection{K=1}
Due to the special case we start with (\ref{Endresult e-Op K=1}). We always have $\tilde{J}=\frac{1}{2}$:
\[\begin{array}{lc@{~}|clllll}

  \lefteqn{\big<\,a_2'(j_1\,j_2)~J'(a_2'\,j_3)\!=\!1~M'\,
           \big|\,\lo{(r)}e^k_{K=1}\,\big|\,
	   a_2(j_1\,j_2)~J(a_2\,j_3)\!=\!0~M\!=\!0\,\big>=}
	   \\
	   \\
	   &\multicolumn{2}{c}{=}&
	   -\displaystyle\sum_{AC} \big[\overline{\tau}_k\big]_{CA}
	   ~\delta_{M'M\!+\!C\!-\!A}
	   ~\sum_{\tilde{j}_1=j_1\pm\frac{1}{2}} \sum_{\tilde{a}_2' \tilde{a}_2}
	   &\displaystyle\sum_{g_2'(j_2\,j_3)} C^{J'\!=\!1~j_1}_{\tilde{J}\!=\!\frac{1}{2}\,~\tilde{j}_1}(A,M',g_2')
	   &\big<~\tilde{a}_2'(\tilde{j}_1\,j_2)~\tilde{J}(\tilde{a}_2'\,j_3)~\big|
	                  ~g_2'(j_2\,j_3)~\tilde{J}(g_2'\,\tilde{j}_1)~ \big>
	   \\[-3mm]
	   &&&&&\big<~{a}_2'({j}_1\,j_2)~{J'}({a}_2'\,j_3)~\big|
	                  ~g_2'(j_2\,j_3)~{J'}(g_2'\,{j}_1)~ \big>
		
	  \\[3mm]&&&&\times\displaystyle\sum_{g_2(j_2\,j_3)} C^{J\!=\!0~j_1}_{\tilde{J}\!=\!\frac{1}{2}\,~\tilde{j}_1}(C,M,g_2)
	   &\big<~\tilde{a}_2(\tilde{j}_1\,j_2)~\tilde{J}(\tilde{a}_2\,j_3)~\big|
	                  ~g_2(j_2\,j_3)~\tilde{J}(g_2\,\tilde{j}_1)~ \big>
	   \\[-3mm]
	   &&&&&\big<~{a}_2({j}_1\,j_2)~{J}({a}_2\,j_3)~\big|
	                  ~g_2(j_2\,j_3)~{J}(g_2\,{j}_1)~ \big>

	   \\[5mm]
	   &&&&\multicolumn{2}{l}{\times 
	    \big<\,\tilde{a}_2'(\tilde{j}_1\,j_2)~\tilde{J}(\tilde{a}_2'\,j_3)\!=\!\frac{1}{2}~M'\!+\!A~
	    \big|~\hat{V}^r~\big|\,
	    \tilde{a}_2(\tilde{j}_1\,j_2)~\tilde{J}(\tilde{a}_2\,j_3)\!=\!\frac{1}{2}~M\!+\!C~
	    \big>
            }

	   \\[10mm]
	   
	   &\multicolumn{2}{c}{=}&
	   -\displaystyle\sum_{AC} \big[\overline{\tau}_k\big]_{CA}
	   ~\delta_{M'M\!+\!C\!-\!A}
	   ~\sum_{\tilde{j}_1=j_1\pm\frac{1}{2}} \sum_{\tilde{a}_2' \tilde{a}_2}
	   
	   \\&&\\
	   &&
	   &\multicolumn{3}{l}{\times \displaystyle\sum_{g_2'={\tiny\left\{\begin{array}{l}
	                                      j_1\!-\!1 \\ j_1 \\j_1\!+\!1 
					   \end{array}\right.}} C^{J'\!=\!1~j_1}_{\tilde{J}\!=\!\frac{1}{2}\,~\tilde{j}_1}(A,M',g_2')
	   \sqrt{(2\tilde{a}_2'+1)(2g_2'+1)}~ (-1)^{\tilde{j}_1+g_2'-\tilde{J}}
	                                       (-1)^{\tilde{j}_1+j_2+j_3+\tilde{J}}
					       \left\{\begin{array}{ccc}
					         \tilde{j}_1&j_2&\tilde{a}_2'\\
						 j_3 & \tilde{J} &g_2'
					       \end{array}\right\}}
					       
	   \\[-3mm]
	   &&&\multicolumn{3}{l}{~\hspace{4cm}~\times 
	   \sqrt{(2{a}_2'+1)(2g_2'+1)}~ (-1)^{{j}_1+g_2'-{J'}}
	                                       (-1)^{{j}_1+j_2+j_3+{J'}}
					       \left\{\begin{array}{ccc}
					         {j}_1&j_2&{a}_2'\\
						 j_3 & {J'} &g_2'
					       \end{array}\right\}}
           %%%%%%%%%%%%%%%%%%%%%%%%%%%%%%%%%%%%%%%%%%%%%%%%%%%%%%%%%%%%%%%%%%%%%%%%%%
	   \\[10mm]
	   
	   &&
	   &\multicolumn{3}{l}{\times ~~~~\displaystyle\sum_{g_2=j_1} C^{J\!=\!0~j_1}_{\tilde{J}\!=\!\frac{1}{2}\,~\tilde{j}_1}(C,M,g_2)
	   \sqrt{(2\tilde{a}_2+1)(2g_2+1)}~ (-1)^{\tilde{j}_1+g_2-\tilde{J}}
	                                       (-1)^{\tilde{j}_1+j_2+j_3+\tilde{J}}
					       \left\{\begin{array}{ccc}
					         \tilde{j}_1&j_2&\tilde{a}_2\\
						 j_3 & \tilde{J} &g_2
					       \end{array}\right\}}
					       
	   \\[3mm]
	   &&&\multicolumn{3}{l}{~\hspace{3.7cm}~\times 
	   \sqrt{(2{a}_2+1)(2g_2+1)}~ (-1)^{{j}_1+g_2-{J}}
	                                       (-1)^{{j}_1+j_2+j_3+{J}}
					       \left\{\begin{array}{ccc}
					         {j}_1&j_2&{a}_2\\
						 j_3 & {J} &g_2
					       \end{array}\right\}}

	   \\[10mm]				       
					       
	   &&&&\multicolumn{2}{l}{\times 
	    \big<\,\tilde{a}_2'(\tilde{j}_1\,j_2)~\tilde{J}(\tilde{a}_2'\,j_3)\!=\!\frac{1}{2}~M'\!+\!A~
	    \big|~\hat{V}^r~\big|\,
	    \tilde{a}_2(\tilde{j}_1\,j_2)~\tilde{J}(\tilde{a}_2\,j_3)\!=\!\frac{1}{2}~M\!+\!C~
	    \big>
            }
 
	 \\[3mm]
	 &&\multicolumn{4}{l}{
	   \fcmt{15}{Note that 
	   \begin{itemize}
	      \item[(i)]{due to integer arguments the $(-1)$-prefactors are equal to 1}
	      
	      \item[(ii)]{with (\ref{explicit V-ME 3-vertex}) we have \fbox{$\tilde{a}_2'\stackrel{!}{=}\tilde{a}_2=j_3\pm\frac{1}{2}$} and have $MEV2D[\tilde{j}_1,j_2,j_3,\tilde{J}=\frac{1}{2},a_2=j_3+\frac{1}{2}]$, \underline{independent} of $\tilde{a}_2$
	      \subitem{$\rightarrow$ We can use the orthogonality relations of the $6j$-symbols (see \cite{Edmonds}, p.96) in order to carry out the remaining sum over $\tilde{a}_2$:
	      \[
	        \displaystyle\sum_{\tilde{a}_2}(2\tilde{a}_2+1) 
		        \left\{\begin{array}{ccc}
			  \tilde{j}_1&j_2&\tilde{a}_2\\
			  j_3 & \tilde{J} & g_2'
			\end{array}\right\}
			\left\{\begin{array}{ccc}
			  \tilde{j}_1&j_2&\tilde{a}_2\\
			  j_3 & \tilde{J} & g_2'
			\end{array}\right\}
	         = \frac{1}{(2g_2+1)}~\delta_{g_2'g_2} ~~\leadsto\fbox{$g_2'\stackrel{!}{=}g_2=j_1$}
	      \]}}
	      
	      \item[(iii)]{For $J=0$ we have
	        \subitem{a)~~$a_2=j_3$}
		
		\subitem{b)~~\[ \left\{\begin{array}{ccc}
		                   j_1&j_2&a_2\\
				   j_3&J&g_2
				\end{array} \right\}
				=
				\left\{\begin{array}{ccc}
		                   j_1&j_2&j_3\\
				   j_3&0&j_1
				\end{array} \right\}
				=
				\left\{\begin{array}{ccc}
		                   j_2&j_1&j_3\\
				   0&j_3&j_1
				\end{array} \right\}
				=\frac{(-1)^{j_1+j_2+j_3}}{\sqrt{(2j_1+1)(2j_3+1)}}
		             \]}}		    
           \end{itemize} }}
	\\&&    
	    
\end{array}\]

\[\begin{array}{lc@{~}|clllll}
     &&\\
     &\multicolumn{2}{c}{=}&
     -\displaystyle\sum_{A~C}~\big[\overline{\tau}_k\big]_{CA}~\delta_{M'\,M\!+\!C\!-\!A}
     (-1)^{j_1+j_2+j_3}~\sqrt{(2j_1+1)(2a_2'+1)} ~
     \left\{\begin{array}{ccc}
        j_2&j_1&a_2'\\
	1&j_3&j_1
     \end{array}\right\}
     ~\hspace{5cm}~ 
     
     \\&&
     \\
     &&&\times
     
     \displaystyle\sum_{\tilde{j}_1=j_1\pm\frac{1}{2}}
     
     C^{J'=1~j_1}_{\tilde{J}=\frac{1}{2}~\,\tilde{j}_1}(A,M',j_1)~
     C^{J=0~j_1}_{\tilde{J}=\frac{1}{2}~\,\tilde{j}_1}(C,0,j_1)
     ~\times~(\ell_P)^{3r}\big|\tilde{Z} \big|^{\frac{r}{2}}\cdot MEV2D\Big[\tilde{j}_1,j_2,j_3,\tilde{J}=\frac{1}{2},j_3+\frac{1}{2}\Big]^{\frac{r}{2}}
     
     \\[8mm]
     &&&\fcmt{10}{Now $C^{J=0~j_1}_{\tilde{J}=\frac{1}{2}~\,\tilde{j}_1}(C,0,j_1)
                      =\left\{\begin{array}{rcl}
		          (-1)^{2j_1+1} \big[\frac{j_1}{2j_1+1}\big]^{\frac{1}{2}}&~~~~~&\tilde{j}_1=j_1-\frac{1}{2}\\
			  
			  (-1)^{2j_1} \big[\frac{j_1+1}{2j_1+1}\big]^{\frac{1}{2}}&~~~~~&\tilde{j}_1=j_1+\frac{1}{2}\\
			  
		       \end{array} \right.$
		       \\
		       independent of $C$ and $(-1)^{3j_1}=(-1)^{-j_1}$}
     \\[18mm]
     &\multicolumn{2}{c}{=}&
     -(\ell_P)^{3r}\big|\tilde{Z} \big|^{\frac{r}{2}}\cdot\displaystyle\sum_{A~C}~\big[\overline{\tau}_k\big]_{CA}~\delta_{M'\,M\!+\!C\!-\!A}
     (-1)^{-j_1+j_2+j_3}~\sqrt{(2a_2'+1)} ~
     \left\{\begin{array}{ccc}
        j_2&j_1&a_2'\\
	1&j_3&j_1
     \end{array}\right\}
     
     \\\\
     
     &\multicolumn{2}{c}{}&\times 
     \Bigg\{-\sqrt{j_1}~~C^{J'=1~j_1}_{\tilde{J}=\frac{1}{2}~\,{j}_1-\frac{1}{2}}(A,M',j_1)
           ~~MEV2D\Big[j_1-\frac{1}{2},j_2,j_3,\frac{1}{2},j_3+\frac{1}{2}\Big]^{\frac{r}{2}} 
     \\\\
     
     &\multicolumn{2}{c}{}& 
     ~~~~+\sqrt{j_1+1}~~C^{J'=1~j_1}_{\tilde{J}=\frac{1}{2}~\,{j}_1+\frac{1}{2}}(A,M',j_1)
           ~~MEV2D\Big[j_1+\frac{1}{2},j_2,j_3,\frac{1}{2},j_3+\frac{1}{2}\Big]^{\frac{r}{2}}~\Bigg\} 

\end{array}\]

\be\label{Gauge inv. 3-vertex: e_{K=1} }\ee

%%%%%%%%%%%%%%%%%%%%%%%%%%%%%%%%%%%%%%%%%%%%%%%%%%%%%%%%%%%%%%%%%%%%%%%%%%%%%%%%%%%%%%%%%%%%%%%%%%%%%%%
\pagebreak
\subsubsection{K=2}
We always have $\tilde{J}=\frac{1}{2}$:
\[\begin{array}{lc@{~}|clllll}

  \lefteqn{\big<\,a_2'(j_1\,j_2)~J'(a_2'\,j_3)\!=\!1~M'\,
           \big|\,\lo{(r)}e^k_{K=2}\,\big|\,
	   a_2(j_1\,j_2)~J(a_2\,j_3)\!=\!0~M\!=\!0\,\big>=}
	   \\
	   \\
	   &\multicolumn{2}{c}{=}&
	   -\displaystyle\sum_{AC} \big[\overline{\tau}_k\big]_{CA}
	   ~\delta_{M'M\!+\!C\!-\!A}
	   ~\sum_{\tilde{j}_2=j_2\pm\frac{1}{2}} \sum_{\tilde{a}_2' \tilde{a}_2}
	   &\displaystyle\sum_{g_2'(j_1\,j_3)} C^{J'\!=\!1~j_2}_{\tilde{J}\!=\!\frac{1}{2}\,~\tilde{j}_2}(A,M',g_2')
	   &\big<~\tilde{a}_2'({j}_1\,\tilde{j}_2)~\tilde{J}(\tilde{a}_2'\,j_3)~\big|
	                  ~g_2'(j_1\,j_3)~\tilde{J}(g_2'\,\tilde{j}_2)~ \big>
	   \\[-3mm]
	   &&&&&\big<~{a}_2'({j}_1\,j_2)~{J'}({a}_2'\,j_3)~\big|
	                  ~g_2'(j_1\,j_3)~{J'}(g_2'\,{j}_2)~ \big>
		
	  \\[3mm]&&&&\times\displaystyle\sum_{g_2(j_1\,j_3)} C^{J\!=\!0~j_2}_{\tilde{J}\!=\!\frac{1}{2}\,~\tilde{j}_2}(C,M,g_2)
	   &\big<~\tilde{a}_2({j}_1\,\tilde{j}_2)~\tilde{J}(\tilde{a}_2\,j_3)~\big|
	                  ~g_2(j_1\,j_3)~\tilde{J}(g_2\,\tilde{j}_2)~ \big>
	   \\[-3mm]
	   &&&&&\big<~{a}_2({j}_1\,j_2)~{J}({a}_2\,j_3)~\big|
	                  ~g_2(j_1\,j_3)~{J}(g_2\,{j}_2)~ \big>

	   \\[5mm]
	   &&&&\multicolumn{2}{l}{\times 
	    \big<\,\tilde{a}_2'({j}_1\,\tilde{j}_2)~\tilde{J}(\tilde{a}_2'\,j_3)\!=\!\frac{1}{2}~M'\!+\!A~
	    \big|~\hat{V}^r~\big|\,
	    \tilde{a}_2({j}_1\,\tilde{j}_2)~\tilde{J}(\tilde{a}_2\,j_3)\!=\!\frac{1}{2}~M\!+\!C~
	    \big>
            }

	   \\[10mm]
	   
	   &\multicolumn{2}{c}{=}&
	   -\displaystyle\sum_{AC} \big[\overline{\tau}_k\big]_{CA}
	   ~\delta_{M'M\!+\!C\!-\!A}
	   ~\sum_{\tilde{j}_2=j_2\pm\frac{1}{2}} \sum_{\tilde{a}_2' \tilde{a}_2}
	   
	   \\&&\\
	   &&
	   &\multicolumn{3}{l}{\times \displaystyle\sum_{g_2'={\tiny\left\{\begin{array}{l}
	                                      j_2\!-\!1 \\ j_2 \\j_2\!+\!1 
					   \end{array}\right.}} C^{J'\!=\!1~j_2}_{\tilde{J}\!=\!\frac{1}{2}\,~\tilde{j}_2}(A,M',g_2')
	   \sqrt{(2\tilde{a}_2'+1)(2g_2'+1)}~ 
	   (-1)^{j_1+\tilde{j}_2-\tilde{a}_2'}
	   (-1)^{\tilde{j}_2+g_2'-\tilde{J}}
	   (-1)^{{j}_1+\tilde{j}_2+j_3+\tilde{J}}
					       \left\{\begin{array}{ccc}
					         \tilde{j}_2&j_1&\tilde{a}_2'\\
						 j_3 & \tilde{J} &g_2'
					       \end{array}\right\}}
					       
	   \\[-3mm]
	   &&&\multicolumn{3}{l}{~\hspace{4cm}~\times 
	   \sqrt{(2{a}_2'+1)(2g_2'+1)}~ 
	   (-1)^{j_1+j_2-a_2'}
	   (-1)^{{j}_2+g_2'-{J'}}
	   (-1)^{{j}_1+j_2+j_3+{J'}}
					       \left\{\begin{array}{ccc}
					         {j}_2&j_1&{a}_2'\\
						 j_3 & {J'} &g_2'
					       \end{array}\right\}}
           %%%%%%%%%%%%%%%%%%%%%%%%%%%%%%%%%%%%%%%%%%%%%%%%%%%%%%%%%%%%%%%%%%%%%%%%%%
	   \\[10mm]
	   
	   &&
	   &\multicolumn{3}{l}{\times ~~~~\displaystyle\sum_{g_2=j_2} C^{J\!=\!0~j_2}_{\tilde{J}\!=\!\frac{1}{2}\,~\tilde{j}_2}(C,M,g_2)
	   \sqrt{(2\tilde{a}_2+1)(2g_2+1)}~ 
	   (-1)^{j_1+\tilde{j}_2-\tilde{a}_2}
	   (-1)^{\tilde{j}_2+g_2-\tilde{J}}
	   (-1)^{{j}_1+\tilde{j}_2+j_3+\tilde{J}}
					       \left\{\begin{array}{ccc}
					         \tilde{j}_2&j_1&\tilde{a}_2\\
						 j_3 & \tilde{J} &g_2
					       \end{array}\right\}}
					       
	   \\[3mm]
	   &&&\multicolumn{3}{l}{~\hspace{3.7cm}~\times 
	   \sqrt{(2{a}_2+1)(2g_2+1)}~ 
	   (-1)^{j_1+j_2-a_2}
	   (-1)^{{j}_2+g_2-{J}}
	   (-1)^{{j}_1+j_2+j_3+{J}}
					       \left\{\begin{array}{ccc}
					         {j}_2&j_1&{a}_2\\
						 j_3 & {J} &g_2
					       \end{array}\right\}}

	   \\[10mm]				       
					       
	   &&&&\multicolumn{2}{l}{\times 
	    \big<\,\tilde{a}_2'({j}_1\,\tilde{j}_2)~\tilde{J}(\tilde{a}_2'\,j_3)\!=\!\frac{1}{2}~M'\!+\!A~
	    \big|~\hat{V}^r~\big|\,
	    \tilde{a}_2({j}_1\,\tilde{j}_2)~\tilde{J}(\tilde{a}_2\,j_3)\!=\!\frac{1}{2}~M\!+\!C~
	    \big>
            }
 
	 \\[3mm]
	 &&\multicolumn{4}{l}{
	   \fcmt{15}{Note that 
	   \begin{itemize}
	      \item[(i)]{due to integer arguments the $(-1)$-prefactors are equal to 
	                 $(-1)^{\tilde{a}_2'-\tilde{a}_2}(-1)^{a_2'-a_2}$}
	      
	      \item[(ii)]{with (\ref{explicit V-ME 3-vertex}) we have \fbox{$\tilde{a}_2'\stackrel{!}{=}\tilde{a}_2=j_3\pm\frac{1}{2}$} and have $MEV2D[{j}_1,\tilde{j}_2,j_3,\tilde{J}=\frac{1}{2},a_2=j_3+\frac{1}{2}]$, \underline{independent} of $\tilde{a}_2$
	      \subitem{$\rightarrow$ We can use the orthogonality relations of the $6j$-symbols (see \cite{Edmonds}, p.96) in order to carry out the remaining sum over $\tilde{a}_2$:
	      \[
	        \displaystyle\sum_{\tilde{a}_2}(2\tilde{a}_2+1) 
		        \left\{\begin{array}{ccc}
			  \tilde{j}_2&j_1&\tilde{a}_2\\
			  j_3 & \tilde{J} & g_2'
			\end{array}\right\}
			\left\{\begin{array}{ccc}
			  \tilde{j}_2&j_1&\tilde{a}_2\\
			  j_3 & \tilde{J} & g_2'
			\end{array}\right\}
	         = \frac{1}{(2g_2+1)}~\delta_{g_2'g_2} ~~\leadsto\fbox{$g_2'\stackrel{!}{=}g_2=j_2$}
	      \]}}
	      
	      \item[(iii)]{For $J=0$ we have
	        \subitem{a)~~$a_2=j_3$}
		
		\subitem{b)~~\[ \left\{\begin{array}{ccc}
		                   j_2&j_1&a_2\\
				   j_3&J&g_2
				\end{array} \right\}
				=
				\left\{\begin{array}{ccc}
		                   j_2&j_1&j_3\\
				   j_3&0&j_2
				\end{array} \right\}
				=
				\left\{\begin{array}{ccc}
		                   j_1&j_2&j_3\\
				   0&j_3&j_2
				\end{array} \right\}
				=\frac{(-1)^{j_1+j_2+j_3}}{\sqrt{(2j_2+1)(2j_3+1)}}
		             \]}}		    
           \end{itemize} }}
	\\&&    
	    
\end{array}\]

\[\begin{array}{lc@{~}|clllll}
     &&\\
     &\multicolumn{2}{c}{=}&
     -\displaystyle\sum_{A~C}~\big[\overline{\tau}_k\big]_{CA}~\delta_{M'\,M\!+\!C\!-\!A}
     (-1)^{j_1+j_2+j_3}~(-1)^{a_2'-j_3}~\sqrt{(2j_2+1)(2a_2'+1)} ~
     \left\{\begin{array}{ccc}
        j_1&j_2&a_2'\\
	1&j_3&j_2
     \end{array}\right\}
     ~\hspace{5cm}~ 
     
     \\&&
     \\
     &&&\times
     
     \displaystyle\sum_{\tilde{j}_2=j_2\pm\frac{1}{2}}
     
     C^{J'=1~j_2}_{\tilde{J}=\frac{1}{2}~\,\tilde{j}_2}(A,M',j_2)~
     C^{J=0~j_2}_{\tilde{J}=\frac{1}{2}~\,\tilde{j}_2}(C,0,j_2)
     ~\times~(\ell_P)^{3r}\big|\tilde{Z} \big|^{\frac{r}{2}}\cdot MEV2D\Big[{j}_1,\tilde{j}_2,j_3,\tilde{J}=\frac{1}{2},j_3+\frac{1}{2}\Big]^{\frac{r}{2}}
     
     \\[8mm]
     &&&\fcmt{10}{Now $C^{J=0~j_2}_{\tilde{J}=\frac{1}{2}~\,\tilde{j}_1}(C,0,j_2)
                      =\left\{\begin{array}{rcl}
		          (-1)^{2j_2+1} \big[\frac{j_2}{2j_2+1}\big]^{\frac{1}{2}}&~~~~~&\tilde{j}_2=j_2-\frac{1}{2}\\
			  
			  (-1)^{2j_2} \big[\frac{j_2+1}{2j_2+1}\big]^{\frac{1}{2}}&~~~~~&\tilde{j}_2=j_2+\frac{1}{2}\\
			  
		       \end{array} \right.$
		       \\
		       independent from $C$ and $(-1)^{3j_2}=(-1)^{-j_2}$}
     \\[18mm]
     &\multicolumn{2}{c}{=}&
     -(\ell_P)^{3r}\big|\tilde{Z} \big|^{\frac{r}{2}}\cdot \displaystyle\sum_{A~C}~\big[\overline{\tau}_k\big]_{CA}~\delta_{M'\,M\!+\!C\!-\!A}
     (-1)^{j_1-j_2+a_2'}~\sqrt{(2a_2'+1)} ~
     \left\{\begin{array}{ccc}
        j_1&j_2&a_2'\\
	1&j_3&j_2
     \end{array}\right\}
     
     \\\\
     
     &\multicolumn{2}{c}{}&\times 
     \Bigg\{-\sqrt{j_2}~~C^{J'=1~j_2}_{\tilde{J}=\frac{1}{2}~\,{j}_2-\frac{1}{2}}(A,M',j_2)
           ~~MEV2D\Big[j_1,j_2-\frac{1}{2},j_3,\frac{1}{2},j_3+\frac{1}{2}\Big]^{\frac{r}{2}} 
     \\\\
     
     &\multicolumn{2}{c}{}& 
     ~~~~+\sqrt{j_2+1}~~C^{J'=1~j_2}_{\tilde{J}=\frac{1}{2}~\,{j}_2+\frac{1}{2}}(A,M',j_2)
           ~~MEV2D\Big[j_1,j_2+\frac{1}{2},j_3,\frac{1}{2},j_3+\frac{1}{2}\Big]^{\frac{r}{2}}~\Bigg\} 

\end{array}\]

\be\label{Gauge inv. 3-vertex: e_{K=2} }\ee

 \pagebreak
%%%%%%%%%%%%%%%%%%%%%%%%%%%%%%%%%%%%%%%%%%%%%%%%%%%%%%%%%%%%%%%%%%%%%%%%%%%%%%%%%%%%%%%%%%%%%%%%%%%%%
\subsubsection{K=3}%
%%%%%%%%%%%%%%%%%%%%
In this special case (since $N=3$ we have to consider the general expression for $K=N$) we start from (\ref{Endresult e-Op K=N}). Again $\tilde{J}=\frac{1}{2}$: 

\[\begin{array}{lc@{~}|clllll}

  \lefteqn{\big<\,a_2'(j_1\,j_2)~J'(a_2'\,j_3)\!=\!1~M'\,
           \big|\,\lo{(r)}e^k_{K=3}\,\big|\,
	   a_2(j_1\,j_2)~J(a_2\,j_3)\!=\!0~M\!=\!0\,\big>=}
	   \\
	   \\
	   &\multicolumn{2}{c}{=}&
	   -\displaystyle\sum_{AC} \big[\overline{\tau}_k\big]_{CA}
	   ~\delta_{M'M\!+\!C\!-\!A}
	   ~\sum_{\tilde{j}_3=j_3\pm\frac{1}{2}} 
	   &
	    C^{J'\!=\!1~j_3}_{\tilde{J}\!=\!\frac{1}{2}\,~\tilde{j}_3}(A,M',a_2')
	   ~ C^{J\!=\!0~j_3}_{\tilde{J}\!=\!\frac{1}{2}\,~\tilde{j}_3}(C,M,a_2)
	   &\big<~\tilde{a}_2({j}_1\,\tilde{j}_2)~\tilde{J}(\tilde{a}_2\,j_3)~\big|
	                  ~g_2(j_1\,j_3)~\tilde{J}(g_2\,\tilde{j}_2)~ \big>
	   \\[5mm]
	   &&&&\multicolumn{2}{l}{\times 
	    \big<\,{a}_2'({j}_1\,{j}_2)~\tilde{J}({a}_2'\,\tilde{j}_3)\!=\!\frac{1}{2}~M'\!+\!A~
	    \big|~\hat{V}^r~\big|\,
	    {a}_2({j}_1\,{j}_2)~\tilde{J}({a}_2\,\tilde{j}_3)\!=\!\frac{1}{2}~M\!+\!C~
	    \big>
            }
        \\&&
	\\
	&&\multicolumn{4}{l}{\fcmt{17}{Note that
	  \begin{itemize}
	     \item[(i)]{with (\ref{explicit V-ME 3-vertex}) we have \fbox{${a}_2'\stackrel{!}{=}{a}_2=j_3$}}
	     
	     \item[(ii)]{the Hilbertspace structure is as follows: Because we must have $|a_2-\tilde{j}_3|=\frac{1}{2}$ the matrix element of the volume operator has to be taken according to $\tilde{j}_3$:
	                 
			 \[\begin{array}{lclllll}
			      \tilde{j}_3=j_3-\frac{1}{2}&~~\rightarrow~~& 
			           a_2=\left\{\begin{array}{l}
			                        j_3-1\\j_3
					      \end{array}\right.
			      &~~~~&\hat{q}=\left(\begin{array}{cc}
			                        0&-\mb{i}A_1\\
						\mb{i}A_1&0
					    \end{array} \right)
			      &~~~~&
			      A_1=A_1\Big[j_1,j_2,j_3-\frac{1}{2},\tilde{J}=\frac{1}{2},a_2=j_3\Big]
			      \\\\
			      \tilde{j}_3=j_3+\frac{1}{2}&~~\rightarrow~~& 
			           a_2=\left\{\begin{array}{l}
			                        j_3\\j_3+1
					      \end{array}\right. 
			      &~~~~&\hat{q}=\left(\begin{array}{cc}
			                        0&-\mb{i}A_1\\
						\mb{i}A_1&0
					    \end{array} \right)
			      &~~~~&
			      A_1=A_1\Big[j_1,j_2,j_3+\frac{1}{2},\tilde{J}=\frac{1}{2},a_2=j_3+1\Big]  
			  \end{array}\]
			 }
	     
	     \item[(iii)]{$C^{J=0~j_3}_{\tilde{J}=\frac{1}{2}~\,\tilde{j}_3}(C,0,j_3)
                      =\left\{\begin{array}{rcl}
		          (-1)^{2j_3+1} \big[\frac{j_3}{2j_3+1}\big]^{\frac{1}{2}}&~~~~~&\tilde{j}_3=j_3-\frac{1}{2}\\
			  
			  (-1)^{2j_3} \big[\frac{j_3+1}{2j_3+1}\big]^{\frac{1}{2}}&~~~~~&\tilde{j}_3=j_3+\frac{1}{2}\\
			  
		       \end{array} \right.$}
	  \end{itemize} }}

   \\&&\\&&\\&&
   \\
    &\multicolumn{2}{l}{=}&\multicolumn{3}{l}{
      -(\ell_P)^{3r}\big|\tilde{Z} \big|^{\frac{r}{2}}\cdot
      \displaystyle\sum_{AC} \big[\overline{\tau}_k\big]_{CA}
      ~\delta_{a_2'a_2}~\delta_{M'M\!+\!C\!-\!A}
      ~\frac{(-1)^{2j_3}}{\sqrt{2j_3+1}}
      ~\Bigg\{ -\sqrt{j_3}~C^{J'\!=\!1~j_3}_{\tilde{J}\!=\!\frac{1}{2}\,~{j}_3-\frac{1}{2}}(A,M',j_3)
                          ~MEV2D\Big[j_1,j_2,j_3-\frac{1}{2},\frac{1}{2},j_3 \Big]^{\frac{r}{2}}
    }
    \\
    &\multicolumn{2}{l}{ }&\multicolumn{3}{l}{\hspace{6.1cm}
    +\sqrt{j_3+1}~C^{J'\!=\!1~j_3}_{\tilde{J}\!=\!\frac{1}{2}\,~{j}_3+\frac{1}{2}}(A,M',j_3)
                          ~MEV2D\Big[j_1,j_2,j_3+\frac{1}{2},\frac{1}{2},j_3+1 \Big]^{\frac{r}{2}}
    ~~ \Bigg\}}
    			  
\end{array}\] \be\label{Gauge inv. 3-vertex: e_{K=3} }\ee

\pagebreak

\subsubsection{The expressions $q_{IL}(\frac{1}{2})=\delta_{il}~~\lo{(\frac{1}{2})}e^i_I~~
                                                                 \lo{(\frac{1}{2})}e^l_L~~$}

Given the explicit expressions
(\ref{Gauge inv. 3-vertex: e_{K=1} }), (\ref{Gauge inv. 3-vertex: e_{K=2} }), 
(\ref{Gauge inv. 3-vertex: e_{K=3} }) we can now calculate the matrix elements of $q_{IL}$-operators as prescribed in 
(\ref{Definition of the q_{IL} at the 3-vertex}):

\ba
   <0\,|\,q_{IL}\big(\frac{1}{2}\big) \,|\,0> 
   =:Q_{IL}&=&
   \sum_{|\,1\,>}  <0\,|\, \lo{(\frac{1}{2})}e^i_I \,|\,1\,>\,<1\,|\, \lo{(\frac{1}{2})}e^l_L\,|\,0>~~\delta_{il} \nonumber\\
   &=&\delta_{il}
   \sum_{a_2'=\left\{ \tiny\begin{array}{l} \!\!j_3\!-\!1\\\!\!j_3\\\!\!j_3\!+\!1\end{array}\right.}
   \sum_{M'=\left\{ \tiny\begin{array}{r} \!\!-1\\\!\!0\\\!\!1\end{array}\right.}~~ \overline{\big<\,a_2'(j_1\,j_2)~J'(a_2'\,j_3)\!=\!1~M'\,
   \big|\,\lo{(\frac{1}{2})}e^i_I\,\big| 
   \,a_2(j_1\,j_2)\!=\!j_3~J(a_2\,j_3)\!=\!0~M\!=\!0\,\big>} \nonumber\\[-5mm]
   &&\hspace{3cm} \times\big<\,a_2'(j_1\,j_2)~J'(a_2\,j_3)\!=\!1~M'\,
   \big|\,\lo{(\frac{1}{2})}e^l_L\,\big| 
   \,a_2(j_1\,j_2)\!=\!j_3~J(a_2\,j_3)\!=\!0~M\!=\!0\,\big> \nonumber\\
\ea

We have done the calculation with MATHEMATICA. Here we have introduced the shorthands\footnote{During the calculation it turned out, that it is much faster if we use placeholders for the matrix elements of the volume operator due to the fact that MATHEMATICA tries to simplify expressions during evaluation.}

\begin{footnotesize}

\[\begin{array}{lcl}
    V_{1A}=MEV2D\big[\tilde{j}_1=j_1-\frac{1}{2},~j_2,~j_3,
                            ~\tilde{J}=\frac{1}{2},~j_3+\frac{1}{2} \big]
    &=&\big[(-j_1+j_2+j_3+1)(j_1-j_2+j_3)(j_1+j_2-j_3)(j_1+j_2+j_3+1) \big]^{\frac{1}{2}}
    
    \\
    V_{1B}=MEV2D\big[\tilde{j}_1=j_1+\frac{1}{2},~j_2,~j_3,
                            ~\tilde{J}=\frac{1}{2},~j_3+\frac{1}{2} \big]
    &=&\big[(-j_1+j_2+j_3)(j_1-j_2+j_3+1)(j_1+j_2-j_3+1)(j_1+j_2+j_3+2) \big]^{\frac{1}{2}}
			    
    \\
    
    V_{2A}=MEV2D\big[j_1,~\tilde{j}_2=j_2-\frac{1}{2},~j_3,
                            ~\tilde{J}=\frac{1}{2},~j_3+\frac{1}{2} \big]
    &=&\big[(-j_1+j_2+j_3)(j_1-j_2+j_3+1)(j_1+j_2-j_3)(j_1+j_2+j_3+1) \big]^{\frac{1}{2}}

    \\
   
   V_{2B}=MEV2D\big[j_1,~\tilde{j}_2=j_2+\frac{1}{2},~j_3,
                            ~\tilde{J}=\frac{1}{2},~j_3+\frac{1}{2} \big]
       &=&\big[(-j_1+j_2+j_3)(j_1-j_2+j_3)(j_1+j_2-j_3+1)(j_1+j_2+j_3+2) \big]^{\frac{1}{2}}

    \\
    
    V_{3A}=MEV2D\big[j_1,~j_2,~\tilde{j}_3=j_3-\frac{1}{2},
                            ~\tilde{J}=\frac{1}{2},~j_3\big]
        &=&\big[(-j_1+j_2+j_3)(j_1-j_2+j_3)(j_1+j_2-j_3+1)(j_1+j_2+j_3+1) \big]^{\frac{1}{2}}
			    
    \\			    
    V_{3B}=MEV2D\big[j_1,~j_2,~\tilde{j}_3=j_3+\frac{1}{2},
                            ~\tilde{J}=\frac{1}{2},~j_3+1 \big]
        &=&\big[(-j_1+j_2+j_3+1)(j_1-j_2+j_3+1)(j_1+j_2-j_3)(j_1+j_2+j_3+2) \big]^{\frac{1}{2}}

\end{array}\]
\end{footnotesize}

Moreover we use the abbreviation \fbox{$A_K=j_K(j_K+1)$}. Note that we now specify to \fbox{$r=\frac{1}{2}$}~ and use the abbreviation \fbox{$Q_{IL}:=<0\,|\,q_{IL}\big(\frac{1}{2}\big) \,|\,0>$}.
\\\\
The result is

\[\begin{array}{rcl}
   Q_{11}&=&(\ell_P)^{3}\big|\tilde{Z}\big|^\frac{1}{2}\cdot
   \frac{4\,{j_1}\,\left( 1 + {j_1} \right) \,{\left( {{{\Mvariable{V}}_{1A}}}^{\frac{1}{4}} - 
        {{{\Mvariable{V}}_{1B}}}^{\frac{1}{4}} \right) }^2}{{\left( 1 + 2\,{j_1} \right) }^2}
	
   \\
   \\
   Q_{12}=Q_{21}&=&(\ell_P)^{3}\big|\tilde{Z}\big|^\frac{1}{2}\cdot
    \frac{-2\,\left( 2 + {\left( -1 \right) }^{2\,\left( {j_1} + {j_2} \right) } \right) \,
    \left( A_1+A_2-A_3  \right) \,
    \left( {{{\Mvariable{V}}_{1A}}}^{\frac{1}{4}} - {{{\Mvariable{V}}_{1B}}}^{\frac{1}{4}} \right) \,
    \left( {{{\Mvariable{V}}_{2A}}}^{\frac{1}{4}} - {{{\Mvariable{V}}_{2B}}}^{\frac{1}{4}} \right) }{3\,
    \left( 1 + 2\,{j_1} \right) \,
    \left( 1 + 2\,{j_2} \right) }
    
    \\
    \\
    Q_{23}=Q_{32}&=&(\ell_P)^{3}\big|\tilde{Z}\big|^\frac{1}{2}\cdot
    \frac{-2\,\left( 2 + {\left( -1 \right) }^{2\,(j_2+{j_3})} \right) \,
    \left( -A_1+A_2+A_3 \right) \,
    \left( {{{\Mvariable{V}}_{2A}}}^{\frac{1}{4}} - {{{\Mvariable{V}}_{2B}}}^{\frac{1}{4}} \right) \,
    \left( {{{\Mvariable{V}}_{3A}}}^{\frac{1}{4}} - {{{\Mvariable{V}}_{3B}}}^{\frac{1}{4}} \right) }{3\,
    \,\left( 1 + 2\,{j_2} \right) \,\left( 1 + 2\,{j_3} \right) }
    \\
    \\
    Q_{13}=Q_{31}&=&(\ell_P)^{3}\big|\tilde{Z}\big|^\frac{1}{2}\cdot
    \frac{-2\,\left( 2 + {\left( -1 \right) }^{2\,(j_1+{j_3})} \right) \,
    \left( A_1-A_2+A_3 \right) \,
    \left( {{{\Mvariable{V}}_{1A}}}^{\frac{1}{4}} - {{{\Mvariable{V}}_{1B}}}^{\frac{1}{4}} \right) \,
    \left( {{{\Mvariable{V}}_{3A}}}^{\frac{1}{4}} - {{{\Mvariable{V}}_{3B}}}^{\frac{1}{4}} \right) }{3\,
    \,\left( 1 + 2\,{j_1} \right) \,\left( 1 + 2\,{j_3} \right) }
    
    \\
    \\
    Q_{22}&=&(\ell_P)^{3}\big|\tilde{Z}\big|^\frac{1}{2}\cdot
    \frac{4\,A_1 \,{\left( {{{\Mvariable{V}}_{2A}}}^{\frac{1}{4}} - 
        {{{\Mvariable{V}}_{2B}}}^{\frac{1}{4}} \right) }^2}{{\left( 1 + 2\,{j_2} \right) }^2}    
    
    \\
    \\
    Q_{33}&=&(\ell_P)^{3}\big|\tilde{Z}\big|^\frac{1}{2}\cdot
    \frac{4\,A_3 \,{\left( {{{\Mvariable{V}}_{3A}}}^{\frac{1}{4}} - 
        {{{\Mvariable{V}}_{3B}}}^{\frac{1}{4}} \right) }^2}{{\left( 1 + 2\,{j_3} \right) }^2}
     
\end{array}\] \be\label{Explicit q_{IJ} matrix elements}\ee

With the explicit expressions in (\ref{Explicit q_{IJ} matrix elements}) we are now able to evaluate (\ref{Definition of e'(B)^2}): 

\ba
   < 0 \,| \,[e']^2\,|\, 0\, >
   &=&
       \epsilon^{IJK}\epsilon^{LMN}~<0\,|\,  
   q_{IL}\big(\frac{1}{2}\big)~  q_{JM}\big(\frac{1}{2}\big)~q_{KN}\big(\frac{1}{2}\big)\,|\,0> \nonumber \\
   &=&
     \epsilon^{IJK}\epsilon^{LMN}~
     <0\,|\,q_{IL}\big(\frac{1}{2}\big) \,|\,0>\,
     <0\,|\,q_{JM}\big(\frac{1}{2}\big) \,|\,0>\,
     <0\,|\,q_{KN}\big(\frac{1}{2}\big) \,|\,0>
   \nonumber\\
   &=&
     \epsilon^{IJK}\epsilon^{LMN}~Q_{IL} ~Q_{JM}~ Q_{KN}
\ea

The result is

\fcmt{17}{\[ \begin{array}{llll}
     \\
    < 0 \,| \,[e']^2\,|\, 0\, >
   &=&\multicolumn{2}{l}{
    \displaystyle\frac{32\,(\ell_P)^{9}\big|\tilde{Z}\big|^\frac{3}{2}}{9\,
    {\left( 1 + 2\,{j_1} \right) }^2\,{\left( 1 + 2\,{j_2} \right) }^2\,{\left( 1 + 2\,{j_3} \right) }^2}}
    
    \\\\
    &&\times&\big[ 108\,A_1 \,A_2 \,
    A_3
      
    \\
    &&&
   - 3\,{\left( 2\,{\left( -1 \right) }^{2\,{j_1}} + 
           {\left( -1 \right) }^{2\,{j_3}} \right) }^2\,A_2 \,
       {\left( A_1 -A_2 + A_3 \right) }^2 
      
    \\
    &&& 
       
       -3\,{\left( 2\,{\left( -1 \right) }^{2\,{j_2}} + {\left( -1 \right) }^{2\,{j_3}} \right) }^2\,
       
       A_1  \,{\left( -A_1 + A_2 + A_3 \right) }^2
       
    \\
    &&&
      
       -3\,{\left( 1 + 2\,{\left( -1 \right) }^{2\,\left( {j_1} + {j_2} \right) } \right) }^2\,
       A_3 \,{\left( A_1 + A_2 - 
           A_3  \right) }^2 \ 
    \\
    &&&  
       -~ \left( 1 + 2\,{\left( -1 \right) }^{2\,\left( {j_1} + {j_2} \right) } \right) \,
       \left( 2\,{\left( -1 \right) }^{2\,{j_1}} + {\left( -1 \right) }^{2\,{j_3}} \right) \,
       \left( 2\,{\left( -1 \right) }^{2\,{j_2}} + {\left( -1 \right) }^{2\,{j_3}} \right) \,
    \\
    &&&~~~~~~~\times
       \left( -A_1 +A_2 +A_3 \right) \,
       \left( A_1 -A_2 + A_3 \right) \,
       \left( A_1 + A_2 - A_3  \right)  
       
     \big]  
     \\\\&&\times&	   
     \,
    {\left( {V_{1A}}^{\frac{1}{4}} - {V_{1B}}^{\frac{1}{4}} \right) }^2\,
    {\left( {V_{2A}}^{\frac{1}{4}} - {V_{2B}}^{\frac{1}{4}} \right) }^2\,
    {\left( {V_{3A}}^{\frac{1}{4}} - {V_{3B}}^{\frac{1}{4}} \right) }^2

\end{array}\] \be\label{e'(B)^2 Endresultat eichinvarianter 3-Vertex} \ee }
\\\\
where all quantities are defined as on the last page and $\tilde{Z}$ is a numerical constant dependent on the regularization of the volume operator and on the Immirzi parameter (see \cite{TT:Closed ME of V in LQG}\footnote{There $\tilde{Z}$ is found to be $\tilde{Z}=\frac{\mb{i}}{4}\Big(\frac{3}{4}\Big)^3$},\cite{Volume_Article_I} for details). The $9^{th}$ power of $\ell_P$ gives together with the $\hbar^{-6}\kappa^{-6}=(\ell_P)^{-12}$ coming from the quantization of the gravitational part $\frac{1}{\sqrt{\det q}}$  of the Hamiltonian in (\ref{quantization step I}) the correct $\frac{1}{(\ell_P)^3}$ prefactor.  
Note the remarkable symmetry of  (\ref{e'(B)^2 Endresultat eichinvarianter 3-Vertex}).
We will illustrate the non-trivial behaviour of (\ref{e'(B)^2 Endresultat eichinvarianter 3-Vertex}) in the following (neglecting the prefactors $(\ell_P)^9|\tilde{Z}|^\frac{3}{2}$) in some examples. 
\vfill
\pagebreak
%%%%%%%%%%%%%%%%%%%%%%%%%%%%%%%%%%%%%%%%%%%%%%%%%%%%%%%%%%%%%%%%%%%%%%%%%%%%%%%%%%%%%%%%%%%%%%%%%%%%%%
\subsection{Different Configurations}
%%%%%%%%%%%%%%%%%%%%%%%%%%%%%%%%%%%%%%%%%%%%%%%%%%%%%%%%%%%%%%%%%%%%%%%%%%%%%%%%%%%%%%%%%%%%%%%%%%%%%%
%%%%%%%%%%%%%%%%%%%%%%%%%%%%%%%%%%%%%%%%%%%%%%%%%%%%%%%%%%%%%%%%%%%%%%%%%%%%%%%%%%%%%%%%%%%%%%%%%%%%%%
\subsubsection{''Oscillating''}
%%%%%%%%%%%%%%%%%%%%%%%%%%%%%%%%%%%%%%%%%%%%%%%%%%%%%%%%%%%%%%%%%%%%%%%%%%%%%%%%%%%%%%%%%%%%%%%%%%%%%%
%%%%%%%%%%%%%%%%%%%%%%%%%%%%%%%%%%%%%%%%%%%
\paragraph{\fbox{$j_1=j_2=\frac{j_3}{2}$}}%
%%%%%%%%%%%%%%%%%%%%%%%%%%%%%%%%%%%%%%%%%%%
If we set $j_1=j_2=\frac{j_3}{2}$ where $j_3 \in \mb{N}$, we get:

\be
< 0 \,| \,[e']^2\,|\, 0\, >
   \propto
\frac{128\,{\sqrt{2}}\,\left( -1 + {\left( -1 \right) }^{{j_3}} \right) \,{{j_3}}^4\,
    \left( -3\,\left( 2 + {\left( -1 \right) }^{{j_3}} \right)  + 
      \left( 1 + {\left( -1 \right) }^{3\,{j_3}} \right) \,{j_3} \right) }{9\,\left( 1 + {j_3} \right) \,
    {\left( 1 + 2\,{j_3} \right) }^{\frac{7}{4}}}
\ee

\begin{figure}[htbp!]
         \cmt{8}{
	 \psfrag{e2}{$[e']^2$}
	 \psfrag{j3}{$j_3$}
	 \includegraphics[height=5cm]{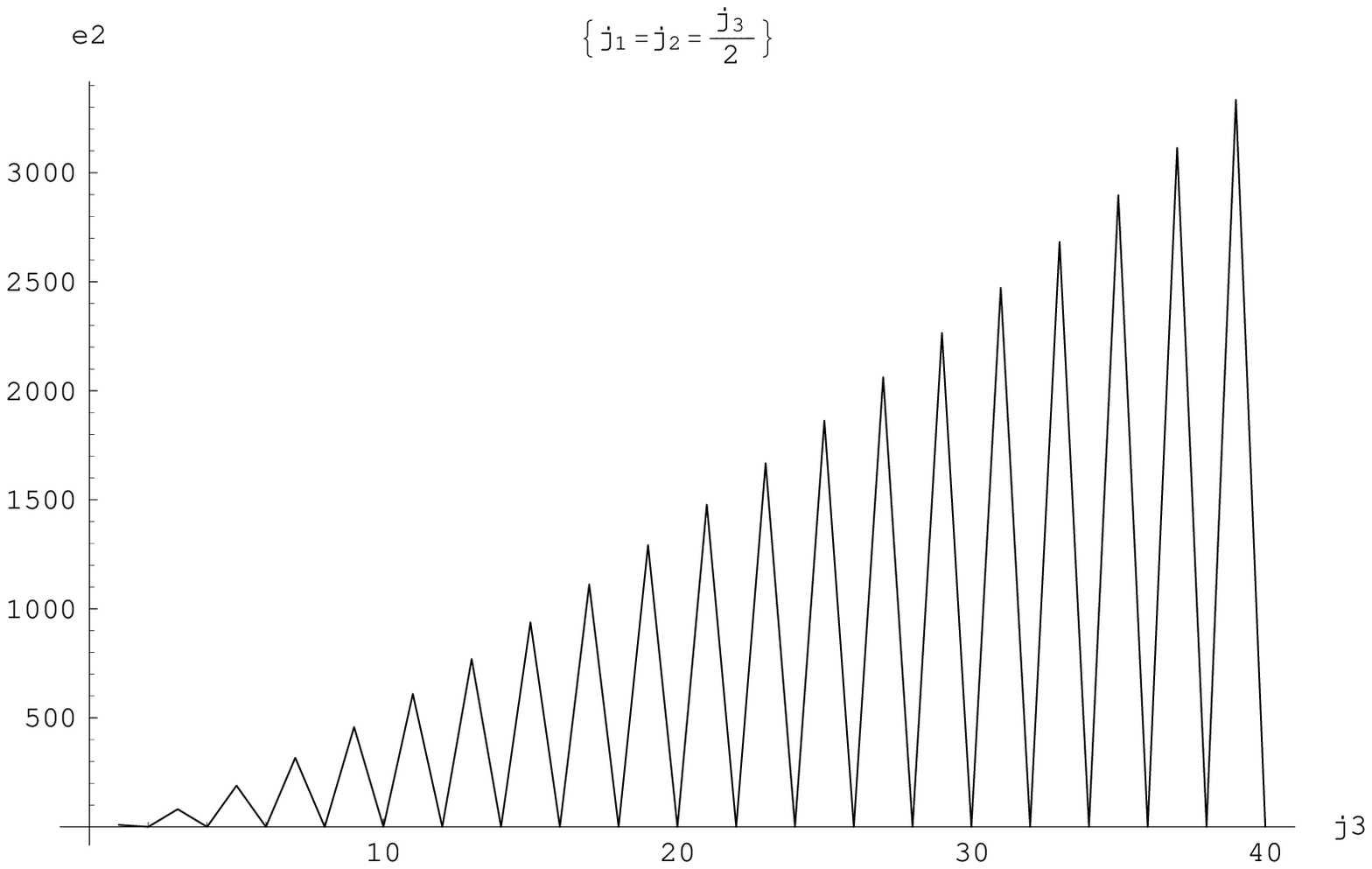}
          \caption{Plot for $j_1=j_2=\frac{j_3}{2}$ where $j_3 \in \mb{N}$ with $1\le j_3 \le 40$. The graph oscillates between 0 (if $j_3$ even) and an increasing value (if $j_3$ odd)} }
 
 \vspace{-6cm}\hspace{8.5cm}
 \cmt{8}{Asymtotically this increases as
  \[\begin{array}{lcl}
     < 0 \,| \,[e']^2\,|\, 0\, >
    & \propto&
    \frac{128\cdot 6 \cdot\sqrt{2} \cdot {j_3}^4}{9(1+j_3)(1+2j_3)^{\frac{11}{4}}}\nonumber\\
    &\stackrel{j_3 \rightarrow \infty}{\propto}&
    35.9 \cdot {j_3}^{\frac{5}{4}}
  \end{array}\]}
  \vspace{4.5cm}
\end{figure}

\vspace{-0.8cm}
%%%%%%%%%%%%%%%%%%%%%%%%%%%%%%%%%%%%%%%%%%%
\paragraph{\fbox{$j_1=j_2=\frac{j_3}{2}+\frac{1}{2}$}}%
%%%%%%%%%%%%%%%%%%%%%%%%%%%%%%%%%%%%%%%%%%%
If we set $j_1=j_2=\frac{j_3}{2}+\frac{1}{2}$ where $j_3 \in \mb{N}$, we get:

\[\begin{array}{lcl}
< 0 \,| \,[e']^2\,|\, 0\, >
   &\propto&\displaystyle
    \frac{1}{9\,
    {\left( 2 + {j_3} \right) }^4\,{\left( 1 + 2\,{j_3} \right) }^2}
    
   \\\\
   &&\times~    
    128\,{\sqrt{2}}\,\left( 1 + {\left( -1 \right) }^{{j_3}} \right) \,
    {{j_3}}^2\,{\left( 1 + {j_3} \right) }^3\,\left( -3\,
       \left( -7 + 3\,{\left( -1 \right) }^{{j_3}} + {\left( -1 \right) }^{3\,{j_3}} \right)  + 
      \left( -1 + {\left( -1 \right) }^{3\,{j_3}} \right) \,{j_3} \right) \,
    \\
    &&\times
    {\left( -\left( 2^{\frac{1}{4}}\,{\left( {{j_3}}^2\,\left( 1 + {j_3} \right)  \right) }^{\frac{1}{8}} \right)
            + {\left( {\left( 1 + {j_3} \right) }^2\,\left( 3 + 2\,{j_3} \right)  \right) }^{\frac{1}{8}} \right)
        }^2\,
	
    \\&&
    \times{\left( {\left( {j_3}\,{\left( 1 + {j_3} \right) }^2 \right) }^{\frac{1}{8}} - 
        {\left( {j_3}\,\left( 3 + 5\,{j_3} + 2\,{{j_3}}^2 \right)  \right) }^{\frac{1}{8}} \right) }^4	
\end{array}\]

\vspace{-0.4cm}

\begin{figure}[htbp!]
         \cmt{8}{
	 \psfrag{e2}{$[e']^2$}
	 \psfrag{j3}{$j_3$}
	 \includegraphics[height=5cm]{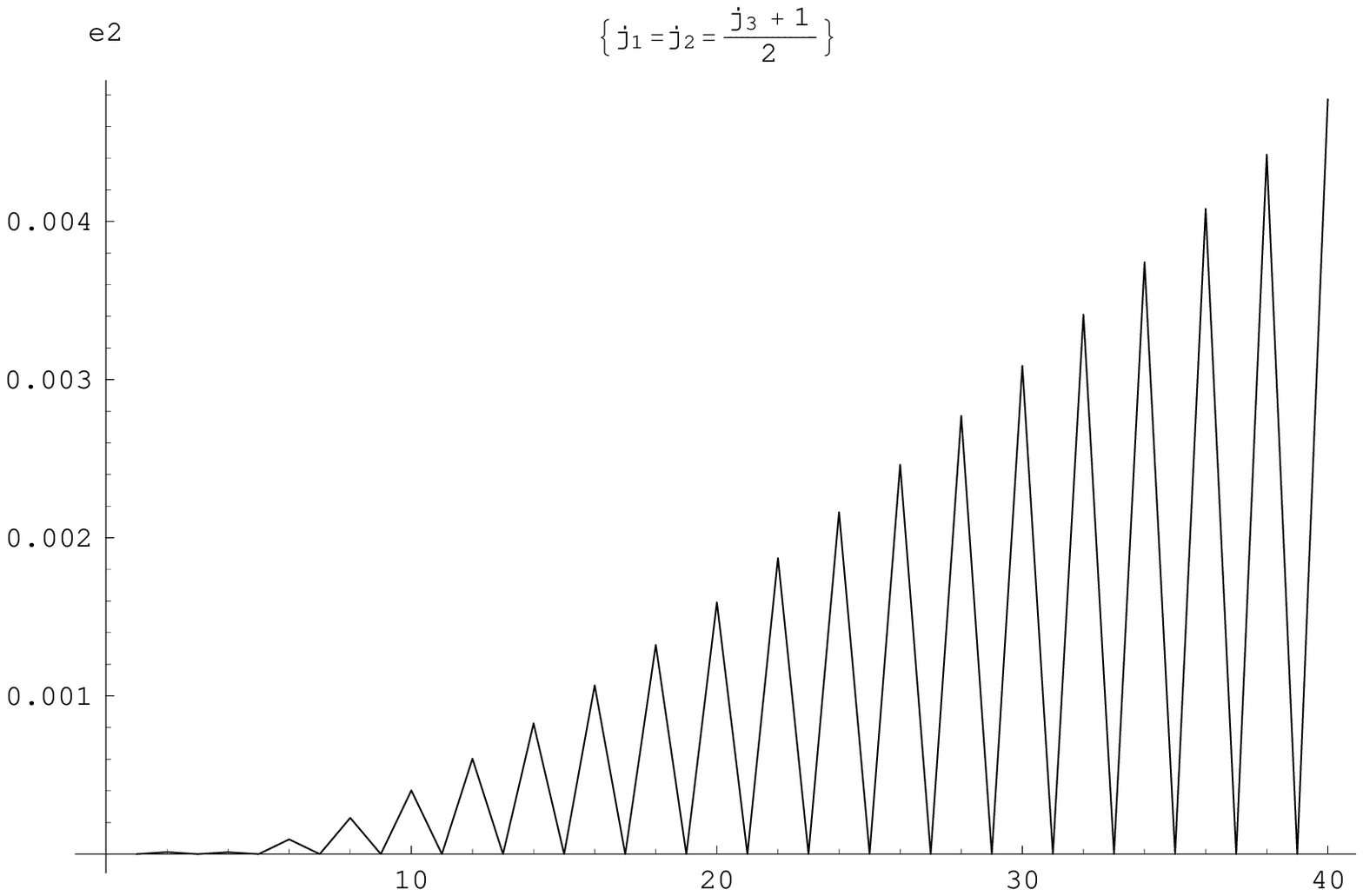}
          \caption{Plot for $j_1=j_2=\frac{j_3+1}{2}$ where $j_3 \in \mb{N}$ with $1\le j_3 \le 40$. The graph oscillates between 0 (if $j_3$ odd) and an increasing value (if $j_3$ even)} }
 
 \vspace{-6cm}\hspace{8.5cm}
 \cmt{8}{Asymtotically this increases as
  \[\begin{array}{lcl}
     < 0 \,| \,[e']^2\,|\, 0\, >
    & \propto&5.9\cdot 10^{-5}~{j_3}^{\frac{5}{4}}
    
  \end{array}\]}
  \vspace{4.5cm}
\end{figure}

\subsubsection{Increasing}
%%%%%%%%%%%%%%%%%%%%%%%%%%%%%%%%%%%%%%%%%%%
\paragraph{\fbox{$j_1=\frac{3}{2}~ j_2=j_3+\frac{1}{2}$}}%
%%%%%%%%%%%%%%%%%%%%%%%%%%%%%%%%%%%%%%%%%%%
If we set $j_1=\frac{3}{2}$, $j_2=j_3+\frac{1}{2}$ where $j_3 \in \mb{N}$, we get:

\[\begin{array}{lcl}
< 0 \,| \,[e']^2\,|\, 0\, >
   &\propto&\displaystyle

   	\frac{1}{3\,\left( 1 + {j_3} \right) \,
    {\left( 1 + 2\,{j_3} \right) }^2}
   \\&&\times
   4\,{j_3}\,\left( -9 + 21\,{j_3} + 14\,{{j_3}}^2 \right) \,
    {\left( -\left( 3^{\frac{1}{8}}\,{\left( {j_3}\,\left( 2 + {j_3} \right)  \right) }^{\frac{1}{8}} \right)
            + {\left( -3 + 4\,{j_3} + 4\,{{j_3}}^2 \right) }^{\frac{1}{8}} \right) }^2
    \\&&\times\,
    {\left( -2\,{\left( {j_3}\,\left( 2 + {j_3} \right)  \right) }^{\frac{1}{8}} + 
        {\sqrt{2}}\,3^{\frac{1}{8}}\,{\left( -3 + 4\,{j_3} + 4\,{{j_3}}^2 \right) }^{\frac{1}{8}} \right) }^2\,
    {\left( {\left( {j_3}\,\left( 3 + 2\,{j_3} \right)  \right) }^{\frac{1}{8}} - 
        {\left( -6 + 9\,{j_3} + 6\,{{j_3}}^2 \right) }^{\frac{1}{8}} \right) }^2

\end{array}\]

\vspace{-0.4cm}

\begin{figure}[htbp!]
         \cmt{8}{
	 \psfrag{e2}{$[e']^2$}
	 \psfrag{j3}{$j_3$}	 
	 \includegraphics[height=5cm]{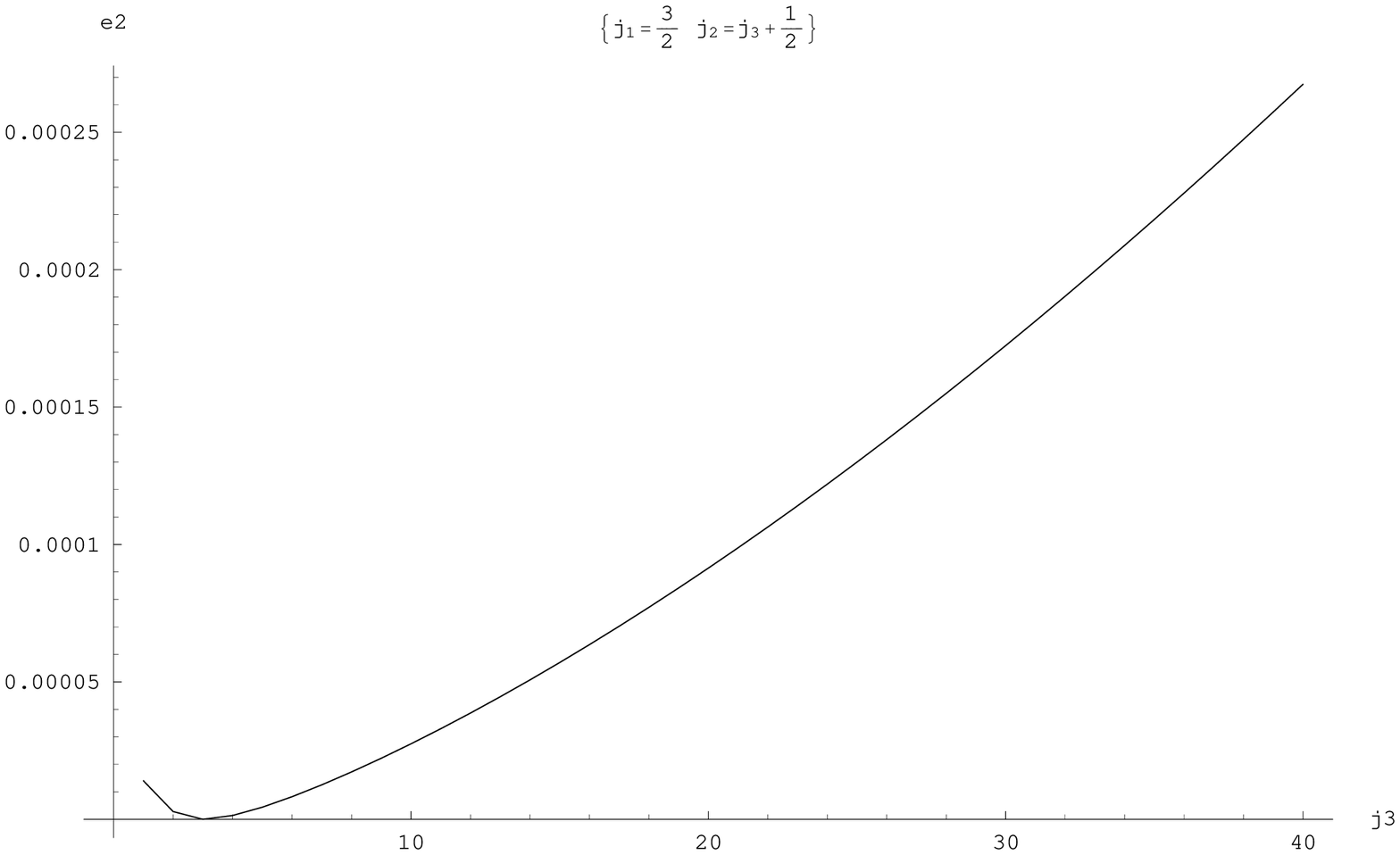}
          \caption{Plot for $j_1=\frac{3}{2}$, $j_2=j_3+\frac{1}{2}$ where $j_3 \in \mb{N}$ with $1\le j_3 \le 40$. The graph first decreases for $1\le j_3 <3 $ and is 0 for $j_3=3$ . It decreases for $j_3>3$} }
 
 \vspace{-6cm}\hspace{8.5cm}
 \cmt{8}{Asymtotically this increases as
  \[\begin{array}{lcl}
     < 0 \,| \,[e']^2\,|\, 0\, >
    & \propto& 1.06\cdot 10^{-6}~{j_3}^{\frac{3}{2}}
    
  \end{array}\]}
  \vspace{4.5cm}
\end{figure}

%%%%%%%%%%%%%%%%%%%%%%%%%%%%%%%%%%%%%%%%%%%%%%%%%%%%%%%%%%%%%%%%%%%%%%%%%%%%%%%%%%%%%%%%%%%%%%%%%%%%%
\subsubsection{\label{Nullkonfigurationen}Identical 0}
If we set $j_1=j_2=j_3$ and more generally $j_1,j_2,j_3\in \mb{N}$ (all spins integer numbers) in (\ref{e'(B)^2 Endresultat eichinvarianter 3-Vertex})  then 

\[\begin{array}{lcl}
< 0 \,| \,[e']^2\,|\, 0\, >
   &=&0

\end{array}\]

\subsection{General Configurations}

Using the general result (\ref{e'(B)^2 Endresultat eichinvarianter 3-Vertex}) (without the prefactors $(\ell_P)^9|Z|^\frac{3}{2}$)  we use the quantity

\[
  Q=\left\{\begin{array}{lcl}
                             30+\ln{[[e']^2(j_1,j_2,j_3)]}&&|j_1-j_2|\le j_3 \le j_1+j_2~~and~~j_1+j_2+j_3~is~integer~~\\
			     &&and~~ [e']^2(j_1,j_2,j_3)\ne 0
			     \\\\
			     0 &~~~&else
                         \end{array}\right.
\]
\\
and do a three dimensional plot\footnote{The numerical constant 30 is added for technical reasons only.} (in the range $\frac{1}{2}\le j_1 \le j_{max}$, $\frac{1}{2}\le j_2 \le j_{max}$ for each fixed value $5\le j_3 \le \frac{15}{2}$ :

It turns out that the non-zero configurations are grouped symetrically along lines parallel to the $j_1=j_2$ axis. The reason for this is of course the integer requirement $j_1+j_2+j_3\stackrel{!}{=}integer$.
Therefore we will get contributions on the $j_1=j_2$-axis only if $j_3$ is integer.
Because (\ref{e'(B)^2 Endresultat eichinvarianter 3-Vertex}) is symmetric with respect to the interchange of $j_1 \leftrightarrow j_2$ we may restrict  ourselves to the range $j_1\ge j_2$\footnote{The rapid increase of the curves for small $j_1$ is due to the fact that the first non zero values has been connected by a line to the last 0.}.

Additionally we extract for each of those 3D-plot a 2 dimensional plot along the lines

\[
  j_2=j_1 - l~~~~~~\mbox{with}~~0\le l\le \min[j_3,j_{max}-\frac{1}{2}]~,~~~ l+\frac{1}{2}\le j_1 \le j_{max}=25
\]

The restriction for the parameter $l$ is a result from the requirements $|j_1-j_2|\le j_3 \le j_1+j_2$ from which for $j_1>j_2$ we may remove the modulus. 
In order to give a better impression we have joined only non-vanishing values of $Q$ along the lines described above\footnote{For integer $j_3$ also every second configuration on the lines gives a 0 due to section \ref{Nullkonfigurationen} and by joining all the points the plots oscillate between 0 and the plotted curves and it would be hardly possible to see anything useful from them in this case if we would join all points.}. 
\vfill
\pagebreak
\begin{figure}[htbp!]
         \cmt{8}{
	 \psfrag{j_1}{$j_1$}
	 \psfrag{j_2}{$j_2$}
	 \psfrag{ln[eB^2]}{$Q$}
	 \psfrag{j_3=5}{}
	 \includegraphics[height=6cm]{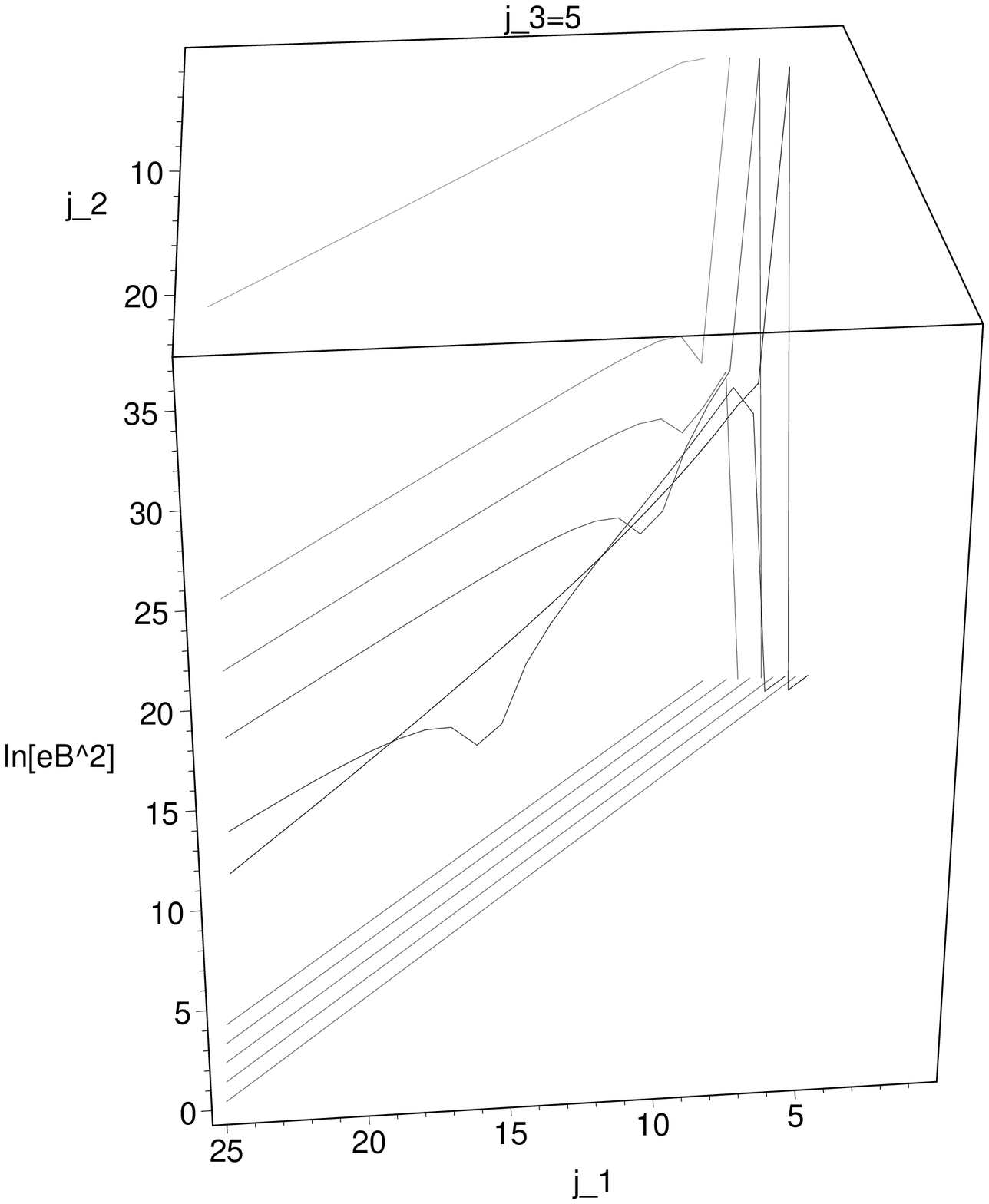}
          \caption{Plot for $j_3=5$} }
	 \cmt{8}{
	 \psfrag{j_1}{$j_1$}
	 \psfrag{j_2}{$j_2$}
	 \psfrag{ln[eB^2]}{$Q$}
	 \psfrag{j_3=5}{}
	 
	 \includegraphics[height=6cm]{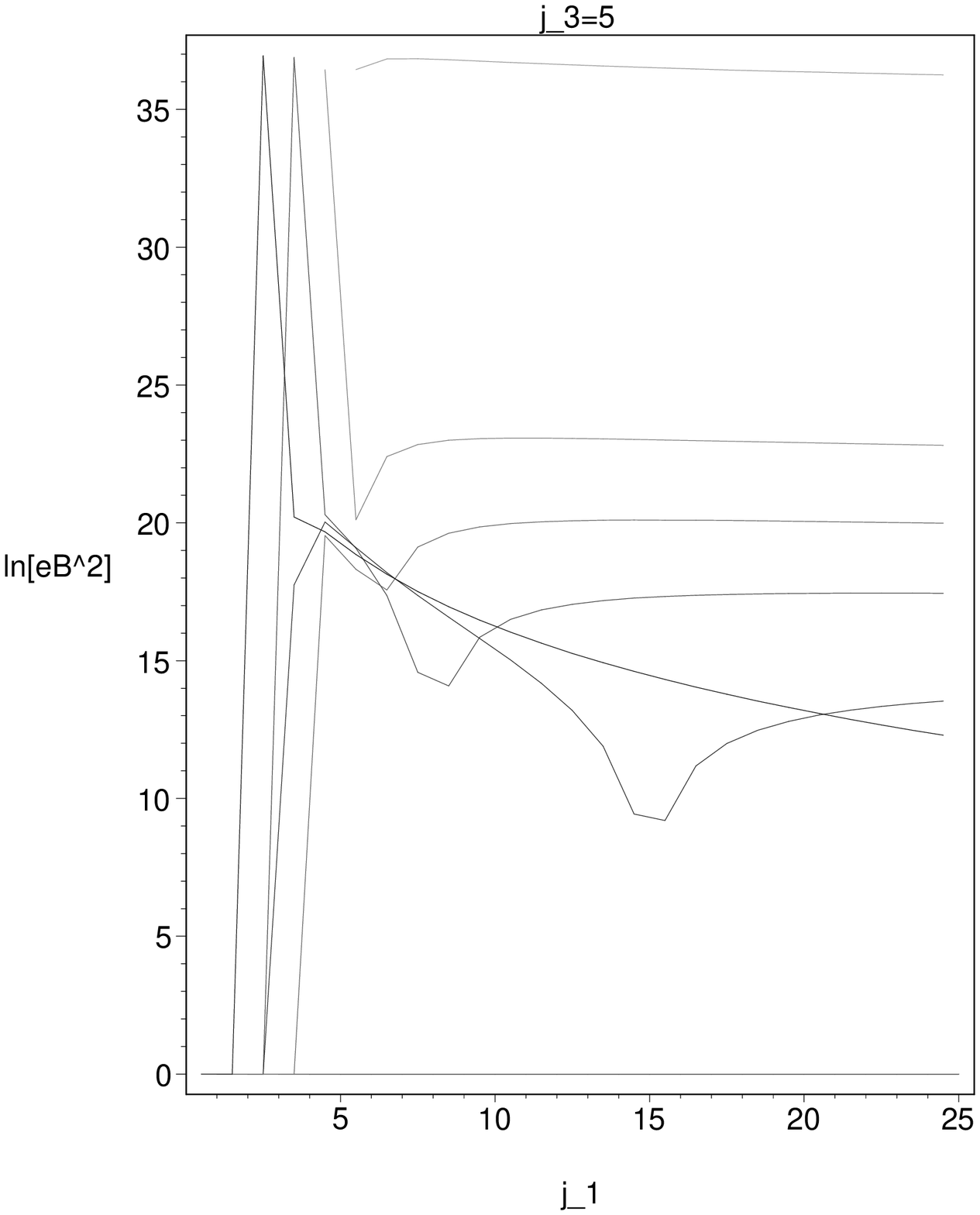}
          \caption{Plot for $j_3=5$} }
 
\end{figure}

\begin{figure}[htbp!]
         \cmt{8}{
	 \psfrag{j_1}{$j_1$}
	 \psfrag{j_2}{$j_2$}
	 \psfrag{ln[eB^2]}{$Q$}
	 \psfrag{j_3=11/2}{}
	 
	 \includegraphics[height=6cm]{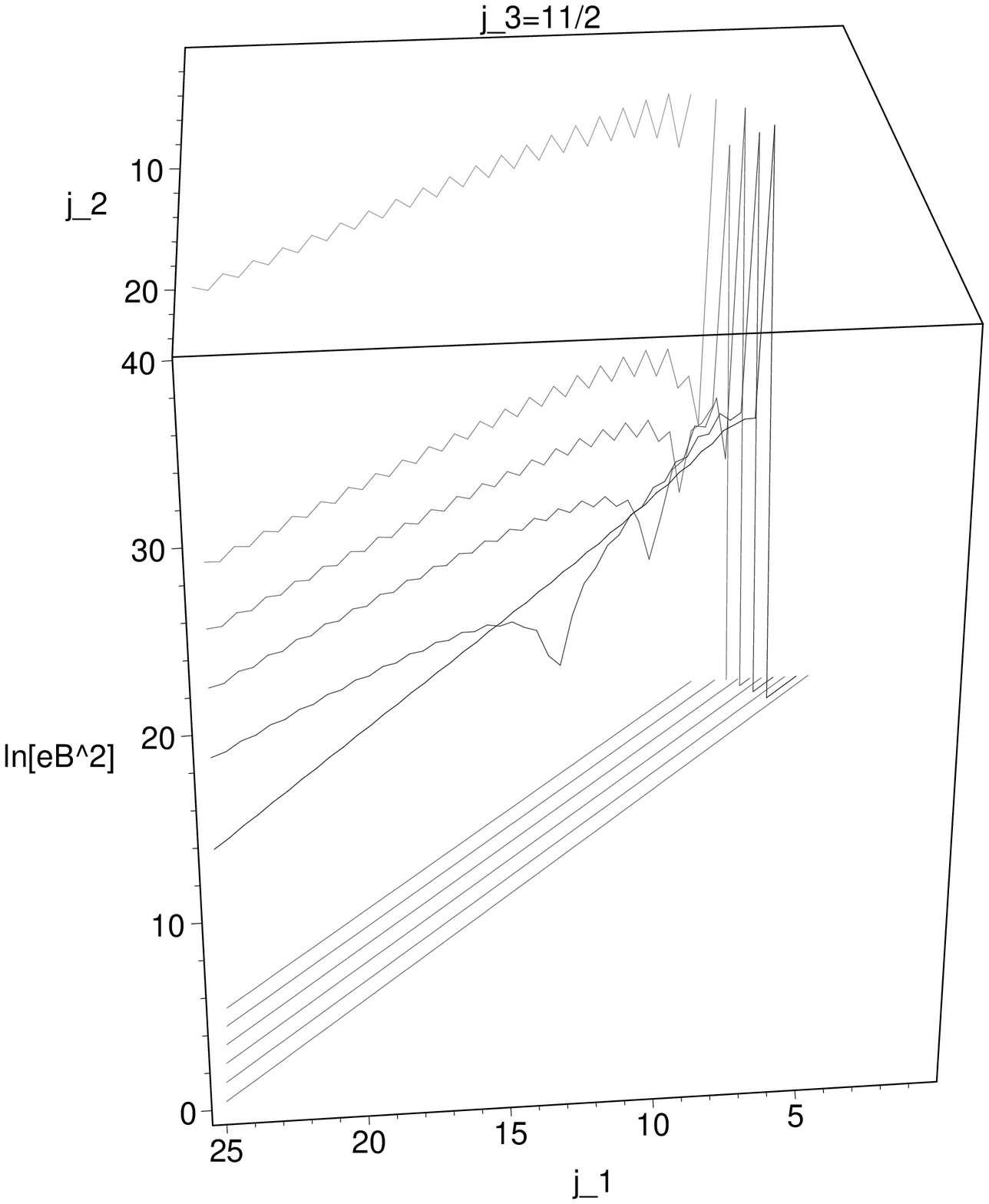}
          \caption{Plot for $j_3=\frac{11}{2}$} }
	 \cmt{8}{
	 \psfrag{j_1}{$j_1$}
	 \psfrag{j_2}{$j_2$}
	 \psfrag{ln[eB^2]}{$Q$}
	 \psfrag{j_3=11/2}{}
	 \includegraphics[height=6cm]{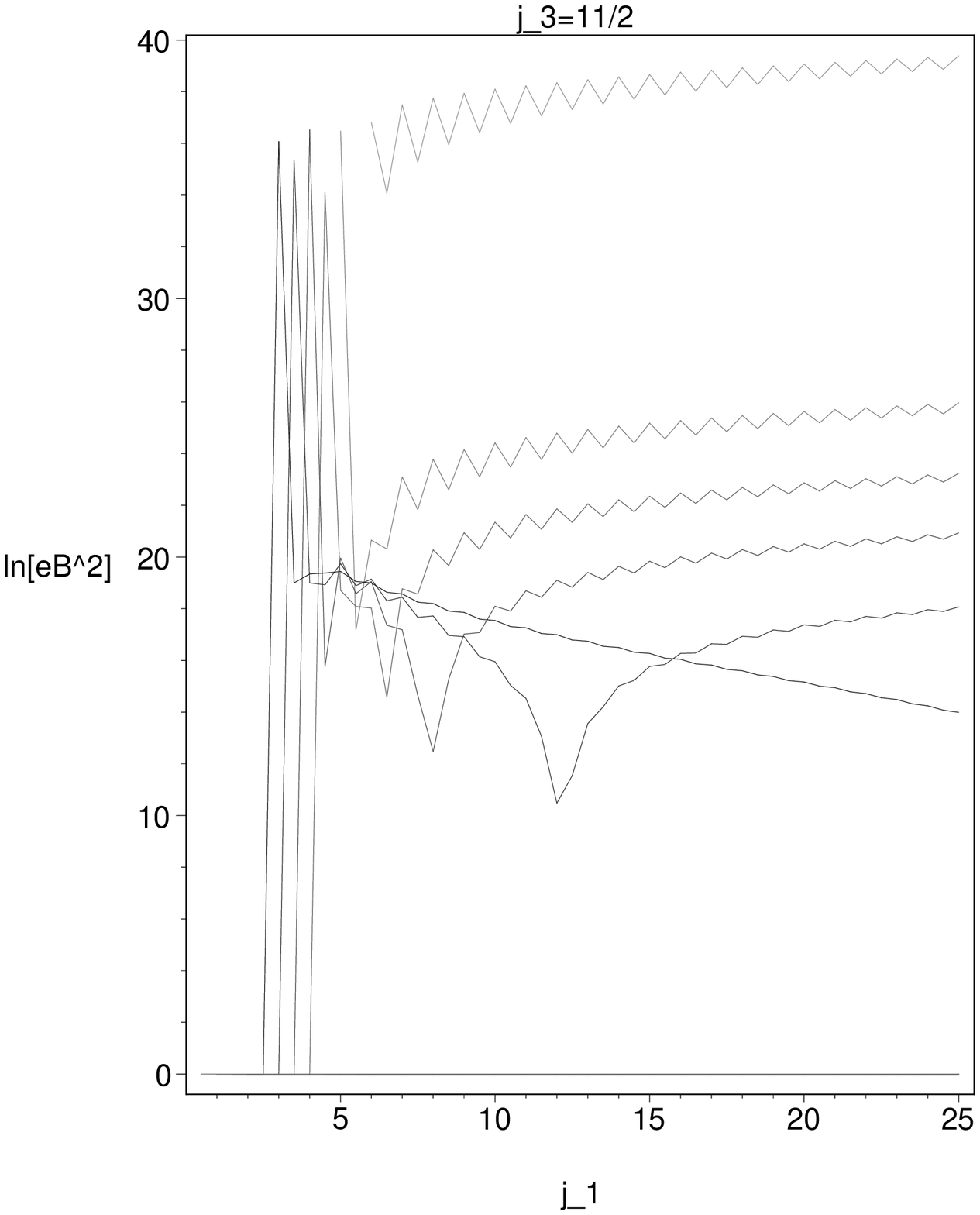}
          \caption{Plot for $j_3=\frac{11}{2}$, $j_2=j_1-l$. The different curves are (bottom to top):
	  $l=\frac{1}{2},\frac{3}{2},\ldots,\frac{11}{2}$} }
 
\end{figure}

\begin{figure}[htbp!]
         \cmt{8}{
	 \psfrag{j_1}{$j_1$}
	 \psfrag{j_2}{$j_2$}
	 \psfrag{ln[eB^2]}{$Q$}
	 \psfrag{j_3=6}{}
	 \includegraphics[height=6cm]{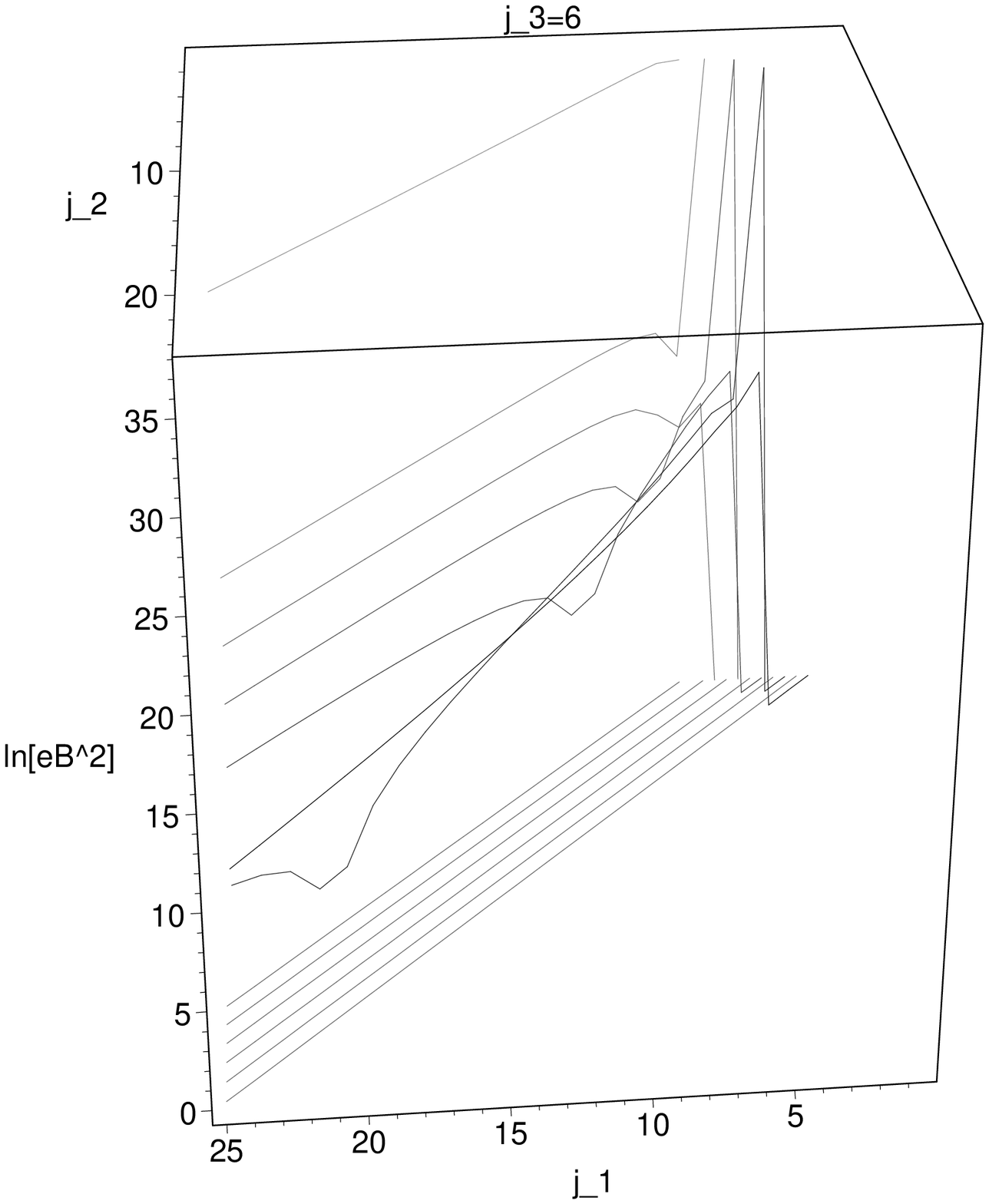}
          \caption{Plot for $j_3=6$} }
	 \cmt{8}{
	 \psfrag{j_1}{$j_1$}
	 \psfrag{j_2}{$j_2$}
	 \psfrag{ln[eB^2]}{$Q$}
	 \psfrag{j_3=6}{}
	 \includegraphics[height=6cm]{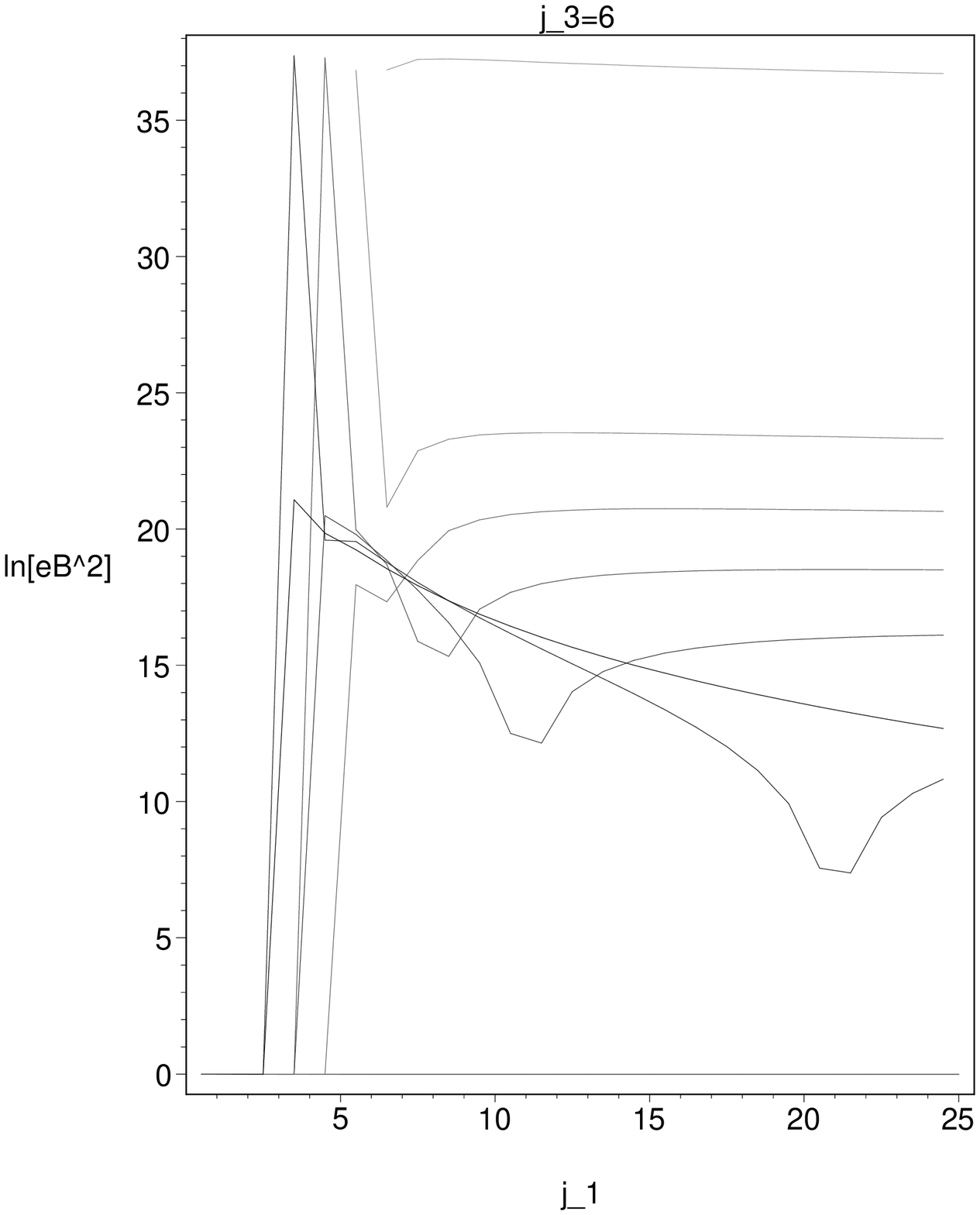}
          \caption{Plot for $j_3=6$} }
 
\end{figure}

\begin{figure}[htbp!]
         \cmt{8}{
	 \psfrag{j_1}{$j_1$}
	 \psfrag{j_2}{$j_2$}
	 \psfrag{ln[eB^2]}{$Q$}
	 \psfrag{j_3=13/2}{}
	 \includegraphics[height=6cm]{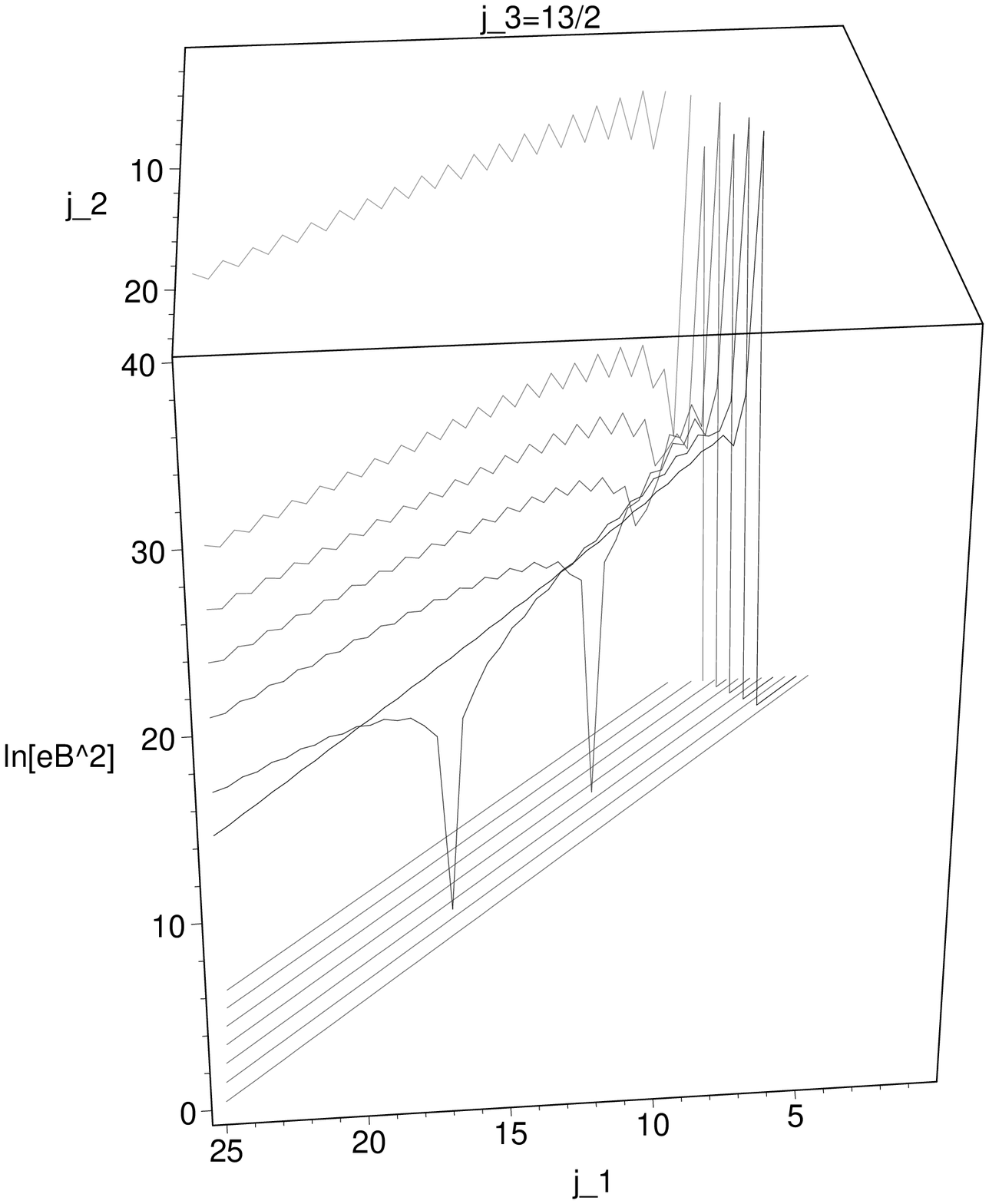}
          \caption{Plot for $j_3=\frac{13}{2}$} }
	 \cmt{8}{
	 \psfrag{j_1}{$j_1$}
	 \psfrag{j_2}{$j_2$}
	 \psfrag{ln[eB^2]}{$Q$}
	 \psfrag{j_3=13/2}{}
	 \includegraphics[height=6cm]{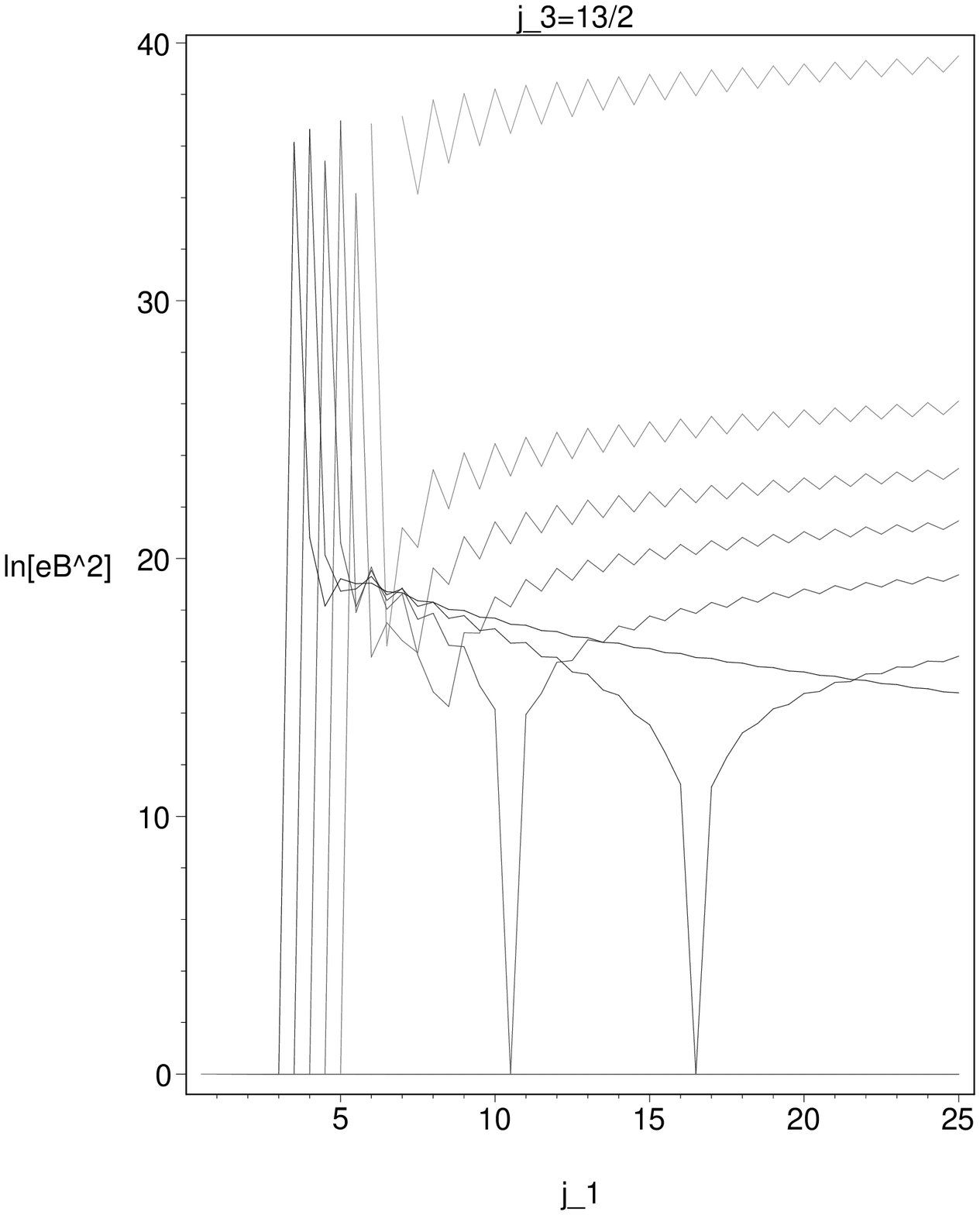}
          \caption{Plot for $j_3=\frac{13}{2}$, $j_2=j_1-l$. The different curves are (bottom to top):
	  $l=\frac{1}{2},\frac{3}{2},\ldots,\frac{13}{2}$} }
 \end{figure}
\begin{figure}[htbp!]
         \cmt{8}{
	 \psfrag{j_1}{$j_1$}
	 \psfrag{j_2}{$j_2$}
	 \psfrag{ln[eB^2]}{$Q$}
	 \psfrag{j_3=7}{}
	 \includegraphics[height=6cm]{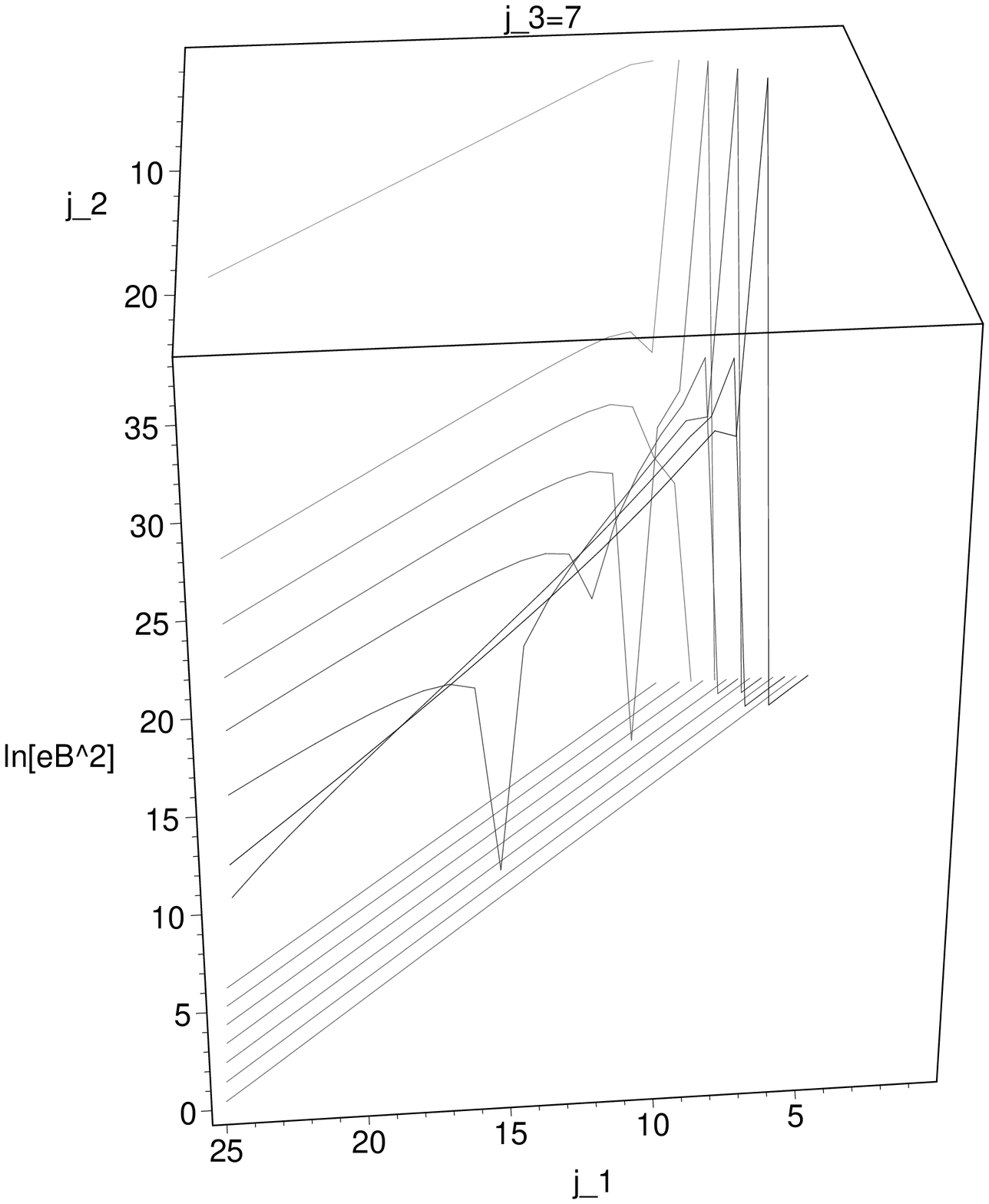}
          \caption{Plot for $j_3=7$} }
	 \cmt{8}{
	 \psfrag{j_1}{$j_1$}
	 \psfrag{j_2}{$j_2$}
	 \psfrag{ln[eB^2]}{$Q$}
	 \psfrag{j_3=7}{}
	 \includegraphics[height=6cm]{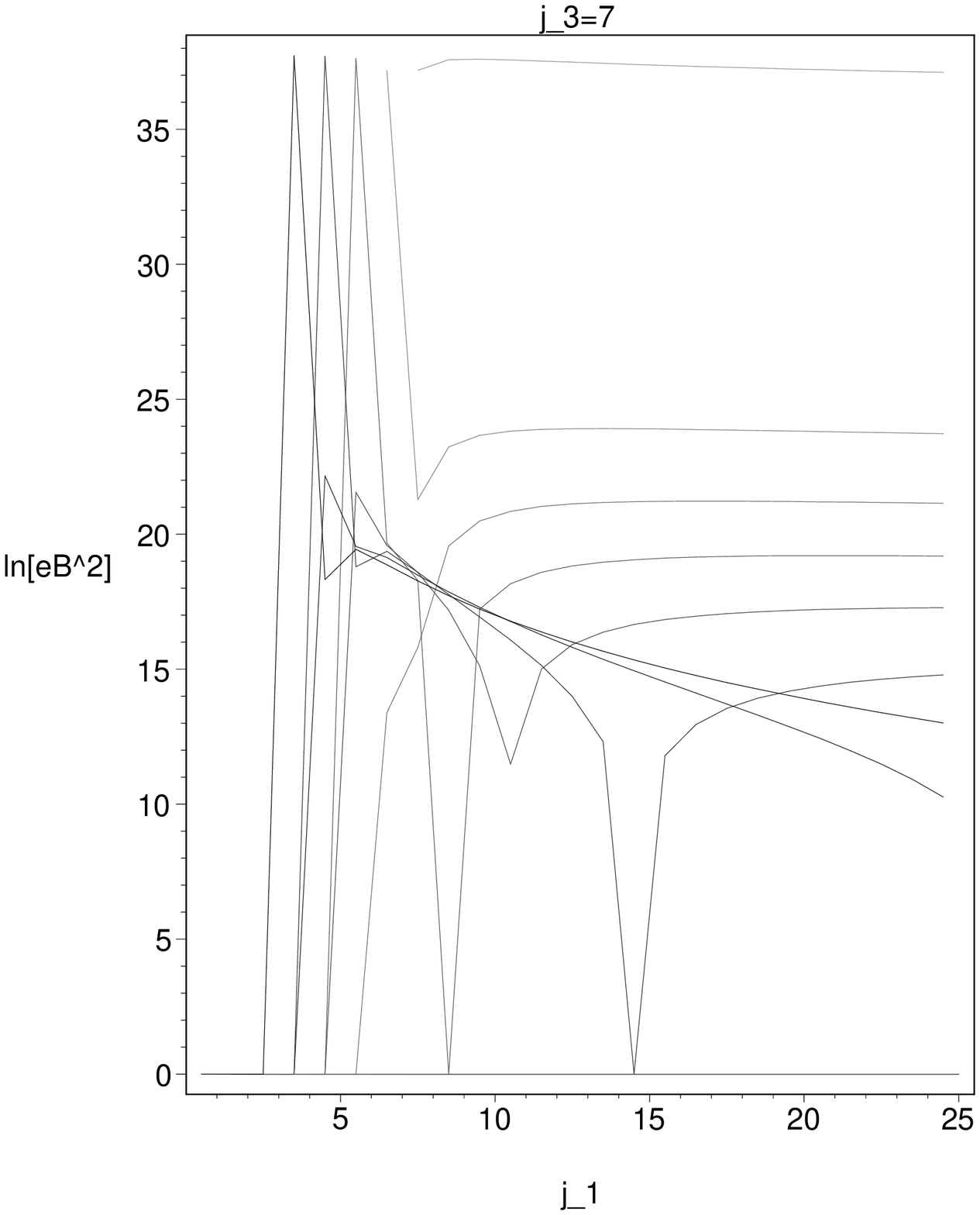}
          \caption{Plot for $j_3=7$} }	   
\end{figure}	
\begin{figure}[htbp!]
         \cmt{8}{
	 \psfrag{j_1}{$j_1$}
	 \psfrag{j_2}{$j_2$}
	 \psfrag{ln[eB^2]}{$Q$}
	 \psfrag{j_3=15/2}{}
	 \includegraphics[height=6cm]{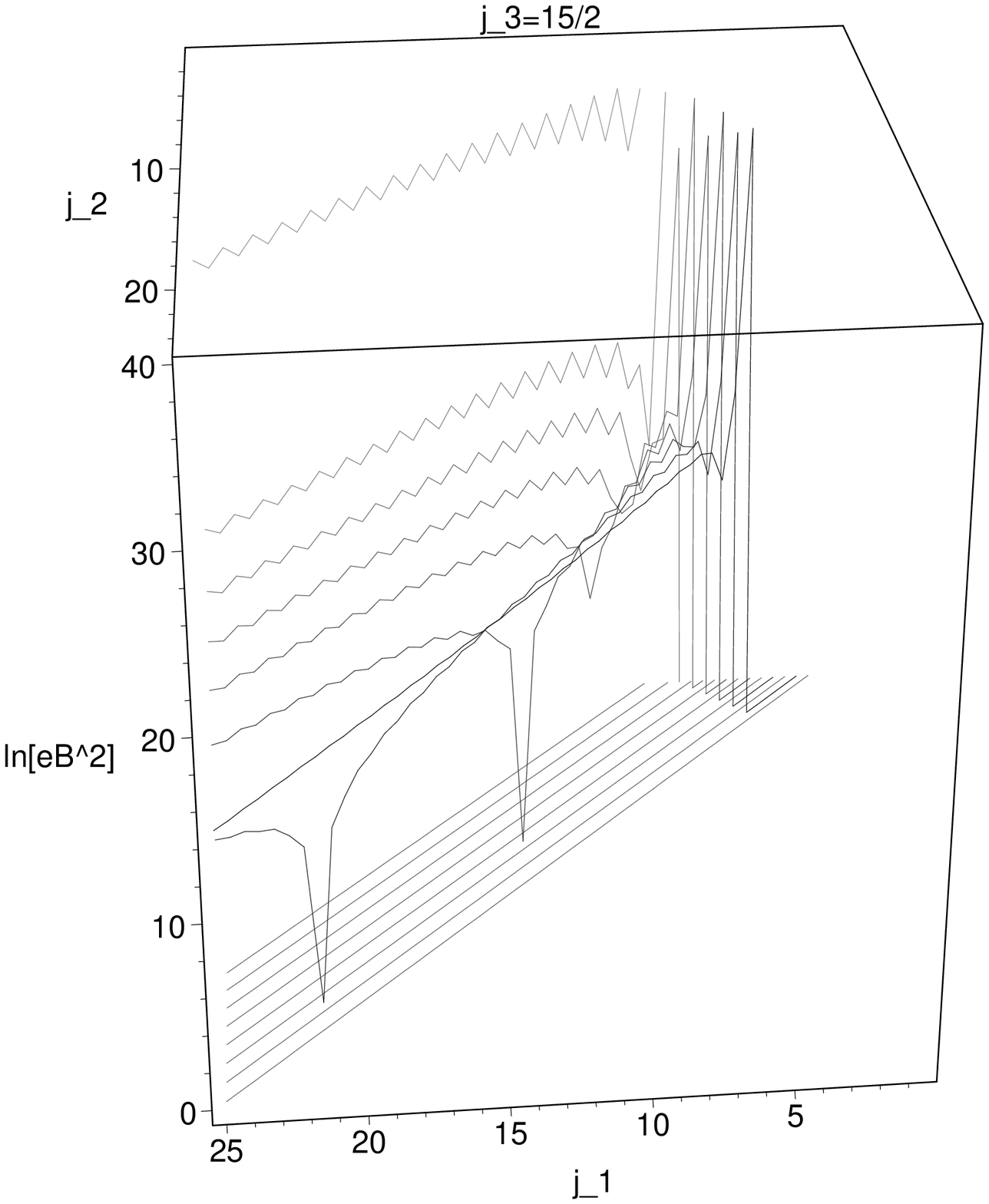}
          \caption{Plot for $j_3=\frac{15}{2}$} }
	 \cmt{8}{
	 \psfrag{j_1}{$j_1$}
	 \psfrag{j_2}{$j_2$}
	 \psfrag{ln[eB^2]}{$Q$}
	 \psfrag{j_3=15/2}{}
	 \includegraphics[height=6cm]{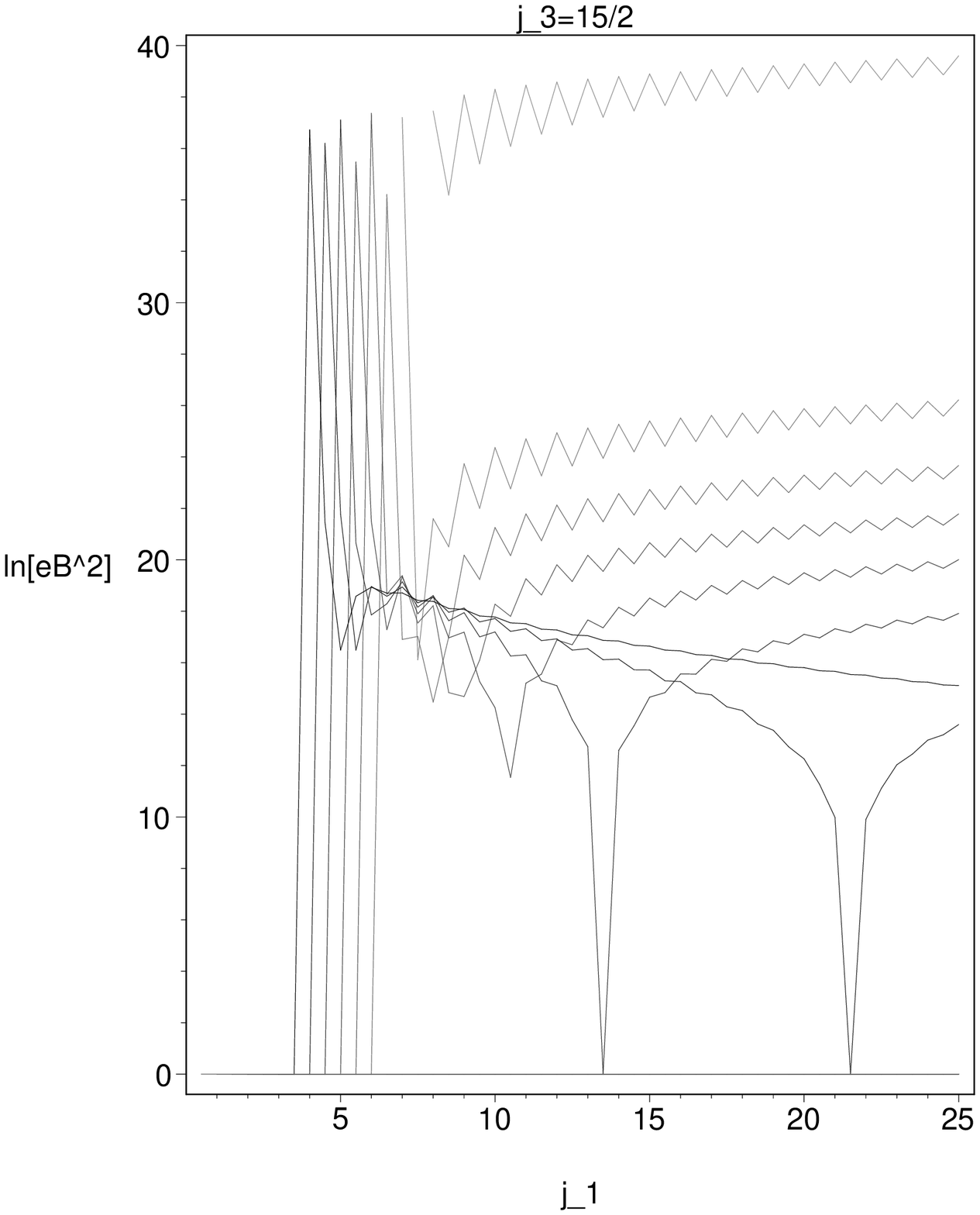}
          \caption{Plot for $j_3=\frac{15}{2}$, $j_2=j_1-l$. The different curves are (bottom to top):
	  $l=\frac{1}{2},\frac{3}{2},\ldots,\frac{15}{2}$} } 
\end{figure}

\pagebreak
%%%%%%%%%%%%%%%%%%%%%%%%%%%%%%%%%%%%%%%%%%%%%%%%%%%%%%%%%%%%%%%%%%%%%%%%%%%%%%%%%%%%%%%%%%%%%%%%%%%%%%
%%%%%%%%%%%%%%%%%%%%%%%%%%%%%%%%%%%%%%%%%%%%%%%%%%%%%%%%%%%%%%%%%%%%%%%%%%%%%%%%%%%%%%%%%%%%%%%%%%%%%%%
\section{Expectation Values in $U(1)^3$ CS} 
%%%%%%%%%%%%%%%%%%%%%%%%%%%%%%%%%%%%%%%%%%%%%%%%%%%%%%%%%%%%%%%%%%%%%%%%%%%%%%%%%%%%%%%%%%%%%%%%%%%%%%%
Note, that for the construction of the $U(1)^3$ coherent states  and for the definitions concerning the volume operator in $U(1)^3$ coherent states we follow \cite{QFT on CST II}. Note however that we will use a slightly different construction in order to find (\ref{Def U(1)^3 CS III}). 
%%%%%%%%%%%%%%%%%%%
\subsection{\label{Rechung mit CS} Setup}
%%%%%%%%%%%%%%%%%%%

Let us have $M$ edges on the vertex $v\in V(\gamma)$ 
Then we find for the expectation value of a polynomial of the operators $\hat{q}^{j_k}_{I_k}$ in our coherent states (\ref{Def U(1)^3 CS III}):

\ba\label{upper bound q-Polynom Eigenwert: Start}
   \Big<~\Psi_{m,\gamma}^{(v)}(A)~\Big|~\prod_{k=1}^N \hat{q}^{j_k}_{I_k} (r)~\Big|~\Psi_{m,\gamma}^{(v)}(A)~ \Big>&=&
   \frac{(2\pi)^{3M}}{\big\| \Psi_{m,\gamma}^{(v)}\big\|^2} \sum_{\{n^j_I \}\in \mb{Z}}
   \mb{ e}^{\sum_{I,j} [-t(I) [n_I^j]^2 + 2 p_I^j(m) n^j_I ] }
   \prod_{k=1}^N \lambda^r\big(\{n^{j_k}_{I_k}  \}\big)\\
   &&\fcmt{9}{where $I \equiv e_I$ now labels the edges $e\in E(v)$ and we recall  
      \[\begin{array}{lcl}
          \lambda^r\big(\{n^{j_k}_{I_k}  \}\big) &\stackrel{(\ref{Aktion von V auf konjugiertes Charge-Netzwerk})}{:=}&
	   \lambda^r\big(\{n^{j}_{I}  \}\big) 
	   -\lambda^r\big(\{n^{j}_{I} +\delta_{II_k} \delta^{jj_k}  \}\big) \\\\
	   
	  \lambda^r\big(\{n^{j}_{I}  \}\big)&\stackrel{(\ref{Def eigenwert V in U(1)^3 charge networks})}{=}& (\ell_P)^{3r}|Z|^{\frac{r}{2}}\Big[\big| \sum\limits_{IJK}\epsilon_{jkl}~ \epsilon(IJK)~ n^j_In^k_Jn^l_K \big|\Big|^{\frac{r}{2}}
	\end{array} \]}\nonumber
\ea
Note, that this expectation value only depends on the pair $m=(A,E)$ the coherent state is peaked on via $p^j_I(m)=p^j_I(A,E)$ according to (\ref{Definition p^j_e}). 
Using the upper bound (\ref{upper bound Gesamteigenwert}) we can estimate (\ref{upper bound q-Polynom Eigenwert: Start}) as

\ba\label{upper bound q-Polynom Eigenwert: 1. Abschaetzung}
   \Big<~\cdot~ \Big>
   &\le&
   \frac{(2\pi)^{3M}(\ell_P)^{3rN}(9M)^N|Z|^\frac{rN}{2}}{\big\| \Psi_{m,\gamma}^{(v)}\big\|^2} \sum_{\{n^j_I \}\in \mb{Z}}
   \mb{ e}^{\sum_{I,j} [-t(I) [n_I^j]^2 + 2 p_I^j(m) n^j_I ] }
   \Big[\sum_{J,j}[n^j_J]^2\Big]^N
\ea

Now define $T_I:=\sqrt{t(I)}$, $x^{j}_{I} := T_I~ n^{j}_{I}$ and perform a Poisson resummation according to (\ref{our PRS form}) to arrive at (Note that we replace at the same time 
$n^j_I \rightarrow \frac{1}{T_I}~ x^j_I$):
 
\ba\label{upper bound q-Polynom Eigenwert: Poisson-Resummation}
   \Big<~\cdot~ \Big>
   &\le&
   \frac{(2\pi)^{3M}(\ell_P)^{3rN}(9M)^N|Z|^\frac{rN}{2}}{\big\| \Psi_{m,\gamma}^{(v)}\big\|^2\prod_{I=1}^M(T_I)^3} \sum_{\{N^j_I \}\in \mb{Z}}~
   \int\limits_{\mb{R}^{3M}}d^{3M}x~
   \mb{ e}^{\sum_{I,j}\big[-[x^j_I]^2+\frac{2}{T_I}x^j_I[p^j_I(m)-\mb{i}\pi N^j_I] \big]}
   \Big[\sum_{J,j}\frac{1}{(T_J)^2}[x^j_J]^2\Big]^N~~~~~~~
\ea

As shown in \cite{QFT on CST II}  we may complete the square in the exponent of the integrals by introducing \linebreak $X_I^j=x^j_I-\frac{1}{T_I}\big(p^j_I(m)-\mb{i}\pi N^j_I \big)$ and rewrite
\ba \label{Erwartungswert nach quadratischer Ergaenzung}
   \Big< \cdot \Big>
    \le&\displaystyle\frac{(2\pi)^{3M}(\ell_P)^{3rN}(9M)^N|Z|^\frac{rN}{2}}{\big\| \Psi_{m,\gamma}^{(v)}\big\|^2\prod_{I=1}^M(T_I)^3} 
     &\displaystyle\sum_{\{N^j_I \}\in \mb{Z}}
     \mb{e}^{\sum_{I,j}\frac{1}{{T_I}^2}\big[[p^j_I(m)]^2-2\mb{i}\pi p^j_I(m)N^j_I -\pi^2[N^j_I]^2\big]}
     \times
     \nonumber
     \\ 
     &&~~~~~~~~\displaystyle\times\int\limits_{\mb{R}^{3M}} d^{3M}X~
     \mb{ e}^{-\sum_{I,j}[X_I^j]^2 }
     \Big[\sum_{J,j}\frac{1}{(T_J)^2}\big|X^j_J+\frac{1}{T_J}\big(p^j_J(m)-\mb{i}\pi N^j_J \big)\big|^2\Big]^N
  \nonumber\\
\ea

For the norm $\|\Psi_{m,\gamma}^{(v)}\|^2$ we find
\ba \label{Norm CS U(1)^3}
   \|\Psi_{m,\gamma}^{(v)}\|^2 
   &=& \prod_{e_I\in E(v) \atop j=1\ldots 3} \|\Psi_{m,I}^j\|^2
   \nonumber\\
   &=& \prod_{e_I\in E(v) \atop j=1\ldots 3}
       2\pi \sqrt{\pi}\frac{1}{T_I}~ 
       \mb{e}^{\frac{1}{{T_I}^2}[p_I^j(m)]^2} 
       \sum_{N^j_I \in \mb{Z}}~\mb{ e}^{-\frac{\pi}{{T_I}^2}\big[\pi[N_I^j]^2 + 2\mb{i}~ N_I^j p_I^j(m)\big] }
   \nonumber\\
   &=&(2\pi)^{3M} \frac{(\sqrt{\pi})^{3M}}{\prod_{I=1}^M (T_I)^3}~
       \mb{e}^{\sum_{I,j}\frac{1}{t_I}[p_I^j]^2} 
       \prod_{I,j}\underbrace{\sum_{\{N^j_I \}\in \mb{Z}}
       \mb{ e}^{-\frac{\pi}{{T_I}^2}\big[\pi[N_I^j]^2 + 2\mb{i}~ N_I^j p_I^j(m)\big] }}
       _{~~~~~~\displaystyle \big[1 + K_{t(I)}^{(j)} \big]}
\ea
here again $K_{t(I)}^{(j)} = \mathcal{O}(t(I)^{\infty})$.

If we now divide (\ref{Erwartungswert nach quadratischer Ergaenzung}) by the norm (\ref{Norm CS U(1)^3}) we find for the expectation value in normalized  $U(1)^3$ coherent states at a $M$-valent vertex $v$:
\ba \label{Erwartungswert normiert}
  \Big< \cdot \Big>
    \le&\displaystyle\underbrace{\frac{(\ell_P)^{3rN}(9M)^N|Z|^\frac{rN}{2}}{\sqrt{\pi}^{3M}
    \prod_{I,i}[1+K_{t(I)}^{i}]}}_{\displaystyle \tilde{\mathcal{K}} } 
     &\displaystyle\sum_{\{N^j_I \}\in \mb{Z}}
     \mb{e}^{-2\pi \mb{i}\sum_{I,j}\frac{1}{(T_I)^2}p^j_I(m)N^j_I }
     \mb{e}^{\sum_{I,j}\frac{\pi^2}{(T_I)^2}[N^j_I]^2}
     \times
     \nonumber
     \\[-7mm] 
     &&~~~~~~~~\displaystyle\times\int\limits_{\mb{R}^{3M}} d^{3M}X~
     \mb{ e}^{-\sum_{I,j}[X_I^j]^2 }
     \Big[\sum_{J,j}\frac{1}{(T_J)^2}\big|X^j_J+\frac{1}{T_J}\big(p^j_J(m)-\mb{i}\pi N^j_J \big)\big|^2\Big]^N
  \nonumber
  \\[8mm]
  \le&\displaystyle\tilde{\mathcal{K}}  
     &\displaystyle\sum_{\{N^j_I \}\in \mb{Z}}
     \Bigg|\mb{e}^{-2\pi \mb{i}\sum_{I,j}\frac{1}{(T_I)^2}p^j_I(m)N^j_I }\Bigg|
     \Bigg|\mb{e}^{\sum_{I,j}\frac{\pi^2}{(T_I)^2}[N^j_I]^2}\Bigg|
     \times
     \nonumber
     \\
     &&~~~~~~~~\displaystyle\times\Bigg|\int\limits_{\mb{R}^{3M}} d^{3M}X~
     \mb{ e}^{-\sum_{I,j}[X_I^j]^2 }
     \Big[\sum_{J,j}\frac{1}{(T_J)^2}\big|X^j_J+\frac{1}{T_J}\big(p^j_J(m)-\mb{i}\pi N^j_J \big)\big|^2\Big]^N\Bigg|
     \nonumber
     \\
\ea

Here we have introduced the abbreviation  $\tilde{\mathcal{K}}$ for the prefactor.
Now we estimate

\ba
   \big|X^j_J+\frac{1}{T_J}\big(p^j_J(m)-\mb{i}\pi N^j_J \big)\big|^2
   &=&\Big(X^j_J +\frac{1}{T_J}p^j_J(m) \Big)^2 +\frac{\pi^2}{(T_J)^2} [N^j_J]^2
   \nonumber
   \\
   &\le&[X^j_J]^2+\frac{2}{T_J}X^j_J|p^j_J(m)|+\frac{1}{(T_J)^2}[p^j_J(m)]^2 +\frac{\pi^2}{(T_J)^2}[N^j_J]^2
   \nonumber
   \\
   &\le&
   \underbrace{\Big[1+\frac{2}{T_J}|p^j_J(m)|\Big]}_{\displaystyle A^j_J}[X^j_J]^2
   +\frac{1}{2}\frac{1}{T_J}|p^j_J(m)|\Big[1+\frac{2}{T_J}|p^j_J(m)| \Big]+\frac{\pi^2}{(T_J)^2}[N^j_J]^2
   \nonumber
   \\
   &=& A^j_J [X^j_J]^2 +\frac{1}{2} \frac{1}{T_J}|p^j_J(m)|A^j_J +\frac{\pi^2}{(T_J)^2}[N^j_J]^2     
   \nonumber
\ea

where we have used that $x\le x^2 +\frac{1}{4}$ $\forall x\in\mb{R}$. Additionally taking into account that $\Big|\mb{e}^{-2\pi \mb{i}\sum_{I,j}\frac{1}{(T_I)^2}p^j_I(m)N^j_I }\Big|\le 1$ and the fact that the remaining two moduli in (\ref{Erwartungswert normiert}) are real numbers we continue with (\ref{Erwartungswert normiert})\footnote{As can be seen from (\ref{Definition p^j_e}), $p^j_J$ is a real number. We only take its modulus here in order to be independent of construction conventions (signatures) of the coherent states.}:

\ba \label{Erwartungswert normiert 2}
  \Big< \cdot \Big>
    &\le&\tilde{\mathcal{K}}\cdot\displaystyle\sum_{\{N^j_I \}\in \mb{Z}}
     \mb{e}^{\sum_{I,j}\frac{\pi^2}{(T_I)^2}[N^j_I]^2}
     \displaystyle\int\limits_{\mb{R}^{3M}} d^{3M}X~
     \mb{ e}^{-\sum_{I,j}[X_I^j]^2 }
     \Big[\sum_{J,j}\frac{1}{(T_J)^2}\Big(A^j_J[X^j_J]^2 +\frac{1}{2T_J}|p^j_J(m)|A^j_J+\frac{\pi^2}{(T_J)^2}[N^j_J]^2 \Big)\Big]^N
  \nonumber\\
\ea

Now let us introduce $X^i_J=:\sqrt{2}~Y^i_J$ and $T:=\min_I T_I$, $p:=\max_{I,j}|p^j_I(m)|$, $A:=1+\frac{p}{T}$. Then an upper bound for  (\ref{Erwartungswert normiert 2}) is given by:

\ba \label{Erwartungswert normiert 3}
  \Big< \cdot \Big>
    &\le&\tilde{\mathcal{K}}\cdot\sqrt{2}^{3M}\displaystyle\sum_{\{N^j_I \}\in \mb{Z}}
     \mb{e}^{\sum_{I,j}\frac{\pi^2}{(T_I)^2}[N^j_I]^2}
     \displaystyle\int\limits_{\mb{R}^{3M}} d^{3M}Y~
     \mb{ e}^{-2\sum_{I,j}[Y_I^j]^2 }
     \Big[\frac{2A}{T^2}\sum_{J,j}[Y^j_J]^2 +\frac{A}{2T^3}3Mp +\frac{\pi^2}{T^4}\sum_{J,j}[N^j_J]^2 \Big]^N
     \nonumber
     \\
     &=&\displaystyle\underbrace{\frac{(\ell_P)^{3rN}(9M)^N|Z|^\frac{rN}{2}\big[\frac{2A}{T^2}\big]^N}
    {\prod_{I,i}[1+K_{t(I)}^{i}]}}_{\displaystyle \mathcal{K} }
    \displaystyle\sum_{\{N^j_I \}\in \mb{Z}}
     \mb{e}^{\sum_{I,j}\frac{\pi^2}{(T_I)^2}[N^j_I]^2}  \times
    \nonumber
    \\[-5mm]
     &&\hspace{4cm}\displaystyle\times \sqrt{\frac{2}{\pi}}^{3M}\int\limits_{\mb{R}^{3M}} d^{3M}Y~
     \mb{ e}^{-2\sum_{I,j}[Y_I^j]^2 }
     \Big[\sum_{J,j}[Y^j_J]^2 +\underbrace{\frac{3Mp}{4T} +\frac{\pi^2}{2AT^2}\sum_{J,j}[N^j_J]^2}_{\displaystyle D} \Big]^N~~~~~~~~~
\ea

If we now define $\|Y\|^2:=\sum\limits_{L,l}[Y^l_L]^2$ we may first rewrite (\ref{Erwartungswert normiert 3}) and then finish with:

\ba\label{Erwartungswert normiert 4}
    \Big< \cdot \Big>
    &\le&
    \mathcal{K}\sum\limits_{\{N^j_I \}\ne \{0\}}\hspace{-3mm}
     \mb{e}^{-\sum\limits_{I,j}\frac{\pi^2}{{T_I}^2}[N^j_I]^2} 
      \sqrt{\frac{2}{\pi}}^{3M}\int\limits_{\mb{R}^{3M}} d^{3M}Y~
     \mb{ e}^{-2\|Y\|^2}
     \Big(\|Y\|^2+D\Big)^N
    \nonumber
    \\
    &&\fcmt{10}{Polynomial theorem: 
              $\big(x+a)^{n_0}=\displaystyle\sum_{n=0}^{n_0}\left( {n_0 \atop n}\right) x^n a^{n_0-n}$}
    \nonumber
    \\[5mm]
    &=&
    \mathcal{K}\sum\limits_{\{N^j_I \}\ne \{0\}}\hspace{-3mm}
     \mb{e}^{-\sum\limits_{I,j}\frac{\pi^2}{{T_I}^2}[N^j_I]^2} 
     \sum\limits_{n=0}^{N}\frac{N!}{(N-n)!~n!} D^{N-n} 
      \sqrt{\frac{2}{\pi}}^{3M}\int\limits_{\mb{R}^{3M}} d^{3M}Y~
     \mb{ e}^{-2\|Y\|^2}
     \|Y\|^{2n}
    \nonumber
    \\
    &&\fcmt{15.5}{Now according to (\ref{Produktformel fuer sphaerische Integrale}) we have $\displaystyle\sqrt{\frac{2}{\pi}}^{3M}\int\limits_{\mb{R}^{3M}} d^{3M}Y~
     \mb{ e}^{-2\|Y\|^2}
     \|Y\|^{2n}=I_{n}=\prod\limits_{l=1}^n\frac{3M+2(l-1)}{4}$,~~~ $I_0=1$}
    \nonumber
    \\[5mm]
    &=&\mathcal{K}\sum\limits_{\{N^j_I \}\ne \{0\}}\hspace{-3mm}
     \mb{e}^{-\sum\limits_{I,j}\frac{\pi^2}{{T_I}^2}[N^j_I]^2} 
     \sum\limits_{n=0}^{N}\frac{N!}{(N-n)!~n!} D^{N-n}
     I_n
    \nonumber
    \\
    &=&\mathcal{K}\sum\limits_{\{N^j_I \}\ne \{0\}}\hspace{-3mm}
     \mb{e}^{-\sum\limits_{I,j}\frac{\pi^2}{{T_I}^2}[N^j_I]^2} 
     \Bigg[D^N+\sum\limits_{n=1}^{N}\frac{N!}{(N-n)!~n!} D^{N-n}
     \prod_{l=1}^n\bigg[\frac{3M+2(l-1)}{4}\bigg]\Bigg] 
 \ea   

In analogy to \cite{QFT on CST II} we see that all sum terms in (\ref{Erwartungswert normiert 4}) are due to the prefactors  $\mb{e}^{-\sum\limits_{I,j}\frac{\pi^2}{{T_I}^2}[N^j_I]^2}$ of order $t^\infty$ as long as $\{N^j_I\}\ne 0$. So one gets (non neglectible) contributions only from the $\{N^j_I\}= \{0\}$ term.

Finally we find\footnote{Neglecting all term of order $t^\infty$.} 

\ba\label{finale obere Schranke fuer Erwartungswert}
   \lefteqn{\Big<~\Psi_{m,\gamma}^{(v)}(A)~\Big|~\prod_{k=1}^N \hat{q}^{j_k}_{I_k} (r)~\Big|~\Psi_{m,\gamma}^{(v)}(A)~ \Big>\le}~~~~~~~~~
   \nonumber\\
   &\le&
     \frac{(\ell_P)^{3rN}(9M)^N|Z|^\frac{rN}{2}\big[\frac{2A}{T^2}\big]^N}
     {\prod_{I,i}[1+K_{t(I)}^{i}]}
     \Bigg[\bigg[\frac{3Mp}{4T}\bigg]^N+\sum\limits_{n=1}^{N}\frac{N!}{(N-n)!~n!} \bigg[\frac{3Mp}{4T}\bigg]^{N-n}
     \prod_{l=1}^n\bigg[\frac{3M+2(l-1)}{4}\bigg]\Bigg]~~~~~~~~~ 
\ea

with 
\[\begin{array}{lllclc}
                                T:=\min\limits_I\{T_I\} &~~~~&\rightarrow~~
				&A&:=1+\frac{p}{T}
				\\
				p:=\max\limits_{I,j}\{|p^j_I(m)|\}
				&&&          
                                
			     \end{array} \]

Concerning the result (\ref{finale obere Schranke fuer Erwartungswert}) some comments are appropriate.
\begin{itemize}
   \item{The upper bound (\ref{finale obere Schranke fuer Erwartungswert}) is robust with respect to the classical configuration the coherent state $\Psi_{m,\gamma}^{(v)}(A)$ is peaked on. Only $p$ has to be bounded from above. 
   In particular at the classically singular configuration of the big bang where $\sqrt{\det(q)}=0$ and thus $E^b_j=\sqrt{\det{(q)}}~e^b_j=0$ which implies $p=0$ nothing special happens.}
   \item{Unfortunately (\ref{finale obere Schranke fuer Erwartungswert}) scales with the number of edges $M$ as $M^{2N}$. So one could essentially take two points of view:
   \subitem{1) We are working on the kinematical level only, therefore we would expect some high valent vertex suppression in the physical theory.}
   \subitem{2) Maybe the approximations used here are too rough, but we see that the only possibility in order to avoid a dependence on the edge number $M$ is to have the integrals (\ref{Produktformel fuer sphaerische Integrale}) independent from their dimension $3M$. 
   Only the integral $I_0$ is naturally independent of $M$. Hence one would like to derive a better estimate so that the integrand becomes independent of $M$. One could expect that this might be possible by taking the sign factor $\epsilon(I,J,K)$ into account which may lead to cancellations. However, as we will show explicitely in \ref{Pathologische Konfiguration} one can construct edge configurations so that (for any $M$) all $\epsilon(I,J,K)$ have equal sign.  }
   \\\\
   The ultimate answer has to be left open as a future task at the moment. It might provide us with some criteria physical states should fulfill.}
\end{itemize}

In any case, for the special case (\ref{quantization step II}) we thus get as upper bound for the expectation value of the gravitational part evaluated at a single vertex $v\in V(\gamma)$		     

\[\begin{array}{lllll}
   \displaystyle\hat{H}_{kin}^{(grav)}:=\widehat{\frac{1}{\sqrt{\det q}}}
   
   &=&
   
   \displaystyle\frac{P}{\kappa^6\hbar^6}~
   \sum\limits_{v\in V(\gamma)}&
   \displaystyle \frac{1}{E(v)} \sum\limits_{IJK}
   \epsilon(IJK)~\epsilon^{IJK}~\epsilon_{ijk}
   ~\lo{(\frac{1}{2})\,}\hat{e}^i_I(v)
   ~\lo{(\frac{1}{2})\,}\hat{e}^j_J(v)
   ~\lo{(\frac{1}{2})\,}\hat{e}^k_K(v)
   \\
   &&&\displaystyle \frac{1}{E(v)}  \sum\limits_{LMN}
   \epsilon(LMN)~\epsilon^{LMN}~\epsilon_{lmn}
   ~\lo{(\frac{1}{2})\,}\hat{e}^l_L(v)
   ~\lo{(\frac{1}{2})\,}\hat{e}^m_M(v)
   ~\lo{(\frac{1}{2})\,}\hat{e}^n_N(v)

\end{array}\]

of $\hat{H}_{kin}$ with $N=6$, $r=\frac{1}{2}$, $E(v)=\left(M\atop 3\right)=$, $P$ being the prefactor as defined in (\ref{quantization step II})\footnote{Note, that the denominator of the gravitational part is equipped with a prefactor $\frac{1}{\hbar^6\kappa^6}=\frac{1}{(\ell_P)^{12}}$}, ($\kappa\hbar=\ell_P^2$) and $\hat{H}_{kin}^{(grav)}$ evaluated at one vertex $v$ only :				     
\ba\label{Expectation value of gravitational part of H_{kin}}
   \lefteqn{\Big<~\Psi_{m,\gamma}^{(v)}(A)~\Big|~\widehat{\frac{1}{\sqrt{\det q}}}~\Big|~\Psi_{m,\gamma}^{(v)}(A)~ \Big>\le}~~~~
   \nonumber\\
   &\le&\displaystyle\frac{P}{\kappa^6\hbar^6}\sum_{v\in V(\gamma)}
    (36)^2\frac{(\ell_P)^{9}(9M)^6|Z|^\frac{3}{2}\big[\frac{2A}{T^2}\big]^6}
     {\prod_{I,i}[1+K_{t(I)}^{i}]}
     \Bigg[\bigg[\frac{3Mp}{4T}\bigg]^6+\sum\limits_{n=1}^{6}\frac{6!}{(6-n)!~n!} \bigg[\frac{3Mp}{4T}\bigg]^{6-n}
     \prod_{l=1}^n\bigg[\frac{3M+2(l-1)}{4}\bigg]\Bigg]~~~~~~~~~   
\ea

Here we again have estimated all sign factors by 1 and also neglected the signs contributed from the $\epsilon_{ijk}, \epsilon_{lmn}$-symbols (which gives the $36^2=(3!)^4$ prefactor).
\vfill

~\\
\\
\\
\section*{Acknowledgements}

J.B. thanks the Gottlieb Daimler-- and Karl Benz--Foundation and the DAAD 
(German Academic 
Exchange Service) for financial support. This work was supported in part
by a grant from NSERC of Canada to the Perimeter Institute for Theoretical 
Physics.

%%%%%%%%%%%%%%%%%%%%%%%%%%%%%%%%%%%%%%%%%%%%%%%%%%%%%%%%%%%%%%%%%%%%%%%%%%%%%%%%%%%%%%%%%%%%%%%%%%%%%%%%
\pagebreak
\begin{appendix}

%%%%%%%%%%%%%%%%%%%%%%%%%%%%%%%%%%%%%%%%%%%%%%%%%%%%%%%%%%%%%%%%%%%%%%%%%%%%%%%%%%%%%%%%%%%%%%%%%%%%%%%%
\section{\label{SU(2) properties}$SU(2)$ properties (definig representation)}
%%%%%%%%%%%%%%%%%%%%%%%%%%%%%%%%%%%%%%%%%%%%%%%%%%%%%%%%%%%%%%%%%%%%%%%%%%%%%%%%%%%%%%%%%%%%%%%%%%%%%%%%

  We have the $\tau$-matrices given by $\tau_k:=-\mb{i}\sigma_k$ with $\sigma_k$ the Pauli-matrices:
  \[\begin{array}{ccccc}
      \sigma_1=\left(\begin{array}{cc}
                     0 & 1 \\
		     1 & 0
                 \end{array}\right)
    &&\sigma_2=\left(\begin{array}{cc}
                     0 & -\mb{i}~ \\
		     \mb{i} & 0
                 \end{array}\right)		 
    &&\sigma_3=\left(\begin{array}{cc}
                     1 & 0 \\
		      0 & -1
                 \end{array}\right)	~~~~~~~\mbox{with}~~~~~\big[\sigma_i,\sigma_j \big]=2\mb{i}\,\epsilon^{ijk}\sigma_k	 
        \\\\
  
      \tau_1=\left(\begin{array}{cc}
                     ~0 & -\mb{i} \\
		     -\mb{i}~ & 0
                 \end{array}\right)
    &&\tau_2=\left(\begin{array}{cc}
                     0 & -1~ \\
		     1 & 0
                 \end{array}\right)		 
    &&\tau_3=\left(\begin{array}{cc}
                     -\mb{i}~ & 0~ \\
		      0 & \mb{i}
                 \end{array}\right)	~~~~~~~\mbox{with}~~~~~\big[\tau_i,\tau_j \big]=2\,\epsilon^{ijk}\tau_k	 
  \end{array}\]

  Additionally we use 
  
  \[\begin{array}{cc}\epsilon=-\tau_2=\left(\begin{array}{cc}
                                 	0 & 1\\
				       -1~ & 0   
                                \end{array} \right) 
				~~~~~\mbox{with the obvious properties}~~~~~
				\epsilon^{-1}=\epsilon^T=-\epsilon=\left(\begin{array}{cc}
                                 	0 & -1\\
				        1 &  0   
                                \end{array} \right)
		   \end{array}\]
  
  In the defining rep. of SU(2) we have for a group element $h\in G$:
  \[\begin{array}{rclll}
  
      h&=&\left(\begin{array}{cc}
                    ~a & b \\
		    -\overline{b}~ & \overline{a}
		\end{array}\right)~~~~~~~~~~~~~  \mbox{with}~~\det{h}=|a|^2+|b|^2=1 
      
      \\\\
	
      h^{-1}&=&\epsilon~h^T~\epsilon^{-1} = \left(\begin{array}{cc}
                                                  \overline{a} & -b \\
		                                  \overline{b}~ & ~a
		                                  \end{array}\right) 
      \\\\	

      \overline{h}&=&\big[h^{-1}\big]^T
                   =\big[\epsilon~h^T~\epsilon^T\big]^T 
		   =\epsilon~h~\epsilon^T
		   =\epsilon~h~\epsilon^{-1}
   \end{array}\]
   
   For the $\tau_k$'s we additionally have
   \[
     -\overline{\tau_k}^T = \tau_k
   \]
   Here always the overline means complex conjugation of the matrix elements and $~^T$ means transpose. 
  
\pagebreak

%%%%%%%%%%%%%%%%%%%%%%%%%%%%%%%%%%%%%%%%%%%%%%%%%%%%%%%%%%%%%%%%%%%%%%%%%%%%%%%%%%%%%%%%%%%%%%%%%%%%%%%%
\section{Action of a Holonomy on a Recoupling Scheme \label{Action of a Holonomy on a Recoupling Scheme}}
%%%%%%%%%%%%%%%%%%%%%%%%%%%%%%%%%%%%%%%%%%%%%%%%%%%%%%%%%%%%%%%%%%%%%%%%%%%%%%%%%%%%%%%%%%%%%%%%%%%%%%%%

%%%%%%%%%%%%%%%%%%%%%%%%%%%%%%%%%%%%%%%%%%%%%%%%%%%%%%%%%%%%%%%%%%%%%%%%%%%%%%%%%%%%%%%%%%%%%%%%%%%%%%%%
\subsection{Representation Matrix Elements of $SU(2)$ Elements}
%%%%%%%%%%%%%%%%%%%%%%%%%%%%%%%%%%%%%%%%%%%%%%%%%%%%%%%%%%%%%%%%%%%%%%%%%%%%%%%%%%%%%%%%%%%%%%%%%%%%%%%%

For the matrix element $\big[\pi_j(h)\big]_{mn}$ of an irreducible representation with weight $j$ of a holonomy $h \in SU(2)$ we have:
\ba \label{Formel Darstellungsmatrixelement}
   \big[\pi_j(h)\big]_{mn} 
   &=& \sqrt{\left( 2j \atop j+m \right)}
       \sqrt{\left( 2j \atop j+n \right)}
       ~h_{(A_1\,B_1}h_{A_2\,B_2}\ldots h_{A_n)\,B_n}
   \nonumber\\
   &=& \underbrace{\Big[\frac{(2j)!}{(j-m)!~(j+m)!}\Big]^{\frac{1}{2}}}_{\displaystyle c_{jm}}
       \underbrace{\Big[\frac{(2j)!}{(j-n)!~(j+n)!}\Big]^{\frac{1}{2}}}_{\displaystyle c_{jn}}~
       h_{(A_1\,B_1}h_{A_2\,B_2}\ldots h_{A_N)\,B_N}
\ea

where the round bracket means symmetrization wrt. the $A_k$'s and $h_{A_kB_k}$ is a matrix element of the defining ($j=\frac{1}{2}$) representation:
\be
  h_{A\,B} := \big[\pi_{j=\frac{1}{2}}\big]_{AB} \mbox{~~~with~~~}
          h= \left(\begin{array}{cc}
	                   a & b\\
			   -\bar{b} & \bar{a}\\
                        \end{array}\right) ~~~~~ \det{h}=h_{11}h_{22}-h_{21}h_{12}=|a|^2+|b|^2=1
\ee

Here we use the conventions:
\[\begin{array}{ccccccccccc}
    h_{1\,1}&=&h_{-\frac{1}{2}~-\frac{1}{2}}& =& a
    &~~~~~~~&
    h_{1\,2}&=&h_{-\frac{1}{2}~\frac{1}{2}}&=& b\\
    h_{2\,1}&=&h_{\frac{1}{2}~-\frac{1}{2}}&=& -\bar{b}
    &~~~~~~~&
    h_{2\,2}&=&h_{\frac{1}{2}~\frac{1}{2}}&=&\bar{a}
\end{array}\] \\[-14mm]\be \label{Konvention 1} \ee

We have
\[\begin{array}{ccccccccc}
      N=2j&~~~~~~&m&=&A_1+\ldots+A_n  &~~~~~~&j+m:& \#A_k = +\frac{1}{2} \\
                  &&n&=&B_1+\ldots+B_n&~~~~~~&j-m:& \#A_k = -\frac{1}{2}\\\\
                                        &&&&&&j+n:& \#B_k = +\frac{1}{2} \\
                                        &&&&&&j-n:& \#B_k = -\frac{1}{2}\\
\end{array}\] \\[-22mm]\be \label{Konvention 2} \ee \\[4mm]

We can explicitely check the representation property:
\[\begin{array}{ccllll}
   \displaystyle\sum_{l=-j}^j \big[\pi_j(h)\big]_{ml} \big[\pi_j(g)\big]_{ln}
   &=& c_{jm}~c_{jn}~
   {\displaystyle\sum_{l=B_1+\ldots+B_N}} 
   (c_{jl})^2~\cdot~
   h_{(A_1B_1}h_{A_2B_2}\ldots h_{A_N)B_N}
   \cdot~
   g_{(B_1C_1}g_{B_2C_2}\ldots g_{B_N)C_N} 
   
   \\\\
   
   &=&c_{jm}~c_{jn}~
   {\displaystyle\sum_{l=B_1+\ldots+B_N}} 
   (c_{jl})^2~\cdot~
   h_{(A_1B_1}h_{A_2B_2}\ldots h_{A_N)B_N}
   \cdot~
   g_{B_1(C_1}g_{B_2C_2}\ldots g_{B_NC_N)} 
   
   \\\\ 
   
   &=&c_{jm}~c_{jn}~
   {\displaystyle\sum_{l=B_1+\ldots+B_N}} 
   \left(2j \atop j+l \right)~\cdot~
   h_{(A_1B_1}h_{A_2B_2}\ldots h_{A_N)B_N}
   \cdot~
   g_{B_1(C_1}g_{B_2C_2}\ldots g_{B_NC_N)} 
   
   \\\\

   &&\fcmt{12}{ Note, that we sum here over one fixed but arbitrary combination $B_1,\ldots,B_n$.            
	    Because of the symmetrization in the $A_k$'s and the $C_k$'s each such arbitrary combination is equivalent to summing over all index-combinations in the tensor space that fulfill 
	    $l=B_1+\ldots+B_N$ (with $j+l:~~\#B_k=+\frac{1}{2}$~~~ $ j-l:~~\#B_k=-\frac{1}{2}$)
				
      Multiplication in the tensor space requires to perform a sum  $\sum_{B_1, \ldots,B_N =\pm \frac{1}{2}}$ over all possible $B_k$-combinations.
      Now this summation contains  
      $\left(2j \atop j+l \right)$ configurations with $B_1+\ldots+B_N=l$ each giving the same contribution to the sum due to the symmetrization in the $A_k$'s and the $C_k$'s.}
   
   \\\\   
      
   &=&c_{jm}~c_{jn}~
   {\displaystyle\sum_{B_1,\ldots,B_N=\pm \frac{1}{2}}} 
   ~
   h_{(A_1B_1}h_{A_2B_2}\ldots h_{A_N)B_N}
   \cdot~
   g_{B_1(C_1}g_{B_2C_2}\ldots g_{B_NC_N)} 
   
   \\\\ 

   &=&c_{jm}~c_{jn}~
      \big[hg\big]_{(A_1(C_1}\big[hg\big]_{A_2C_2}\ldots \big[hg\big]_{A_N)C_N)}
      
   \\\\ 

   &=&c_{jm}~c_{jn}~
      \big[hg\big]_{(A_1C_1}\big[hg\big]_{A_2C_2}\ldots \big[hg\big]_{A_N)C_N}
      
   \\\\ 

   &=&\big[\pi_j(hg)\big]_{mn}

\end{array}\]

%%%%%%%%%%%%%%%%%%%%%%%%%%%%%%%%%%%%%%%%%%%%%%%%%%%%%%%%%%%%%%%%%%%%%%%%%%%%%%%%%%%%%%%%%%%%%%%%%%%%%%%%
\subsection{Adding one more Irreducible Representation - Tensorporduct}
%%%%%%%%%%%%%%%%%%%%%%%%%%%%%%%%%%%%%%%%%%%%%%%%%%%%%%%%%%%%%%%%%%%%%%%%%%%%%%%%%%%%%%%%%%%%%%%%%%%%%%%%
 
\[\begin{array}{cccccccc}
    \lefteqn{h_{A_0B_0}~h_{(A_1B_1}h_{A_2B_2}\ldots h_{A_N)B_N} =}~~~~~\\\\
    &=&&\displaystyle\frac{1}{N!} \sum_{\pi(1\ldots N)}&& 
       h_{A_0B_0}~h_{A_{\pi(1)}B_1}h_{A_{\pi(2)}B_2}\ldots h_{A_{\pi(N)}B_N}  
       
    \\\\
       
    &=&\displaystyle\frac{1}{N+1}&\displaystyle\frac{1}{N!} \sum_{\pi(1\ldots N)}&\Big\{& 
        h_{A_0B_0}~h_{A_{\pi(1)}B_1}h_{A_{\pi(2)}B_2}\ldots h_{A_{\pi(N)}B_N}
    &+&h_{A_{\pi(1)}B_0}~h_{A_0B_1}h_{A_{\pi(2)}B_2}\ldots h_{A_{\pi(n)}B_n}
    \\
    &&&&&&+&h_{A_{\pi(2)}B_0}~h_{A_{\pi(1)}B_1}h_{A_0B_2}\ldots h_{A_{\pi(N)}B_N}
    \\
    &&&&&&\vdots&\\
    &&&&&&+&h_{A_{\pi(N)}B_0}~h_{A_{\pi(1)}B_1}h_{A_{\pi(2)}B_2}\ldots h_{A_0B_N}
    
    \\\\\\\
    
    &&&&+&h_{A_0B_0}~h_{A_{\pi(1)}B_1}h_{A_{\pi(2)}B_2}\ldots h_{A_{\pi(N)}B_N}
       &-&h_{A_{\pi(1)}B_0}~h_{A_0B_1}h_{A_{\pi(2)}B_2}\ldots h_{A_{\pi(N)}B_N}
    \\\\
    &&&&+&h_{A_0B_0}~h_{A_{\pi(1)}B_1}h_{A_{\pi(2)}B_2}\ldots h_{A_{\pi(N)}B_N}
       &-&h_{A_{\pi(2)}B_0}~h_{A_{\pi(1)}B_1}h_{A_0B_2}\ldots h_{A_{\pi(N)}B_N}
       
    \\
    &&&&\vdots&&\vdots
    \\
    &&&&+&h_{A_0B_0}~h_{A_{\pi(1)}B_1}h_{A_{\pi(2)}B_2}\ldots h_{A_{\pi(N)}B_N}
       &-&h_{A_{\pi(N)}B_0}~h_{A_{\pi(1)}B_1}h_{A_{\pi(2)}B_2}\ldots h_{A_0B_N} ~~\Big\}

    \\
    
    &&\multicolumn{6}{l}{
      \fcmt{14.5}{
            Note, that the first terms in this sum are the symmetrization
	    \[
	      \frac{1}{(N+1)!} \sum_{\pi(0\ldots N)}
	      h_{A_{\pi(0)}B_0}~h_{A_{\pi(1)}B_1}h_{A_{\pi(2)}B_2}\ldots h_{A_{\pi(N)}B_N} 
	    \]
	   
	   Furthermore note, that\\[-5mm]	
	   \[
	      h_{A_1B_1} h_{A_2B_2} - h_{A_2B_1} h_{A_1B_2} = \epsilon_{A_1A_2} \epsilon_{B_1B_2} \\
	   \]
	   for the conventions\\[-6mm]   
	   \[    
	       \epsilon=\left(\begin{array}{cc}
	                   0 & 1\\
			   -1 & 0\\
                  \end{array}\right)~~~~~~~~
	        h= \left(\begin{array}{cc}
	                   a & b\\
			   -\bar{b} & \bar{a}\\
                  \end{array}\right) ~
          \]
	  
	  since by definition ~~ $\det{h}=h_{11}h_{22}-h_{21}h_{12}=|a|^2+|b|^2=1 $ 
          }} 
      
      \\\\
      
      &=&\multicolumn{6}{l}{
            h_{(A_0B_0}h_{A_1B_1}h_{A_2B_2}\ldots h_{A_N)B_N}
         } 
	 
      \\\\
      
    &&+~\displaystyle\frac{1}{N+1}&\displaystyle\frac{1}{N!} \sum_{\pi(1\ldots N)}&\Big\{& 
        
     \multicolumn{3}{l}{\epsilon_{A_0A_{\pi(1)}}\epsilon_{B_0B_1}~
     \underbrace{h_{A_{\pi(2)}B_2}h_{A_{\pi(3)}B_3}\ldots h_{A_{\pi(N)}B_N}
     }_{\displaystyle R_1}}
     \\
    &&&&+&\multicolumn{3}{l}{\epsilon_{A_0A_{\pi(2)}}\epsilon_{B_0B_2}~
    \underbrace{h_{A_{\pi(1)}B_1}h_{A_{\pi(3)}B_3}\ldots h_{A_{\pi(N)}B_N}
    }_{\displaystyle R_2}}
       
    \\
    &&&&\vdots&\vdots&
    \\
    &&&&+&\multicolumn{3}{l}{\epsilon_{A_0A_{\pi(N)}}\epsilon_{B_0B_N}~
    \underbrace{h_{A_{\pi(1)}B_1}h_{A_{\pi(2)}B_2}\ldots h_{A_{\pi(N-1)}B_{N-1}}
        }_{\displaystyle R_N}
    ~~\Big\} } 
          
\end{array}\]\\[-12mm] \be\label{Holonomie Multiplikation 1} \ee

In (\ref{Holonomie Multiplikation 1}) we have the following contributions (using the conventions (\ref{Konvention 1}), (\ref{Konvention 2})):
\begin{enumerate}
   \item{$(j\mp n)$ rows will contribute for $ B_0=\pm\frac{1}{2}$ due to $\epsilon_{B_0B_k}$ }
   \item{if we carry out the sum over all permutations $\pi(1\ldots N)$ then in each of the contributing rows we will find $(j\mp m)$ non-zero combinations $\epsilon_{A_0A_{\pi(k)}}$ for $A_0=\pm \frac{1}{2}$ and fixed $A_{\pi(k)}$}
   \item{for each such fixed combination there remain $(2j-1)!=(N-1)!$ permutations of within the $R_k$-terms}
   \item{The contributing $R_k$-terms have $m_{new}^{(i)}=\sum_kA_k (R_i)=m\!+\!A_0$ , $n_{new}^{(i)}=\sum_{k}B_k(R_i)=n\!+\!B_0$ since the contraction with the $\epsilon$'s 'anihilates' one $A_k=\mp\frac{1}{2}$ resp. $B_k=\mp\frac{1}{2}$ for $A_0=\pm\frac{1}{2}$ resp. $B_0=\pm\frac{1}{2}$.  }
   \item{Each $\epsilon$ contributes  a $\pm$ sign}
\end{enumerate}

Now we can discuss the contribution of the second term in (\ref{Holonomie Multiplikation 1}) dependent on the values of $A_0$, $B_0$. Using the conventions (\ref{Konvention 1}), (\ref{Konvention 2}) and the observations made we can write:
\\

\renewcommand{\arraystretch}{1.5}
\begin{tabular}{|c|c||c|c||c|c||c|c|c|c|}\hline
                  && \multicolumn{2}{|c||}{contributing rows}
		  &\multicolumn{2}{|c||}{\cmt{3}{contributing terms in each of them\\}}&&&&  \\ \cline{3-6} 
    $A_0$ & $B_0$ & $\epsilon_{B_0B_k}\ne 0$ iff       & \cmt{1}{valid for $\#~ R_i$}
		  &$\epsilon_{A_0A_{\pi(k)}}\ne 0$ iff&\cmt{1}{valid for terms}
		   &\cmt{2.3}{$\#$ permutations inside $R_i$ }
		   &$m_{new}^{(i)}$&$n_{new}^{(i)}$&prefactor\\\hline\hline 
       		   
	$1 \Leftrightarrow -\frac{1}{2}$    &   $1 \Leftrightarrow -\frac{1}{2}$ 
       &$B_k\stackrel{!}{=}2\Leftrightarrow +\frac{1}{2}$ & $(j+n)$
       &$A_{\pi(k)}\stackrel{!}{=}2\Leftrightarrow +\frac{1}{2}$ & $(j+m)$ 
       & $(2j -1)!$ 
       &$m-\frac{1}{2}$&$n-\frac{1}{2}$&$\epsilon_{12}\epsilon_{12}=1$
       \\\hline
       
	$1 \Leftrightarrow -\frac{1}{2}$    &   $2 \Leftrightarrow +\frac{1}{2}$ 
       &$B_k\stackrel{!}{=}1\Leftrightarrow -\frac{1}{2}$ & $(j-n)$
       &$A_{\pi(k)}\stackrel{!}{=}2\Leftrightarrow +\frac{1}{2}$ & $(j+m)$ 
       & $(2j -1)!$ 
       &$m-\frac{1}{2}$&$n+\frac{1}{2}$&$\epsilon_{12}\epsilon_{21}=-1$
       \\\hline

       	$2 \Leftrightarrow +\frac{1}{2}$    &   $1 \Leftrightarrow -\frac{1}{2}$ 
       &$B_k\stackrel{!}{=}2\Leftrightarrow +\frac{1}{2}$ & $(j+n)$
       &$A_{\pi(k)}\stackrel{!}{=}1\Leftrightarrow -\frac{1}{2}$ & $(j-m)$ 
       & $(2j -1)!$ 
       &$m+\frac{1}{2}$&$n-\frac{1}{2}$&$\epsilon_{21}\epsilon_{12}=-1$
       \\\hline

	$2 \Leftrightarrow +\frac{1}{2}$    &   $2 \Leftrightarrow +\frac{1}{2}$ 
       &$B_k\stackrel{!}{=}1\Leftrightarrow -\frac{1}{2}$ & $(j-n)$
       &$A_{\pi(k)}\stackrel{!}{=}1\Leftrightarrow -\frac{1}{2}$ & $(j-m)$ 
       & $(2j -1)!$ 
       &$m+\frac{1}{2}$&$n+\frac{1}{2}$&$\epsilon_{21}\epsilon_{21}=1$
       \\\hline     
\end{tabular}
\\[5mm]

Thus the sum in (\ref{Holonomie Multiplikation 1}) can then be rewritten as ($N=2j$, using $A_0,B_0=\pm\frac{1}{2}$):
\[\begin{array}{cclc}
    \lefteqn{h_{A_0B_0}~h_{(A_1B_1}h_{A_2B_2}\ldots h_{A_N)B_N} =}~~~~~\\\\
    
      &=& h_{(A_0B_0}h_{A_1B_1}h_{A_2B_2}\ldots h_{A_N)B_N} \\
          
      &&	+~4~A_0B_0~\displaystyle\frac{1}{2j+1}\displaystyle\frac{1}{(2j)!}~(2j-1)!~(j-2A_0\,m)~(j-2B_0\,n)  
        ~~\Big[ h_{(\tilde{A}_1\tilde{B}_1} h_{\tilde{A}_2\tilde{B}_2}\ldots h_{\tilde{A}_{N-1})\tilde{B}_{N-1}}\Big]\bigg|_{\tilde{A}_1+\ldots +\tilde{A}_{N-1}=m+A_0
	                                          \atop
						  \tilde{B}_1+\ldots +\tilde{B}_{N-1}=n+B_0} 
      \\\\
      &=& h_{(A_0B_0}h_{A_1B_1}h_{A_2B_2}\ldots h_{A_N)B_N}\\ 
      && +~4~A_0B_0~\displaystyle\frac{(j-2A_0\,m)~(j-2B_0\,n)}{(2j+1)~2j}
        ~~\Big[ h_{(\tilde{A}_1\tilde{B}_1} h_{\tilde{A}_2\tilde{B}_2}\ldots h_{\tilde{A}_{N-1})\tilde{B}_{N-1}}\Big]\bigg|_{\tilde{A}_1+\ldots +\tilde{A}_{N-1}=m+A_0
	                                          \atop
						  \tilde{B}_1+\ldots +\tilde{B}_{N-1}=n+B_0}

\end{array}\]

Using  (\ref{Formel Darstellungsmatrixelement}) this can be finally rewritten as :

\[\begin{array}{cclc}
    \lefteqn{~\big[\pi_{\frac{1}{2}}(h)\big]_{A_0B_0}~
	     \big[\pi(h)_j\big]_{mn}=}~~~~~~~~\\\\
    
     &=& \frac{(2j)!}{\sqrt{(j+m)!\,(j-m)!\,(j+n)!\,(j-n)!}}~
         \frac{\sqrt{(j+\frac{1}{2}+m+A_0)!\,(j+\frac{1}{2}-m-A_0)!\,
	                          (j+\frac{1}{2}+n+B_0)!\,(j+\frac{1}{2}-n-B_0)!}}{(2(j+\frac{1}{2}))!} 
          
	  ~\big[\pi_{j+\frac{1}{2}}(h)\big]_{m+A_0~n+B_0}\\\\ 
      
      && +\frac{(2j)!}{\sqrt{(j+m)!\,(j-m)!\,(j+n)!\,(j-n)!}}
         \frac{\sqrt{(j-\frac{1}{2}+m+A_0)!\,(j-\frac{1}{2}-m-A_0)!\,
	                          (j-\frac{1}{2}+n+B_0)!\,(j-\frac{1}{2}-n-B_0)!}}{(2(j-\frac{1}{2}))!}
				  \times\\ 

      &&~~~~~~~~~~~~~~~~~~~~\times~4~A_0B_0~\displaystyle\frac{(j-2A_0\,m)~(j-2B_0\,n)}{(2j+1)~2j}
        
         ~\big[\pi_{j-\frac{1}{2}}(h)\big]_{m+A_0~n+B_0}

    \\\\&&\fcmt{5.5}{just set $A_0,B_0=\pm\frac{1}{2}$ and carefully cancel according terms}

\end{array}\] 

\fcmt{17}{ \ba \label{Endformel Holonomiemultiplikation j=1/2}
              \big[\pi_{\frac{1}{2}}(h)\big]_{A_0B_0}~
	           \big[\pi(h)_j\big]_{mn}
	      &=& \displaystyle\frac{\sqrt{(j+2A_0m +1)(j+2B_0n +1)}}{(2j+1)}          
	         ~\big[\pi_{j+\frac{1}{2}}(h)\big]_{m+A_0~n+B_0}~\nonumber\\  
               &&+~4~A_0B_0~\displaystyle\frac{\sqrt{(j-2A_0\,m)~(j-2B_0\,n)}}{(2j+1)}
                  ~\big[\pi_{j-\frac{1}{2}}(h)\big]_{m+A_0~n+B_0} 
	   \ea }

 \fbox{where $-j\le m,n\le j$ and $A_0,B_0 = \pm \frac{1}{2}$}

%%%%%%%%%%%%%%%%%%%%%%%%%%%%%%%%%%%%%%%%%%%%%%%%%%%%%%%%%%%%%%%%%%%%%%%%%%%%%%%%%%%%%%
\subsection{Action on a SNF}
%%%%%%%%%%%%%%%%%%%%%%%%%%%%%%%%%%%%%%%%%%%%%%%%%%%%%%%%%%%%%%%%%%%%%%%%%%%%%%%%%%%%%%%
We have to work out the exact action of a holonomy acting on the gauge invariant SNF:
We will use the general expression (\ref{Endformel Holonomiemultiplikation j=1/2}).

If we use the correspondence for matrix elements of a representation $\sqrt{2j+1}~ \big[\pi_{j}(g)\big]_{m n} = \big|~j~m~;~n~\big>$~~~\footnote{For clarity of the notation we will be a bit sloppy in our notation here. Correctly this correspondence has to be written as\linebreak
$\sqrt{2j+1}~ \big[\pi_{j}(g)\big]_{m n} =\big<~g~ \big|~j~m~;~n~\big>$ and thus expressing the orthonormality of the spin states:\\ 
$\big<\,j'\,m'\,;n'\,\big|\,j\,m\,;\,n\,\big>
  =\displaystyle\int\limits_{SU(2)}d\mu_H(g) \big<\,j'\,m'\,;\,n'\, \big|\,g\,\big>\big<\,g\,\big|\,j\,m\,;\,n\,\big>
  =(2j+1)\cdot\displaystyle\int\limits_{SU(2)}d\mu_H(g) \overline{\big(\big[\pi_{j'}(g)\big]_{m'n'}\big)}\big[\pi_{j}(g)\big]_{mn}
  =\delta_{j'j}\,\delta_{m'm}\,\delta_{n'n}$ as stated by the $Peter\&Weyl~ theorem$. 
  So within our calculations we will use $\sqrt{2j+1}~\big[\pi_{j}(g)\big]_{m n}$ in order to express the function $\sqrt{2j+1}~\big[\pi_{j}(\cdot)\big]_{m n}=:\big|~j~m~;~n~\big>$. } 
  we can rewrite
(\ref{Endformel Holonomiemultiplikation j=1/2}) as:
\ba\label{einfache Holonomiemultiplikation 2}
  \big[\pi_{\frac{1}{2}}(g)\big]_{A_0 B_0} \big|~j~m~;~n~\big>
  &=&
  \underbrace{\Bigg[\frac{\big[(j+2A_0m+1)(j+2B_0n+1)}{(2j+1)(2j+2)}\Bigg]^{\frac{1}{2}}}_{\displaystyle C_{(+)}}
   \big|~j+\frac{1}{2}~m+A_0~;~n+B_0~\big>\nonumber\\
  && \oplus ~\underbrace{4~A_0 B_0 \Bigg[\frac{\big[(j-2A_0m)(j-2B_0n)}{2j~(2j+1)} \Bigg]^{\frac{1}{2}}}_{\displaystyle C_{(-)}}
   \big|~j-\frac{1}{2}~m+A_0~;~n+B_0~\big>
\ea  
We can furthermore rewrite (\ref{einfache Holonomiemultiplikation 2}) by realizing that parts of the coefficients $C_{(+)}$ an $C_{(-)}$ correspond to Clebsch Gordan coefficients:
\ba
   \Bigg[\frac{\big[(j+2A_0m+1)}{(2j+1)}\Bigg]^{\frac{1}{2}}
   &=&
   \big<~j~m~;~\frac{1}{2}~A_0~\big|~j\!+\!\frac{1}{2}~m\!+\!A_0~\big>
   \nonumber\\
   -2~A_0~\Bigg[\frac{\big[(j-2A_0m)}{(2j+1)}\Bigg]^{\frac{1}{2}}
   &=&
   \big<~j~m~;~\frac{1}{2}~A_0~\big|~j\!-\!\frac{1}{2}~m\!+\!A_0~\big>
\ea

So we can finally write 
\ba \label{einfache Holonomiemultiplikation 3}
  \big[\pi_{\frac{1}{2}}(g)\big]_{A_0 B_0} \big|~j~m~;~n~\big> 
  &=&
  \overbrace{\Bigg[\frac{j+2B_0n+1}{2~(j+1)}\Bigg]^{\frac{1}{2}}}^{\displaystyle C^j_{\tilde{j}}(B_0,n)\big|_{\tilde{j}=j+\frac{1}{2}}}
  \big<~j~m~;~\frac{1}{2}~A_0~\big|~j\!+\!\frac{1}{2}~m\!+\!A_0~\big>\cdot
   \big|~j\!+\!\frac{1}{2}~m\!+\!A_0~;~n\!+\!B_0~\big>
  \nonumber
  \\
  &&\oplus~\underbrace{(-2)~B_0 \Bigg[\frac{j-2B_0n}{2~j} \Bigg]^{\frac{1}{2}}}_{\displaystyle C^j_{\tilde{j}}(B_0,n)\big|_{\tilde{j}=j-\frac{1}{2}}}
   \big<~j~m~;~\frac{1}{2}~A_0~\big|~j\!-\!\frac{1}{2}~m\!+\!A_0~\big>\cdot
   \big|~j\!-\!\frac{1}{2}~m\!+\!A_0~;~n\!+\!B_0~\big>
   \nonumber
   \\
   &=&\bigoplus_{\tilde{j}=j\pm\frac{1}{2}} C^j_{\tilde{j}}(B_0,n)~
   \big<~j~m~;~\frac{1}{2}~A_0~\big|~\tilde{j}~m\!+\!A_0~\big>\cdot
   \big|~\tilde{j}~m\!+\!A_0~;~n\!+\!B_0~\big> 
\ea
where we have introduced the coefficients $C^j_{\tilde{j}}(B_0,n)$ \\\\
\fcmt{17}{So we notice a fundamental difference between {\it internal}
(tensorproduct of representations of the same edge-holonomy) and {\it external} (tensorproduct of representations of holonomies of different edges) recoupling !} \vfill

\pagebreak
%%%%%%%%%%%%%%%%%%%%%%%%%%%%%%%%%%%%%%%
\subsection{N-Vertex}
%%%%%%%%%%%%%%%%%%%%%%%%%%%%%%%%%%%%%%%

\subsubsection{Recoupling Schemes}
In this section we will briefly review the definitions of recoupling schemes. We will closely follow \cite{Volume_Article_I}, \cite{TT:Closed ME of V in LQG}.

In what follows we will frequently use \fbox{$\tilde{m}_k:=m_1+m_2+\ldots+m_k$}.
\\

\paragraph{A general (standard-)recoupling scheme} is defined as follows:
\begin{footnotesize}

\[\begin{array}{ccl}
   \lefteqn{\big|\,\vec{a}(12)~J~M~;~\vec{n}\, \big>=}\\\\
   && \multicolumn{1}{l}{\hspace{1mm}=\big|\,a_2(j_1\,j_2)~a_3(a_2\,j_3)\ldots
       a_{K-1}(a_{K-2}\,j_K)~a_K(a_{K-1}\,j_k)~a_{K+1}(a_K\,j_{K-1})\ldots 
       a_{N-1}(a_{N-2}\,j_{N-1})~J(a_{N-1}\,j_N)~M~;~n_1\ldots n_N\, \big>}\\
   &&\left.\begin{array}{cl}
             =~~\displaystyle\sum_{m_1+\ldots+m_N=M}&
	       \big<~j_1~m_1~;~j_2~m_2 ~\big|~a_2(j_1\,j_2)~\tilde{m}_2 ~\big>\\[-3mm]
	     & \big<~a_2~\tilde{m}_2~;~j_3~m_3~\big|~a_3(a_2\,j_3)~\tilde{m}_3~\big>\\
	     & ~~\vdots\\
	     &\big<~a_{K-2}~\tilde{m}_{K-2}~;~j_{K-1}~m_{K-1}~\big|
	           ~a_{K-1}(a_{K-2}\,j_{K-1})~\tilde{m}_{K-1}~\big>\\
             &\big<~a_{K-1}~\tilde{m}_{K-1}~;~j_{K}~m_{K}~\big|
	           ~a_{K}(a_{K-1}\,j_{K})~\tilde{m}_{K}~\big>\\
             &\big<~a_{K}~\tilde{m}_{K}~;~j_{K+1}~m_{K+1}~\big|
	           ~a_{K+1}(a_{K}\,j_{K+1})~\tilde{m}_{K+1}~\big>\\		   
	     & ~~\vdots\\
             &\big<~a_{N-2}~\tilde{m}_{N-2}~;~j_{N-1}~m_{N-1}~\big|
	           ~a_{N-1}(a_{N-2}\,j_{N-1})~\tilde{m}_{N-1}~\big>\\
             &\big<~a_{N-1}~\tilde{m}_{N-1}~;~j_{N}~m_{N}~\big|
	           ~J(a_{N-1}\,j_{N})~M~\big>\\		   
	     
           \end{array}\right\} \fbox{N-1 terms} \\\\
    
     &&\hspace{2.7cm}	   	   
       \big|\,j_1\,m_1\,;\,n_1 \,\big> \,\otimes
       \big|\,j_2\,m_2\,;\,n_2 \,\big> \,\otimes\,\ldots\,\otimes
       \big|\,j_{K-1}\,m_{K-1}\,;\,n_{K-1} \,\big> \,\otimes
       \big|\,j_K\,m_K\,;\,n_K \,\big> \,\otimes
       \big|\,j_{K+1}\,m_{K+1}\,;\,n_{K+1} \,\big>\,\otimes \\
     &&\hspace{64.5mm}  \,\otimes\,\ldots\,\otimes
       \big|\,j_{N-1}\,m_{N-1}\,;\,n_{N-1} \,\big> \,\otimes
       \big|\,j_N\,m_N\,;\,n_N \,\big> 
         
\end{array}\] \\[-8mm]\be\label{Definition recoupling scheme} \ee\\[-8mm]
\end{footnotesize}

%%%%%%%%%%%%%%%%%%%%%%%%%%%%%%%%%%%%%%%%%%%%%%%%%%%%%%%%%%%%%%%%%%%%%%%%%%%%%%%%%%%%%%%%%%%%%%%%%%%%%
\paragraph{Orthogonality Relations Between Recoupling Schemes} For the scalar product of two recoupling schemes we have\footnote{We supress the quantum numbers $n_1,\ldots,n_N$ (in general we have $\delta_{\vec{n}'\,\vec{n}}$) and use the reality of the Clebsch Gordan coefficients}

\begin{footnotesize}
\[\begin{array}{cc@{~}|crclc}
   \lefteqn{\big<\,a_2'~a_3'~\ldots~a_{N-1}'~J'~M' \,\big|\,a_2~a_3~\ldots~a_{N-1}~J~M \,\big>=}\\\\
   
   &\multicolumn{2}{c}{=}&&\displaystyle\sum_{m_1+\ldots+m_N=M \atop m_1'+\ldots+m_N'=M'}
    &\big<\,j_1'\,m_1'~;~j_2'\,m_2' \,\big|\,a_2'\,\tilde{m}_2' \,\big>~
      \big<\,j_1\,m_1~;~j_2\,m_2 \,\big|\,a_2\,\tilde{m}_2 \,\big>
     \\[-5mm]  
  &&&&&\big<\,a_2'\,\tilde{m}_2'~;~j_3'\,m_3' \,\big|\,a_3'\,\tilde{m}_3' \,\big>~
      \big<\,a_2\,\tilde{m}_2~;~j_3\,m_3 \,\big|\,a_3\,\tilde{m}_3 \,\big>
     
     \\
  &&&&&~~\vdots
     \\
  
  &&&&&\big<\,a_{N-2}'\,\tilde{m}_{N-2}'~;~j_{N-1}'\,m_{N-1}' \,
      \big|\,a_{N-1}'\,\tilde{m}_{N-1}' \,\big>~
      \big<\,a_{N-2}\,\tilde{m}_{N-2}~;~j_{N-1}\,m_{N-1} \,\big|\,a_{N-1}\,\tilde{m}_{N-1} \,\big>
     \\  
  &&&&&\big<\,a_{N-1}'\,\tilde{m}_{N-1}'~;~j_N'\,m_N' \,\big|\,J'\,M' \,\big>~
      \big<\,a_{N-1}\,\tilde{m}_{N-1}~;~j_N\,m_N \,\big|\,J\,M\,\big>
     \\&&
     \\
  &&&&&\underbrace{\big<\,j_1'\,m_1'\,\big|\,j_1\,m_1\,\big>}
                _{\displaystyle\delta_{j_1'j_1}\delta_{m_1'm_1}} \cdot
      \underbrace{\big<\,j_2'\,m_2'\,\big|\,j_2\,m_2\,\big>}
                _{\displaystyle\delta_{j_2'j_2}\delta_{m_2'm_2}}~ \cdot
      ~\ldots~\cdot				    
      \underbrace{\big<\,j_{N-1}'\,m_{N-1}'\,\big|\,j_{N-1}\,m_{N-1}\,\big>} 
                _{\displaystyle\delta_{j_{N-1}'j_{N-1}}\delta_{m_{N-1}'m_{N-1}}} \cdot
      \underbrace{\big<\,j_2'\,m_2'\,\big|\,j_2\,m_2\,\big>}
                _{\displaystyle\delta_{j_N'j_N}\delta_{m_N'm_N}}
		
  \\&&\\
  
   &\multicolumn{2}{c}{=}&\delta_{MM'} &\displaystyle\sum_{m_1+\ldots+m_N=M }
    &\big<\,j_1\,m_1~;~j_2\,m_2 \,\big|\,a_2'\,\tilde{m}_2 \,\big>~
      \big<\,j_1\,m_1~;~j_2\,m_2 \,\big|\,a_2\,\tilde{m}_2 \,\big>
     \\[-3mm]  
  &&&&&\big<\,a_2\,\tilde{m}_2~;~j_3\,m_3 \,\big|\,a_3\,\tilde{m}_3 \,\big>~
      \big<\,a_2'\,\tilde{m}_2~;~j_3\,m_3 \,\big|\,a_3'\,\tilde{m}_3 \,\big>
     
     \\
  &&&&&~~\vdots
     \\
  
  &&&&&\big<\,a_{N-2}'\,\tilde{m}_{N-2}~;~j_{N-1}\,m_{N-1} \,
      \big|\,a_{N-1}'\,\tilde{m}_{N-1} \,\big>~
      \big<\,a_{N-2}\,\tilde{m}_{N-2}~;~j_{N-1}\,m_{N-1} \,\big|\,a_{N-1}\,\tilde{m}_{N-1} \,\big>
     \\  
  &&&&&\big<\,a_{N-1}'\,\tilde{m}_{N-1}~;~j_N\,m_N \,\big|\,J'\,M \,\big>~
      \big<\,a_{N-1}\,\tilde{m}_{N-1}~;~j_N\,m_N \,\big|\,J\,M\,\big>

  \\&&
  \\%%%%%%%%%%%%%%%%%%%%%%%%%%%%%%%%%%%%%%%%%%%%%%%%%%%%%%%%
  &&&\multicolumn{3}{l}{\fcmt{5}{Introduce:
                                $m_2=\tilde{m_2}-m_1$}}
  \\&&%%%%%%%%%%%%%%%%%%%%%%%%%%%%%%%%%%%%%%%%%%%%%%%%%%%%%%
  \\
   
  \end{array}\]

\[\begin{array}{cc@{~}|lcll}

   &\multicolumn{2}{c}{=}&\delta_{MM'} \displaystyle\sum_{m_1}&\displaystyle\sum_{\tilde{m}_2+m_3+\ldots+m_N=M }
    &\big<\,j_1\,m_1~;~j_2\,\tilde{m}_2-m_1 \,\big|\,a_2\,\tilde{m}_2 \,\big>~
      \big<\,j_1\,m_1~;~j_2\,\tilde{m}_2-m_1 \,\big|\,a_2'\,\tilde{m}_2 \,\big>
     \\[-3mm]  
  &&&&&\big<\,a_2\,\tilde{m}_2~;~j_3\,m_3 \,\big|\,a_3\,\tilde{m}_3 \,\big>~
      \big<\,a_2'\,\tilde{m}_2~;~j_3\,m_3 \,\big|\,a_3'\,\tilde{m}_3 \,\big>
     
     \\
  &&&&&~~\vdots
     \\
  
  &&&&&\big<\,a_{N-2}'\,\tilde{m}_{N-2}~;~j_{N-1}\,m_{N-1} \,
      \big|\,a_{N-1}'\,\tilde{m}_{N-1} \,\big>~
      \big<\,a_{N-2}\,\tilde{m}_{N-2}~;~j_{N-1}\,m_{N-1} \,\big|\,a_{N-1}\,\tilde{m}_{N-1} \,\big>
     \\  
  &&&&&\big<\,a_{N-1}'\,\tilde{m}_{N-1}~;~j_N\,m_N \,\big|\,J'\,M \,\big>~
      \big<\,a_{N-1}\,\tilde{m}_{N-1}~;~j_N\,m_N \,\big|\,J\,M\,\big>
  
  \\&&
  \\%%%%%%%%%%%%%%%%%%%%%%%%%%%%%%%%%%%%%%%%%%%%%%%%%%%%%%%%%%%%%%%%%%%%%%%%%%%%%%%%%%%%%%%%%%%%%%%
  &&&\multicolumn{3}{l}{\fcmt{12}{Use unitarity of the Clebsch Gordan coefficients in order to carry out the sum over $m_1$}}
  \\&&%%%%%%%%%%%%%%%%%%%%%%%%%%%%%%%%%%%%%%%%%%%%%%%%%%%%%%%%%%%%%%%%%%%%%%%%%%%%%%%%%%%%%%%%%%%%%%%
  \\

   &\multicolumn{2}{c}{=}&\delta_{MM'} \delta_{a_2'a_2}&\displaystyle\sum_{\tilde{m}_2+m_3+\ldots+m_N=M }
    &\big<\,a_2\,\tilde{m}_2~;~j_3\,m_3 \,\big|\,a_3\,\tilde{m}_3 \,\big>~
      \big<\,a_2\,\tilde{m}_2~;~j_3\,m_3 \,\big|\,a_3'\,\tilde{m}_3 \,\big>
     
     \\[-3mm]
  &&&&&~~\vdots
     \\
  
  &&&&&\big<\,a_{N-2}'\,\tilde{m}_{N-2}~;~j_{N-1}\,m_{N-1} \,
      \big|\,a_{N-1}'\,\tilde{m}_{N-1} \,\big>~
      \big<\,a_{N-2}\,\tilde{m}_{N-2}~;~j_{N-1}\,m_{N-1} \,\big|\,a_{N-1}\,\tilde{m}_{N-1} \,\big>
     \\  
  &&&&&\big<\,a_{N-1}'\,\tilde{m}_{N-1}~;~j_N\,m_N \,\big|\,J'\,M \,\big>~
      \big<\,a_{N-1}\,\tilde{m}_{N-1}~;~j_N\,m_N \,\big|\,J\,M\,\big>

  \\&&
  \\%%%%%%%%%%%%%%%%%%%%%%%%%%%%%%%%%%%%%%%%%%%%%%%%%%%%%%%%%%%%
  &&&\multicolumn{3}{l}{\fcmt{5}{Introduce:
                                $m_3=\tilde{m_3}-\tilde{m}_2$}}
  \\&&%%%%%%%%%%%%%%%%%%%%%%%%%%%%%%%%%%%%%%%%%%%%%%%%%%%%%%%%%%%%
  \\

   &\multicolumn{2}{c}{=}&\delta_{MM'} \delta_{a_2'a_2}\displaystyle\sum_{\tilde{m}_2}
   &\displaystyle\sum_{\tilde{m}_3+m_4+\ldots+m_N=M }
    &\big<\,a_2\,\tilde{m}_2~;~j_3\,\tilde{m}_3-\tilde{m}_2 \,\big|\,a_3\,\tilde{m}_3 \,\big>~
      \big<\,a_2\,\tilde{m}_2~;~j_3\,\tilde{m}_3-\tilde{m}_2 \,\big|\,a_3'\,\tilde{m}_3 \,\big>
     
     \\[-3mm]
  &&&&&~~\vdots
     \\
  
  &&&&&\big<\,a_{N-2}'\,\tilde{m}_{N-2}~;~j_{N-1}\,m_{N-1} \,
      \big|\,a_{N-1}'\,\tilde{m}_{N-1} \,\big>~
      \big<\,a_{N-2}\,\tilde{m}_{N-2}~;~j_{N-1}\,m_{N-1} \,\big|\,a_{N-1}\,\tilde{m}_{N-1} \,\big>
     \\  
  &&&&&\big<\,a_{N-1}'\,\tilde{m}_{N-1}~;~j_N\,m_N \,\big|\,J'\,M \,\big>~
      \big<\,a_{N-1}\,\tilde{m}_{N-1}~;~j_N\,m_N \,\big|\,J\,M\,\big>

      \\&&
       \\%%%%%%%%%%%%%%%%%%%%%%%%%%%%%%%%%%%%%%%%%%%%%%%%%%%%%%%%%%%%%%%%%%%%%%%%%%%%%%%%%%%%%%%%%%%%%%%%
  &&&\multicolumn{3}{l}{\fcmt{12}{Use unitarity of the Clebsch Gordan coefficients in order to carry out the sum over $\tilde{m}_2$}}
  \\&&%%%%%%%%%%%%%%%%%%%%%%%%%%%%%%%%%%%%%%%%%%%%%%%%%%%%%%%%%%%%%%%%%%%%%%%%%%%%%%%%%%%%%%%%%%%%%%%%
  \\&&
  \\ 

   &\multicolumn{2}{c}{=}&\delta_{MM'} \delta_{a_2'a_2}\delta_{a_3'a_3}&\displaystyle\sum_{\tilde{m}_3+m_4+\ldots+m_N=M }
    &\big<\,a_3\,\tilde{m}_3~;~j_4\,m_4 \,\big|\,a_4\,\tilde{m}_4 \,\big>~
      \big<\,a_3\,\tilde{m}_3~;~j_4\,m_4 \,\big|\,a_4'\,\tilde{m}_4 \,\big>
     
     \\[-3mm]
  &&&&&~~\vdots
     \\
  
  &&&&&\big<\,a_{N-2}'\,\tilde{m}_{N-2}~;~j_{N-1}\,m_{N-1} \,
      \big|\,a_{N-1}'\,\tilde{m}_{N-1} \,\big>~
      \big<\,a_{N-2}\,\tilde{m}_{N-2}~;~j_{N-1}\,m_{N-1} \,\big|\,a_{N-1}\,\tilde{m}_{N-1} \,\big>
     \\  
  &&&&&\big<\,a_{N-1}'\,\tilde{m}_{N-1}~;~j_N\,m_N \,\big|\,J'\,M \,\big>~
      \big<\,a_{N-1}\,\tilde{m}_{N-1}~;~j_N\,m_N \,\big|\,J\,M\,\big>
  \\
  &\multicolumn{2}{c}{\vdots}&\\    
  &\multicolumn{2}{c}{\vdots}&\multicolumn{3}{l}{\fcmt{6}{This process continues $N-2$ times}}\\
  &\multicolumn{2}{c}{\vdots}&\\
  &&\\
  
  &\multicolumn{2}{c}{=}&
   \multicolumn{2}{l}{\delta_{MM'}\delta_{a_2a_2'}\ldots\delta_{a_{N-1}a_{N-1}'}
   \displaystyle\sum_{\tilde{m}_{N-1}+m_N=M}}
   &\big<\,a_{N-1}\,\tilde{m}_{N-1}~;~j_N\,m_N \,\big|\,J'\,M \,\big>~
      \big<\,a_{N-1}\,\tilde{m}_{N-1}~;~j_N\,m_N \,\big|\,J\,M\,\big>
  \\[-3mm]&&
  \\&\multicolumn{2}{c}{=}&
  \multicolumn{3}{l}{\delta_{MM'}
                        \delta_{a_2a_2'}\delta_{a_3a_3'}\ldots
			\delta_{a_{N-1}a_{N-1}'}\delta_{JJ'}
			\delta_{j_1'j_1}\ldots \delta_{j_N'j_N}
			\delta_{n_1'n_1}\ldots\delta{n_N'n_N}}

\end{array}\]
\end{footnotesize}
In the last line we have reintroduced the $n$-quantum numbers for completeness.
Note, that it is also possible, to rearrange the Clebsch Gordan coefficients and to start from the last line to the top (see \cite{Edmonds}).

The result is written here only to demonstrate the nice properties of the Clebsch Gordan coefficients. 
\linebreak It can easily be understood by recalling the definition of a recoupling scheme \linebreak
$\big|~a_2(j_1~j_2)~a_3(a_2~j_3)\ldots a_{N-1}(a_{N-2}~j_{N-1})~J(a_{N-1}~j_N)~M>$ as the simultaneous eigenstate for the operators $(G_2)^2=(J_1+J_2)^2, (G_3)^2=(G_2+J_3)^2, \ldots , (G_{N-1})^2=(G_{N-2}+J_{N-1})^2, J^2=(G_{N-1}+J_N)^2=(J_1+\ldots+J_N)^2 $ with eigenvalues $a_2(a_2+1), a_3(a_3+1),\ldots, g_{N-1}(g_{N-1}+1), J(J+1)$ .

\vfill
\pagebreak
\paragraph{Partial Orthogonality Relations Between Recoupling Schemes}

The same argument can also be applied to cases, where we have to calculate the scalar product of two recoupling schemes of different recoupling order. For illustration let us consider two recoupling schemes 
\begin{footnotesize}

\ba
   \big|~\vec{a}~J~M~\big>
   &=&\big|\,a_2(j_1\,j_2)~a_3(a_2\,j3)\ldots a_{K\!-\!1}(a_{K\!-\!2}\,j_{K\!-\!1}~a_K(a_{K\!-\!1}\,j_K)\ldots 
   a_L(a_{L\!-\!1}\,j_L)~a_{L\!+\!1}(a_{L}\,j_{L\!+\!1} \ldots a_{N\!-\!1}(a_{N\!-\!2}\,j_{N\!-\!1})~J(a_{N\!-\!1}\,j_N)~M\big>
   \nonumber
   \\
   \nonumber
   \\
   \big|~\vec{g}~J~M~\big>
   &=&\big|\,g_2(j_1\,j_2)~g_3(a_2\,j3)\ldots g_{K\!-\!1}(g_{K\!-\!2}\,j_{K\!-\!1}~g_K(g_{K\!-\!1}\,j_P)\ldots 
   g_L(g_{L\!-\!1}\,j_R)~ g_{L\!+\!1}(g_{L}\,j_{L\!+\!1})~ \ldots g_{N\!-\!1}(g_{N\!-\!2}\,j_{N\!-\!1})~ J(g_{N\!-\!1}\,j_N)~M\big>
  \nonumber
  \\\nonumber  \\
\ea  
\end{footnotesize}

Here from $2\ldots K-1$ the spins $j_1,j_2 \ldots j_{K-1}$ are couples in $\vec{a}$ and $\vec{g}$ an the same order. Then $j_K \ldots j_L$ are coupled in the standard way to $\vec{a}$ but in a different order to $\vec{g}$. After that $j_{L+1} \ldots j_{N}$ are successively coupled to each scheme again. 

Now it is clear that $\vec{a}$ and $\vec{g}$ simultaneously diagonalize not only $(G_2)^2=(J_1+J_2)^2, (G_3)^2=(G_2+J_3)^2 \ldots,$ $ (G_{K-1})^2=(G_{K-2}+J_{K-1})^2$ but also $(G_L)^2 \ldots (G_{N-1})^2,J^2=(J_1+\ldots+j_N)^2$. Therefore we can write down immediately

\begin{footnotesize}

\ba\label{Partial orthogonal recoupling schemes}
   \big<~\vec{a}~J~M~\big|~\vec{g}~J~M~\big>
   &=&
   \big< a_K(a_{K\!-\!1}\,j_K)\ldots 
   a_{L\!-\!1}(a_{L\!-\!2}\,j_{L\!-\!1})~ a_L(a_{L\!-\!1}\,j_L)~ \big|~g_K(a_{K\!-\!1}\,j_P)\ldots g_{L\!-\!1}(g_{L\!-\!2}\,j_Q)~a_L(g_{L\!-\!1}\,j_R)~\big>~ 
   \\
   \nonumber
   &&\times
   \delta_{a_2g_2}\ldots\delta_{a_{K-1}g_{K-1}}~\times~\delta_{a_Lg_L}\ldots \delta_{a_{N-1}g_{N-1}}~\times
   ~\delta_{JJ'} ~\times~\delta_{MM'}       
\ea  
\end{footnotesize}

Note, that (\ref{Partial orthogonal recoupling schemes}) is according to lemma 5.1 and 5.2 in \cite{TT:Closed ME of V in LQG}.
\\\\
\paragraph{After the Action of a holonomy} we have according to (\ref{einfache Holonomiemultiplikation 3}) : 

\begin{footnotesize}

\[\begin{array}{ccl}
   \lefteqn{\big[\pi_{\frac{1}{2}}(h_K) \big]_{AB}~\big|\,\vec{a}(12)~J~M~;~\vec{n}\, \big>=}\\\\
   &&\begin{array}{cl}
             =\displaystyle\bigoplus_{\tilde{j}_K=j_K\pm\frac{1}{2}}~\sum_{m_1+\ldots+m_N=M}&
	       \big<~j_1~m_1~;~j_2~m_2 ~\big|~a_2(j_1\,j_2)~\tilde{m}_2 ~\big>\\[-3mm]
	     & \big<~a_2~\tilde{m}_2~;~j_3~m_3~\big|~a_3(a_2\,j_3)~\tilde{m}_3~\big>\\
	     & ~~\vdots\\
	     &\big<~a_{K-1}~\tilde{m}_{K-1}~;~j_{K}~m_{K}~\big|
	           ~a_{K}(a_{K-1}\,j_{K})~\tilde{m}_{K}~\big>~~
	      \underbrace{\big<\,j_K\,m_K~;~\frac{1}{2}\,A~\big|~\tilde{j}_K\,m_K+A~\big>}_{\cmt{3}{this will spoil the use of the unitarity trick!}}~~
	      C^{j_K}_{\tilde{j}_K}(B,n_K)\\[-10mm]
             & ~~\vdots\\
             &\big<~a_{N-1}~\tilde{m}_{N-1}~;~j_{N}~m_{N}~\big|
	           ~J(a_{N-1}\,j_{N})~M~\big>\\		   
	     
           \end{array} \\\\
    
     &&\hspace{4cm}	   	   
       \big|\,j_1\,m_1\,;\,n_1 \,\big> \,\otimes\,\ldots\,\otimes
       \big|\,\tilde{j}_K\,m_K+A\,;\,n_K+B \,\big> 
       \,\otimes\,\ldots\,\otimes\,
       \big|\,j_N\,m_N\,;\,n_N \,\big> 
         
\end{array}\] \be\label{Holonomie auf Kopplungsschema}\ee
\end{footnotesize}

Now the action (\ref{Holonomie auf Kopplungsschema}) of a holonomy clearly prevents us from directly using the unitarity properties of the Clebsch Gordan coefficients from $j_K$ on, since then we have different Clebsch Gordan coefficients and the unitarity condition cannot be used any more!
\\\\

\pagebreak
%%%%%%%%%%%%%%%%%%%%%%%%%%%%%%%%%%%%%%%%%%%%%%%%%%%%%%%%%%%%%%%%%%%%%%%%%%%%%%%%%%%%%%%%%%%%%%%%%%%
\paragraph{Cure: Expansion into another Recoupling Scheme\\\\}
%%%%%%%%%%%%%%%%%%%%%%%%%%%%%%%%%%%%%%%%%%%%%%%%%%%%%%%%%%%%%%%%%%%%%%%%%%%%%%%%%%%%%%%%%%%%%%%%%%%

In order to get rid of all the $m$-summations and to obtain a closed expression we will expand the recoupling scheme (\ref{Definition recoupling scheme}) as follows:  

\begin{footnotesize}
\[\begin{array}{l}
     \big|\,a_2(j_1\,j_2)~\ldots~   a_{K-1}(a_{K-2}\,j_{K-1})~a_K(a_{K-1}\,j_k)~a_{K+1}(a_K\,j_{K+1})~\ldots~
          a_{N-1}(a_{N-2}\,j_{N-1})~J(a_{N-1}\,j_N)~M\,\vec{n}\,\big>=  \\\\
     
     =\displaystyle\sum_{\vec{g}}\underbrace{\big<~\vec{g}~J~M~\vec{n}\big|~\vec{a}~J~M~\vec{n}~\big>~}_{\displaystyle 3nj\mbox{-symbol}}~
     
     \big|\,g_2(j_1\,j_2)~\ldots~   g_{K-1}(g_{K-2}\,j_{K-1})~g_K(g_{K-1}\,j_{k+1})~g_{K+1}(g_K\,j_{K+2})~\ldots~
          g_{N-1}(g_{N-2}\,j_{N})~J(g_{N-1}\,j_K)~M\,\vec{n}\,\big>.  

\end{array}\]
\end{footnotesize}
\\[-8mm]
\be\label{Expansion II}\ee
       
In the following calculation we always have $M'=M$ and $J'=J$, since otherwise the scalar product would va-nish. \fbox{We supress the quantum numbers $M$ and $\vec{n}=(n_1\ldots n_N)$}. Note, that we will frequently use the reality of the $3nj$-symbols, that is
    $\big<~\vec{g}~J~M~\vec{n}\big|~\vec{a}~J~M~\vec{n}~\big>
    =
    \big<~\vec{a}~J~M~\vec{n}\big|~\vec{g}~J~M~\vec{n}~\big>$ and (\ref{Partial orthogonal recoupling schemes}). Moreover $\big<~\%~\big|$ means a bra-vector containing the same arguments as its companion ket-vector ($\big|\cdot\big>$).  
We have
\begin{footnotesize}
\noindent

\[\begin{array}{cc@{~}|llll}

    \lefteqn{\big<\,\vec{a}~J\,\big|\,\vec{g}~J\,\big>:=}
    \\\\    
     &\multicolumn{2}{c}{=}&\multicolumn{3}{l}{
     \big<\,a_2(j_1\,j_2)~\ldots~   a_{K-1}(a_{K-2}\,j_{K-1})~a_K(a_{K-1}\,j_K)~a_{K+1}(a_K\,j_{K+1})~\ldots~
          a_{N-1}(a_{N-2}\,j_{N-1})~J(a_{N-1}\,j_N)
	  \,\big|~~~~~~}
     \\
	  &&&\multicolumn{2}{c}{~~~~~~\big|\,
	  g_2(j_1\,j_2)~\ldots~
          g_{K-1}(g_{K-2}\,j_{K-1})~g_K(g_{K-1}\,j_{k+1})~g_{K+1}(g_K\,j_{K+2})~\ldots~
          g_{N-1}(g_{N-2}\,j_{N})~J(g_{N-1}\,j_K)\,\big>} 
      \\&&
      \\%%%%%%%%%%%%%%%%%%%%%%%%%%%%%%%%%%%%%
      &\multicolumn{2}{c}{=}&
      \delta_{a_2\,g_2} \ldots \delta_{a_{K-1}\, g_{K-1}} ~\times
      \\
      &&&\multicolumn{2}{l}{\times \big<a_K(a_{K\!-\!1}\,j_K)~a_{K\!+\!1}(a_K\,j_{K\!+\!1})\ldots
          a_{N\!-\!1}(a_{N\!-\!2}\,j_{N\!-\!1})~J(a_{N\!-\!1}\,j_N)
	  \,\big|\,
	  g_K(g_{K\!-\!1}\,j_{K\!+\!1})~g_{K\!+\!1}(g_K\,j_{K\!+\!2})\ldots
          g_{N\!-\!1}(g_{N\!-\!2}\,j_{N})~J(g_{N\!-\!1}\,j_K)\,\big>}& 
      \\&&
      \\%%%%%%%%%%%%%%%%%%%%%%%%%%%%%%%%%%%%%
      &&&\multicolumn{3}{l}{\fcmt{14}{Insert:~~~
           $\mb{1}=\displaystyle\sum_{\vec{h}}~
	   \big|\,h_K(a_{K-1}\,j_{K+1})~h_{K+1}(h_K\,j_K)~h_{K+2}(h_{K+1}\,j_{K+2})~\ldots~ J(h_{N-1}\,j_N)~\big>~\big<~~\% ~~\big|$}}
      \\&&
      \\%%%%%%%%%%%%%%%%%%%%%%%%%%%%%%%%%%%%% 
      &\multicolumn{2}{c}{=}&
      \multicolumn{2}{l}{\delta_{a_2\,g_2} \ldots \delta_{a_{K-1}\, g_{K-1}} ~\displaystyle\sum_{\vec{h}}~\bigg\{}
      \\
      &&&\multicolumn{2}{l}{\big<a_K(a_{K\!-\!1}\,j_K)~\overbrace{a_{K\!+\!1}(a_K\,j_{K\!+\!1})\ldots
          a_{N\!-\!1}(a_{N\!-\!2}\,j_{N\!-\!1})~J(a_{N\!-\!1}\,j_N)
	  \,\big|\,
	  h_K(a_{K-1}\,j_{K+1})~h_{K+1}(h_K\,j_K)~h_{K+2}(h_{K+1}\,j_{K+2})\ldots J(h_{N-1}\,j_N)}^{\displaystyle \delta_{a_{K+1}h_{K+1}}~\ldots~\delta_{a_Nh_N}}
	  
	  \,\big>}
      \\
      &&&\multicolumn{2}{l}{
	  \big<\,h_K(a_{K-1}\,\underbrace{j_{K+1})~h_{K+1}(h_K\,j_K)~h_{K+2}(h_{K+1}\,j_{K+2})\ldots 
	  J(h_{N-1}\,j_N)\,
	  \big|\,g_K(g_{K\!-\!1}}_{\displaystyle\delta_{h_K g_K}}
	  \,j_{K\!+\!1})~g_{K\!+\!1}(g_K\,j_{K\!+\!2})\ldots
          g_{N\!-\!1}(g_{N\!-\!2}\,j_{N})~J(g_{N\!-\!1}\,j_K)\,\big>\bigg\}}& 
      \\&&
      \\%%%%%%%%%%%%%%%%%%%%%%%%%%%%%%%%%%%%% 
      &\multicolumn{2}{c}{=}&
      \delta_{a_2\,g_2} \ldots \delta_{a_{K-1}\, g_{K-1}}
      &
      \big<a_K(a_{K\!-\!1}\,j_K)~a_{K\!+\!1}(a_K\,j_{K\!+\!1})
	  \,\big|\,
	  g_K(a_{K-1}\,j_{K+1})~a_{K+1}(g_K\,j_K)\,\big>
      \\
      &&&\multicolumn{2}{c}{
	  \big<\,a_{K+1}(g_K\,j_K)~a_{K+2}(a_{K+1}\,j_{K+2})\ldots 
	  J(a_{N-1}\,j_N)\,
	  \big|\,g_{K+1}(g_{K}\,j_{K\!+\!2})~g_{K\!+\!2}(g_{K+1}\,j_{K\!+\!3})\ldots
          g_{N\!-\!1}(g_{N\!-\!2}\,j_{N})~J(g_{N\!-\!1}\,j_K)\,\big>} 
      \\&&
      \\%%%%%%%%%%%%%%%%%%%%%%%%%%%%%%%%%%%%%
      &&&\multicolumn{3}{l}{\fcmt{17}{Insert:~~~
           $\mb{1}=\displaystyle\sum_{\vec{h}}~
	   \big|\,h_{K+1}(g_{K}\,j_{K+2})~h_{K+2}(h_{K+1}\,j_K)~h_{K+3}(h_{K+2}\,j_{K+3})~\ldots~ h_{N-1}(h_{N-2}\,j_{N-1})~J(h_{N-1}\,j_N)~\big>~\big<~~\% ~~\big|$}}
      \\&&
      \\%%%%%%%%%%%%%%%%%%%%%%%%%%%%%%%%%%%%% 
      &\multicolumn{2}{c}{=}&
      \delta_{a_2\,g_2} \ldots \delta_{a_{K-1}\, g_{K-1}}
      &
      \big<a_K(a_{K\!-\!1}\,j_K)~a_{K\!+\!1}(a_K\,j_{K\!+\!1})
	  \,\big|\,
	  g_K(a_{K-1}\,j_{K+1})~a_{K+1}(g_K\,j_K)\,\big>
      \\
      &&&\multicolumn{2}{l}{\displaystyle\sum_{\vec{h}}
	  \big<\,a_{K+1}(g_K\,j_K)~\overbrace{a_{K+2}(a_{K+1}\,j_{K+2})\ldots 
	  J(a_{N-1}\,j_N)\,
	  \big|\,h_{K+1}(g_{K}\,j_{K+2})~h_{K+2}(h_{K+1}\,j_K)~h_{K+3}(h_{K+2}\,j_{K+3})~\ldots~ ~J(h_{N-1}\,j_N)}^{\displaystyle\delta_{a_{K+2}h_{K+2}~\ldots~\delta_{a_Nh_N}}}
	  ~\big>}
       \\[-3mm]
       &&&\multicolumn{2}{l}{~~~~~  
	  \big<\,h_{K+1}(g_{K}\,\underbrace{j_{K+2})~h_{K+2}(h_{K+1}\,j_K)~h_{K+3}(h_{K+2}\,j_{K+3})~\ldots~ ~J(h_{N-1}\,j_N)
	  ~\big|\,
	  g_{K+1}(g_{K}\,}_{\displaystyle\delta_{h_{K+1}g_{K+1}}}
	  j_{K\!+\!2})~g_{K\!+\!2}(g_{K+1}\,j_{K\!+\!3})\ldots
          ~J(g_{N\!-\!1}\,j_K)~\big>} 
      \\&&
      \\%%%%%%%%%%%%%%%%%%%%%%%%%%%%%%%%%%%%%
      &\multicolumn{2}{c}{=}&
      \delta_{a_2\,g_2} \ldots \delta_{a_{K-1}\, g_{K-1}}
      &
      \big<a_K(a_{K\!-\!1}\,j_K)~a_{K\!+\!1}(a_K\,j_{K\!+\!1})
	  \,\big|\,
	  g_K(a_{K-1}\,j_{K+1})~a_{K+1}(g_K\,j_K)\,\big>
      \\
      &&&&\big<a_{K\!+\!1}(g_{K}\,j_K)~a_{K\!+\!2}(a_{K\!+\!1}\,j_{K\!+\!2})
	  \,\big|\,
	  g_{K\!+\!1}(g_{K}\,j_{K+2})~a_{K+2}(g_{K\!+\!1}\,j_K)\,\big>
      \\
      &&&\multicolumn{2}{c}{
	  \big<\,a_{K+2}(g_{K+1}\,j_K)~a_{K+3}(a_{K+2}\,j_{K+2})\ldots 
	  J(a_{N-1}\,j_N)\,
	  \big|\,g_{K+2}(g_{K+1}\,j_{K+3})~g_{K+3}(g_{K+2}\,j_{K+4})\ldots
          g_{N-1}(g_{N-2}\,j_{N})~J(g_{N-1}\,j_K)\,\big>} 
      \\&&
      
      %%%%%%%%%%%%%%%%%%%%%%%%%%%%%%%%%%%%%
\end{array}\]

\[\begin{array}{cc@{~}|llll}
      
      &\multicolumn{2}{c}{\vdots}&&
      \\%%%%%%%%%%%%%%%%%%%%%%%%%%%%%%%%%%%%%
      &\multicolumn{2}{c}{\vdots}&\multicolumn{3}{l}{\fcmt{6}{This can be continued until we finally get:}}
      \\&\multicolumn{2}{c}{\vdots}&
      \\%%%%%%%%%%%%%%%%%%%%%%%%%%%%%%%%%%%%% 
      &\multicolumn{2}{c}{=}&
      \delta_{a_2\,g_2} \ldots \delta_{a_{K-1}\, g_{K-1}}
      &
      \big<a_K(a_{K\!-\!1}\,j_K)~a_{K\!+\!1}(a_K\,j_{K\!+\!1})
	  \,\big|\,
	  g_K(a_{K-1}\,j_{K+1})~a_{K+1}(g_K\,j_K)\,\big>
      \\
      &&&&\big<a_{K\!+\!1}(g_{K}\,j_K)~a_{K\!+\!2}(a_{K\!+\!1}\,j_{K\!+\!2})
	  \,\big|\,
	  g_{K\!+\!1}(g_{K}\,j_{K+2})~a_{K+2}(g_{K\!+\!1}\,j_K)\,\big>
      \\
      &&&&~~\vdots
      \\
      &&&&\big<a_{N\!-\!3}(g_{N-4}\,j_K)~a_{N\!-\!2}(a_{N\!-\!3}\,j_{N\!-\!2})
	  \,\big|\,
	  g_{N\!-\!3}(g_{N-4}\,j_{N-2})~a_{N-2}(g_{N\!-\!3}\,j_K)\,\big>
      \\&&
      \\
      &&&\multicolumn{2}{c}{
	  \big<\,a_{N-2}(g_{N-3}\,j_K)~a_{N-1}(a_{N-2}\,j_{N-1}) 
	  ~J(a_{N-1}\,j_N)\,
	  \big|\,g_{N-2}(g_{N-3}\,j_{N-1})~
          g_{N-1}(g_{N-2}\,j_{N})~J(g_{N-1}\,j_K)\,\big>} &
      \\&&
      \\%%%%%%%%%%%%%%%%%%%%%%%%%%%%%%%%%%%%%
      &&&\multicolumn{3}{l}{\fcmt{17}{Insert:~~~
           $\mb{1}=\displaystyle\sum_{\vec{h}}~
	   \big|\,h_{N-2}(g_{N-3}\,j_{N-1})~h_{N-1}(h_{N-2}\,j_K)~J(h_{N-1}\,j_N)=h_N~\big>~\big<~~\% ~~\big|$}}
      \\&&
      \\%%%%%%%%%%%%%%%%%%%%%%%%%%%%%%%%%%%%% 
      &\multicolumn{2}{c}{=}&
      \delta_{a_2\,g_2} \ldots \delta_{a_{K-1}\, g_{K-1}}
      &
      \big<a_K(a_{K\!-\!1}\,j_K)~a_{K\!+\!1}(a_K\,j_{K\!+\!1})
	  \,\big|\,
	  g_K(a_{K-1}\,j_{K+1})~a_{K+1}(g_K\,j_K)\,\big>
      \\
      &&&&\big<a_{K\!+\!1}(g_{K}\,j_K)~a_{K\!+\!2}(a_{K\!+\!1}\,j_{K\!+\!2})
	  \,\big|\,
	  g_{K\!+\!1}(g_{K}\,j_{K+2})~a_{K+2}(g_{K\!+\!1}\,j_K)\,\big>
      \\
      &&&&~~\vdots
      \\
      &&&&\big<a_{N\!-\!3}(g_{N-4}\,j_K)~a_{N\!-\!2}(a_{N\!-\!3}\,j_{N\!-\!2})
	  \,\big|\,
	  g_{N\!-\!3}(g_{N-4}\,j_{N-2})~a_{N-2}(g_{N\!-\!3}\,j_K)\,\big>
      \\&&
      \\
      &&&\multicolumn{3}{c}{
	  \displaystyle\sum_{\vec{h}}~\big<\,a_{N-2}(g_{N-3}\,j_K)~\overbrace{a_{N-1}(a_{N-2}\,j_{N-1}) 
	  ~J(a_{N-1}\,j_N)\,
	  \big|\,h_{N-2}(g_{N-3}\,j_{N-1})~h_{N-1}(h_{N-2}\,j_K)~J(h_{N-1}\,j_N)
	  }^{\displaystyle\delta_{a_N-1h_N-1}\delta_{a_Nh_N}}\,\big>}
      \\[-3mm]	   
      &&&\multicolumn{3}{c}{
	  ~~~~~\big<\,h_{N-2}(g_{N-3}\,\underbrace{j_{N-1})~h_{N-1}(h_{N-2}\,j_K)~J(h_{N-1}\,j_N)\,
	  \big|\,g_{N-2}(g_{N-3}}_{\displaystyle\delta_{h_{N-2}g_{N-2}}}
	  \,j_{N-1})~
          g_{N-1}(g_{N-2}\,j_{N})~J(g_{N-1}\,j_K)\,\big>} 
      
      \\&&
      \\%%%%%%%%%%%%%%%%%%%%%%%%%%%%%%%%%%%%%
      &\multicolumn{2}{c}{=}&
      \delta_{a_2\,g_2} \ldots \delta_{a_{K-1}\, g_{K-1}}
      &
      \big<a_K(a_{K\!-\!1}\,j_K)~a_{K\!+\!1}(a_K\,j_{K\!+\!1})
	  \,\big|\,
	  g_K(a_{K-1}\,j_{K+1})~a_{K+1}(g_K\,j_K)\,\big>
      \\
      &\multicolumn{2}{c}{~}&&\big<a_{K\!+\!1}(g_{K}\,j_K)~a_{K\!+\!2}(a_{K\!+\!1}\,j_{K\!+\!2})
	  \,\big|\,
	  g_{K\!+\!1}(g_{K}\,j_{K+2})~a_{K+2}(g_{K\!+\!1}\,j_K)\,\big>
      \\
      &\multicolumn{2}{c}{~}&&~~\vdots
      \\
      &\multicolumn{2}{c}{~}&&\big<a_{N\!-\!3}(g_{N-4}\,j_K)~a_{N\!-\!2}(a_{N\!-\!3}\,j_{N\!-\!2})
	  \,\big|\,
	  g_{N\!-\!3}(g_{N-4}\,j_{N-2})~a_{N-2}(g_{N\!-\!3}\,j_K)\,\big>
      \\
      &\multicolumn{2}{c}{~}&&\big<a_{N\!-\!2}(g_{N-3}\,j_K)~a_{N\!-\!1}(a_{N\!-\!2}\,j_{N\!-\!1})
	  \,\big|\,
	  g_{N\!-\!2}(g_{N-3}\,j_{N-1})~a_{N-1}(g_{N\!-\!2}\,j_K)\,\big>
      \\
      &\multicolumn{2}{c}{~}&&\big<a_{N\!-\!1}(g_{N-2}\,j_K)~J(a_{N\!-\!1}\,j_{N})
	  \,\big|\,
	  g_{N\!-\!1}(g_{N-2}\,j_{N})~J(g_{N\!-\!1}\,j_K)\,\big>
 
 \end{array}\] \be\label{3nj Expansion I} \ee   

\end{footnotesize}

Note, that all expressions in (\ref{3nj Expansion I}) can be rewritten as $6j$-symbols!
Moreover for the {\bf special case \fbox{$K=1$}} we have to start from

\begin{footnotesize}

\[\begin{array}{cc@{~}|llll}
      
   \big|~\vec{g}~J~M~\big>&\multicolumn{2}{c}{=}&
       \big|~g_2(j_2\,j_3)~g_3(g_2\,j_4) \ldots g_{N-1}(g_{N-2}\,j_N)~J(g_{N-1}\,j_1)~M~\big>

\end{array}\]
\end{footnotesize}

and obtain a result similar to (\ref{3nj Expansion I}) which (structurally) differs only by the recoupling-order of the first two spins: 

\begin{footnotesize}

\[\begin{array}{cc@{~}|llll}

    \big<\,\vec{a}~J\,\big|\,\vec{g}~J\,\big>      
    
    &\multicolumn{2}{c}{:=}&
      
      &
      \big<a_2(j_1\,j_2)~a_{3}(a_2\,j_{3})
	  \,\big|\,
	  g_2(j_{2}\,j_{3})~a_{3}(g_2\,j_1)\,\big>
      \\
      &\multicolumn{2}{c}{~}&&
      \big<a_{3}(g_{2}\,j_1)~a_{4}(a_{3}\,j_{4})
	  \,\big|\,
	  g_{3}(g_{2}\,j_{4})~a_{4}(g_{3}\,j_1)\,\big>
      \\
      &\multicolumn{2}{c}{~}&&
      \big<a_{4}(g_{3}\,j_1)~a_{5}(a_{4}\,j_{5})
	  \,\big|\,
	  g_{4}(g_{3}\,j_{5})~a_{5}(g_{4}\,j_1)\,\big>
      \\
      &\multicolumn{2}{c}{~}&&~~\vdots
      \\
      &\multicolumn{2}{c}{~}&&
      \big<a_{N\!-\!3}(g_{N-4}\,j_1)~a_{N\!-\!2}(a_{N\!-\!3}\,j_{N\!-\!2})
	  \,\big|\,
	  g_{N\!-\!3}(g_{N-4}\,j_{N-2})~a_{N-2}(g_{N\!-\!3}\,j_1)\,\big>
      \\
      &\multicolumn{2}{c}{~}&&
      \big<a_{N\!-\!2}(g_{N-3}\,j_1)~a_{N\!-\!1}(a_{N\!-\!2}\,j_{N\!-\!1})
	  \,\big|\,
	  g_{N\!-\!2}(g_{N-3}\,j_{N-1})~a_{N-1}(g_{N\!-\!2}\,j_1)\,\big>
      \\
      &\multicolumn{2}{c}{~}&&
      \big<a_{N\!-\!1}(g_{N-2}\,j_1)~J(a_{N\!-\!1}\,j_{N})
	  \,\big|\,
	  g_{N\!-\!1}(g_{N-2}\,j_{N})~J(g_{N\!-\!1}\,j_1)\,\big>
 
\end{array}\]\be\label{3nj Expansion: special case K=1} \ee 

\end{footnotesize}

%%%%%%%%%%%%%%%%%%%%%%%%%%%%%%%%%%%%%%%%%%%%%%%%%%%%%%%%%%%%%%%%%%%%%%%%%%%%%%%%%%%%%%%%%%%%%%%%%%%%%%%

\pagebreak
\subsubsection{Discussion: Action of a holonomy $\big[\pi_{\frac{1}{2}}(h_{e_K})\big]_{AB}$ on a Recoupling Scheme\\}
If we use the expansion (\ref{Expansion II}) with the coefficients explicitely obtained in (\ref{3nj Expansion I}), and the abbreviations

\begin{footnotesize}
\ba
   \big|\,\vec{a}\,J\,M\,;\,\vec{n}\,\big>&:=&\big|\,a_2(j_1\,j_2)~a_3(a_2\,j_3)\ldots
                    a_{K\!-\!1}(a_{K\!-\!2}\,j_{K\!-\!1})~a_K(a_{K\!-\!1}\,j_K)~a_{K\!+\!1}(a_K\,j_{K\!+\!1})\ldots
		    a_{N\!-\!1}(a_{N\!-\!2}\,j_{N\!-\!1})~a_N(a_{N\!-\!1}\,j_N)\!=\!J~M~;~\vec{n}\,\big>
   \nonumber\\
   \big|\,\vec{a}'\,J'\,M'\,;\,\vec{n}'\,\big>&:=&\big|\,a_2'(j_1'\,j_2')~a_3'(a_2'\,j_3')\ldots
                    a_{K\!-\!1}'(a_{K\!-\!2}'\,j_{K\!-\!1}')~a_K'(a_{K\!-\!1}'\,j_K')~a_{K\!+\!1}'(a_K'\,j_{K\!+\!1}')\ldots
		    a_{N\!-\!1}'(a_{N\!-\!2}'\,j_{N\!-\!1}')~a_N'(a_{N\!-\!1}'\,j_N')\!=\!J'~M'~;~\vec{n}'\,\big>
   \nonumber\\\nonumber\\
   \big|\,\vec{g}\,J\,M\,;\,\vec{n}\,\big>&:=&\big|\,g_2(j_1\,j_2)~g_3(g_2\,j_3)~\ldots~
                    g_{K\!-\!1}(g_{K\!-\!2}\,j_{K\!-\!1})~g_K(g_{K\!-\!1}\,j_{K\!+\!1})~g_{K\!+\!1}(g_K\,j_{K\!+\!2})\ldots
		    g_{N\!-\!1}(g_{N\!-\!2}\,j_{N})~g_N(g_{N\!-\!1}\,j_K)\!=\!J~M~;~\vec{n}\,\big>
   \nonumber\\
   \big|\,\vec{g}'\,J'\,M'\,;\,\vec{n}'\,\big>&:=&\big|\,g_2'(j_1'\,j_2')~g_3'(g_2'\,j_3')\ldots
                    g_{K\!-\!1}'(g_{K\!-\!2}'\,j_{K\!-\!1}')~g_K'(g_{K\!-\!1}'\,j_{K\!+\!1}')~g_{K\!+\!1}'(g_K'\,j_{K\!+\!2}')~\ldots~
		    g_{N\!-\!1}'(g_{N\!-\!2}'\,j_{N}')~g_N'(g_{N\!-\!1}'\,j_K')\!=\!J'~M'~;~\vec{n}'\,\big>
  \nonumber		    
\ea
\end{footnotesize}
\be \label{Definitions for the action of holonomy on a recoupling scheme} \ee\\[-13mm]

we can write:\\[-10mm]

\ba  \label{Entwicklung in anderes Recoupling Scheme}
   \big<\,\vec{a}'\,J'\,M'\,;\,\vec{n}'\,\big|\,
    \big[\pi_{\frac{1}{2}}(h_K) \big]_{AB}
    \,\big|\,\vec{a}\,J\,M\,;\,\vec{n}\,\big> 
    =&
    \displaystyle\sum_{\vec{g}'\,\vec{g}}&
    \big<\,\vec{a}'\,J'\,M'\,;\,\vec{n}'\,\big|\,\vec{g}'\,J'\,M'\,;\,\vec{n}'\,\big>
    \nonumber\\[-3.5mm]
    &&
    \big<\,\vec{g}'\,J'\,M'\,;\,\vec{n}'\,\big|\,
    \big[\pi_{\frac{1}{2}}(h_K) \big]_{AB}\,
    \big|\,\vec{g}\,J\,M\,;\,\vec{n}\,\big>
    \nonumber\\
    &&
    \big<\,\vec{g}\,J\,M\,;\,\vec{n}\,\big|
    \,\vec{a}\,J\,M\,;\,\vec{n}\,\big>
\ea

Let us introduce the shorthands $\tilde{m}_l=m_1+\ldots+m_l$ , that is the sum of all $m$'s according to the recoupling order in $\vec{g}$. Especially we have
\ba
   \tilde{m}_{l<K}&=&m_1+\ldots+m_l \nonumber\\
   \tilde{m}_{l=K}&=&m_1+\ldots+m_{K-1}+m_{K+1} \nonumber\\   
   \tilde{m}_{l> K}&=&m_1+\ldots+m_{K-1}+m_{K+1}+\ldots+m_l \nonumber\\
   \tilde{m}_{N-1}&=&m_1+\ldots+m_{K-1}+m_{K+1}+\ldots+m_N \nonumber\\    
   M=\tilde{m}_{N}&=&m_1+\ldots+m_{K-1}+m_{K+1}+\ldots+m_N+m_K  
\ea

Using these conventions and the coefficient definition of (\ref{einfache Holonomiemultiplikation 3}) we can write\footnote{If we consider the gauge behaviour of a recoupling state of total angular momentum $J$ after the action of a holonomy of \linebreak weight $\frac{1}{2}$ we can easily see that the resulting state transforms under gauge transformations according to $J\otimes\frac{1}{2}= \bigoplus\limits_{\tilde{J}=J\pm\frac{1}{2}}\tilde{J}$.}:
\begin{footnotesize}

\[\begin{array}{cc@{~}|lllll}
    \lefteqn{ \big<\,\vec{g}'\,J'\,M'\,;\,\vec{n}'\,\big|\,
              \big[\pi_{\frac{1}{2}}(h_K) \big]_{AB}\,
              \big|\,\vec{g}\,J\,M\,;\,\vec{n}\,\big>~=~}
    \\
    \\%%%%%%%%%%%%%%%%%%%%%%%%%%%%%%%%%%%%%
    &\multicolumn{2}{c}{=}&\displaystyle\sum_{\tilde{j}_K=j_K\pm\frac{1}{2}\atop \tilde{J}=J\pm\frac{1}{2}}&
                           \displaystyle\sum_{m_1'\!+\!\ldots\!+m_N'\!+\!\ldots\!+m_K'\!=\!M' 
			                \atop m_1\!+\!\ldots\!+m_N\!+\!\ldots\!+m_K\!=\!M}
    &\big<\,j_1'\,m_1'\,;\,j_2'\,m_2'\,\big|\,g_2'(j_1'\,j_2')~\tilde{m}_2'\,\big>
    &\big<\,j_1\,m_1\,;\,j_2\,m_2\,\big|\,g_2(j_1\,j_2)~\tilde{m}_2\,\big>
    \\[-5mm]%%%%%%%%%%%%%%%%%%%%%%%%%%%%%%%%%%%%%
    &&&&
    &\big<\,g_2'\,\tilde{m}_2'\,;\,j_3'\,m_3'\,\big|\,g_3'(g_2'\,j_3')~\tilde{m}_3'\,\big>
    &\big<\,g_2\,\tilde{m}_2\,;\,j_3\,m_3\,\big|\,g_3(g_2\,j_3)~\tilde{m}_3\,\big>
    \\%%%%%%
    &&&&&~~\vdots&~~\vdots
    \\
    &&&&\multicolumn{2}{l}{\big<\,g_{K-2}'\,\tilde{m}_{K-2}'\,;\,j_{K-1}'\,m_{K-1}'\,
     \big|\,g_{K-1}'(g_{K-2}'\,j_{K-1}')~\tilde{m}_{K-1}'\,\big>}
    &\big<\,g_{K-2}\,\tilde{m}_{K-2}\,;\,j_{K-1}\,m_{K-1}\,
     \big|\,g_{K-1}(g_{K-2}\,j_{K-1})~\tilde{m}_{K-1}\,\big>
    \\
    &&&&\multicolumn{2}{l}{\big<\,g_{K-1}'\,\tilde{m}_{K-1}'\,;\,j_{K+1}'\,m_{K+1}'\,
     \big|\,g_{K}'(g_{K-1}'\,j_{K+1}')~\tilde{m}_{K}'\,\big>}
     &
     \big<\,g_{K-1}\,\tilde{m}_{K-1}\,;\,j_{K+1}\,m_{K+1}\,
     \big|\,g_{K}(g_{K-1}\,j_{K+1})~\tilde{m}_{K}\,\big>
    \\
    &&&&\multicolumn{2}{l}{\big<\,g_{K}'\,\tilde{m}_{K}'\,;\,j_{K+2}'\,m_{K+2}'\,
     \big|\,g_{K+1}'(g_{K}'\,j_{K+2}')~\tilde{m}_{K+1}'\,\big>}
     &
     \big<\,g_{K}\,\tilde{m}_{K}\,;\,j_{K+2}\,m_{K+2}\,
     \big|\,g_{K+1}(g_{K}\,j_{K+2})~\tilde{m}_{K+1}\,\big>
    \\%%%%%%
    &&&&&~~\vdots&~~\vdots
    \\
    &&&&\multicolumn{2}{l}{\big<\,g_{N-2}'\,\tilde{m}_{N-2}'\,;\,j_{N}'\,m_{N}'\,
     \big|\,g_{N-1}'(g_{N-2}'\,j_{N}')~\tilde{m}_{N-1}'\,\big>}
    &\big<\,g_{N-2}\,\tilde{m}_{N-2}\,;\,j_{N}\,m_{N}\,
     \big|\,g_{N-1}(g_{N-2}\,j_{N})~\tilde{m}_{N-1}\,\big>
    \\
    &&&&\multicolumn{2}{l}{\big<\,g_{N-1}'\,\tilde{m}_{N-1}'\,;\,j_{K}'\,m_{K}'\,
     \big|\,g_{N}'(g_{N-1}'\,j_{K}')=J'~M'\,\big>}
     &
     \big<\,g_{N-1}\,\tilde{m}_{N-1}\,;\,j_{K}\,m_{K}\,
     \big|\,g_{N}(g_{N-1}\,j_{K})=J~M\,\big>~\times
     \\&&
     \\&&&&\multicolumn{2}{l}{\times~C^{j_K}_{\tilde{j}_K}(B,n_K)~~ \big<\,j_K\,m_K\,;\,\frac{1}{2}\,A \,\big|\,\tilde{j}_K\,m_K\!+\!A \,\big>~\times}
     \\&&
     \\&&
     \\
     &&&&\multicolumn{3}{l}{
         \times~\underbrace{\big<\,j_1'\,m_1'\,;\,n_1'\,\big|\,j_1\,m_1\;\,n_1 \,\big>}
	 _{\displaystyle\delta_{j_1'j_1}\delta_{m_1'm_1}\delta_{n_1'n_1}}
	 ~\ldots~
	 \underbrace{\big<\,j_{K-1}'\,m_{K-1}'\,;\,n_{K-1}' \,\big|\,j_{K-1}\,m_{K-1}\,;\,n_{K-1}  \,\big>}
	 _{\displaystyle\delta_{j_{K-1}'j_{K-1}}\delta_{m_{K-1}'m_{K-1}}\delta_{n_{K-1}'n_{K-1}}}
	 \underbrace{\big<\,j_K'\,m_K'\,;\,n_K' \,\big|\,\tilde{j}_K\,m_K\!+\!A\,;\,n_K\!+\!B \,\big>}
	 _{\displaystyle\delta_{j_{K}'\tilde{j}_{K}}\delta_{m_{K}'m_{K}+A}\delta_{n_{K}'n_{K}+B}} ~\times}
     \\&&
     \\
     &&&&\multicolumn{3}{l}{
	 \times~\underbrace{\big<\,j_{K+1}'\,m_{K+1}'\,;\,n_{K+1}' \,
	 \big|\,j_{K+1}\,m_{K+1}\,;\,n_{K+1} \,\big>}
	 _{\displaystyle\delta_{j_{K+1}'j_{K+1}}\delta_{m_{K+1}'m_{K+1}}\delta_{n_{K+1}'n_{K+1}}}
	 ~\ldots~
	 \underbrace{\big<\,j_N'\,m_N'\,;\,n_N' \,\big|\,j_N\,m_N\,;\,n_N \,\big>}
	 _{\displaystyle\delta_{j_{N}'j_{N}}\delta_{m_{N}'m_{N}}\delta_{n_{N}'n_{N}}}~\times~\delta_{J'\tilde{J}} }
\end{array}\]

%%%%%%%%%%%%%%%%%%%%%%%%%%%%%%%%%%%%%%%%%%%%%%%%%%%%%%%%%%%%%%%%%%%%%%%%%%%%%%%%%%%%%%%%%%%%%%%%%%%%
%%%%%%%%%%%%%%%%%%%%%%%%%%%%%%%%%%%%%%%%%%%%%%%%%%%%%%%%%%%%%%%%%%%%%%%%%%%%%%%%%%%%%%%%%%%%%%%%%%%%

\[\begin{array}{cc@{~}|lllll}
    &\multicolumn{2}{c}{=}&\displaystyle\sum_{\tilde{j}_K=j_K\pm\frac{1}{2}\atop \tilde{J}=J\pm\frac{1}{2}}&
                           \displaystyle\sum_{m_1\!+\!\ldots\!+m_N\!+\!\ldots\!+m_K\!=\!M}
    &\big<\,j_1\,m_1\,;\,j_2\,m_2\,\big|\,g_2'(j_1\,j_2)~\tilde{m}_2\,\big>
    &\big<\,j_1\,m_1\,;\,j_2\,m_2\,\big|\,g_2(j_1\,j_2)~\tilde{m}_2\,\big>
    \\[-5mm]%%%%%%%%%%%%%%%%%%%%%%%%%%%%%%%%%%%%%
    &&&&
    &\big<\,g_2'\,\tilde{m}_2\,;\,j_3\,m_3\,\big|\,g_3'(g_2'\,j_3)~\tilde{m}_3\,\big>
    &\big<\,g_2\,\tilde{m}_2\,;\,j_3\,m_3\,\big|\,g_3(g_2\,j_3)~\tilde{m}_3\,\big>
    \\%%%%%%
    &&&&&~~\vdots&~~\vdots
    \\
    &&&&\multicolumn{2}{l}{\big<\,g_{K-2}'\,\tilde{m}_{K-2}\,;\,j_{K-1}\,m_{K-1}\,
     \big|\,g_{K-1}'(g_{K-2}'\,j_{K-1})~\tilde{m}_{K-1}\,\big>}
    &\big<\,g_{K-2}\,\tilde{m}_{K-2}\,;\,j_{K-1}\,m_{K-1}\,
     \big|\,g_{K-1}(g_{K-2}\,j_{K-1})~\tilde{m}_{K-1}\,\big>
    \\
    &&&&\multicolumn{2}{l}{\big<\,g_{K-1}'\,\tilde{m}_{K-1}\,;\,j_{K+1}\,m_{K+1}\,
     \big|\,g_{K}'(g_{K-1}'\,j_{K+1})~\tilde{m}_{K}\,\big>}
     &
     \big<\,g_{K-1}\,\tilde{m}_{K-1}\,;\,j_{K+1}\,m_{K+1}\,
     \big|\,g_{K}(g_{K-1}\,j_{K+1})~\tilde{m}_{K}\,\big>
    \\
    &&&&\multicolumn{2}{l}{\big<\,g_{K}'\,\tilde{m}_{K}\,;\,j_{K+2}\,m_{K+2}\,
     \big|\,g_{K+1}'(g_{K}'\,j_{K+2})~\tilde{m}_{K+1}\,\big>}
     &
     \big<\,g_{K}\,\tilde{m}_{K}\,;\,j_{K+2}\,m_{K+2}\,
     \big|\,g_{K+1}(g_{K}\,j_{K+2})~\tilde{m}_{K+1}\,\big>
    \\%%%%%%
    &&&&&~~\vdots&~~\vdots
    \\
    &&&&\multicolumn{2}{l}{\big<\,g_{N-2}'\,\tilde{m}_{N-2}\,;\,j_{N}\,m_{N}\,
     \big|\,g_{N-1}'(g_{N-2}'\,j_{N})~\tilde{m}_{N-1}\,\big>}
    &\big<\,g_{N-2}\,\tilde{m}_{N-2}\,;\,j_{N}\,m_{N}\,
     \big|\,g_{N-1}(g_{N-2}\,j_{N})~\tilde{m}_{N-1}\,\big>
    \\
    &&&&\multicolumn{2}{l}{\big<\,g_{N-1}'\,\tilde{m}_{N-1}\,;\,\tilde{j}_{K}\,m_{K}\!+\!A\,
     \big|\,g_{N}'(g_{N-1}'\,\tilde{j}_{K})=J'~M\!+\!A\,\big>}
     &
     \big<\,g_{N-1}\,\tilde{m}_{N-1}\,;\,j_{K}\,m_{K}\,
     \big|\,g_{N}(g_{N-1}\,j_{K})=J~M\,\big>~\times
     \\&&
     \\&&&&\multicolumn{2}{l}{\times~C^{j_K}_{\tilde{j}_K}(B,n_K)~ \big<\,j_K\,m_K\,;\,\frac{1}{2}\,A \,\big|\,\tilde{j}_K\,m_K\!+\!A \,\big>~\times}
     \\&&
     
     \\
     &&&&\multicolumn{3}{l}{
         \times~\Big[\displaystyle\prod_{L=1 \atop L\ne K}^N  \delta_{n_L'n_L}\delta_{j_L'j_L}\Big]
	         \times\delta_{n_K'n_K+B}~\delta_{j_K'\tilde{j}_K}~\times~\delta_{J'\tilde{J}}}
     \\&&
     \\&&&\multicolumn{4}{l}{\fcmt{10}{Now we can use the unitarity properties of the Clebsch Gordan Coefficients to carry out all but the last sum (starting from the first line) over the $m$'s}}
     \\&&%%%%%%%%%%%%%%%%%%%%%%%%%%%%%%%%%%%%%%%%%%%%%%%%%%%
     \\&\multicolumn{2}{c}{=}&\multicolumn{4}{l}{
              \displaystyle\sum_{\tilde{j}_K=j_K\pm\frac{1}{2} \atop \tilde{J}=J\pm\frac{1}{2}}
                               C^{j_K}_{\tilde{j}_K}(B,n_K) ~~~~~~
                           \displaystyle\sum_{\tilde{m}_{N-1}+m_K=M}
                                \big<\,g_{N-1}'\,\tilde{m}_{N-1}\,;\,\tilde{j}_{K}\,m_{K}\!+\!A\,
                                \big|\,J'(g_{N-1}'\,\tilde{j}_{K})~M\!+\!A\,\big>}
     
     \\[-3.5mm]
     &&&&\multicolumn{3}{l}{\hspace{4.26cm}                           \big<\,g_{N-1}\,\tilde{m}_{N-1}\,;\,j_{K}\,m_{K}\,
                                \big|\,J(g_{N-1}\,j_{K})~M\,\big>~}
     \\
     &&&&\multicolumn{3}{l}{\hspace{4.26cm}
                \big<\,j_K\,m_K\,;\,\frac{1}{2}\,A \,\big|\,\tilde{j}_K\,m_K\!+\!A \,\big>}

     \\&&
     \\
     &&&&   \multicolumn{3}{l}{\hspace{4.26cm}
                \times~\Big[\displaystyle\prod_{L=1 \atop L\ne K}^N  \delta_{n_L'n_L}\delta_{j_L'j_L}\Big]
	         \times\delta_{n_K'n_K+B}~\delta_{j_K'\tilde{j}_K}}
		 		
     \\
     &&&&   \multicolumn{3}{l}{\hspace{4.26cm}
                \times\delta_{M'M+A}~\delta_{J'\tilde{J}}~\displaystyle\prod_{L=1}^{N-1}\delta_{g_{L}'g_{L}}}		
     \\&&%%%%%%%%%%%%%%%%%%%%%%%%%%%%%%%%%%%%%%%%%%%%%%%%%%%%
     \\&\multicolumn{2}{c}{=}&\multicolumn{4}{l}{\displaystyle\sum_{\tilde{j}_K=j_K\pm\frac{1}{2}\atop \tilde{J}=J\pm\frac{1}{2}}
                               C^{j_K}_{\tilde{j}_K}(B,n_K)~~~~~~
                           \displaystyle\sum_{\tilde{m}_{N-1}}
                   \big<\,g_{N-1}\,\tilde{m}_{N-1}\,;\,\tilde{j}_{K}\,M\!-\!\tilde{m}_{N-1}\!+\!A\,
                                \big|\,\tilde{J}(g_{N-1}\,\tilde{j}_{K})~M\!+\!A\,\big>}
     
     \\[-3.5mm]
     &&&&\multicolumn{3}{l}{\hspace{3.14cm}                           \big<\,g_{N-1}\,\tilde{m}_{N-1}\,;\,j_{K}\,M\!-\!\tilde{m}_{N-1}\,
                                \big|\,J(g_{N-1}\,j_{K})~M\,\big>~}
     \\
     &&&&\multicolumn{3}{l}{\hspace{3.14cm}
                \big<\,j_K\,M-\tilde{m}_{N-1}\,;\,\frac{1}{2}\,A \,\big|\,\tilde{j}_K\,M-\tilde{m}_{N-1}\!+\!A \,\big>}

     \\&&
     \\
     
     &&&&   \multicolumn{3}{l}{\hspace{3.14cm}
                \times~\Big[\displaystyle\prod_{L=1 \atop L\ne K}^N  \delta_{n_L'n_L}\delta_{j_L'j_L}\Big]
	         \times\delta_{n_K'n_K+B}~\delta_{j_K'\tilde{j}_K}}
		 		
     \\
     &&&&   \multicolumn{3}{l}{\hspace{3.14cm}
                \times\delta_{M'M+A}~\delta_{J'\tilde{J}}~\displaystyle\prod_{L=1}^{N-1}\delta_{g_{L}'g_{L}}}		
     \\&&%%%%%%%%%%%%%%%%%%%%%%%%%%%%%%%%%%%%%%%%%%%%%%%%%%%%
     \\&\multicolumn{2}{c}{=}&\multicolumn{4}{l}{\displaystyle\sum_{\tilde{j}_K=j_K\pm\frac{1}{2}\atop\tilde{J}=J\pm\frac{1}{2}}
                               C^{j_K}_{\tilde{j}_K}(B,n_K)~~
                               C^{J~j_K}_{\tilde{J}~ \tilde{j}_K}(A,M,g_{N-1})~\times~
			       
    \Big[\displaystyle\prod_{L=1 \atop L\ne K}^N  \delta_{n_L'n_L}\delta_{j_L'j_L}\Big]
	         \times\delta_{n_K'n_K+B}~\delta_{j_K'\tilde{j}_K}
		 		
    \times\delta_{M'M+A}~\delta_{J'\tilde{J}}~\displaystyle\prod_{L=1}^{N-1}\delta_{g_{L}'g_{L}}}		
			       
\end{array}\] 
\end{footnotesize} \be \label{Holonomieaktion auf allgemeine Kante}\ee

Here in the last line we have introduced $C^{J~j_K}_{\tilde{J}~ \tilde{j}_K}(A,M,g_{N-1})$ as can be seen from the context or explicitely in (\ref{Abkuerzungsdefinition}).

\pagebreak

Using (\ref{3nj Expansion I}) and (\ref{Holonomieaktion auf allgemeine Kante}) we can now complete the expansion (\ref{Entwicklung in anderes Recoupling Scheme})

\begin{footnotesize}

\[\begin{array}{cc@{~}|ccllcccc} 
   \lefteqn{\big<\,\vec{a}'\,J'\,M'\,;\,\vec{n}'\,\big|\,
    \big[\pi_{\frac{1}{2}}(h_K) \big]_{AB}
    \,\big|\,\vec{a}\,J\,M\,;\,\vec{n}\,\big>
    =:\big<\,T'_{J'}\,\big|\,\big[\pi_{\frac{1}{2}}(h_K)\big]_{AB}\,\big|\,T_J\,\big>=}
    
    \\
    \\ 
    
    &\multicolumn{2}{c}{=}& 
    \displaystyle\sum_{\vec{g}'\,\vec{g}}&
     \multicolumn{4}{l}{
    \big<\,\vec{a}'\,J'\,M'\,;\,\vec{n}'\,\big|\,\vec{g}'\,J'\,M'\,;\,\vec{n}'\,\big>
    ~
    \big<\,\vec{g}'\,J'\,M'\,;\,\vec{n}'\,\big|\,
    \big[\pi_{\frac{1}{2}}(h_K) \big]_{AB}\,
    \big|\,\vec{g}\,J\,M\,;\,\vec{n}\,\big>
    ~
    \big<\,\vec{g}\,J\,M\,;\,\vec{n}\,\big|
    \,\vec{a}\,J\,M\,;\,\vec{n}\,\big>}
    \\[-3mm]&&
    \\%%%%%%%%%%%%%%%%%%%%%%%%%%%%%%%%%%%%%
    
      &\multicolumn{2}{c}{=}&
      \displaystyle\sum_{\tilde{j}_K=j_K\pm\frac{1}{2}\atop \tilde{J}=J\pm\frac{1}{2}}~~\sum_{g'_K\ldots g_{N\!-\!1}' \atop g_K\ldots g_{N\!-\!1}}&\bigg\{
      
      \\
      &&&\multicolumn{2}{l}{
      \big<a_K'(a_{K\!-\!1}'\,j_K')~a_{K\!+\!1}'(a_K'\,j_{K\!+\!1}')
	  \,\big|\,
	  g_K'(a_{K-1}'\,j_{K+1})~a_{K+1}'(g_K'\,j_K)\,\big>}
      &
      \big<a_K(a_{K\!-\!1}\,j_K)~a_{K\!+\!1}(a_K\,j_{K\!+\!1})
	  \,\big|\,
	  g_K(a_{K-1}\,j_{K+1})~a_{K+1}(g_K\,j_K)\,\big>&&&~	  
      \\
      
      &&&\multicolumn{2}{l}{\big<a_{K\!+\!1}'(g_{K}'\,j_K')~a_{K\!+\!2}'(a_{K\!+\!1}'\,j_{K\!+\!2}')
	  \,\big|\,
	  g_{K\!+\!1}'(g_{K}'\,j_{K+2}')~a_{K+2}'(g_{K\!+\!1}'\,j_K')\,\big>}
      &\big<a_{K\!+\!1}(g_{K}\,j_K)~a_{K\!+\!2}(a_{K\!+\!1}\,j_{K\!+\!2})
	  \,\big|\,
	  g_{K\!+\!1}(g_{K}\,j_{K+2})~a_{K+2}(g_{K\!+\!1}\,j_K)\,\big>
	  	  
      \\
      
      &&&\vdots&&~~~~~~~~~\vdots
      
      \\
      
      &&&\multicolumn{2}{l}{\big<a_{N\!-\!3}'(g_{N-4}'\,j_K')~a_{N\!-\!2}'(a_{N\!-\!3}'\,j_{N\!-\!2}')
	  \,\big|\,
	  g_{N\!-\!3}'(g_{N-4}'\,j_{N-2}')~a_{N-2}'(g_{N\!-\!3}'\,j_K')\,\big>}
      &\big<a_{N\!-\!3}(g_{N-4}\,j_K)~a_{N\!-\!2}(a_{N\!-\!3}\,j_{N\!-\!2})
	  \,\big|\,
	  g_{N\!-\!3}(g_{N-4}\,j_{N-2})~a_{N-2}(g_{N\!-\!3}\,j_K)\,\big>
	  	  
      \\
      
      &&&\multicolumn{2}{l}{\big<a_{N\!-\!2}'(g_{N-3}'\,j_K')~a_{N\!-\!1}'(a_{N\!-\!2}'\,j_{N\!-\!1}')
	  \,\big|\,
	  g_{N\!-\!2}'(g_{N-3}'\,j_{N-1}')~a_{N-1}'(g_{N\!-\!2}'\,j_K')\,\big>}
      &\big<a_{N\!-\!2}(g_{N-3}\,j_K)~a_{N\!-\!1}(a_{N\!-\!2}\,j_{N\!-\!1})
	  \,\big|\,
	  g_{N\!-\!2}(g_{N-3}\,j_{N-1})~a_{N-1}(g_{N\!-\!2}\,j_K)\,\big>
	  	  
      \\
      
      &&&\multicolumn{2}{l}{\big<a_{N\!-\!1}'(g_{N-2}'\,j_K')~J'(a_{N\!-\!1}'\,j_{N}')
	  \,\big|\,
	  g_{N\!-\!1}'(g_{N-2}'\,j_{N}')~J'(g_{N\!-\!1}'\,j_K')\,\big>}
      &\big<a_{N\!-\!1}(g_{N-2}\,j_K)~J(a_{N\!-\!1}\,j_{N})
	  \,\big|\,
	  g_{N\!-\!1}(g_{N-2}\,j_{N})~J(g_{N\!-\!1}\,j_K)\,\big>
	  	  
      \\
      
      &&&\multicolumn{2}{l}{\delta_{a_2'\,g_2'} \ldots \delta_{a_{K-1}'\, g_{K-1}'}}
      &\delta_{a_2\,g_2} \ldots \delta_{a_{K-1}\, g_{K-1}}  

     \\&& 
     \\&&&\multicolumn{5}{l}{C^{j_K}_{\tilde{j}_K}(B,n_K)~~
                               C^{J~j_K}_{\tilde{J}~ \tilde{j}_K}(A,M,g_{N-1})~\times~
			       
    \Big[\displaystyle\prod_{L=1 \atop L\ne K}^N  \delta_{n_L'n_L}\delta_{j_L'j_L}\Big]
	         \times\delta_{n_K'n_K+B}~\delta_{j_K'\tilde{j}_K}
		 		
    \times\delta_{M'M+A}~\delta_{J'\tilde{J}}~\displaystyle\prod_{L=1}^{N-1}\delta_{g_{L}'g_{L}}
    ~~~~\bigg\}}

     \\&&  

    \\%%%%%%%%%%%%%%%%%%%%%%%%%%%%%%%%%%%%%
    
      &\multicolumn{2}{c}{=}&
      \displaystyle\sum_{\tilde{j}_K=j_K\pm\frac{1}{2}\atop \tilde{J}=J\pm\frac{1}{2}}~~\sum_{g_{K}\ldots g_{N\!-\!1}}&\bigg\{
      
      \\
      &&&\multicolumn{2}{l}{
      \big<a_K'(a_{K\!-\!1}\,\tilde{j}_k)~a_{K\!+\!1}'(a_K'\,j_{K\!+\!1})
	  \,\big|\,
	  g_K(a_{K-1}\,j_{K+1})~a_{K+1}'(g_K\,\tilde{j}_K)\,\big>}
      &
      \big<a_K(a_{K\!-\!1}\,j_K)~a_{K\!+\!1}(a_K\,j_{K\!+\!1})
	  \,\big|\,
	  g_K(a_{K-1}\,j_{K+1})~a_{K+1}(g_K\,j_K)\,\big>&&&~	  
      \\
      
      &&&\multicolumn{2}{l}{\big<a_{K\!+\!1}'(g_{K}\,\tilde{j}_K)~a_{K\!+\!2}'(a_{K\!+\!1}'\,j_{K\!+\!2})
	  \,\big|\,
	  g_{K\!+\!1}(g_{K}\,j_{K+2})~a_{K+2}'(g_{K\!+\!1}\,\tilde{j}_K)\,\big>}
      &\big<a_{K\!+\!1}(g_{K}\,j_K)~a_{K\!+\!2}(a_{K\!+\!1}\,j_{K\!+\!2})
	  \,\big|\,
	  g_{K\!+\!1}(g_{K}\,j_{K+2})~a_{K+2}(g_{K\!+\!1}\,j_K)\,\big>
	  	  
      \\
      
      &&&\vdots&&~~~~~~~~~\vdots
      
      \\
      
      &&&\multicolumn{2}{l}{\big<a_{N\!-\!3}'(g_{N-4}\,\tilde{j}_K)~a_{N\!-\!2}'(a_{N\!-\!3}'\,j_{N\!-\!2})
	  \,\big|\,
	  g_{N\!-\!3}(g_{N-4}\,j_{N-2})~a_{N-2}'(g_{N\!-\!3}\,\tilde{j}_K)\,\big>}
      &\big<a_{N\!-\!3}(g_{N-4}\,j_K)~a_{N\!-\!2}(a_{N\!-\!3}\,j_{N\!-\!2})
	  \,\big|\,
	  g_{N\!-\!3}(g_{N-4}\,j_{N-2})~a_{N-2}(g_{N\!-\!3}\,j_K)\,\big>
	  	  
      \\
      
      &&&\multicolumn{2}{l}{\big<a_{N\!-\!2}'(g_{N-3}\,\tilde{j}_K)~a_{N\!-\!1}'(a_{N\!-\!2}'\,j_{N\!-\!1})
	  \,\big|\,
	  g_{N\!-\!2}(g_{N-3}\,j_{N-1})~a_{N-1}'(g_{N\!-\!2}\,\tilde{j}_K)\,\big>}
      &\big<a_{N\!-\!2}(g_{N-3}\,j_K)~a_{N\!-\!1}(a_{N\!-\!2}\,j_{N\!-\!1})
	  \,\big|\,
	  g_{N\!-\!2}(g_{N-3}\,j_{N-1})~a_{N-1}(g_{N\!-\!2}\,j_K)\,\big>
	  	  
      \\
      
      &&&\multicolumn{2}{l}{\big<a_{N\!-\!1}'(g_{N-2}\,\tilde{j}_K)~J'(a_{N\!-\!1}'\,j_{N})
	  \,\big|\,
	  g_{N\!-\!1}(g_{N-2}\,j_{N})~J'(g_{N\!-\!1}\,\tilde{j}_K)\,\big>}
      &\big<a_{N\!-\!1}(g_{N-2}\,j_K)~J(a_{N\!-\!1}\,j_{N})
	  \,\big|\,
	  g_{N\!-\!1}(g_{N-2}\,j_{N})~J(g_{N\!-\!1}\,j_K)\,\big>

     \\&& 
     \\&&&\multicolumn{5}{l}{C^{j_K}_{\tilde{j}_K}(B,n_K)~~
                               C^{J~j_K}_{\tilde{J}~ \tilde{j}_K}(A,M,g_{N-1})~\times~
			       
    \Big[\displaystyle\prod_{L=1 \atop L\ne K}^N  \delta_{n_L'n_L}\delta_{j_L'j_L}\Big]
	         \times\delta_{n_K'n_K+B}~\delta_{j_K'\tilde{j}_K}
		 		
    \times\delta_{M'M+A}~\delta_{J'\tilde{J}}~\times~

    \prod_{R=2}^{K-1}\delta_{a_R'a_R} ~~~~\bigg\}  }

     \\&&  
    \\%%%%%%%%%%%%%%%%%%%%%%%%%%%%%%%%%%%%%
    
      &\multicolumn{2}{c}{\stackrel{\cdot}{=}}&\multicolumn{5}{l}{ \displaystyle\sum_{\tilde{j}_K=j_K\pm\frac{1}{2}\atop \tilde{J}=J\pm\frac{1}{2}}
                 C^{j_K}_{\tilde{j}_K}(B,n_K)
       \displaystyle\sum_{g_{N-1}} 
                 C^{J~j_K}_{\tilde{J}~ \tilde{j}_K}(A,M,g_{N-1})
       \displaystyle\sum_{g_K\ldots g_{N-2}}
		 
                 C^{\vec{a}~J~j_K}_{\vec{a}'\tilde{J}~\tilde{j}_K}(g_K,\ldots,g_{N-1})
	}
     \\&&&
     
     \multicolumn{3}{c}{	 
		 \times~
			       
    \Big[\displaystyle\prod_{L=1 \atop L\ne K}^N  \delta_{n_L'n_L}\delta_{j_L'j_L}\Big]
	         \times\delta_{n_K'n_K+B}~\delta_{j_K'\tilde{j}_K}
		 		
    \times\delta_{M'M+A}~\delta_{J'\tilde{J}}~\times~
    
    \prod_{R=2}^{K-1}\delta_{a_R'a_R}} 
      %%%%%%%%%%%%%%%%%%%%%%%%%%%%%%%%%%%%%%%%%%%%%%%%%%%%%%%%%%%%%%%%%%%%%%%%%%%%
    \\
    &\multicolumn{2}{c}{\stackrel{\cdot}{=}}&C^{T_J}_{T'_{J'}}(K,A,B)

    \\[-5mm]
      
\end{array}\]  
\end{footnotesize}   
\be\label{Holonomieaktion auf allgemeine Kante Endresultat} \ee
\begin{footnotesize}       
      \\[2mm]
      \fcmt{12}{
       We have used the following abbreviations ($A,B=\pm\frac{1}{2}$) :
        \ba\label{Abkuerzungsdefinition} 
           C^{j_K}_{\tilde{j}_K}(B,n_K)
	   &=&
	      \left\{ \begin{array}{rcl}
	               -2B~\bigg[\displaystyle\frac{j_K-2Bn_K}{2j_K} \bigg]^{\frac{1}{2}}
		       &~~~~~&\mbox{if}~~~\tilde{j}_K=j_K-\frac{1}{2} 
		       
		       \\\\
		       
		       \bigg[\displaystyle\frac{j_K+2Bn_K+1}{2(j_K+1)} \bigg]^{\frac{1}{2}}
		       &~~&\mbox{if}~~~\tilde{j}_K=j_K+\frac{1}{2}
	              \end{array} \right.
	   \nonumber
	   \\
	   \nonumber
	   \\[5mm] 
           C^{J~j_K}_{\tilde{J}~ \tilde{j}_K}(A,M,g_{N-1})
	   &=&\displaystyle\sum_m
               \big<\,g_{N\!-\!1}\,m\,;\,j_K\,M\!-\!m\,\big|\,J(g_{N\!-\!1}\,j_K)~M\big>
	    \nonumber\\ [-1.5mm]
            &&
	       ~~~~~\big<\,g_{N\!-\!1}\,m\,;\,\tilde{j}_K\,M\!-\!m\!+\!A_0\,\big|\,J'(g_{N\!-\!1}\,\tilde{j}_K)~M\!+\!A\big> 
           \nonumber\\
            &&
	        ~~~~~\big<~j_K~M\!-\!m~;~\frac{1}{2}~A~\big|~\tilde{j}_K~M\!-m\!+\!A~\big>
           \ea 
           }
\end{footnotesize}
~
\\\\[1cm]
Now the reason for our expansions becomes clear: for arbitrary $K$ the action of a holonomy $\pi_j(h_K)$ on a standard recoupling scheme can be expressed as a sum of standard recoupling schemes, where the expansion coefficients have a modular structure, which enables us to give explicit general equations since all the expressions in (\ref{Holonomieaktion auf allgemeine Kante Endresultat}) can be calculated separately. 

Finally we have achieved our goal to express the state resulting from the action of a holonomy of an edge on a (standrad) recoupling scheme as a sum over recoupling schemes:

\fcmt{16}{\ba
    \big[\pi_{\frac{1}{2}}(h_K) \big]_{AB}
    \,\big|\,T_J\,\big>
    &=&
    \sum_{T'_{J'}}~ C^{T_J}_{T'_{J'}}(K,A,B)~
    \big|\,T'_{J'}\,\big>\,
\ea   }
\\[5mm]
%%%%%%%%%%%%%%%%%%%%%%%%%%%%%%%%%%%%%%%%%%%%%%%%%%%%%%%%%
where $T'_{J'}$ is a multilabel containig all the quantum numbers $\vec{a}',J',\vec{j}',\vec{m}',\vec{n}'$\footnote{e.g. $\vec{n}=\{n_1,\ldots,n_N\}$}
\paragraph{Note again the range of the variables involved,}in order to get a non vanishing expansion coefficient:\\\\
%%%%%%%%%%%%%%%%%%%%%%%%%%%%%%%%%%%%%%%%%%%%%%%%%%%%%%%%%
\fcmt{16}{\[\begin{array}{cclccclccclcl}
	  \tilde{j_K}&=&j_K\pm\frac{1}{2}&~~&m_K'&=&m_K+A&~~~~&n_K'&=&n_K+B&~~&\mbox{for the action of the holonomy $\hat{h}_K:=\big[\pi_{\frac{1}{2}}(h_K)\big]_{AB}$}\\
	  j_l'&=&j_l&&m_l'&=&m_l&&n_l'&=&n_l&&\forall~~l\ne k\\\cline{1-11}\\
	  
	  M'&=&M+A &&J'&=&J\pm\frac{1}{2}

	  \end{array}\]}

\vfill

\pagebreak

%%%%%%%%%%%%%%%%%%%%%%%%%%%%%%%%%%%%%%%%%%%%%%%%%%%%%%%%%%%%%%%%%%%%%%%%%%%%%%%%%%%%%%%%%%%%%%%%%%%%%%
\section{Definitions and Conventions for $U(1)^3$ Coherent State Calculation}
%%%%%%%%%%%%%%%%%%%%%%%%%%%%%%%%%%%%%%%%%%%%%%%%%%%%%%%%%%%%%%%%%%%%%%%%%%%%%%%%%%%%%%%%%%%%%%%%%%%%%%

In this section we will summarize the construction of  complexifier coherent states for Loop Quantum Gravity. For a more detailed introduction we refer to \cite{TT:review,CGS II,QFT on CST I,QFT on CST II}.

%%%%%%%%%%%%%%%%%%%%%%%%%%%%%%%%%%%
\subsection{General Construction}
%%%%%%%%%%%%%%%%%%%%%%%%%%%%%%%%%%%
Formally a complexifier coherent state can be constructed by
\ba\label{Allgemeinste Def CS}
   \Psi_m(A)
   &=&\Big[\mb{e}^{-\frac{1}{\hbar}C}\delta_{A'} \big](A)_{A'\rightarrow A'^{\mb{C}}(m)}
\ea

Here $\delta_{A'}(A)=\sum\limits_{s\in S}T_s(A')\overline{T_s(A)}$ denotes the $\delta$-distribution on the (quantum)configuration space $\overline{\mathcal{A}}$ with respect to the Ashtekar-Lewandowski-measure $\mu_0$ in the thus built kinematical Hilbert space $\mathcal{H}_{kin}^0=L_2(\overline{\mathcal{A}},d\mu_0)$. The sum has to be extended over all spin network labels $s$ and the $T$'s are the corresponding spin network functions which provide an orthonormal basis. 
$C$ is called complexifier, $A'\rightarrow A'^{\mb{C}}$ indicates that after the action of $\mb{e}^{-\frac{1}{\hbar}C}$ the whole expression has to be analytically continued to all values the complexification $A'^{\mb{C}}$ of the connection $A'$ given by
\ba\label{Iterierte Poissonklammer}
   A'^{\mb{C}}&=&\sum_{n=0}^{\infty}\frac{\mb{i}^n}{n!}\big\{A,C \big\}_{(n)}
\ea 

can take (with the iterated Poisson bracket given by $\big\{F,G \big\}_{(0)}=F$
and $\big\{F,G \big\}_{(n+1)}=\big\{\{F,G \}_{(n)},G \big\}$). This construction works for every compact gauge group $G$.

%%%%%%%%%%%%%%%%%%%%%%%%%%%%%%%%%%%%%%%
\subsection{$G=SU(2)$}
%%%%%%%%%%%%%%%%%%%%%%%%%%%%%%%%%%%%%%%
The Hilbert space $\mathcal{H}_{kin}^0=L_2(\overline{\mathcal{A}},d\mu_0)$ is constructed as an inductive limit of subspaces $\mathcal{H}_{kin}^\gamma=L_2(\overline{\mathcal{A}},d\mu_\gamma)$ consisting of square integrable fuctions $T_\gamma$ cylindrical with respect to graphs $\gamma$ consisting of analytical embedded edges $e\in E(\gamma)$ which intersect in the vertices $v \in V(\gamma)$.
The coherent states are restricted to an arbitrary but fixed graph $\gamma$, because due to the uncountability of the set $s$ in (\ref{Allgemeinste Def CS}) the thus constructed coherent state would not be normalizable. In order to allow distributional connections $A\in \overline{\mathcal{A}}$ one  regularizes the classical Poisson algebra of connections $A^j_a(x)$ and electric fields $E^b_k(y)$
\be\label{Klassische Poissonalgebra}
   \big\{A^j_a(x),A_b^k(y) \big\}=\big\{E_j^a(x),E^b_k(y) \big\}=0
   ~~~~~
   \big\{A^j_a(x),E^b_k(y) \big\}=\kappa\delta^a_b \delta^k_j \delta(x,y)
\ee
by smearing the connection over the one dimensional edges $e$ of a graph in order to obtain holonomies $h_e(A):=\mathcal{P}\mb{e}^{\int_e A}$ and integrating the electric fields over surfaces $S$ in order to get electric fluxes $E_j(S)=\int_S *E_j$. One finds
\ba\label{Klassische Poissonalgebra ausgeschmiert}
   \big\{ h_e,h_{e'} \big\}=\big\{E_j(S),E_k(S') \big\}&=&0~~~~~~\mbox{if $S$, $S'$ do not intersect (as it will be the case in our later considerations)}
   \nonumber\\
   \big\{E_j(S),h_e\big\}&=&\left\{\begin{array}{ccl}
                                     0&~~~~&e\cap S = \emptyset~\mbox{or}~e\cap S = e\\
				     \kappa\sigma(e,S)\frac{\tau_j}{2}h_e&~~~~&e \cap S = u
				     ~~~~~~
				     \mbox{$u\ldots$beginning point of $e$}
                                 \end{array} \right.
\ea
Here the edge $e$ is adapted to the surface $S$, that is $e$ is outgoing from $S$. The orientataion of the tangent $\dot{e}(u)$ of $e$ compared to the surface normal at the intersectiong point $u$ is denoted by $\sigma(e,S)$. For the surface normal pointing up $\sigma(e,S)=1$ if $\dot{e}(u)$ points up,
$\sigma(e,S)=-1$ if $\dot{e}(u)$ points down . 
The Peter\&Weyl theorem is then exployed in order to go to the spin network representation, built from the matrix-element functions of the representation matrices of the holonomies, whose closed linear span provides a basis of $\mathcal{H}_{kin}^\gamma=L_2(\overline{\mathcal{A}},d\mu_\gamma)$.
\ba
   h_e(A) &\mapsto& \sqrt{\dim{\pi_j}} \big[ \pi_j(h_e(A))\big]_{mn}
          = \sqrt{\dim{\pi_j}} \big[ \pi_j\big]_{mn}\big(h_e(A)\big)
	  =: \big<~h_e(A)~\big|~j~m~;~n~\big>
	  \nonumber\\
   T_{\gamma\vec{j}\vec{m}\vec{n}}
   &:=&\prod_{e\in E(\gamma)} \sqrt{\dim\pi_j}\big[ \pi_j(h_e(A))\big]_{mn} 
\ea 
Here $j=0,\frac{1}{2},1,\frac{3}{2},\ldots$ is the weight of the representation and 
$m,n=-j,-j+1,\ldots,j-1,j$ denote its matrix element. The Poisson algebra (\ref{Klassische Poissonalgebra ausgeschmiert}) is then represented on $\mathcal{H}_{kin}^\gamma$ as
\ba\label{Ausgeschmierte Poisson Algebra Quantisiert}
   \hat{\pi}_j\big[h_e\big]_{m_0n_0} T_{\gamma\vec{j}\vec{m}\vec{n}}
   &=&\pi_j\big[h_e\big]_{m_0n_0}\cdot T_{\gamma\vec{j}\vec{m}\vec{n}}
   \nonumber\\
   \hat{E}_k(S_e)~T_{\gamma\vec{j}\vec{m}\vec{n}}
   &=&\mb{i}\hbar \big\{E_k(S_e),h_e \big\}
   =X^R_k(S_e) T_{\gamma\vec{j}\vec{m}\vec{n}}
   \nonumber\\
   &=&
   \tr_{j_e}\Big[(\tau_k~h_e)^T~\frac{\partial}{h_e}\Big] T_{\gamma\vec{j}\vec{m}\vec{n}}
   =\sum_{p,q=-j_e}^{j_e} \pi_{j_e}\big[(\tau_k h_e)^T\big]_{pq}  \frac{\partial}{\pi_{j_e}\big[h_e\big]_{pq}} T_{\gamma\vec{j}\vec{m}\vec{n}} 
\ea
 where $S_e$ denotes that the smearing surface is transversal to the edge $e$ and $X^r_k(S_e)$ is the right invariant vectorfield on $G$, $~^T$ means transpose.

%%%%%%%%%%%%%%%%%%%%%%%%%
\subsection{$G=U(1)^3$}
%%%%%%%%%%%%%%%%%%%%%%%%%
\subsubsection{Charge Networks}

We will specify now to $G=U(1)^3$ instead of $G=SU(2)$ as an Abelianized model of General Relativity. Instead of the 3 quantum numbers $j,m,n$ we will have three copies of $U(1)$ for each edge with three charges $n_1,n_2,n_3 \in \mb{Z}$. The point is now, that the two index sets posses the same cardinality and therefore calculations can be (as a proof of concept) first carried out in $U(1)^3$ theory which as we will show is much simpler. The justification for this is the fact that the results seem to be qualitatively the same.

Let $A_a^k$ be a $U(1)^3$ connection on a three dimensional manifold $\sigma$ and $E$ is a conjugate electric field satisfying the canonical Poisson brackets (\ref{Klassische Poissonalgebra ausgeschmiert}). Again the index $a,b,c,\ldots=1,2,3$ are spatial indices whereas  
$i,j,k,\ldots=1,2,3$ now indicate the copy of $U(1)$

In order to regularize (\ref{Klassische Poissonalgebra}) we again introduce holonomy and flux variables:
\ba\label{E(S) erste Definition}
   h^j_e(A)&:=&\mb{e}^{\mb{i}\int\limits_e A^j_a(x) dx^a}=\mb{e}^{\mb{i}\theta^j_e(A)}
   \nonumber\\[3mm]
   E_j(S)&:=&\int\limits_S *E=\int\limits_S \epsilon_{abc} E^a_j(x)dx^adx^b
   =\int\limits_U du^1du^2\big[\epsilon_{abc} X^b_{S,u^1}X^c_{S,u^2}\big]E^a_j\big(X_s(u)\big)
   \nonumber\\
   &:=&\int\limits_U d^2u~ n_a^S(u)~E^a_j\big(X_S(u)\big)
\ea
Here we have the embeddings of the edge $e: \mb{R} \mapsto \sigma $, $[0,1]\ni t\mapsto e(t)$ and 
of the surface $S$ $X_S: \mb{R}^2 \supset U \mapsto \sigma$ $ [-\frac{1}{2},\frac{1}{2}]^2 \ni u \mapsto X_S(u)$. The quantity $n_a^S(u)$ is called the ''normal'' of $S$.        
Now (\ref{Klassische Poissonalgebra ausgeschmiert}) reads as

\be \label{Klassische U(1)^3 Poissonalgebra ausgeschmiert}
   \big\{ h_e^j,h_{e'}^k \big\}=\big\{E_j(S),E_k(S') \big\}=0
   ~~~~~
   \big\{E_j(S),h_e^k\big\}=\left\{\begin{array}{ccl}
                                     0&~~~~&e\cap S = \emptyset~\mbox{or}~e\cap S = e\\
				     K\cdot\mb{i}\kappa\,\delta^k_j\sigma(e,S)h_e&~~~~&e \cap S = u
                                 \end{array} \right.
\ee
The prefactor $K=\frac{1}{2}$ if $u$ is the beginning or end point of $e$ (which can always be achieved by adapting the edge $e$ to the surface $S$) or 
$K=1$ if $u$ is an internal point of $e$.
This raises the following 
\begin{Definition}
   \begin{itemize}
      \item[]{~\\[-4mm]}
      \item[(i)]{A charge network $c$ is a pair ($\gamma$(c),n(c)) consisting of a graph $\gamma(c)$ together with a coloring of each of its edges $e \in E(\gamma)$ with 3 charges $n^j_e(c) \in \mb{Z}$. A charge network state is the following function on the space $\mathcal{C}=\mathcal{A}$ of smooth connections:
      \be\label{Definition charge network}
         T_c: ~A\mapsto \prod_{e\in\gamma(c) \atop j=1,2,3} \big[h^j_e(A) \big]^{n^j_e(c)}
      \ee
      Note, that if we would work at the gauge invariant level, then at each $v\in V(\gamma)$ for each $j$ the sum of the charges of the edges would have to add up to 0. However, we will not use gauge invariant states to begin with.}
      \item[(ii)]{The Hilbert space $\mathcal{H}_{kin}^\gamma$ is defined as the closed linear span of the charge network functions which form an orthonormal basis, that is 
      \be
         \big< T_c \big| T_{c'} \big>_{kin}=\delta_{c,c'}
      \ee
      and their finite linear span is dense. }
      \item[(iii)]{The representation of (\ref{Klassische U(1)^3 Poissonalgebra ausgeschmiert}) on $\mathcal{H}_{kin}$ is then defined by
      \be\label{Klassische U(1)^3 Poissonalgebra ausgeschmiert und quantisiert}
         \hat{h}^j_eT_c=h^j_eT_c~~~~~~~~~~~~~~~~~~
	 \hat{E}_j(S)T_c=\mb{i}\hbar\big\{E_j(S),T_c \big\}
      \ee}
   \end{itemize}
\end{Definition}

%%%%%%%%%%%%%%%%%%%%%%%%%%%%%%%%%%%%%%%%%%%%%%%%%%%%%%%%%%%%%%%%%%%%%%%%%%%%%%%%%%%%%%%%%%%%%%%%%%%%%%
\subsubsection{Construction of the Coherent States}
%%%%%%%%%%%%%%%%%%%%%%%%%%%%%%%%%%%%%%%%%%%%%%%%%%%%%%%%%%%%%%%%%%%%%%%%%%%%%%%%%%%%%%%%%%%%%%%%%%%%%%%

\begin{figure}[!hbtp] 
  \centering
  \begin{minipage}[t]{6cm} 
      \psfrag{B}{$x^3$}
      \psfrag{D}{$x^1$}
      \psfrag{C}{$x^2$}
      \psfrag{sigma}{surface~$\sigma_t^I$:~~$t_I(x)=const$}
      \includegraphics[height=3.6cm,width=6cm]{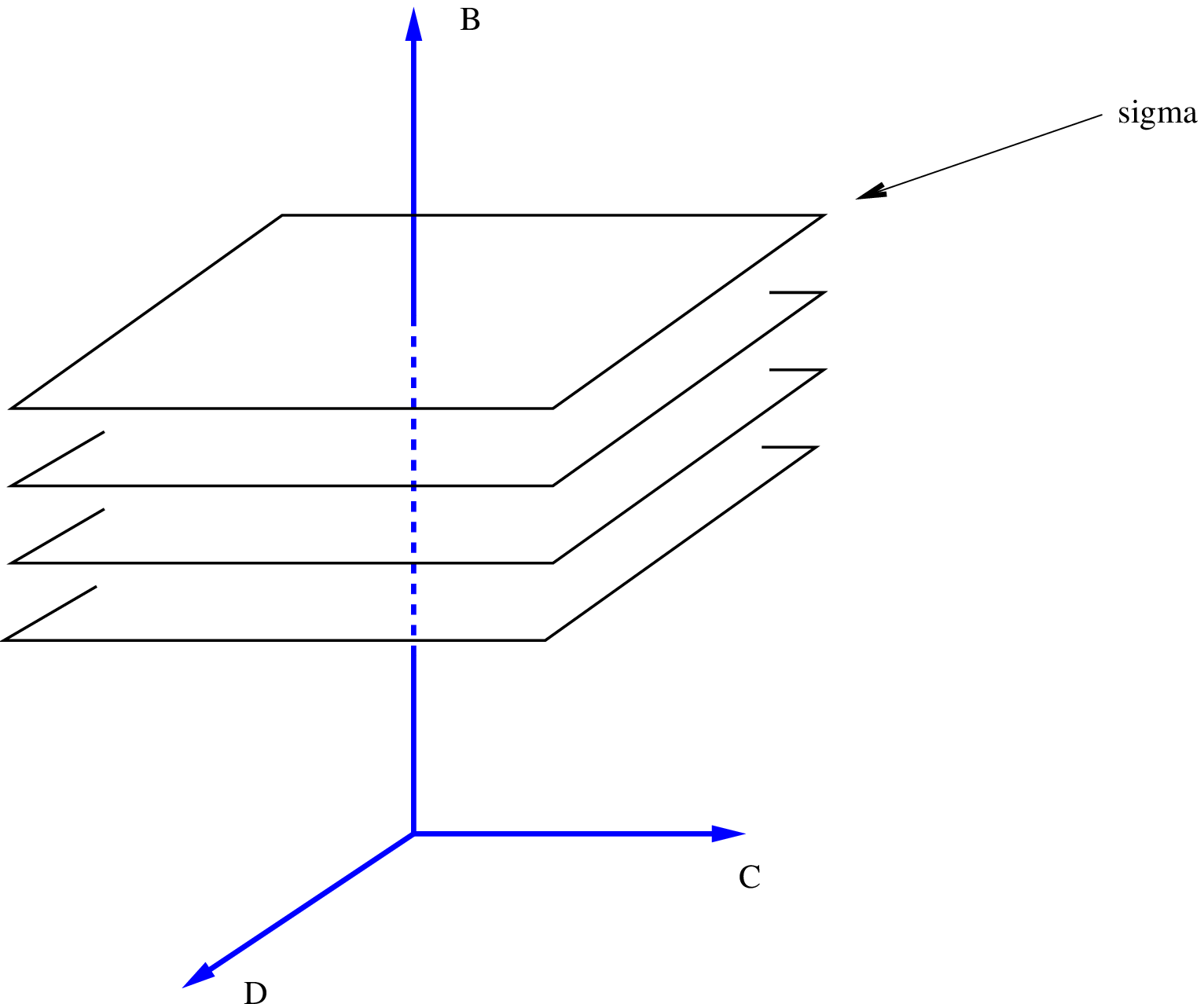}
      \caption{Foliation $\mathcal{F}^I$ ($I=3$ direction) into surfaces $\sigma_t^I$ of constant foliation parameter $t$ . The foliation varies smoothly with $t$.}
      \label{foliation}
 \end{minipage}
 \hspace{2cm}
 \begin{minipage}[t]{6cm} 
      \psfrag{S}{$S$}
      \psfrag{sigma}{$\sigma_t^I$}
      \includegraphics[height=1.5cm,width=6cm]{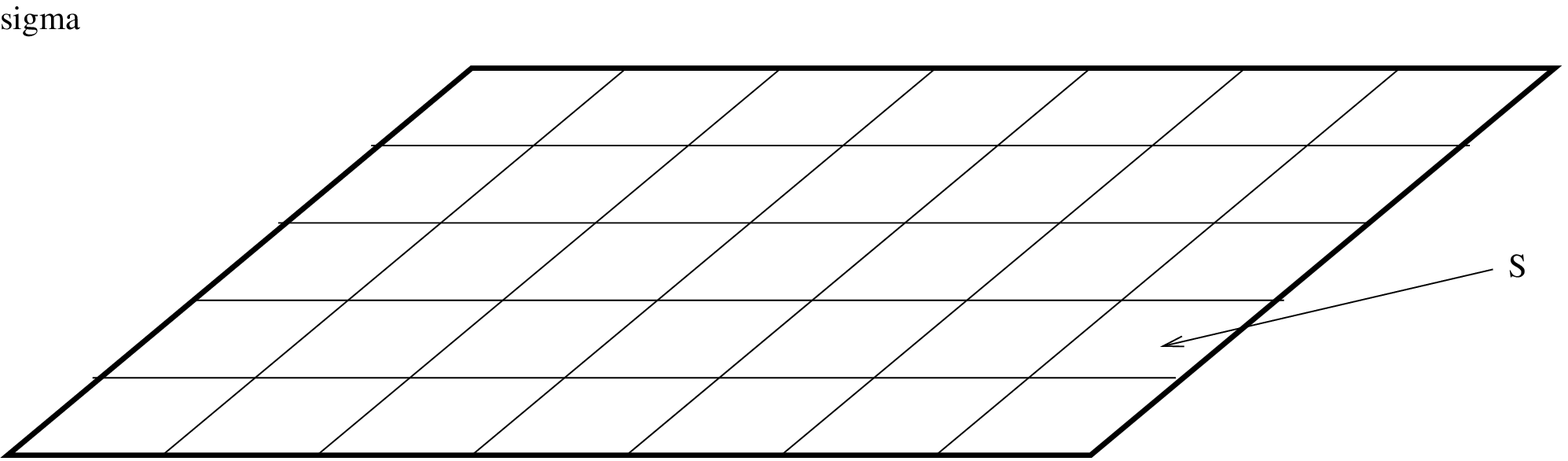}
      \caption{Choosing a parquette $\mathcal{P}^I_t$ $\Leftrightarrow$ Partition of the  $t_I(x)=const$ surface $\sigma_t^I$ into small surfaces $S$. Also $\mathcal{P}^I_t$ varies smoothly with $t_I$. Notice, that for each $I,x$ we get a {\bf unique} surface $S^I_x$ such that $x\in S^I_x$.}
      \label{foliation2}
  \end{minipage} 
\end{figure}

Although our coherent state calculations closely follows \cite{QFT on CST I},\cite{QFT on CST II} we will use a somewhat more general and flexible construction principle here:
\\ 
In order to give a regularized explicit expression for a $U(1)^3$ group coherent state over an arbitrary graph $\gamma$ we introduce 3 foliations $\mathcal{F}_I$ ($I=1,2,3$) of $\sigma$ into 2-dimensional hypersurfaces $\sigma_t^I$ such that two leaves $\sigma^I_t$, $\sigma^J_t$ for $I\ne J$ intersect transversally if they intersect at all (see figure \ref{foliation}). 

In addition choose a parquette $\mathcal{P}^I_t$ (see figure \ref{foliation2}), that is we partition each of the $\sigma^I_t$ for fixed $(I,t)$ in small surfaces $S\subset\sigma^I_t$. 

Note, that each $S^I$ is defined by its embedding $X_{S(t)}:\mb{R}^2 \supset U \mapsto \sigma , 
\big[-\frac{1}{2},\frac{1}{2} \big]^2 \ni u \mapsto X_{S(t)}(u)$, which is $t$-dependend due to the $t$-dependence of $\mathcal{P}^I_t$. 

By construction there is a bijection $Y_I:~~\sigma\mapsto \bigcup\limits_{t\in \mb{R}}(t,\sigma^I_t)$, 
~~$x\mapsto \big(t_I(x),u_I(x)\big)$ with $u_I(x)=\big(u_I^1(x),u_I^2(x)\big)$.

Now we can write down a complexifier $\mathcal{C}$, which depends, of course, on the foliation $\mathcal{F}$:
\be\label{Definition des Complexifiers}
   \mathcal{C}_{\mathcal{F}}=
			             \frac{1}{2\kappa L^{3}}\sum\limits_{I=1}^3 \int\limits_{ \mb{R}} dt
			             \sum\limits_{S\in \mathcal P^I_t} 
				       \delta^{jk} E_j(S) E_k(S)
\ee
Here $L$ is a length parameter we keep unspecified at the moment.

For the complexifier (\ref{Definition des Complexifiers}) we notice that only for $n=0,1$ we get non vanishing iterated Poisson brackets in (\ref{Iterierte Poissonklammer}). This gives rise to the specific result:
\ba\label{complexified connection}
   Z^j_a(x)
   &=&A^j_a(x)-\frac{\mb{i}}{L^3}\sum_{I=1}^3 E_j(S^I_x)~ n_a^{S^I_x}(u^I_x)~k^I(x)
   ~~~\mbox{with}~~k^I(x)=\Bigg[\bigg| \det\bigg( \frac{\partial X_{S^I_x(t)}(u)}{\partial (t,u^1,u^2}\bigg)\bigg|^{-1} \Bigg]_{t=t_I(x)\atop u=u_I(x)}
\ea

Note, that by (\ref{Klassische Poissonalgebra}) we only get a non vanishing contribution for each foliation direction $I$ if $x$ is an element of a per construction unique subsurface $S^I_x$. In this surface $x$ coincides with a unique pair $u^I_x=(u^1(x),u^2(x))$ of embedding parameters. So the positive number $k^I$ is the inverse of the Jacobian of the embedding $X_{S_x^I(t)} $ with respect the foliation parameter $t$ and the embedding parameters $u=(u^1,u^2)$ at the point $t^I_x$ and $u^I_x=(u^1(x),u^2(x))$. According to (\ref{E(S) erste Definition}) the ''normal'' of the surface $S^I_x$ is denoted by $n^{S^I_x}(u^I_x)$, $E_j(S^I_x)$ is defined accordingly.

Note that in the case of 'small' surfaces $S^I_x$ we get the bijection:
\be
   Z^j_a(x)\approx A^j_a(x) -\frac{\mb{i}}{L^3}E^b_j(x)~q_{ab}^0(x) 
   ~~~~~\mbox{with}~~~
    q^0_{ab}(x)=\sum_{I=1}^3 k^I(x) \int\limits_{S^I_x}d^2u~n_a^{S^I_x}(u) n_b^{S^I_x}(u)
\ee 
Hence $m=(A,E) \mapsto Z(A,E)=Z(m)$ becomes a bijection because $q^0_{ab}$ is non-degenerate.

Now we want to construct the holonomies of the complexified connection $A$ according to (\ref{Klassische U(1)^3 Poissonalgebra ausgeschmiert}). The holonomies $h^j_e(A)$ here are simple complex numbers since we work in $U(1)^3$ instead of $SU(2)$:

\ba
   h^j_e(A) 
   &=& \mb{e}^{\mb{i}\int_e A^j_a(e(t))~ \dot{e}^a(t) dt} \nonumber\\
   &=& \mb{e}^{\mb{i}~ \theta_e^j(A)}        
\ea 

By using (\ref{complexified connection}) specified to a pair $m=(A,E)$ we define
\ba \label{Definition p^j_e}
   p^j_e(m)&=& \mb{i} \int\limits_e \big[Z^j_a(e(t))\big|_m - A^j_a(e(t))\big|_m \big]~ \dot{e}^a(t)~ dt
\ea
So the complexified holonomy $h^j_e\big(Z(m)\big)$ is given by
\ba
   h^j_e\big(Z(m)\big)
   &=&\mb{e}^{\mb{i}\int_e Z^j_a(e(t))\big|_m~ \dot{e}^a(t) dt} 
   \nonumber\\
   &=&\mb{e}^{\mb{i}\int_e \big[Z^j_a(e(t))\big|_m - A^j_a(e(t))\big|_m + A^j_a(e(t))\big|_m \big]~
      \dot{e}^a(t)~ dt} 
   \nonumber\\
   &=& \mb{e}^{p_e^j(m)} ~\mb{e}^{\mb{i}\int_e A^j_a(e(t))\big|_m~ \dot{e}^a(t) dt}
   \nonumber\\ 
   &=& \mb{e}^{p_e^j(m)} ~\mb{e}^{\mb{i}~ \theta^j_e(A|_m)}
   \nonumber\\
   &=& \mb{e}^{p_e^j(m)} ~h^j_e(A|_m)
\ea

Using (\ref{Klassische U(1)^3 Poissonalgebra ausgeschmiert}), 
(\ref{Definition charge network}), 
(\ref{Klassische U(1)^3 Poissonalgebra ausgeschmiert und quantisiert}) we can explicitely evaluate the action 
of the electric fluxes $E_j(S)$ on a charge network function $T_c$:
\ba\label{E auf Tc I}
   \hat{E}_j(S)T_c
   &=&\hat{E}_j(S)\prod_{e \in E(\gamma(c))\atop j=1,2,3} \big[h^j_e(A) \big]^{n^j_e(c)}
   \nonumber
   \\
   &=&-\frac{\hbar\kappa}{2} \underbrace{\Big[\sum_{e \in \gamma(c)}\sigma(e,S)n^e_j(c)\Big]} T_c
   \nonumber
   \\
   &=&-\frac{\hbar\kappa}{2}\hspace{1.3cm} \lambda^j_c(S) \hspace{1cm} T_c
\ea
which turn out to be diagonal with eigenvalue $\lambda^j_c(S)$. Next we want to evaluate the action of the operator-version of the complexifier (\ref{Definition des Complexifiers}) itself:
\ba
   \hat{\mathcal{C}}_{\mathcal{F}} T_c 
   &=&\frac{1}{2\kappa L^{3}}\Big[\sum\limits_{I=1}^3 \int\limits_{ \mb{R}} dt
			             \sum\limits_{S\in \mathcal P^I_t} 
				       \delta^{jk} E_j(S) E_k(S)\Big] T_c
   \nonumber
   \\
   &=&\frac{\hbar^2\kappa}{8L^{3}}
       \underbrace{\Big[\sum\limits_{I=1}^3 \int\limits_{ \mb{R}} dt
                \sum\limits_{S\in \mathcal P^I_t} 
		\delta_{jk} \lambda^j_c(S) \lambda^k_c(S)\Big]} T_c
   \nonumber
   \\
   &=&\frac{\hbar^2\kappa}{8L^{3}}\hspace{1.8cm} \lambda_c(\mathcal{F})\hspace{1cm} T_c	       
\ea
Since $\hat{\mathcal{C}}_{\mathcal{F}}$ only consists of electric field operators and our theory is abelian anyway, it is diagonal with eigenvalues $\lambda_c(\mathcal{F})$, which depend on the choice of the foliation $\mathcal{F}$ .

The $\delta$-distribution on $\mathcal{H}_{kin}$ is (formally) given by 
\be\label{formal delta distribution}
   \delta_{A'}(A)=\sum_c T_c(A')\overline{T_c(A)}
\ee 

By invoking the definition (\ref{Allgemeinste Def CS}) of a complexifier coherent state together with (\ref{formal delta distribution}) we are now able to evaluate down coherent states centered at the phase space pair $m=(A,E)$ on the kinematical Hilbert space $\mathcal{H}_{kin}^0$:

\ba\label{Def U(1)^3 CS I}
   \Psi_m(A)
   &=&\Big[\mb{e}^{-\frac{1}{\hbar}\hat{\mathcal{C}}}\delta_{A'} \big](A)_{A'\rightarrow Z(m)}
   \nonumber
   \\
   &=&\sum_c T_c \big(Z(m) \big)~ \mb{e}^{-\frac{\hbar\kappa}{8 L^3} \lambda_c(\mathcal{F})}
           ~\overline{T}_c
\ea
Here the sum extends over all possible graphs and edge charges.
Again, to make (\ref{Def U(1)^3 CS I}) normalizable we have to restrict ourselces to an arbitrary but fixed graph $\gamma$(c). Then we have:
\ba\label{Def U(1)^3 CS II}
   \Psi_{m,\gamma}(A)
   &=&\sum_{c} 
         \mb{e}^{-\frac{\hbar\kappa}{8 L^3} \lambda_c(\mathcal{F})}
           ~T_c^\gamma(Z(m))\overline{T}_c^\gamma
   \nonumber
   \\
   &=&\sum_{\{\vec{n} \}} \mb{e}^{-\frac{\hbar\kappa}{8 L^3} \lambda_{\vec{n}}(\mathcal{F})} 
   \prod_{e \in E(\gamma) \atop j=1,2,3} \big[h^j_e(Z(m))h^j_e(A)^{-1} \big]^{n_e^j} 
\ea 
In the last line the symbol $\vec{n}$ denotes the sum over all possible charge configurations the individual copies uf $U(1)$ at every edge $E\in E(\gamma)$ can take. 
In the last step we will bring (\ref{Def U(1)^3 CS I}) to the simpler form\footnote{This will be the case when the parquettes $\mathcal{P}^I_t$ are fine enough such that to a good approximation $\sigma(e,S)\cdot\sigma(e',S)=0$ iff $e\ne e'$ and the $\big(\lambda^j_c(S)\big)^2$ in (\ref{E auf Tc I})-terms decompose. }:
\be
   \lambda_{\vec{n}}(\mathcal{P})= \sum_{e\in E(\gamma)\atop j=1,2,3} f_e [n^e_j]^2 
\ee
with $f_e=\sum\limits_{I=1}^3\int\limits_\mb{R}dt\sum\limits_{S\in \mathcal{P}^I_t}\big(\sigma(e,S)\big)^2$ being a dimensionful edge specific function, $f_e>0 ~~\forall e\in E(\gamma)$, $\big[f_e\big]=cm^1$.
We can thus introduce the (edge specific) dimensionless classicality parameter
\be
   t(e):=\frac{\hbar\kappa}{4L^3} f_e
\ee
with $L$ being an (at the moment unspecified) parameter of dimension meter,

Then (\ref{Def U(1)^3 CS II}) can be simplified to
\ba\label{Def U(1)^3 CS III}
   \Psi_{m,\gamma}(A)
   &=&\sum_{\vec{n}}  
   \prod_{e \in E(\gamma) \atop j=1,2,3} 
   \mb{e}^{-\frac{t(e)}{2}[n^j_e]^2}  \big[h^j_e(Z(m))h^j_e(A)^{-1} \big]^{n_e^j} 
   \nonumber
   \\
   &=&\prod_{e \in E(\gamma) \atop j=1,2,3} \sum_{n^j_e \in \mb{Z}} 
   \mb{e}^{-\frac{t(e)}{2}[n^j_e]^2}  \big[h^j_e(Z(m))h^j_e(A)^{-1} \big]^{n_e^j}
   \nonumber
   \\
   &=&\prod_{e \in E(\gamma) \atop j=1,2,3} \sum_{n^j_e \in \mb{Z}} 
   \mb{e}^{-\frac{t(e)}{2}[n^j_e]^2}  \big[\mb{e}^{p^j_e(m)}\mb{e}^{\mb{i}\theta_e^j(m)}~~\mb{e}^{-\mb{i}\theta^j_e(A)} \big]^{n_e^j}
   \nonumber
   \\
   &=&\prod_{e \in E(\gamma) \atop j=1,2,3} \sum_{n^j_e \in \mb{Z}} 
   \mb{e}^{-\frac{t(e)}{2}[n^j_e]^2 +n_e^j p^j_e(m)}\big[\mb{e}^{\mb{i}\theta_e^j(m)}~~\mb{e}^{-\mb{i}\theta^j_e(A)} \big]^{n_e^j} 
\ea

Note that the coherent state as defined in (\ref{Def U(1)^3 CS III}) is not yet normalized. We can perform the normalization by employing the Poisson resummation formula (\ref{our PRS form}) in order to obtain its norm. The result is:
\ba \label{Norm CS U(1)^3 general}
   \|\Psi_{m,\gamma}(A)\|^2 
   &=& \prod_{e\in E(\gamma) \atop j=1\ldots 3} \|\Psi_{m,e}^j\|^2
   \nonumber\\
   &=& \prod_{e\in E(\gamma) \atop j=1\ldots 3}
       2\pi \sqrt{\frac{\pi}{t(e)}}~ 
       \mb{e}^{\frac{1}{t(e)}[p_e^j(m)]^2} 
       \underbrace{\sum_{N^j_e \in \mb{Z}}~\mb{ e}^{-\frac{\pi}{t(e)}\big[\pi[N_e^j]^2 + 2\mb{i}~ N_e^j p_e^j(m)\big] } }_{~~~~~~\displaystyle \big[1 + K_{t(e)} \big]}
\ea
Here the $N^j_e$ are new summation variables\footnote{especially they have nothing to do with the charge labels $n^j_e$ as can be seen from (\ref{our PRS form}).} $K_{t(e)} = \mathcal{O}(t(e)^{\infty})$ dentotes a function of order $t(e)^\infty$ symbolizing that $\lim_{t(e)\rightarrow 0}\frac{K_{t(e)}}{t(e)^b}=0$ $\forall b<\infty$ such that for small $t(e)$ this quantity can be neglected because then $K_{t(e)}\ll 1$.

\pagebreak
%%%%%%%%%%%%%%%%%%%%%%%%%%%%%%%%%%%%%%%%%%%%%%%%%%%%%%%%%%%%%%%%%%%%%%%%%%%%%%%%%%%%%%%%%%%%%%%%%%%%%%%
\subsection{The Volume Operator $\hat{V}$ Acting on $U(1)^3$-Charge Networks}
%%%%%%%%%%%%%%%%%%%%%%%%%%%%%%%%%%%%%%%%%%%%%%%%%%%%%%%%%%%%%%%%%%%%%%%%%%%%%%%%%%%%%%%%%%%%%%%%%%%%%%%
\subsubsection{Setup}
%%%%%%%%%%%%%%%%%%%%%%

Following \cite{QFT on CST II} we consider polynomials of the operator
\ba
   \hat{q}_e^j(v,r) &=& \tr{\big[\tau_j \hat{h}_e [(\hat{h}_e)^{-1}, (\hat{V}_v)^r] \big]}~~~~~\mbox{for $SU(2)$}\\
   \hat{q}_e^j(v,r) &=& \hat{h}_e^j [(\hat{h}^j_e)^{-1}, (\hat{V}_v)^r] 
   ~~~~~~~~~~~~~\mbox{for $U(1)^3$}~~~~~~ r \in\mb{Q} \label{U(1)^3 Q}  
\ea

Now the action of the $(U(1)^3$ version (\ref{U(1)^3 Q}) of $\hat{q}_{e_{I_0}}^{j_0}(v,r)$ on a charge network state 
\be
   T_c=\prod_{ e_I \in E(v) \atop j=1,2,3 } \big[h^j_I (A)\big]^{n_I^j(c)}
\ee 
at the vertex $v\in V(\gamma)$ is

\ba
   \hat{q}_{e_{I_0}}^{j_0}(v,r)~ T_c 
   &=&  \Big( \hat{V}^r_v 
                       - \hat{h}_{I_0}^{j_0} \hat{V}^r_v \big[\hat{h}_{I_0}^{j_0}\big]^{-1} \Big)~ T_c
   \nonumber\\
   &=& \Big\{\lambda^r\big(\{n^j_I \} \big) 
                       - \lambda^r\big(\{n^j_I - \delta^{jj_0}\delta_{II_0}\} \big) \Big\} ~ T_c
\ea

With respect to $U(1)^3$ charge networks $T_c$ the volume operator $\hat{V}(R)$ according to the classical expression of the volume of a spatial region $R$
\be
   \mbox{Vol}(R)=\int\limits_R d^3x \sqrt{\big|\det{(E)} \big|}
\ee

is already diagonal and its action on charge network states $T_c$ is given by  
\ba\label{Def eigenwert V in U(1)^3 charge networks}
   \hat{V}(R)~T_c
   &=&\sum_{v \in V(\gamma)}\hat{V}_v ~T_c
   \nonumber\\
   &=& \sum_{v \in V(\gamma)}
      (\ell_P)^3\sqrt{\bigg|Z\cdot \sum_{e_I\cap e_J\cap e_K=v} \epsilon_{jkl}~\epsilon(I,J,K)~ n_I^j n_J^k n_K^l \bigg|} ~T_c
   \nonumber\\
   &=& \sum_{v \in V(\gamma)} \lambda\big(\{n^j_I \} \big) ~T_c        
\ea
Here the sum runs over all triple of edges at the vertex $v$, $\epsilon(I,J,K)=\sgn\big(\epsilon_{abc}~\dot{e}^a_I(v)\dot{e}^b_J(v)\dot{e}^c_K(v) \big)$gives the sign of the determinant of the tangents $\dot{e}_L(v)$ of the edges $e_L$ evaluated at the vertex $v$ and zero if the tangents are linearly dependend, which is the case in particular if at least two of the labels $I,J,K$ are equal or if one edge in the triple is the analytic continuation of another. Furthermore $\ell_P$ is the Planck length, and $Z$ a constant prefactor dependent on the regularization of the volume operator and the Immirzi parameter (see  \cite{Volume_Article_I}, \cite{TT:Closed ME of V in LQG}, for details\footnote{In \cite{TT:Closed ME of V in LQG} $Z$ is found to be $Z=\frac{1}{3!}\Big(\frac{3}{4} \Big)^3$}. 
Note, that due to the construction of the coherent states (\ref{Def U(1)^3 CS III}) we will evaluate 
$\hat{V}$ on the conjugated charge network states 
\be
   \bar{T}_c=\prod_{ e_I \in E(v) \atop j=1,2,3 } \big[(h^j_I (A))^{-1}\big]^{n_I^j(c)}
            =\prod_{ e_I \in E(v) \atop j=1,2,3 } \big[h^j_I (A)\big]^{-n_I^j(c)}
\ee 
according to 
\ba\label{Aktion von V auf konjugiertes Charge-Netzwerk}
   \hat{q}_{e_{I_0}}^{j_0}(v,r)~ \bar{T}_c 
   &=&  \Big( \hat{V}^r_v 
                     - \hat{h}_{I_0}^{j_0} \hat{V}^r_v \big[\hat{h}_{I_0}^{j_0}\big]^{-1} \Big)~ \bar{T}_c
   \nonumber\\
   &=& \Big\{\lambda^r\big(\{n^j_I \} \big) 
                       - \lambda^r\big(\{n^j_I + \delta^{jj_0}\delta_{II_0}\} \big) \Big\} ~ \bar{T}_c 
   \nonumber\\
   &=:&\lambda^r\big(\{n^{j_0}_{I_0} \} \big)~ 
                        \bar{T}_c 
\ea

 \pagebreak
%%%%%%%%%%%%%%%%%%%%%%%%%%%%%%%%%%%%%%%%%%%%%%%%%%%%%%%%%%%%%%%%%%%%%%%%%%%%%%%%%%%%%%%%%%%%%%
\subsubsection{Upper Bound for the Eigenvalues $\lambda^r\big(\{n^{j_0}_{I_0} \}\big)$ of $\hat{q}_{e_{I_0}}^{j_0}(v,r)$}
%%%%%%%%%%%%%%%%%%%%%%%%%%%%%%%%%%%%%%%%%%%%%%%%%%%%%%%%%%%%%%%%%%%%%%%%%%%%%%%%%%%%%%%%%%%%%%

We will derive an upper bound for the modulus of the eigenvalues 
$\big|\lambda^r\big(\{n^{j_0}_{I_0} \}\big)\big|= \Big|\lambda^r\big(\{n^j_I \} \big) 
                       - \lambda^r\big(\{n^j_I + \delta^{jj_0}\delta_{II_0}\} \big) \Big|$
of the operators $\hat{q}_{e_{I_0}}^{j_0}(v,r)$ as evaluated on a (conjugated) charge network on a single vertex $v$ with an arbitrary number $M$ of outgoing edges $e\in E(v)$, $E(v)$ being the set of edges of $v$.\newline
Here $\lambda\big(\big\{n_I^j\big\}\big) 
=\ell_P^{3}\sqrt{\big|Z\cdot\sum_{I,J,K} \epsilon_{ijk}~\epsilon(I,J,K)~ n_I^in_J^jn_K^k \big|}$ are the eigenvalues of the volume operator as defined in (\ref{Def eigenwert V in U(1)^3 charge networks}),
(\ref{Aktion von V auf konjugiertes Charge-Netzwerk}) where now the edge labels $I,J,K =1,2,\ldots,M$ and again $i,j,k=1,2,3$ for the three copies of $U(1)$. Recall that the charges $n^j_J\in \mb{Z}$ are integer numbers.   
\\
Let us first consider the eigenvalues of the volume operator:

\[\begin{array}{rlll}
  \lambda^r\big(\big\{n_I^i\big\}\big)  &=&
  (\ell_P)^{3r} |Z|^{\frac{r}{2}}\bigg|\displaystyle\sum_{I,J,K} \epsilon(I,J,K)~\epsilon_{ijk}~ n_I^in_J^jn_K^k \bigg|^{\frac{r}{2}}
  \\
  &=:&(\ell_P)^{3r} |Z|^{\frac{r}{2}}\big|\rho\big|^{\frac{r}{2}}
  \\
  &=:&(\ell_P)^{3r} |Z|^{\frac{r}{2}}~ a^{\frac{r}{2}}

  \\
  \\

  \lambda^r\big(\big\{n_I^i+\delta^{ii_0}\delta_{II_0}\big\}\big) 
  &=&
  (\ell_P)^{3r}|Z|^{\frac{r}{2}}\bigg|\displaystyle\sum_{I,J,K} \epsilon(I,J,K)~\epsilon_{ijk}~ n_I^in_J^jn_K^k +
  \\
  &&+\displaystyle\sum_{J,K} \epsilon(I_0,J,K)~\epsilon_{i_0jk}~ n_J^jn_K^k
  +\displaystyle\sum_{I,K} \epsilon(I,I_0,K)~\epsilon_{ii_0k}~ n_I^in_K^k
  +\displaystyle\sum_{I,J} \epsilon(I,J,I_0)~\epsilon_{iji_0}~ n_I^in_J^j\bigg|^{\frac{r}{2}}
  \\
  
  &=&
  (\ell_P)^{3r}|Z|^{\frac{r}{2}} \bigg|\displaystyle\sum_{I,J,K} \epsilon(I,J,K)~\epsilon_{ijk}~ n_I^in_J^jn_K^k +
  3\cdot\displaystyle\sum_{J,K} \epsilon(I_0,J,K)~\epsilon_{i_0jk}~ n_J^jn_K^k
  \bigg|^{\frac{r}{2}}
  \\
  &=:&(\ell_P)^{3r}|Z|^{\frac{r}{2}}\big|\rho + \sigma \big|^{\frac{r}{2}}
  \\
  &=:&(\ell_P)^{3r}|Z|^{\frac{r}{2}}~b^{\frac{r}{2}}
\end{array}\]\\[-15mm]\be\label{Definition a und b}\ee\\[-2mm]
%\end{footnotesize}

Here in the definition of $b$ we have explicitely decomposed the contribution from the modification $n^{i_0}_{I_0}\rightarrow n^{i_0}_{I_0}+1$ of the charge label $n^{i_0}_{I_0}$ caused by the action of holonomies in the $\hat{q}_{e_{I_0}}^{j_0}(v,r)$ operator. 
Due to the (double) antisymmetry of the $\epsilon(I,J,K)\epsilon_{ijk}$ prefactors in the sum terms and the fact that the remaining summation variables always run over $J,K=1,\ldots,M$ we can factor out the multiplicity factor $3$.
\\
We can therefore write:
\ba\label{Genereller Eigenwert von q}
   \lambda^r\big(\{n^{j_0}_{I_0} \}\big)&=& \lambda^r\big(\{n^j_I \} \big) 
                       - \lambda^r\big(\{n^j_I + \delta^{jj_0}\delta_{II_0}\} \big) 
   =(\ell_P)^{3r}|Z|^{\frac{r}{2}}\big(a^\frac{r}{2}-b^\frac{r}{2} \big)
   =(\ell_P)^{3r}|Z|^{\frac{r}{2}}\big(\big|\rho\big|^\frac{r}{2}-\big|\rho+\sigma\big|^\frac{r}{2} \big) 	       
\ea
\\
Now let $\frac{r}{2}=\frac{K}{L}<1~~$ be a rational number with $K<L$ and $K,L\in\mb{N}$. By invoking the (generalized) binomial theorem $\frac{x^N-y^N}{x-y}=\sum\limits_{k=0}^{N-1}x^ky^{N-1-k}$ ($N\in \mb{N}$ and $x\ne y\in\mb{C}$) and the geometric series $\sum\limits_{k=0}^{N-1}x^k=\frac{1-x^N}{1-x}$ ($x\ne 1\in \mb{C}, N\in \mb{N}$) we may write:
\ba
   a^{\frac{K}{L}}-b^{\frac{K}{L}}
   &=&\big[a^\frac{1}{L}\big]^K - \big[b^\frac{1}{L}\big]^K
   \nonumber
   \\
   &=&\big(a^\frac{1}{L} - b^\frac{1}{L} \big) \sum_{k=0}^{K-1}a^\frac{k}{L}b^\frac{K-1-k}{L}
   \nonumber
   \\
   &=&\big(\big[a^\frac{1}{L}\big]^L - \big[b^\frac{1}{L}\big]^L \big) \frac{\sum\limits_{k=0}^{K-1}a^\frac{k}{L}b^\frac{K-1-k}{L}}{\sum\limits_{l=0}^{L-1}a^\frac{l}{L}
   b^\frac{L-1-l}{L}}
   \nonumber
   \\
   &=&\big(a-b \big)~ b^\frac{K-L}{L}~ \frac{\sum\limits_{k=0}^{K-1}\Big[\big(\frac{a}{b}\big)^\frac{1}{L}\Big]^k}
   {\sum\limits_{l=0}^{L-1}\Big[\big(\frac{a}{b}\big)^\frac{1}{L}\Big]^l}
   \nonumber
\ea
With $\beta:=\big(\frac{a}{b}\big)^{\frac{1}{L}}$, $K<L$ this results in 
\ba\label{Umschreibung}
   a^{\frac{K}{L}}-b^{\frac{K}{L}}
   &=&
   \big(a-b \big)~ \frac{1}{b^{1-\frac{K}{L}}}~\frac{1-\beta^K}{1-\beta^L}
\ea
In order to give an upper bound for the modulus of (\ref{Umschreibung}) first notice, that $\big|x\big|-\big|y\big|\le \big|x-y\big|~~~\forall x,y\in\mb{C}$. Secondly $\frac{1-\beta^K}{1-\beta^L}\le 1~~~\forall \beta\in\mb{R}$ if $K<L$. Thirdly  $\frac{1}{b^{1-\frac{K}{L}}}\le 1 ~~~\forall~ b\ge 1$. 
Now let us discuss the special cases
\begin{itemize}
   \item[\fbox{$\beta=1$}]{ but then $a=b$  and thus $a^{\frac{K}{L}}-b^{\frac{K}{L}}=0$}
   \item[\fbox{$b<1$}]{ By definition $a,b\ge 0$ may only vary in integer steps, since the edge charges $n^i_I\in\mb{Z}$ are integer numbers, therefore \fbox{the only possible value $b<1$ is $b=0$}. But if $b=0$ we must have $\rho=-\sigma$ in (\ref{Definition a und b}) and therefore\linebreak $a=|\rho|\ge 0$. Then eigther $a=0$ and thus $a^{\frac{K}{L}}-b^{\frac{K}{L}}=0$ or $a\ge 1$ and thus 
   $a^{\frac{K}{L}}-b^{\frac{K}{L}}=a^{\frac{K}{L}}\le a =|\rho|=|\sigma| $ }
\end{itemize}
So we can give the general upper bound (note, that we sum over the indices $j,k$,~ $M$ is the number of edges at the vertex):
\ba
   \big|a^{\frac{K}{L}}-b^{\frac{K}{L}}\big|
   &\le&
   \big(a-b \big)=\Big|\big|\rho\big|-\big|\rho+\sigma \big| \Big| 
   \nonumber
   \\
   &\le& \big|\rho-\rho-\sigma\big|=\big|\sigma\big|
   = \Big|3\cdot\displaystyle\sum_{J,K} \epsilon(I_0,J,K)~\epsilon_{i_0jk}~ n_J^jn_K^k\Big|
   \nonumber
   \\
   &\le&3\cdot\displaystyle\sum_{J,K} \Big|\epsilon(I_0,J,K)~\epsilon_{i_0jk}~ n_J^jn_K^k\Big|
   \nonumber
   \\
   &=&3\cdot\displaystyle\sum_{J,K} \Big|\epsilon_{i_0jk}~ n_J^jn_K^k\Big|
   \nonumber
   \\
   &\le&3\cdot\displaystyle\sum_{J,K}\sum_{j,k=1,2,3} \Big|n_J^jn_K^k\Big|
   \nonumber
   \\
   &\le&3\cdot\displaystyle\sum_{J,K}\sum_{j,k=1,2,3} \frac{1}{2}\Big(|n_J^j|^2+|n_K^k|^2\Big)
   \nonumber
   \\
   &=&\frac{3}{2}\cdot 3M\cdot\Big(\displaystyle\sum_{J}\sum_{j=1,2,3}|n_J^j|^2
                          +\displaystyle\sum_{K}\sum_{k=1,2,3}|n_K^k|^2\Big)
   \nonumber
   \\
   &=&9M\cdot\displaystyle\sum_{J}\sum_{j=1,2,3}|n_J^j|^2   
\ea 

The final result for an upper bound of the modulus of the eigenvalue $\lambda^r\big(\{n^{j_0}_{I_0} \}\big)$ in (\ref{Genereller Eigenwert von q}) then reads 
\ba\label{upper bound Gesamteigenwert}
   \big|\lambda^r\big(\{n^{j_0}_{I_0} \}\big)\big|
   &=& (\ell_P)^{3r}|Z|^{\frac{r}{2}}\cdot\big|a^\frac{r}{2}-b^\frac{r}{2} \big|
   \nonumber
   \\
   &\le&
   (\ell_P)^{3r}|Z|^{\frac{r}{2}}9M
   \displaystyle\sum_{J}\sum_{j=1,2,3}|n_J^j|^2
\ea

\pagebreak
%%%%%%%%%%%%%%%%%%%%%%%%%%%%%%%%%%%%%%%%%%%%%%%%%%%%%%%%%%%%%%%%%%%%%%%%%%%%%%%%%%%%%%%%%%%%%%%%%%%%%%
\subsection{\label{Pathologische Konfiguration} Explicit Construction of a 'Pathological' Edge Configuration at a Vertex v}
%%%%%%%%%%%%%%%%%%%%%%%%%%%%%%%%%%%%%%%%%%%%%%%%%%%%%%%%%%%%%%%%%%%%%%%%%%%%%%%%%%%%%%%%%%%%%%%%%%%%%%
Here we will give an explicit construction of a configuration of $M$ edges outgoing from a vertex $v$ where each ordered edge triple $e_I,e_J,e_K$, $I<J<K$, contributes with a negative sign factor $\epsilon(I,J,K)$ of its tangents.

Since we are only interested in the sign factor we can make simplifying assumptions, especially we may choose certain numerical values.

Consider the vertex $v$ as the origin of a 3 dimensional coordinate system with axis $x,y,z$. 
Now consider a circle with radius $r=1$ centered at $y=1$, parallel to the $x-z$ -plane. Let every edge tangent $\dot{e}_K$ end on a point on the circle with coordinates 
$\big(\cos\phi_K,1,\sin\phi_K\big)$ and $\phi_K=2\pi\frac{K}{M}$. Now one may check that for each orederd triple $e_I,e_J,e_K$ with $I<J<K\le M$ we have:
\be
   \det\big({\dot{e}_I,\dot{e}_J,\dot{e}_K}\big)
   = -4\cdot\sin\Big[\pi\frac{K-I}{M}\Big]
            \sin\Big[\pi\frac{K-J}{M}\Big]
	    \sin\Big[\pi\frac{J-I}{M}\Big]
\ee
			 
Since for all arguments $x$ of the $sin$-functions we have $0<x\le\pi$ all of these functions  are $\ge 0$ and therefore we get $\epsilon(I,J,K)=\sgn\Big(\det\big({\dot{e}_I\dot{e}_J,\dot{e}_K}\big)\Big)=-1$ for all ordered edge triples $e_I,e_J,e_K$ with $I<J<K$ at $v$.

\begin{figure}[hbt]
    \center
    \cmt{8}{
    \psfrag{x}{${x}$}
    \psfrag{y}{${y}$}
    \psfrag{z}{${z}$}
    \psfrag{phik}{$\phi_K$}
    \psfrag{r}{$r$}
    \psfrag{v}{$v$}
    \psfrag{e1}{$\dot{e}_1$}
    \psfrag{e2}{$\dot{e}_2$}
    \psfrag{ek}{$\dot{e}_K$}
    \psfrag{eM}{$\dot{e}_M$}
    \includegraphics[height=5cm]{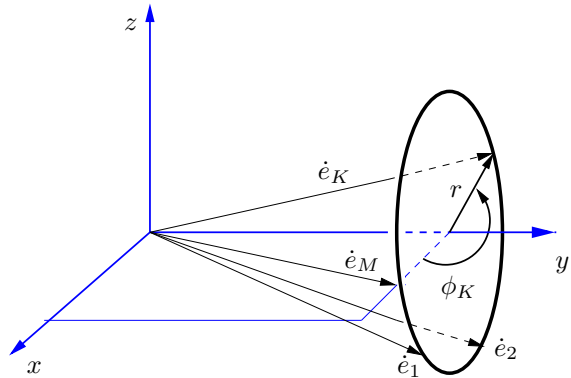} 
    \caption{ 'Pathological' edge configutration}}
\end{figure} 
 
As one can see, this edge configuration is quite special - it is like all edges lying in one octant only (if we rotate the coordinate system). 
Such an edge configuration would appear to look rather 1 than 3 dimensional. This shows that the sign factor cannot be used in general to achieve an $M$-independent bound of the expectation values (\ref{finale obere Schranke fuer Erwartungswert}).     
 \pagebreak                    
%%%%%%%%%%%%%%%%%%%%%%%%%%%%%%%%%%%%%%%%%%%%%%%%%%%%%%%%%%%%%%%%%%%%%%%%%%%%%%%%%%%%%%%%%%%%%%%%%%%%%%
\subsection{Theorems Needed}
%%%%%%%%%%%%%%%%%%%%%%%%%%%%%%%%%%%%%%%%%%%%%%%%%%%%%%%%%%%%%%%%%%%%%%%%%%%%%%%%%%%%%%%%%%%%%%%%%%%%%%%
%%%%%%%%%%%%%%%%%%%%%%%%%%%%%%%%%%%%%%%%%%%%%%%%%%%%%%%%%%%%%%%%%%%%%%%%%%%%%%%%%%%%%%%%%%%%%%%%%%%%%%%%
\subsubsection{\label{Spherical coordinate upper bound} Spherical Coordinates}
%%%%%%%%%%%%%%%%%%%%%%%%%%%%%%%%%%%%%%%%%%%%%%%%%%%%%%%%%%%%%%%%%%%%%%%%%%%%%%%%%%%%%%%%%%%%%%%%%%%%%%%%

For integrals of the form
\ba
   I_k:=\sqrt{\frac{2}{\pi}}^m \int_{\mb{R}^m}d^mx ~\mb{e}^{-2\|x\|^2}\|x\|^{2k}
   ~~~\mbox{with}~~k,m\in\mb{N}
\ea 
there is a recursion relation
\ba
   I_k=\frac{m+2(k-1)}{4}I_{k-1} ~~~~\mbox{with}~~~I_0=1
\ea
such that we can write for $k\ge 1$ , $I_0=1$
\ba\label{Produktformel fuer sphaerische Integrale}
   I_{k}=\prod_{l=1}^k\frac{m+2(l-1)}{4}
\ea
or explicitly
\ba
   \fbox{\mbox{m even}}&~~~~&I_k=\frac{\big(\frac{m}{2}+k-1\big)!}{2^k\big(\frac{m}{2} \big)!}
   \\
   \fbox{\mbox{m odd}}&~~~~&I_k=\frac{\big(m-1+2k \big)!\big(\frac{m-1}{2}\big)!}{8^k\big(m-1\big)!\big(\frac{m-1}{2}+k \big)!}
\ea
~\\[1cm]
%%%%%%%%%%%%%%%%%%%%%%%%%%%%%%%%%%%%%%%%%%%%%%%%%%%%%%%%%%%%%%%%%%%%%%%%%%%%%%%%%%%%%%%%%%%%%%%%%%%%%%%%
\subsubsection{Poisson Resummation Formula}
%%%%%%%%%%%%%%%%%%%%%%%%%%%%%%%%%%%%%%%%%%%%%%%%%%%%%%%%%%%%%%%%%%%%%%%%%%%%%%%%%%%%%%%%%%%%%%%%%%%%%%%%
\begin{Theorem}[Poisson Summation Formula] \label{our PRS form}~\\
Let $f$ be an $L_1(\mb{R},dx)$ function such that the series
$$ \phi(y)=\sum_{n=-\infty}^\infty f(y+ns)$$ 
is absolutely and uniformly convergent for $y\in [0,s],s>0$.
Then
\be \label{3.9}
\sum_{n=-\infty}^\infty f(ns)=\frac{2\pi}{s}
\sum_{N=-\infty}^\infty \tilde{f}(\frac{2\pi N}{s})
\ee
where $\tilde{f}\big(\frac{2\pi N}{s}\big):=\displaystyle\int_{\mb{R}}f(x)\mb{e}^{-2\pi\mb{i}\frac{N}{s} x} dx$~is the Fourier transform of $f$~and $x=s\cdot k$~~.
\end{Theorem}
The proof of this theorem can be found in any textbook on Fourier series,
see e.g. the classical book by Bochner \cite{Bochner}. \\
The importance of this remarkable theorem for our purposes is that it 
converts a 
slowly converging series $\sum_n f(ns)$ as $s\to 0$ into a 
possibly rapidly converging series $\frac{1}{s}\sum_N \tilde{f}(2\pi 
N/s)$ of which in our case almost only the term with $N=0$ will be 
relevant.

\end{appendix}
\pagebreak
%%%%%%%%%%%%%%%%%%%%%%%%%%%%%%%%%%%%%%%%%%%%%%%%%%%%%%%%%%%%%%%%%%%%%%%%%%%%%%%%%%%%%%%%%%%%%%%%%%%

%%%%%%%%%%%%%%%%%%%%%%%%%%%%%%%%%%%%%%%%%%%%%%%%%%%%%%%%%%%%%%%%%%%%%%%%%%%%%%%%%%%%%%%%%%%%%%%%%%%%%%%

%%%%%%%%%%%%%%%%%%%%%%%%%%%%%%%%%%%%%%%%%%%%%%%%%%%%%%%%%%%%%%%%%%%%%%%%%%%%%%%%%%%%%%%%%%%%%%%%%%%%%%%

\end{document}